\documentclass[12pt]{article}
\usepackage{amsmath}
\usepackage{amsthm}
\usepackage{amssymb}
\usepackage{amsfonts}
\usepackage{epic}
\usepackage{eepic}
 \usepackage{times}
\usepackage[matrix,arrow]{xy}
\usepackage{macros}
\usepackage{epsfig}
\usepackage{graphicx}
\usepackage{bm}
\theoremstyle{plain}

\newtheorem{propn}{Proposition}

\theoremstyle{remark}

\newtheorem{claim}{Claim}

\newcommand{\sst}{\scriptscriptstyle}

\newcommand{\Li}{{\rm Li}}
\newcommand{\vt}{\vartheta}

\newcommand{\ka}{\kappa}
\renewcommand{\1}{\one}
\renewcommand{\2}{\two}
\newcommand{\3}{\three}
\newcommand{\4}{{\mathfrak 4}}
\newcommand{\5}{{\mathfrak 5}}
\newcommand{\6}{{\mathfrak 6}}

\newcommand{\id}{{\rm id}}

\newcommand{\ii}{{\textup i}}
\newcommand{\pa}{\partial}
\newcommand{\ot}{\otimes}
\newcommand{\ra}{\to}

\newcommand{\fr}[2]{{\textstyle \frac{#1}{#2} }}

\newcommand{\fsl}{{\mathfrak s}{\mathfrak l}}

\renewcommand{\=}[1]{\stackrel{(\ref{#1})}{=}}
\newcommand{\df}{\equiv}

\DeclareMathOperator*{\Res}{Res}
\newcommand{\al}{\alpha}

\newcommand{\be}{\beta}
\newcommand{\ga}{\gamma}
\newcommand{\Ga}{\Gamma}
\newcommand{\de}{\delta}
\newcommand{\De}{\Delta}
\newcommand{\ep}{\epsilon}
\newcommand{\la}{\lambda}
\newcommand{\La}{\Lambda}
\newcommand{\om}{\omega}
\newcommand{\Om}{\Omega}
\newcommand{\si}{\sigma}
\newcommand{\up}{\Upsilon}
\newcommand{\vf}{\varphi}

\newcommand{\BBE}{E}

\newcommand{\bq}{\bar{q}}

\newcommand{\bz}{{\bar{z}}}

\newcommand{\CA}{{\mathcal A}}
\newcommand{\CB}{{\mathcal B}}

\newcommand{\CC}{{\mathcal C}}
\newcommand{\CD}{{\mathcal D}}
\newcommand{\CE}{{\mathcal E}}
\newcommand{\CF}{{\mathcal F}}
\newcommand{\CG}{{\mathcal G}}
\newcommand{\CH}{{\mathcal H}}

\newcommand{\CI}{{\mathcal I}}
\newcommand{\CJ}{{\mathcal J}}
\newcommand{\CK}{{\mathcal K}}
\newcommand{\CL}{{\mathcal L}}
\newcommand{\CM}{{\mathcal M}}
\newcommand{\CN}{{\mathcal N}}
\newcommand{\CO}{{\mathcal O}}
\newcommand{\CP}{{\mathcal P}}
\newcommand{\CQ}{{\mathcal Q}}
\newcommand{\CR}{{\mathcal R}}
\newcommand{\CS}{{\mathcal S}}
\newcommand{\CT}{{\mathcal T}}

\newcommand{\CU}{{\mathcal U}}
\newcommand{\CV}{{\mathcal V}}
\newcommand{\CW}{{\mathcal W}}

\newcommand{\CZ}{{\mathcal Z}}

\newcommand{\SA}{{\mathsf A}}
\newcommand{\SB}{{\mathsf B}}
\newcommand{\SC}{{\mathsf C}}
\newcommand{\SD}{{\mathsf D}}

\newcommand{\SF}{{\mathsf F}}

\newcommand{\SH}{{\mathsf H}}

\newcommand{\sk}{{\mathsf k}}
\newcommand{\SL}{{\mathsf L}}

\newcommand{\SM}{{\mathsf M}}

\newcommand{\SO}{{\mathsf O}}
\newcommand{\SP}{{\mathsf P}}
\newcommand{\SQ}{{\mathsf Q}}

\renewcommand{\SS}{{\mathsf S}}
\newcommand{\ST}{{\mathsf T}}
\newcommand{\SU}{{\mathsf U}}
\newcommand{\SV}{{\mathsf V}}
\newcommand{\SW}{{\mathsf W}}

\newcommand{\SZ}{{\mathsf Z}}

\newcommand{\fe}{{\mathfrak e}}
\newcommand{\fg}{{\mathfrak g}}

\newcommand{\fv}{{\mathfrak v}}

\newcommand{\fw}{{\mathfrak w}}

\newcommand{\sa}{{\mathsf a}}

\newcommand{\sfc}{{\mathsf c}}
\newcommand{\sd}{{\mathsf d}}

\newcommand{\sh}{{\mathsf h}}
\newcommand{\sll}{{\mathsf l}}

\newcommand{\so}{{\mathsf o}}
\newcommand{\sq}{{\mathsf q}}
\newcommand{\spp}{{\mathsf p}}

\newcommand{\mss}{{\mathsf s}}

\newcommand{\su}{{\mathsf u}}

\newcommand{\sx}{{\mathsf x}}
\newcommand{\sy}{{\mathsf y}}
\newcommand{\sz}{{\mathsf z}}

\newcommand{\Loc}{{\SL\so\sfc}}

\newcommand{\BD}{{\mathbb D}}

\newcommand{\homsl}{\CM^{\BC}_{\rm char}(C)}
\newcommand{\homslr}{\CM^{\BR}_{\rm char}(C)}
\newcommand{\upc}{\Upsilon_{\rm cl}}

\newcommand{\FV}{{\mathfrak V}}

\newcommand{\zero}{{\mathfrak 0}}
\newcommand{\0}{{\mathfrak 0}}

\newcommand{\one}{{\mathfrak 1}}
\newcommand{\two}{{\mathfrak 2}}
\newcommand{\three}{{\mathfrak 3}}

\newcommand{\BF}{{\mathbb F}}

\newcommand{\BR}{{\mathbb R}}

\newcommand{\BL}{{\mathbb L}}
\newcommand{\BC}{{\mathbb C}}

\newcommand{\BP}{{\mathbb P}}
\newcommand{\BS}{{\mathbb S}}

\newcommand{\BU}{{\mathbb U}}
\newcommand{\BZ}{{\mathbb Z}}

\newcommand{\rf}[1]{(\ref{#1})}

\newcommand{\Fus}[6]{F_{#5#6}^{}\big[\,{}^{#3}_{#4}\;{}^{#2}_{#1}\,\big]} 
 
\newcommand{\Fusan}[6]{F_{#5#6}^{\rm\sst PT}\big[\,{}^{#3}_{#4}\;{}^{#2}_{#1}\,\big]} 
\newcommand{\FusC}[6]{F_{#5#6}^{\rm\sst L}\big[\,{}^{#3}_{#4}\;{}^{#2}_{#1}\,\big]} 
 
\newcommand{\fus}[6]{f_{#5#6}^{}\big[\,{}^{#3}_{#4}\;{}^{#2}_{#1}\,\big]} 
\newcommand{\dus}[6]{d_{#5#6}^{(k)}\big[\,{}^{#3}_{#4}\;{}^{#2}_{#1}\,\big]} 
\newcommand{\nc}{\newcommand}

\nc{\rnc}{\renewcommand} \nc{\beq}{\begin{equation}}
\nc{\eeq}{\end{equation}} \nc{\beqa}{\begin{eqnarray}}
\nc{\eeqa}{\end{eqnarray}}

\begin{document}
\title{Supersymmetric gauge theories,
quantization of $\CM_{\rm flat}$, and conformal field theory}
\author{J. Teschner and G. S. Vartanov}
\address{
DESY Theory, Notkestr. 85, 22603 Hamburg, Germany}
\maketitle

\begin{quote}
\centerline{\bf Abstract}
{\small  We will propose a derivation of 
the correspondence between certain gauge theories with $N=2$
supersymmetry and conformal field theory discovered by Alday,
Gaiotto and Tachikawa in the spirit of Seiberg-Witten theory.
Based on certain results from the literature we argue that the quantum theory 
of the moduli spaces of flat $SL(2,\BR)$-connections
represents a non-perturbative ``skeleton'' of the gauge theory,
protected by supersymmetry. It follows that instanton partition
functions can be characterized as solutions to a Riemann-Hilbert 
type problem. In order to solve it, 
we describe the quantization of  the moduli spaces of 
flat connections explicitly in terms of two natural sets of 
Darboux coordinates.
The kernel describing the relation between the two pictures 
represents the solution to the Riemann Hilbert problem,
and is naturally identified with the Liouville conformal blocks.
}\end{quote}


\tableofcontents

\newpage

\section{Introduction}
\setcounter{equation}{0}

This work is motivated by the discovery \cite{AGT} of 
remarkable relations between certain $N=2$ supersymmetric gauge theories
and conformal field theories. The defining data for the
relevant class of gauge theories, nowadays often called class $\CS$,  
can be encoded in certain geometrical structures
associated to Riemann surfaces $C$ of genus $g$ with $n$ punctures \cite{G09}. 
We will restrict attention to the case where the 
gauge group is $[SU(2)]^{3g-3+n}$, for which the corresponding 
conformal field theory is the Liouville theory.
The gauge theory corresponding to a Riemann surface $C$
will be denoted $\CG_C$. 

The authors of \cite{AGT} discovered 
relations between the instanton partition functions 
$\CZ^{\rm inst}(a,m,\tau,\ep_\1,\ep_\2)$  
defined in \cite{N}\footnote{Based in parts on earlier work  
\cite{MNS1,MNS2,LNS} in this direction.} for some gauge theories 
$\CG_C$ of class $\CS$ on the one hand, 
and the conformal blocks of the Liouville conformal field 
theory \cite{Teschner:2001rv}
on the other hand. Using this observation one may furthermore
use the variant of the localization technique developed   
in \cite{Pe} to find relations
between expectation values of Wilson loops in $\CG_C$ and 
certain Liouville correlation functions on $C$. The results
of \cite{Pe,AGT} were further developed and generalized in particular
in \cite{GOP,HH}, and the results of \cite{AFLT} prove the validity of
these relations for the cases where the Riemann surface $C$ has genus 
zero or one, and
arbitrary number of punctures. 

This correspondence can be used as a powerful tool for the study 
of non-perturbative effects in N=2 gauge theories. As an example 
let us note that techniques from the
study of Liouville theory \cite{Teschner:2001rv} can be used to effectively
resum the instanton expansions, leading to highly nontrivial
quantitative checks of the strong-weak coupling conjectures
formulated in \cite{G09} for gauge theories of class $\CS$.  
However, gaining a deeper understanding of the origin of the
relations between N=2 gauge theories and conformal field theories
discovered in \cite{AGT}
seems highly desirable.

We will propose a derivation of the relations discovered in \cite{AGT} 
based on certain physically motivated assumptions. 
We will in particular make the following assumptions:
\begin{itemize}
\item 
The instanton partition functions $\CZ^{\rm inst}(a,m,\tau,\ep_\1,\ep_\2)$
are holomorphic in the UV gauge couplings $\tau$, and can be analytically 
continued over the gauge theory coupling constant space. 
Singularities are in one-to-one correspondence with weakly-coupled 
Lagrangian descriptions of $\CG_C$.
\item Electric-magnetic duality exchanges 
Wilson- and 't Hooft loops.
\end{itemize}
Our approach works for all $g$ and $n$.
One may observe an analogy with the reasoning used by
Seiberg and Witten in their derivations of the prepotentials
for certain examples of gauge theories from this class \cite{SW1,SW2}.
This is not completely surprising, as the prepotential can be recovered from the
instanton partition functions $\CZ^{\rm inst}(a,m,\tau,\ep_\1,\ep_\2)$  
in the limit $\ep_1,\ep_2\ra 0$.

A basic observation underlying our approach is that
the instanton partition functions $\CZ^{\rm inst}$ 
can be interpreted as certain wave-functions $\Psi_\tau(a)$ representing
states in subspaces $\CH_0$ of the Hilbert spaces $\CH$ defined by
studying $\CG_C$ on suitable four-manifolds.
Indeed, the localization methods used in
\cite{Pe,HH} show that the path integrals representing 
Wilson loop expectation values, for example, localize to the 
quantum mechanics of the scalar zero modes of $\CG_C$. The
instanton partition functions represent certain wave-functions
in the zero mode quantum mechanics the path integral localizes to.

Supersymmetric versions of the Wilson- and 't Hooft loop operators
act naturally on the zero mode
Hilbert space $\CH_0$, generating a sub-algebra 
$\CA_{\ep_1,\ep_2}$
of the algebra of operators. 
A key information needed as input for our approach is 
contained in the statement that the algebra 
$\CA_{\ep_1,\ep_2}$ is isomorphic to the quantized algebra
of functions on the moduli space $\CM_{\rm flat}(C)$ of flat 
$SL(2,\BR)$-connections on $C$. A derivation of this
fact, applicable to all theories $\CG_C$,
was proposed in \cite{NW}. It is strongly supported by the explicit
calculations performed for certain theories from 
class $\CS$ in \cite{Pe,AGT,GOP,IOT}. A more direct
way to understand why the algebra $\CA_{\ep_1,\ep_2}$ 
is related to the quantization of the moduli spaces 
$\CM_{\rm flat}(C)$ can propbably be based on the work \cite{GMN3}
which relates the algebra of the loop operators
to the quantization of the Darboux coordinates from \cite{GMN1}.

We view the algebra of supersymmetric
loop operators $\CA_{\ep_1\ep_2}$ and its representation
on $\CH_0$ as a non-perturbative "skeleton" 
of the gauge theory $\CG_C$ which is protected by some unbroken
supersymmetry.
This structure determines the low-energy 
physics of $\CG_C$ and its finite-size corrections on certain
supersymmetric backgrounds, as follows from the localization
of the path integral studied in \cite{Pe,GOP,HH}

The instanton partition functions $\CZ^{\rm inst}(a,m,\tau,\ep_\1,\ep_\2)$ 
may then be characterized as wave-functions of
joint eigenstates of the Wilson loop operators whose eigenvalues
are given by the Coulomb branch parameters $a$.
It follows from our assumptions above that the 
instanton partition functions $\CZ^{\rm inst}_{2}(a_2,m,\tau_\2,\ep_\1,\ep_\2)$ 
and
$\CZ^{\rm inst}_{1}(a_1,m,\tau_\1,\ep_\1,\ep_\2)$ 
associated to two different weakly-coupled Lagrangian descriptions
must be related linearly as
\begin{equation}\label{S-duality}
\begin{aligned}
&\CZ^{\rm inst}_{2}(a_2,m,\tau_\2,\ep_\1,\ep_\2)\,=\,\\
&\quad= f(m,\tau_\2,\ep_\1,\ep_\2)\int da_1\;K(a_2,a_1;m;\ep_\1,\ep_\2)
\CZ^{\rm inst}_{1}(a_1,m,\tau_\1(\tau_2),\ep_\1,\ep_\2)\,.
\end{aligned}\end{equation}
The $a_2$-independent prefactor $f(m,\tau_\2,\ep_\1,\ep_\2)$ 
describes a possible change of regularization scheme used in the
definition of the instanton partition functions.
Knowing the relation between the algebra 
$\CA_{\ep_1,\ep_2}$ and the quantum theory of $\CM_{\rm flat}(C)$
will allow us to determine the kernels 
$K(a_2,a_1;m;\ep_\1,\ep_\2)$ in \rf{S-duality} explicitly.
These are the main pieces of data needed for the formulation of 
a generalized Riemann-Hilbert problem characterizing 
the  instanton partition functions. 

The resulting mathematical problem is not of standard
Riemann-Hilbert type in two respects: One is, on the one hand, 
dealing with
infinite dimensional representations of the relevant monodromy groups,
here the mapping class groups of the Riemann surfaces $C$. We will,
on the other hand, find that the
$a_2$-independent prefactors $f(m,\tau_\2,\ep_\1,\ep_\2)$ 
in \rf{S-duality} can not be eliminated in general\footnote{This is 
the case for surfaces
of higher genus. The prefactors could be eliminated for the cases studied 
in \cite{AGT}, and some generalizations like the so-called linear quiver
theories.}. Their appearance is closely related to the fact that
the representation of the mapping class group of $C$ 
described by the kernels $K(a_2,a_1;m;\ep_\1,\ep_\2)$ is found to 
be {\it projective}. Without prefactors $f(m,\tau_\2,\ep_\1,\ep_\2)$
which, roughly speaking, cancel the projectiveness there could not
exist any solution to our generalized Riemann-Hilbert problem.

Working out the kernels $K(a_2,a_1;m;\ep_\1,\ep_\2)$ is the 
content of Part II of this paper, containing a detailed study of
the quantum theory of the relevant connected 
component $\CM_{\rm flat}^0(C)$ of $\CM_{\rm flat}(C)$.
In Part III we describe how the Riemann-Hilbert problem for
$\CZ^{\rm inst}(a,m,\tau,\ep_\1,\ep_\2)$ is solved by Liouville theory. 
We explain how Liouville theory is related to the quantum
theory of $\CM_{\rm flat}^0(C)$, which is equivalent to the quantum theory
of the Teichm\"uller spaces $\CT(C)$. The relation between
Liouville theory and the quantization of $\CM_{\rm flat}^0(C)$,
combined with the connection between instanton partition functions 
$\CZ^{\rm inst}(a,m,\tau,\ep_\1,\ep_\2)$ and wave-functions in the quantum 
theory of $\CM_{\rm flat}^0(C)$ yields a way to 
derive and explain the correspondence found in \cite{AGT}.
One of the main technical problems addressed in Part III is the proper
characterization of the prefactors $f(m,\tau_\2,\ep_\1,\ep_\2)$ 
in \rf{S-duality} which are related to the projective line bundle 
whose importance for conformal field theory was emphasized 
by Friedan and Shenker \cite{FS}.

There is an alternative approach towards proving the AGT-correspondence,
which relates the series expansion 
of $\CZ^{\rm inst}(a,m,\tau,\ep_\1,\ep_\2)$
defined from the equivariant cohomology of
instanton moduli spaces more directly to the 
definition of the conformal blocks of Liouville theory obtained 
from the representation theory of the Virasoro algebra.
Important progress has been made along these
lines. A first proof of the AGT-correspondence
for a subset of gauge theories $\CG_C$ from class $\CS$ 
was obtained in \cite{AFLT} by
finding closed formulae for the coefficients appearing 
in the series expansions of the Liouville conformal blocks
that directly match the formulae known for the expansion 
coefficients of   $\CZ^{\rm inst}(a,m,\tau,\ep_\1,\ep_\2)$ from the instanton 
calculus.
An important step towards a more conceptual
explanation was taken by identifying the Virasoro algebra
as a symmetry of the equivariant cohomology of the
instanton moduli spaces \cite{SchV,MO}. A physical approach to these results
was described in \cite{Tan}.
 
This approach may be seen as complementary to the
one used in this paper: 
It elucidates the mathematical structure of the perturbative
expansion of $\CZ^{\rm inst}(a,m,\tau,\ep_\1,\ep_\2)$ as defined
from a given Lagrangian description for $\CG_C$.
The arguments presented here
relate the non-perturbative "skeleton" of $\CG_C$ to global
objects on $C$ instead.

The results in Parts II and III of this paper are of independent
interest. Part II describes the quantization of $\CM_{\rm flat}^0(C)$
using the Darboux variables which were recently used in a
related context in \cite{NRS}.\footnote{Partial results in this direction were previously obtained in \cite{DG}.} 
These results give an  
alternative representation for the
quantum theory of the Teichm\"uller spaces which is based 
on pants decompositions instead of triangulations of $C$,
as is important for understanding the relation to Liouville theory.
Our approach
is related to the one pioneered in \cite{F97,CF,Ka1} by a nontrivial
unitary transformation that we construct explicitly.

In Part III we extend the relation between quantization of
the Teichm\"uller spaces and Liouville theory found in \cite{T03}
for surfaces of genus $0$ to arbitrary genus. 
An important subtlety is to properly take into account 
the projective line bundle over moduli space whose relevance
for conformal field theory was first emphasized in \cite{FS}.
This allows us to find the appropriate way to cancel
the central extension of the canonical connection on the space of 
conformal blocks defined by the energy-momentum tensor. Doing this 
is crucial for having a solution of the Riemann-Hilbert problem
of our interest at all. 

The results of Part III also seem to be interesting from 
a purely mathematical perspective. They amount to an 
interpretation  
of conformal field theory in terms of the harmonic analysis 
on the Teichm\"uller spaces, which can be seen
as symmetric spaces for the 
group ${\rm Diff}_0(S^1)$.

Our work realizes part of a larger picture  outlined in \cite{T10} 
relating 
the quantization of the Hitchin moduli spaces, integrable models 
and conformal field theory.
In order to get a connection to supersymmetric
gauge theories extending the connections discussed here 
one needs to consider insertions
of surface operators on the gauge theory side. This is currently under
investigation \cite{FGT}.

\noindent
{\bf Acknowledgements:}
We would like to thank T. Dimofte, S. Gukov, R. Kashaev and
S. Shatashvili for useful discussions on related topics.

\newpage

\section{Riemann surfaces: Some basic definitions and results}

\setcounter{equation}{0}

Let us introduce some basic definitions
concerning Riemann surfaces that will be used throughout the paper.

\subsection{Complex analytic gluing construction}
\label{sec:glueing}

A convenient family of particular choices for coordinates on 
$\CT(C)$ is produced from the complex-analytic 
gluing construction of Riemann surfaces $C$ from three punctured
spheres \cite{Ma,HV}. Let us  briefly review this construction.

Let $C$ be a (possibly disconnected) Riemann surface. 
Fix a complex number $q$ with $|q|<1$, and
pick two points $Q_1$ and $Q_2$ on $C$
together with coordinates $z_i(P)$ in a neighborhood of 
$Q_i$, $i=1,2$, such that $z_i(Q_i)=0$, and such that the discs $D_i$,
\[
D_i\,:=\,\{\,P_i\in C_i\,;\,|z_i(P_i)|<|q|^{-\frac{1}{2}}\,\}\,,
\]
do not intersect.
One may define the annuli $A_i$,
\[
A_i\,:=\,\{\,P_i\in C_i\,;\,|q|^{\frac{1}{2}}<|z_i(P_i)|<|q|^{-\frac{1}{2}}\,\}\,.
\]
To glue $A_1$ to $A_2$ let us identify two points $P_1$ and $P_2$ on $A_1$ and $A_2$, 
respectively, iff the coordinates of these two points satisfy the equation
\begin{equation}\label{glueid}
z_1(P_1)z_2(P_2)\,=\,q\,.
\end{equation}
If $C$ is connected one creates an additional handle, and if $C=C_1\sqcup C_2$
has two connected components one gets a single connected component after 
performing the gluing operation. In the limiting case where $q=0$ 
one gets a nodal surface which represents a component of 
the boundary $\pa\CM(C)$ defined by the Deligne-Mumford 
compactification $\overline{\CM}(C)$.

By iterating the gluing operation one may build any 
Riemann surface $C$ of genus $g$ with $n$ punctures 
from three-punctured spheres $C_{0,3}$. 
Embedded into $C$ we naturally get a collection of annuli 
$A_1,\dots,A_h$, where
\begin{equation}
h\,:=\,3g-3+n\,,
\end{equation}
The construction above can be used to define 
an $3g-3+n$-parametric family or Riemann
surfaces, parameterized by a collection $q=(q_1,\dots,q_h)$ 
of complex parameters. These parameters can be taken
as complex-analytic coordinates for a neighborhood of 
a component in the boundary $\pa\CM(C)$ with respect to 
its natural complex structure \cite{Ma}. 

Conversely, assume given a Riemann surface $C$ and a 
cut system, a collection 
$\CC=\{\ga_1,\dots,\ga_h\}$ of 
homotopy classes of non-intersecting simple closed curves
on $C$.  
Cutting along all the curves in $\CC$ produces a pants decompostion,
$C\setminus\CC\simeq\bigsqcup_{v}C_{0,3}^v$, where the $C_{0,3}^v$
are three-holed spheres.

Having glued $C$ from three-punctured spheres defines a
distinguished cut system, defined by a collection
of simple closed curves $\CC=\{\ga_1,\dots,\ga_h\}$ 
such that $\ga_r$ can be embedded into the annulus $A_r$
for $r=1,\dots,h$. 

\begin{figure}[htb]
\epsfxsize4cm
\centerline{\epsfbox{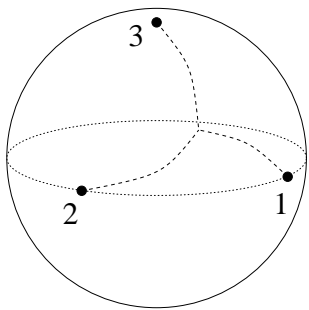}}
\caption{\it A sphere with three punctures, and 
a choice of branch cuts for the definition of 
the logarithms of local coordinates around the punctures.}
\label{threept}\vspace{.3cm}
\end{figure}

An important deformation of the complex structure of $C$ is 
the Dehn-twist: It corresponds to rotating one end of an 
annulus $A_r$ by $2\pi$ before regluing, and can be described by
a change of the local coordinates used in the gluing
construction. The coordinate $q_r$ can not distinguish complex
structures related by a Dehn twist in $A_r$. It is often
useful to replace the coordinates $q_r$
by logarithmic coordinates $\tau_r$ such that $q_r=e^{2\pi i \tau_r}$. 
This corresponds to replacing the gluing identification \rf{glueid} 
by its logarithm. In order to define the logarithms of the coordinates
$z_i$ used in \rf{glueid}, one needs to introduce branch cuts
on the three-punctured spheres, an example being
depicted in Figure \ref{threept}.

By imposing the requirement that the branch cuts chosen on each
three-punctured sphere glue to a connected three-valent 
graph $\Ga$ on $C$, one gets an unambiguous definition of the
coordinates $\tau_r$. We see that the logarithmic versions of
the gluing construction that define the coordinates $\tau_r$
are parameterized by the pair of data $\si=(\CC_\si,\Ga_\si)$,
where $\CC_\si$ is the cut system defined by the gluing construction,
and $\Ga_\si$ is the three-valent graph specifying the choices 
of branch cuts. In order to have a handy terminology we will 
call the pair of data  $\si=(\CC_\si,\Ga_\si)$ a {\it pants decomposition},
and the three-valent graph $\Ga_\si$ will be called the Moore-Seiberg graph, 
or MS-graph associated to a pants decomposition $\si$.

The gluing construction depends on the choices of 
coordinates around the punctures
$Q_i$.  There exists an ample supply of choices
for the coordinates $z_i$ such that the union of the 
neighborhoods $\CU_{\si}$ produces a cover of $\CM(C)$ \cite{HV}. 
For a fixed choice of these coordinates
one produces families of Riemann surfaces fibred over
the multi-discs $\CU_{\si}$ with coordinates $q$.
Changing the coordinates $z_i$ around $q_i$ produces a 
family of Riemann surfaces which
is locally biholomorphic to the initial one \cite{RS}. 

\subsection{The Moore-Seiberg groupoid}

\begin{figure}[t]
\epsfxsize9cm
\centerline{\epsfbox{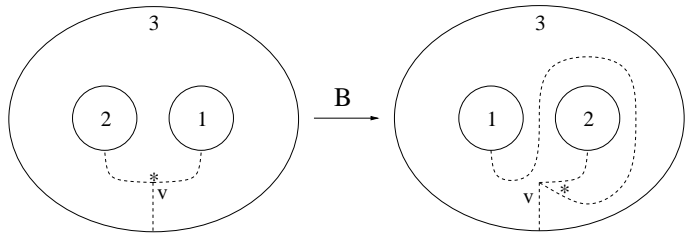}}
\caption{The B-move}\label{bmove}\vspace{-.03cm}
\end{figure}

\begin{figure}[t]
\epsfxsize9cm
\centerline{\epsfbox{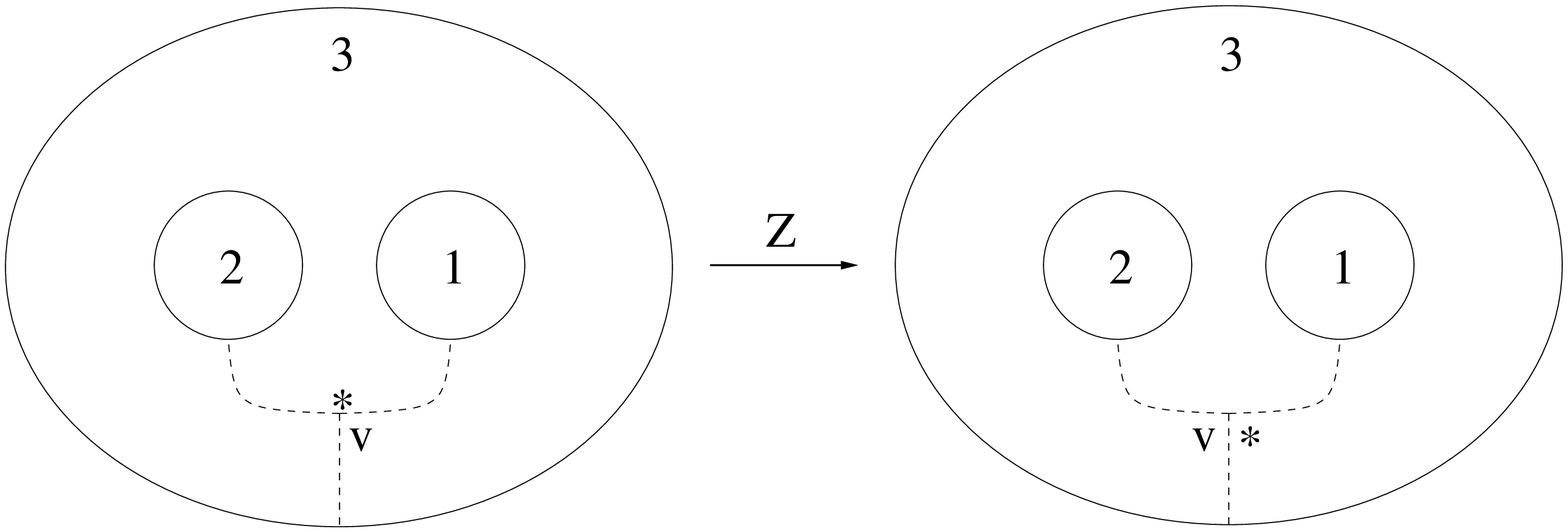}}
\caption{The Z-move}\label{zmove}
\end{figure}

\begin{figure}[t]
\epsfxsize4.5cm
\centerline{\epsfbox{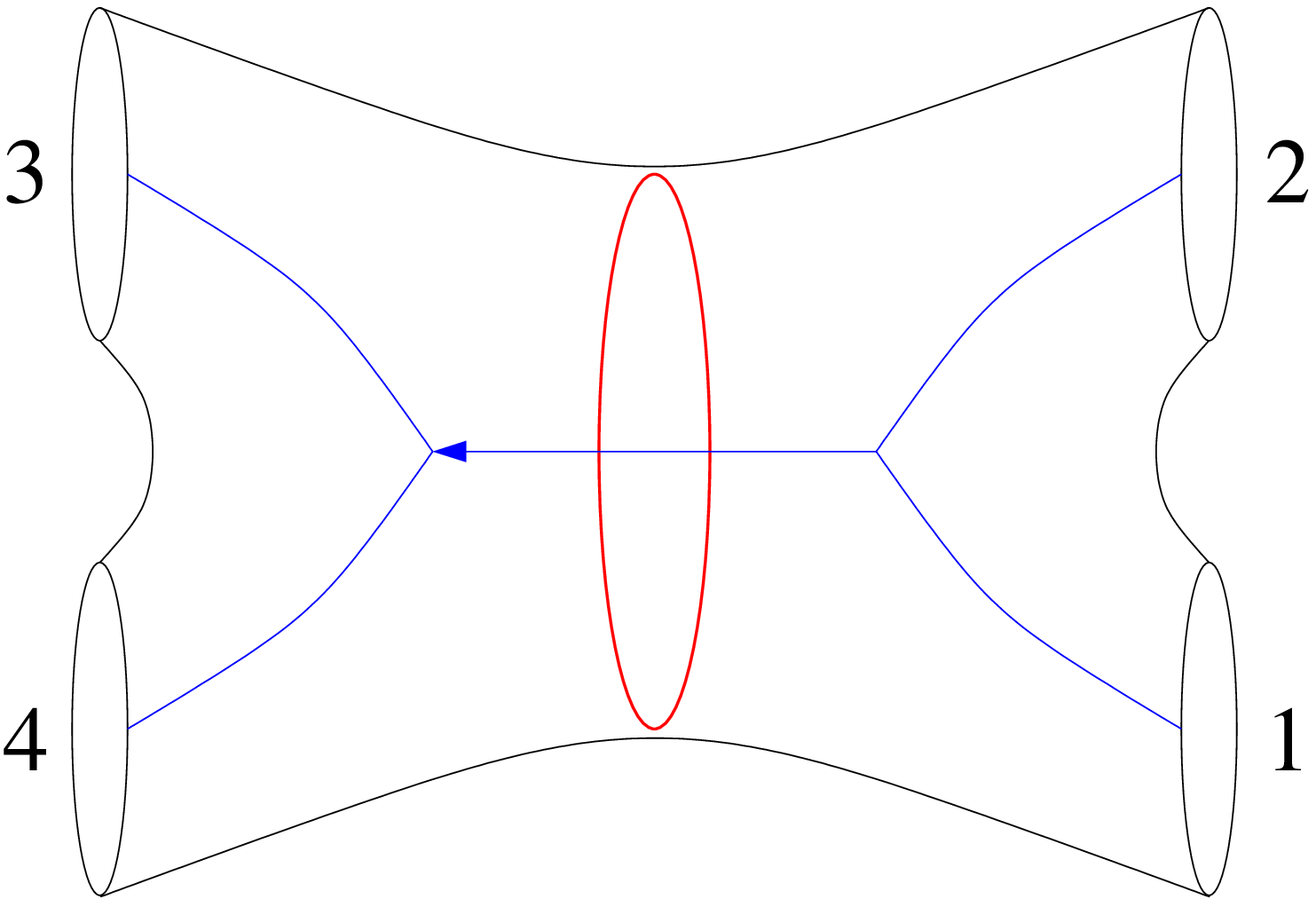}\hspace{.5cm}$\Longrightarrow$\hspace{.5cm}
\epsfxsize4.5cm\epsfbox{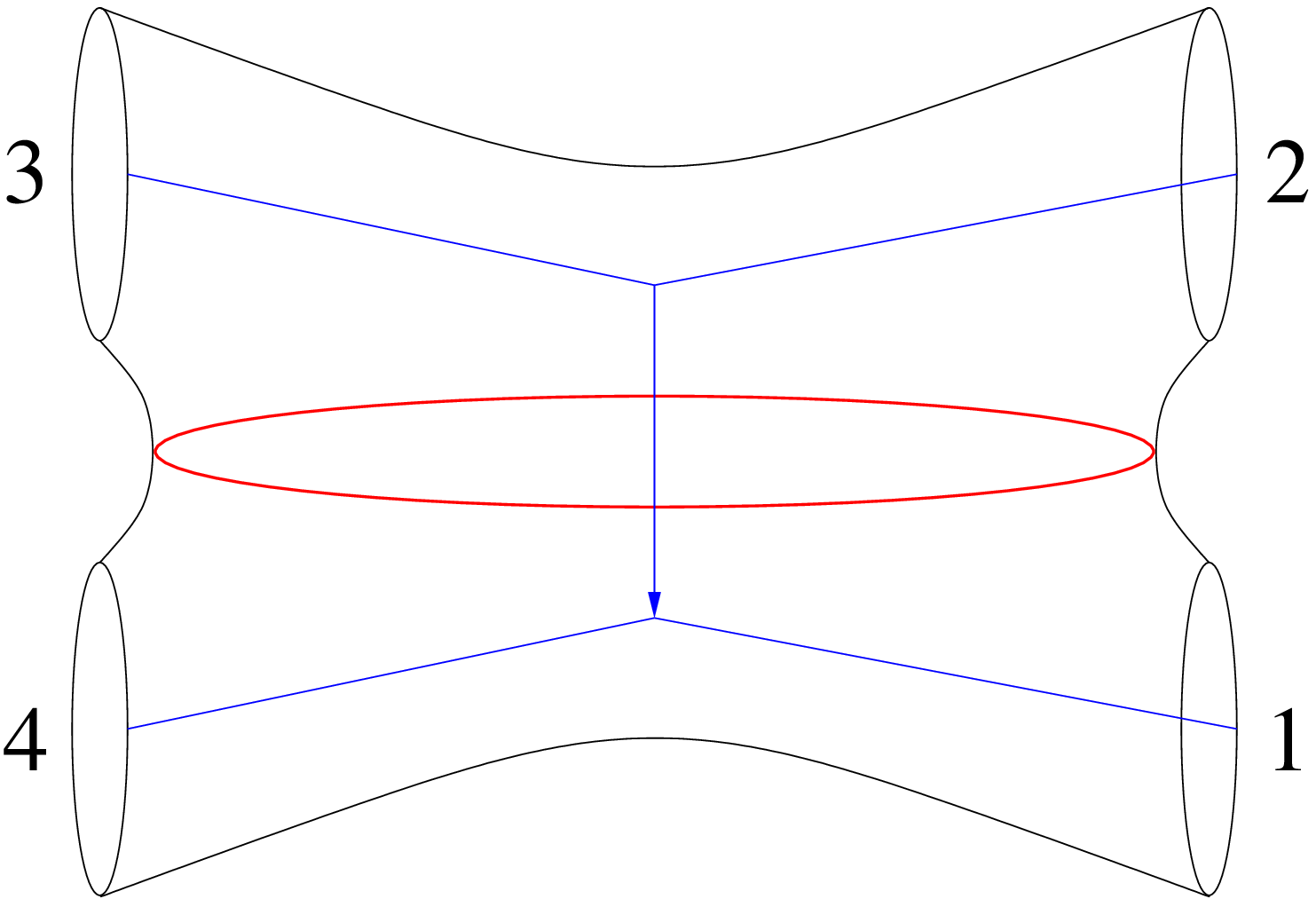}}
\caption{The F-move}\label{fmove}\vspace{.3cm}
\end{figure}

\begin{figure}[t]
\epsfxsize5.5cm
\centerline{\epsfbox{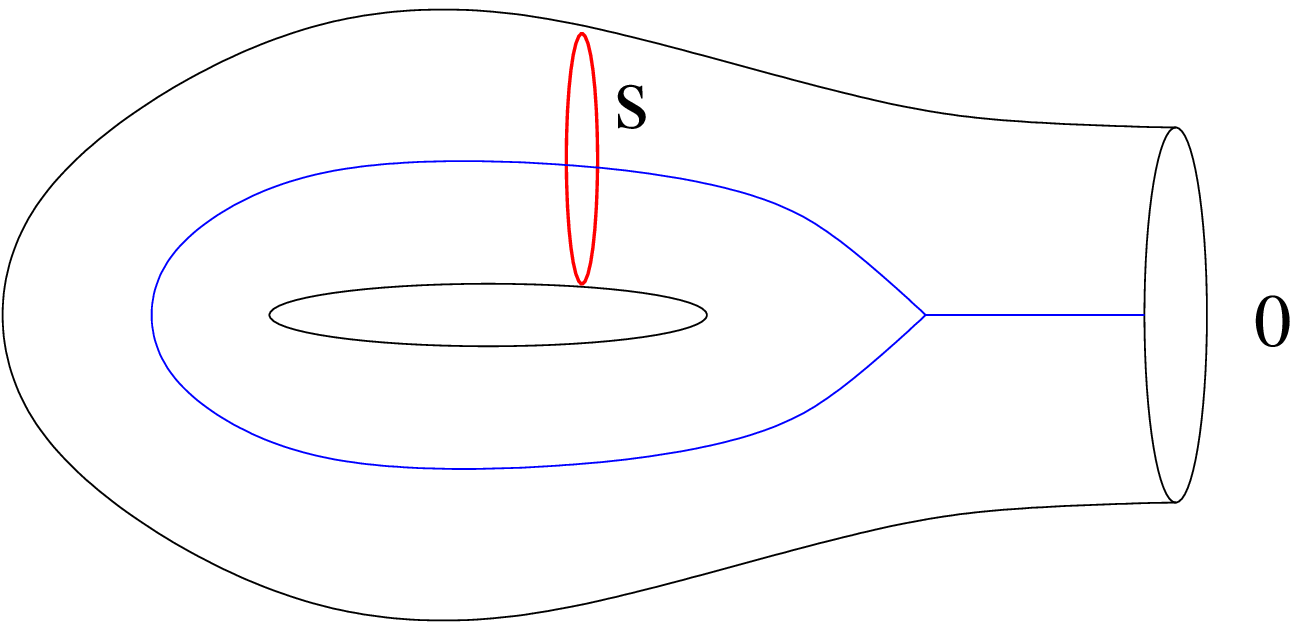}\hspace{.5cm}$\Longrightarrow$\hspace{.5cm}
\epsfxsize5.5cm\epsfbox{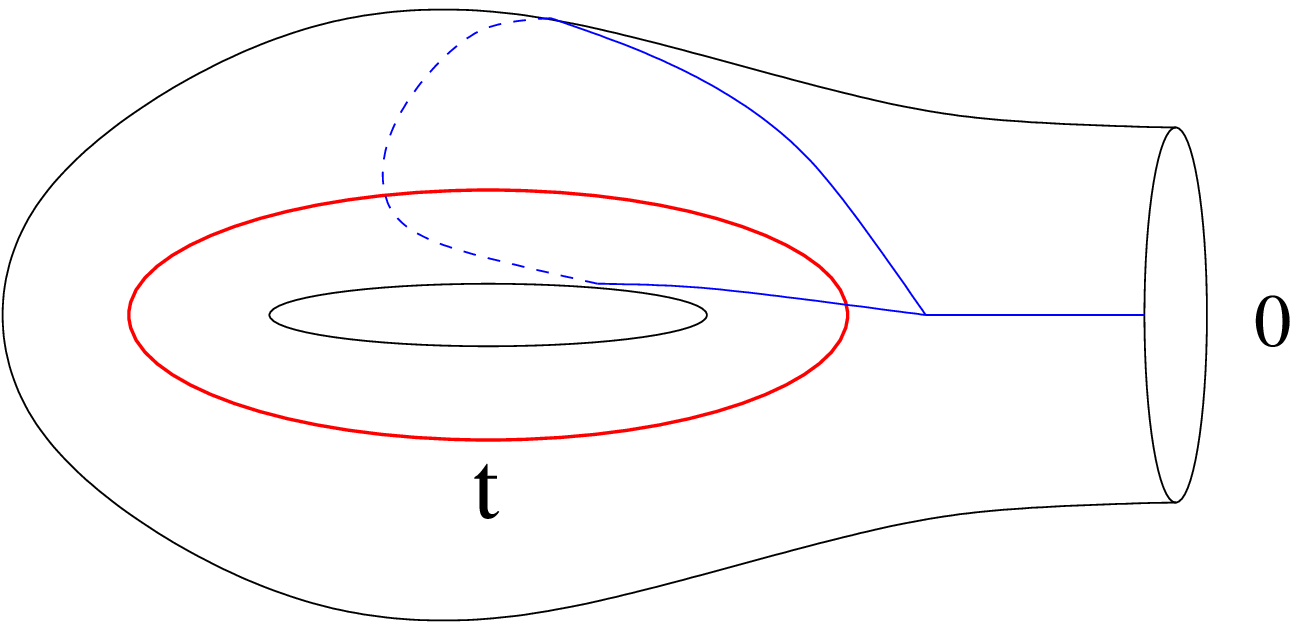}}
\caption{The S-move}\label{smove}\vspace{.3cm}
\end{figure}

Let us note \cite{MS,BK} that
any two 
different pants decompositions 
$\si_2$, $\si_1$ can be connected by a sequence of
elementary moves localized in subsurfaces of $C_{g,n}$ of type
$C_{0,3}$, $C_{0,4}$ and $C_{1,1}$. These will be called the 
$B$, $F$, $Z$ and $S$-moves, respectively.
Graphical representations for the elementary 
moves $B$, $Z$,  $F$, and $S$ are given in 
Figures \ref{bmove},  \ref{zmove},  \ref{fmove},
and \ref{smove}, respectively.

One may formalize
the resulting structure by introducing a two-dimensional 
CW complex $\CM(C)$ with set of vertices $\CM_\0(C)$
given by the
pants decompositions $\si$, and a 
set of edges $\CM_\1(C)$ associated to the elementary
moves. 

The Moore-Seiberg groupoid is defined to be the path groupoid of 
 $\CM(C)$. It can be described in terms of generators and relations,
the generators being associated with the edges of  $\CM(C)$, 
and the relations associated with the faces of $\CM(C)$. 
The classification of the relations was first presented in \cite{MS},
and rigorous mathematical proofs have been presented in \cite{FG,BK}.
The relations are all represented by sequences of moves localized
in subsurfaces $C_{g,n}$ with genus $g=0$ and $n=3,4,5$ punctures,
as well as $g=1$, $n=1,2$. Graphical representations of the
relations can be found in \cite{MS,FG,BK}.

\subsection{Hyperbolic metrics vs. flat connections}

The classical uniformization theorem ensures existence and uniqueness of 
a hyperbolic metric, 
a metric of constant negative curvature, on a Riemann surface $C$.
In a local chart with complex analytic coordinates $y$
one may represent this metric in the form $ds^2=e^{2\vf}dyd\bar y$, with
$\vf$ being a solution to the Liouville equation
$\pa\bar \pa\vf=\mu e^{2\vf}dy d\bar y$.

There is a well-known relation between the Teichm\"uller space $\CT(C)$
and a connected component of the moduli space 
$\CM_{\rm flat}(C)$ of flat ${\rm PSL}(2,\BR)$-connections 
on $C$. The relevant component will be denoted 
as $\CM_{\rm flat}^0(C)$. The relation between $\CT(C)$  and
$\CM_{\rm flat}^0(C)$ may be desrcribed as follows.

To a hyperbolic metric  $ds^2=e^{2\vf}dyd\bar y$ let us 
associate the connection $\nabla=\nabla'+\nabla''$, 
and
\begin{equation}\label{hypconn}
\nabla''=\bar\pa\,,\qquad\nabla'\,=\,\pa+M(y)dy\,, \qquad
M(y)\,=\,\bigg(\,\begin{matrix} 0 & -t \\ 1 & 0 \end{matrix}\,\bigg)\,,
\end{equation}
with $t$ constructed from $\vf(y,\bar y)$ as
\begin{equation}
t:=\,-(\pa_y\vf)^2+\pa_y^2\vf\,.
\end{equation}
This connection is flat since $\pa_y\bar{\pa}_{\bar y}\vf
=\mu e^{2\vf}$
implies $\bar\pa t=0$. The form \rf{hypconn} of $\nabla$
is preserved by changes
of local coordinates if $t=t(y)$ transforms as
\begin{equation}\label{opertrsf-a}
t(y)\;\mapsto\;(y'(w))^2t(y(w))+\frac{1}{2}\{y,w\}\,,
\end{equation}
where the 
Schwarzian derivative $\{y,w\}$ is defined as
\begin{equation}
\{y,w\}\,\equiv\,\left(\frac{y''}{y'}\right)'-
\frac{1}{2}\left(\frac{y''}{y'}\right)^2\,.
\end{equation}
Equation \rf{opertrsf-a}  
is the transformation law characteristic for {\it projective}
connections, which are also called $\fsl_2$-opers, or opers for short.

The hyperbolic metric $ds^2=e^{2\vf}dyd\bar y$ can be constructed from 
the solutions to $\nabla s=0$ which implies
that the component $\chi$ of
$s=(\eta,\chi)$ solves a second order differential equation of the form
\begin{equation}\label{DFuchsian}
(\pa_y^2+t(y))\chi=0\,.
\end{equation}
Picking two linearly independent solutions $\chi_\pm$ of \rf{DFuchsian} with
$\chi_+'\chi_--\chi_-'\chi_+=1$ allows us to represent
$e^{2\vf}$ as $e^{2\vf}=-(\chi_+\bar\chi_--\chi_-\bar\chi_+)^{-2}$.
The  hyperbolic metric $ds^2=e^{2\vf}dyd\bar y$ may now be written
in terms of the quotient $A(y):=\chi_+/\chi_-$ as
\begin{equation}
ds^2\,=\,e^{2\vf}dyd\bar y\,=\,\frac{\pa A\bar\pa {\bar A}}{({\rm Im}(A))^2}\,.
\end{equation}
It follows that $A(y)$ represents a conformal mapping from $C$ to 
a domain $\Omega$ 
in the upper half plane $\BU$ with its standard constant curvature 
metric. $C$ is therefore conformal to $\BU/\Ga$, where
the Fuchsian group $\Ga$ is the monodromy group of the 
connection $\nabla$.

\subsection{Hyperbolic pants decomposition and Fenchel-Nielsen coordinates}\label{sec:FNdef}

Let us consider hyperbolic surfaces $C$ of 
genus $g$ with $n$ holes. We will assume that the holes
are represented by geodesics in the hyperbolic metric.
A pants decomposition of a hyperbolic surface 
$C$ is defined, as before, by a cut system which in this context
may be represented by a collection 
$\CC=\{\ga_1,\dots,\ga_h\}$ of non-intersecting simple closed geodesics
on $C$.
The complement 
$C\setminus\CC$ is a disjoint union $\bigsqcup_{v}C_{0,3}^v$ of
three-holed spheres (trinions).
One may reconstruct $C$ from the resulting collection 
of trinions by pairwise gluing of boundary 
components. 

For given lengths of the three boundary geodesics there is a unique
hyperbolic metric on each trinion $C_{0,3}^v$. 
Introducing a numbering of the boundary geodesics $\ga_i(v)$, $i=1,2,3$,
one gets three distinguished geodesic arcs $\ga_{ij}(v)$, $i,j=1,2,3$
which connect the boundary components pairwise.  
Up to homotopy there are exactly 
two tri-valent graphs $\Ga_{\pm}^v$ on $C_{0,3}^v$ 
that do not intersect any $\ga_{ij}(v)$.
We may assume that these graphs glue to two connected graphs 
$\Ga_\pm$ on $C$. 
The pair of data $\si=(\CC_\si,\Ga_\si)$, where $\Ga_\si$ is one
of the MS graphs $\Ga_{\pm}$ associated 
to a hyperbolic pants decomposition, can be used to distinguish
different pants decompositions in hyperbolic geometry.

The data  $\si=(\CC_\si,\Ga_\si)$ 
can also be used to define the classical Fenchel-Nielsen coordinates
for $\CT(C)$ as follows. 
Note that the edges $e$ of $\Ga_{\si}$ are in one-to-one correspondence
with the curves $\ga_e$ in $\CC_{\si}$. To each edge $e$ let us 
first associate the length $l_e$ of the geodesic $\ga_e$.

\begin{figure}[htb]
\epsfxsize6cm
\centerline{\epsfbox{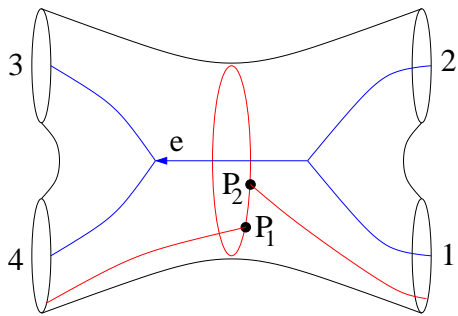}}
\caption{\it A four-holed sphere with MS graph (blue) and
the geodesics used in the definition of the Fenchel-Nielsen
coordinates (red).}
\label{FN}\vspace{.3cm}
\end{figure}

In order to define the Fenchel-Nielsen twist variables we need to consider
two basic cases: Either a given $\ga_e\in\CC$ separates two different
trinions $C_{0,3}^{v_1}$ and $C_{0,3}^{v_2}$, or it is the result 
of the identification of two boundary components
of a single trinion. 
In order to fix a precise prescription in the first case let us
assume that $C$ and 
the edge $e$ are oriented. One may then define
a numbering of the boundary components of the four-holed sphere
$C_{0,4}^{v_{12}}$ obtained by gluing $C_{0,3}^{v_1}$ and $C_{0,3}^{v_2}$:
Number $1$ is assigned to the boundary component intersecting the
next edge of $\Ga_\si$ on the right of the 
tail of the edge $e$, number $4$ to  the boundary component intersecting the
next edge of $\Ga_\si$ to the left of the tip of $e$.
There are geodesic 
arcs $\ga_{4e}(v_2)$ and $\ga_{1e}(v_1)$ on $C_{0,3}^{v_1}$ and $C_{0,3}^{v_2}$
that intersect $\ga_e$ in points $P_1$, and $P_2$, respectively.
This set-up is drawn in Figure \ref{FN}.

The twist variable $k_e$ is then defined to be the geodesic distance between
$P_1$ and $P_2$, and the twist angle $\theta_e=2\pi k_e/l_e$.
The second case (gluing of two holes in one trinion gives
sub-surface $C_e$ of type $C_{1,1}$) is treated similarly.

We see that the role of the MS-graph $\Ga_\si$ is to distinguish
pants decompositions related by Dehn-twists, corresponding 
to $\theta_e\ra\theta_e+2\pi$.

\subsection{Trace coordinates}\label{sec:loops}

Given a flat ${\rm SL}(2,\BC)$-connection $\nabla=d-A$, one may define 
its holonomy $\rho(\ga)$ along a closed loop $\ga$ 
as $\rho(\ga)=\CP\exp(\int_\ga A)$. The assignment
$\ga\mapsto \rho(\ga)$ defines a representation of $\pi_1(C)$ in 
${\rm SL}(2,\BC)$, defining a point in the so-called
character variety 
\begin{equation}
\homsl:={\rm Hom}(\pi_1(C),{\rm PSL}(2,\BC))/{\rm PSL}(2,\BC)\,.
\end{equation}
The Fuchsian groups $\Ga$ represent a 
connected component 
$\CM_{\rm char}^{\BR,0}(C)\simeq \CT(C)$ 
in the {\it real} character variety
\begin{equation}
\homslr:={\rm Hom}(\pi_1(C),{\rm PSL}(2,\BR))/{\rm PSL}(2,\BR)\,.
\end{equation}
which will be of main interest here. $\homslr$ is naturally
identified with the moduli space $\CM_{\rm flat}(C)$ of
flat ${\rm PSL}(2,\BR)$ connections on $C$, and 
$\CM_{\rm char}^{\BR,0}(C)$ represents the so-called 
Teichm\"uller component $\CM_{\rm flat}^0(C)$ within $\CM_{\rm flat}(C)$.

\subsubsection{Topological classification of closed loops}

With the help of pants decompositions one may conveniently classify all
non-selfintersecting closed loops on $C$ up to homotopy. To a loop $\ga$
let us associate the collection of integers $(r_e,s_e)$ 
associated to all edges $e$ of $\Ga_{\si}$ which are defined as follows.
Recall that there is a unique curve $\ga_e\in\CC_\si$ that intersects 
a given edge $e$ on $\Ga_\si$ exactly once, and which does not intersect
any other edge. The integer $r_e$ is defined as 
the number of intersections between $\ga$ and 
the curve $\ga_e$. Having chosen an orientation 
for the edge $e_r$ we will define
$s_e$ to be the intersection index between $e$ and $\ga$. 

Dehn's theorem (see \cite{DMO} for a nice discussion)
ensures that the curve $\ga$ is up to homotopy uniquely classified by the
collection of integers $(r,s)$, subject to the restrictions 
\begin{equation}
\begin{aligned}
{\rm (i)} \quad & 
r_e\geq 0\,,\\ {\rm (ii)} \quad & {\rm if}\;\;r_e=0\;\Rightarrow\;s_e\geq 0\,,\\
{\rm (iii)} \quad &
r_{e_\1}+r_{e_\2}+r_{e_\3}\in 2\BZ\;\,{\rm whenever}\;\,
\ga_{e_\1},\ga_{e_\2},\ga_{e_\3}\;\,\text{bound the same trinion}.
\end{aligned}
\end{equation}
We will use the notation $\ga_{(r,s)}$ for the geodesic which has
parameters $(r,s):e\mapsto (r_e,s_e)$.

\subsubsection{Trace functions}

The trace functions
\begin{equation}\label{tracedef}
L_\ga:=\nu_\ga{\rm tr}(\rho(\ga))\,,
\end{equation}
represent useful coordinate functions for $\homsl$.
The signs $\nu_\ga\in\{+1,-1\}$ in the definition 
\rf{tracedef} will be specified when this becomes relevant.
Real values of the trace functions $L_{\ga}$ 
characterize $\homslr$.

If the representation $\rho$ is the one coming 
from the uniformization 
of $C$, it is an elementary exercise in hyperbolic geometry
to show that the length $l_\ga$ of the geodesic $\ga$ is related
to $L_{\ga}$ by
\begin{equation}
|L_{\ga}|\,=\,2\cosh(l_\ga/2)\,.
\end{equation}
Representing the
points in $\CM_{\rm char}^{\BR,0}(C)$ by representations 
$\rho:\pi_1(C)\ra {\rm SL}(2,\BR)$, we will
always choose the sign $\nu_\ga$ in \rf{tracedef} such
that $L_{\ga}=2\cosh(l_\ga/2)$.
 
We may then analytically 
continue the trace functions $L_{\ga}$ defined thereby
to coordinates on the natural
complexification $\CM_{\rm char}^{\BC,0}(C)\subset 
\CM_{\rm char}^{\BC}(C)$ of $\CM_{\rm char}^{\BR,0}(C)$.
The representations $\rho:\pi_1(C)\ra {\rm PSL}(2,\BC)$ that
are parameterized by $\CM_{\rm char}^{\BC,0}(C)$ are called 
quasi-Fuchsian. It is going to be important for us to have 
coordinates $L_\ga$ that are 
complex analytic on $\CM_{\rm char}^{\BC,0}(C)$ on the one hand,
but {\it positive} (and larger than two) 
when restricted to the real slice
$\CM_{\rm char}^{\BR,0}(C)$ on the other hand.

\subsubsection{Skein algebra}

The well-known relation ${\rm tr}(g){\rm tr}(h)={\rm tr}(gh)+{\rm tr}(gh^{-1})$
valid for any pair of $SL(2)$-matrices $g,h$ implies that the
geodesic length functions satisfy the 
 so-called skein relations,
\begin{equation}\label{skeinrel}
L_{\ga_1} L_{\ga_2}\,=\,L_{S(\ga_1,\ga_2)}\,,
\end{equation}
where $S(\ga_1,\ga_2)$ is the loop obtained from 
$\ga_1$, $\ga_2$ by means of the smoothing operation,
defined as follows. The application of $S$ to a single intersection
point of $\ga_1$, $\ga_2$ is depicted in 
Figure \ref{skeinfig} below.
\begin{figure}[htb]
\epsfxsize8cm
\centerline{\epsfbox{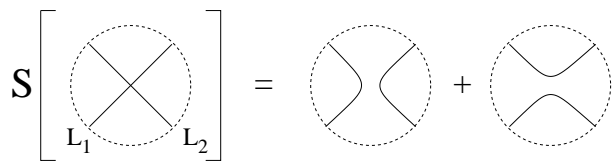}}
\caption{The symmetric 
smoothing operation}\label{skeinfig}\vspace{.3cm}
\end{figure}
The general result is obtained by
applying this rule at each intersection point, and summing the
results.

The coordinate functions $L_\ga$ generate the 
commutative algebra $\CA(C)\simeq {\rm Fun}^{\rm alg}(\CM_{\rm flat}(C))$ 
of functions  on $\CM_{\rm flat}(C)$. 
As set of generators one may take the functions
$L_{(r,s)}\equiv L_{\ga_{(r,s)}}$. 
The skein relations imply various relations among the 
$L_{(r,s)}$. It is not hard to see that these relations
allow one to express arbitrary $L_{(r,s)}$ in terms
of a finite subset of the set of $L_{(r,s)}$.

\subsubsection{Generators and relations} \label{sec:Genrel}

The pants decompositions allow us to describe $\CA(C)$ in terms of generators and
relations. Let us note that 
to each internal\footnote{An internal edge does not end 
in a boundary component of $C$.} edge $e$ of the MS-graph $\Ga_\si$ of $\si$
there corresponds a unique  curve $\ga_e$
in the cut system $\CC_\si$. There is a unique subsurface 
$C_{e}\hookrightarrow C$
isomorphic to either $C_{0,4}$ or $C_{1,1}$ that contains $\ga_e$ in 
the interior of $C_e$. The subsurface $C_e$ has boundary components
labeled by numbers $1,2,3,4$ according to the convention introduced
in Subsection \ref{sec:FNdef} if $C_e\simeq C_{0,4}$, and if $C_e\simeq C_{1,1}$
we will assign to the single boundary component the number $0$.

For each edge $e$ let us introduce the geodesics $\ga^e_t$
which have Dehn parameters $(r^e,0)$, where
$r_{e'}^e=2\de_{e,e'}$ if $C_e\simeq C_{0,4}$ and 
$r_{e'}^e=\de_{e,e'}$ if $C_e\simeq C_{1,1}$. These geodesics
are depicted as red curves on the right halfs of Figures \ref{fmove}
and \ref{smove}, respectively. There furthermore exist unique
geodesics $\ga^e_u$ 
with Dehn parameters $(r^e,s^e)$, where $s_{e'}^e=\de_{e,e'}$.
We will denote
$L_{k}^e\equiv |{\rm tr}(\ga^e_k)|$, where $k\in\{s,t,u\}$.
The set
$\{L_s^e,L_t^e,L_u^e\,;\,\ga_e\in\CC_{\si}\}$ generates $\CA(C)$.

These coordinates are not independent, though. Further relations
follow from the relations in $\pi_1(C)$.
It can be shown (see e.g. \cite{Go09} for a review) 
that any triple of
coordinate functions $L_s^e$, $L_t^e$ and $L_u^e$
satisfies an algebraic
relation of the form 
\begin{equation}
\label{algrel}
P_e(L_s^e,L_t^e,L_u^e)\,=\,0\,.
\end{equation}
The polynomial $P_e$ in \rf{algrel} is
for $C_e\simeq C_{0,4}$ explicitly given
as\footnote{Comparing to \cite{Go09} note
that some signs were absorbed by a suitable choice of the signs
$\nu_\ga$ in \rf{tracedef}.} 
\begin{align}
 P_e(L_s, L_t, L_u) := &-L_s L_t L_u + L_s^2 + L_t^2 + L_u^2 \nonumber \\
& +L_s (L_3L_4 + L_1L_2) + L_t (L_2L_3 + L_1L_4) + L_u (L_1L_3 + L_2L_4)
\nonumber \\ &
-4 + L_1^2+L_2^2+L_3^2+L_4^2+L_1L_2L_3L_4\,,
\label{W04}\end{align}
while for $C_e\simeq C_{1,1}$ we take $P$ to be 
\begin{align}
 P_e(L_s, L_t, L_u) := L_s^2 + L_t^2 + L_u^2 - L_s L_t L_u + L_0-2\,.
\end{align}
In the expressions above we have denoted $L_{i}:=|{\rm Tr}(\rho(\ga_i))|$,
$i=0,1,2,3,4$, where $\ga_0$ is the geodesic representing the 
boundary of $C_{1,1}$, while
$\ga_i$, $i=1,2,3,4$ represent 
the boundary components of $C_{0,4}$, labelled according
to the convention above.   

\subsubsection{Poisson structure}

\begin{figure}[t]
\epsfxsize8cm
\centerline{\epsfbox{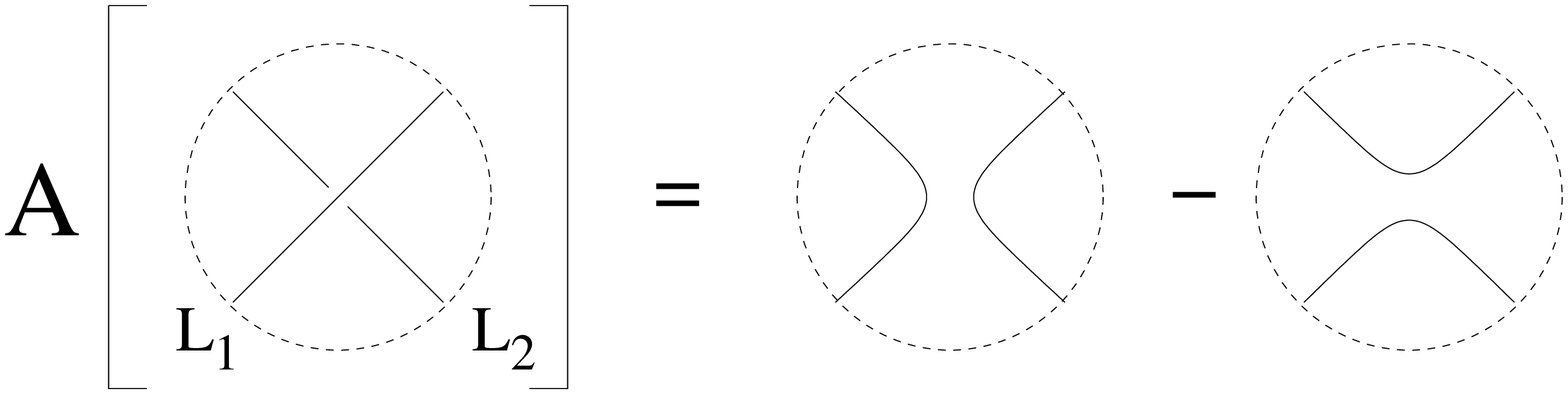}}
\caption{The anti-symmetric smoothing operation}\label{skeinfigas}\vspace{.3cm}
\end{figure}

There is also a natural Poisson bracket on $\CA(C)$ \cite{Go86}, 
defined such that 
\begin{equation}
\{\,L_{\ga_1}\,,\, L_{\ga_2}\,\}\,=\,L_{A(\ga_1,\ga_2)}\,,
\end{equation}
where $A(\ga_1,\ga_2)$ is the loop obtained from 
$\ga_1$, $\ga_2$ by means of the anti-symmetric smoothing operation,
defined as above, but replacing the rule depicted in 
Figure \ref{skeinfig} by the one depicted in Figure 
\ref{skeinfigas}.

The resulting expression for the Poisson bracket $\{\,L_s^e\,,L_t^{e}\,\}$
can be written elegantly in the form
\begin{equation}\label{loopPB}
\{\,L_s^e\,,L_t^{e}\,\}\,=\,\frac{\pa}{\pa L_u^e} P_e(L_s^e,L_t^e,L_u^e)\,.
\end{equation}
It is remarkable that the same polynomial appears both in 
\rf{algrel} and in \rf{loopPB}, which indicates that the symplectic
structure on  $\CM_{\rm flat}$ is compatible with its 
structure as algebraic variety.

This Poisson structure coincides with the Poisson structure coming
from the natural symplectic structure on $\CM_{\rm flat}(C)$ which
was introduced by Atiyah and Bott.

\subsection{Darboux coordinates for $\CM_{\rm flat}(C)$}
\label{sec:2hyp}

One may express $L^e_s$, $L^e_t$ and $L^e_u$
in terms of the Fenchel-Nielsen coordinates 
$l_e$ and $k_e$ \cite{Go09}.
The expressions
are
\begin{subequations}\label{FN-FK}
\begin{align}
& L^e_s\,=\,2\cosh(l_e/2)\,,
\end{align}
and for $C_e\simeq C_{1,1}$,
\begin{align}
& L^e_t\big((L^e_s)^2-4\big)^{\frac{1}{2}}\,=\,2\cosh(k_e/2)
\sqrt{(L_s^e)^2+L_0^e-2}\\
& L^e_u\big((L^e_s)^2-4\big)^{\frac{1}{2}}\,=\,
2\cosh((l_e+k_e)/2)
\sqrt{(L_s^e)^2+L_0^e-2}\,,
\end{align}
while for $C_e\simeq C_{0,4}$,
\begin{align}
& L^e_t\big((L^e_s)^2-4\big)\,=\,
2(L^e_2L^e_3+L^e_1L^e_4)+L^e_s(L^e_1L^e_3+L^e_2L^e_4) \label{cl-'t Hooft}\\
& \hspace{3cm}
+2\cosh(k_e)
\sqrt{c_{12}(L^e_s)c_{34}(L^e_s)}\,,
\notag\\
& L^e_u\big((L^e_s)^2-4\big)\,=\,
L^e_s(L^e_2L^e_3+L^e_1L^e_4)+2(L^e_1L^e_3+L^e_2L^e_4) \label{cl-dyonic}\\
& \hspace{3cm}
+2\cosh((2k_e-l_e)/2)
\sqrt{c_{12}(L^e_s)c_{34}(L^e_s)}\,,
\notag
\end{align}
\end{subequations}
where $L_i^e=2\cosh\frac{l_i^e}{2}$, and $c_{ij}(L_s)$ is defined as
\begin{align}\label{cijdef}
c_{ij}(L_s) & \,=\,L_s^2+L_i^2+L_j^2+L_sL_iL_j-4\ \\ \nonumber & =
2 \cosh \fr{l_s+l_i+l_j}{4} 2 \cosh \fr{l_s+l_i-l_j}{4}
2 \cosh \fr{l_s-l_i+l_j}{4} 2 \cosh \fr{l_s-l_i-l_j}{4}.
\end{align}
These expressions
ensure that the algebraic relations $P_e(L_s,L_t,L_t)=0$
are satisfied.

The coordinates $l_e$ and $k_e$ are known to be
Darboux-coordinates for $\CM_{\rm flat}(C)$, 
having the
Poisson bracket 
\begin{equation}
\{\,l_e\,,\,k_{e'}\,\}\,=\,2\de_{e,e'}\,.
\end{equation}
This was recently observed and exploited in a related context in 
\cite{NRS}.

Other natural sets of 
Darboux-coordinates $(l_e,k_e)$ can be obtained by means of 
canonical transformations $k_e'=k_e+f(l)$. By a suitable 
choice of $f(l)$, one gets Darboux coordinates $(l_e,k_e)$
in which, for example, 
the expression for $L_t^e$ in \rf{FN-FK} is replaced by 
\begin{align}\label{FN-FK'}
L^e_t((L^e_s)^2-4)\,=\,&
\;2({L}_2L^e_3+L^e_1L^e_4)+L^e_s(L^e_1L^e_3+L^e_2L^e_4) \\
& +
2 \cosh \fr{l^e_s+l^e_1-l^e_2}{4} 2 \cosh \fr{l^e_s+l^e_2-l^e_1}{4} 
2 \cosh \fr{l^e_s+l^e_3-l^e_4}{4} 2 \cosh \fr{l^e_s+l^e_4-l^e_3}{4} \;e^{+k_s'}\notag  \\
& +
2 \cosh \fr{l^e_s+l^e_1+l^e_2}{4} 2 \cosh \fr{l^e_s-l^e_1-l^e_2}{4} 
2 \cosh \fr{l^e_s+l^e_3+l^e_4}{4} 2 \cosh \fr{l^e_s-l^e_3-l^e_4}{4} \;e^{-k_s'}
\,. \notag
\end{align}
The Darboux coordinates $(l_e,k_e)$ are equally good to 
represent the Poisson structure of $\CM_G(C_{0,4})$, but they 
have the advantage that the expressions for $L_\ka^e$ do not 
contain square-roots. This remark will later turn out to be useful.


\newpage 

\part{\Large Supersymmetric gauge theories}\label{par:SUSYgauge}

Summary:
\begin{itemize}
\item Review of SUSY gauge theories $\CG_C$ 
of class $\CS$ on $4d$ ellipsoids.
\item The path integrals representing
supersymmetric observables on $4d$ ellipsoids localize to the 
quantum mechanics of the scalar zero modes of $\CG_C$. 
\item The instanton partition functions can be interpreted
as certain wave-functions $\Psi_{\tau}(a)$ in 
the zero mode quantum mechanics.
\item The Wilson and 't Hooft loops act nontrivially on the
wave-functions $\Psi_{\tau}(a)$.
\item Algebra $\CA_{\ep_1,\ep_2}$ generated
by supersymmetric Wilson and 't Hooft loops 
is isomorphic to the quantized algebra of functions on 
a component of $\CM_{\rm flat}(C)$.
\item Physical reality properties of Wilson and 't Hooft loops $\Rightarrow$
Relevant for $\CG_C$ 
is the component $\CM_{\rm flat}^0(C)\subset \CM_{\rm flat}(C)$
isomorphic to the Teichm\"uller space $\CT(C)$.
\item Analyticity + behavior under S-duality $\Rightarrow$ 
Instanton partition functions can be characterized as
solutions to a Riemann-Hilbert type
problem.
\end{itemize}

\section{Quantization of $\CM_{\rm flat}(C)$ from gauge theory}

\setcounter{equation}{0}

To a Riemann surface $C$ of genus $g$ and $n$ punctures 
one may associate \cite{G09} a four-dimensional gauge theory $\CG_C$ with 
$\CN=2$ supersymmetry, gauge group $({\rm SU}(2))^{3g-3+n}$ and
flavor symmetry $({\rm SU}(2))^n$. 
In the cases where $(g,n)=(0,4)$ and $(g,n)=(1,1)$ one would 
get the supersymmetric gauge theories commonly referred to
as $N_f=4$ and $N=2^*$-theory, respectively.
The aim of this introductory section is to review the relation
between $C$ and $\CG_C$ along with recent exact results on  
expectation values of certain supersymmetric observables in $\CG_C$.

\subsection{Supersymmetric gauge theories of class $\CS$}

The gauge theory $\CG_C$ has a Lagrangian description for
each choice of a pants decomposition $\si$. We will now describe the relevant 
parts of the mapping between geometric structures on $C$ and
the defining data of $\CG_C$. 

The field content of $\CG_C$ is determined as follows. To each 
internal edge $e\in\Ga_{\si}$ there is an associated $\CN=2$ 
vector multiplet 
containing a vector field $A^e_{\mu}$, two fermions $\la_e$, $\bar{\la}_e$,
and two real scalars $\phi_e$, $\bar\phi_e$. 
Matter fields are represented by (half-)hypermultiplets
associated to the vertices $v$ of $\Ga_{\si}$. 
They couple only to the gauge fields associated to the edges
that meet at the vertex $v$. There are $n$ mass parameters 
associated to the boundary components of $C$. We refer 
to \cite{HKS2} for a description of the necessary building 
blocks for building the Lagrangian of $\CG_C$ associated to
a pants decomposition $\si$.

The Lagrangian for  $\CG_C$ will include kinetic terms for the gauge fields
$A^e_\mu$
with gauge coupling constants $g_e$, 
and it may include topological
terms
with theta angles $\theta_e$. These parameters 
are related to the gluing parameters $q_e$ as
\begin{equation}\label{qvstau}
q_e:=\,e^{2\pi i \tau_e}\,,\qquad
\tau_e:=\frac{4\pi i}{g_e^2}+\frac{\theta_e}{2\pi}\,.
\end{equation}
In order to define UV couplings constants like $g_e^2$ one generically
needs to fix a particular scheme for calculating amplitudes or expectation
values. Using a different scheme will lead to equivalent results related
by analytic redefinitions of the coupling constants. This ambiguity
will be mapped to the dependence of the coordinates $q_e$ for 
$\CT(C)$ on the choices of local coordinates around the punctures.
Equation \rf{qvstau} describes the relation which holds for a particular 
scheme in $\CG_C$, and a particular choice of local coordinates around
the punctures of $C_{0,3}$.

Different Lagrangian descriptions are related by S-duality. It follows
from the description of the gauge theories $\CG_C$ from class
$\CS$ given in \cite{G09} that
the groupoid of S-duality transformations coincides with the
Moore-Seiberg groupoid for the gauge theories of class $\CS$.

\subsection{Supersymmetric gauge theories on ellipsoids}

It may be extremely useful to 
study quantum field theories on compact Euclidean space-times
or on compact spaces rather than flat $\BR^4$. 
Physical quantities get finite size corrections which 
encode deep information on 
the quantum field theory we study. The zero modes of the
fields become dynamical, and have to be treated quantum-mechanically.

In the case of supersymmetric quantum field theories there are 
not many compact background space-times that allow us to preserve
part of the supersymmetry. A particularly interesting family 
of examples was studied in \cite{HH}, generalizing the seminal work
of Pestun \cite{Pe}.

\subsubsection{The four-dimensional ellipsoid}

Let us consider gauge theories $\CG_C$ on the four-dimensional ellipsoid
\begin{equation}
{\BBE}_{\ep_1,\ep_2}^4:=\,\{\,(x_0,\dots,x_4)\,|\,x_0^2+
\ep_1^2(x_1^2+x_2^2)+\ep_2^2(x_3^2+x_4^2)=1\,\}\,.
\end{equation}
Useful polar coordinates for ${\BBE}_{\ep_1,\ep_2}^4$
are defined as
\begin{equation}x_0\,=\,\sin\rho\,,\qquad
\begin{aligned}
&x_1=\ep_1^{-1}\cos\rho\,\cos\theta\,\cos\vf\,,\\
&x_2=\ep_1^{-1}\cos\rho\,\cos\theta\,\sin\vf\,,\\
&x_3=\ep_2^{-1}\cos\rho\,\sin\theta\,\cos\chi\,,\\
&x_4=\ep_2^{-1}\cos\rho\,\sin\theta\,\sin\chi\,.
\end{aligned}
\end{equation}
It was shown in \cite{Pe,HH} for some
examples of gauge theories $\CG_C$ that one of the supersymmetries  $Q$
is preserved on
 ${\BBE}_{\ep_1,\ep_2}^4$. It seems straightforward to 
generalize the proof of existence of an unbroken
supersymmetry $Q$ 
to all gauge theories $\CG_C$ of class $\CS$.

\subsubsection{Supersymmetric loop operators}

Supersymmetric Wilson loops can be defined as\footnote{We adopt 
the conventiones used in \cite{HH}.}
\begin{subequations}
\label{Wilsondef}
\begin{align}
& W_{e,\1}:={\rm Tr}_F\,\CP\exp\bigg(i\int_{S^1_\1}d\vf\;\Big(
A_{e,\vf}-\frac{1}{\ep_1}(\phi_e+\bar\phi_e)\Big)\bigg)\,,\\
& W_{e,\2}:={\rm Tr}_F\,\CP\exp\bigg(i\int_{S^1_\2}d\chi\;\Big(
A_{e,\chi}-\frac{1}{\ep_2}(\phi_e+\bar\phi_e)\Big)\bigg)\,,
\end{align}
\end{subequations}
with traces taken in the fundamental representation of $SU(2)$, 
and
contours of integration being
\begin{align}
&S^1_1:=\,\{\,(x_0,\dots,x_4)=
(\pi/2,\ep_1^{-1}\cos\vf,\ep_1^{-1}\sin\vf,0,0)\,,\,\vf\in[0,2\pi)\,\}\,,\\
&S^1_2:=\,\{\,(x_0,\dots,x_4)=
(\pi/2,0,0,\ep_2^{-1}\cos\chi,\ep_2^{-1}\sin\chi)\,,\,\chi\in[0,2\pi)\,\}\,,
\end{align}

The 't Hooft loop observables $T_{e,i}$, $i=1,2$, 
can be defined semiclassically for vanishing theta-angles
$\theta_e=0$ 
by the boundary condition 
\begin{equation}
F_e\,\sim\,\frac{B_e}{4}\ep_{ijk}\frac{x^i}{|\vec{x}|^3}dx^k\wedge dx^j\,,
\end{equation}
near the contours $S^1_i$, $i=1,2$. The coordinates $x^i$ are
local coordinates for the space transverse to $S^1_i$, 
and $B_e$ is an element of the Cartan subalgebra
of $SU(2)_e$. In order to get supersymmetric observables
one needs to have a corresponding singularity at $S^1_i$ 
for the scalar
fields $\phi_e$, $\bar\phi_e$. For the details of the definition and the
generalization to $\theta_e\neq 0$ we
refer to \cite{GOP}.

It is shown in \cite{Pe,HH,GOP} that these observables 
are left invariant by the supersymmetry $Q$ preserved on 
${\BBE}_{\ep_1,\ep_2}^4$.

\subsubsection{Expectation values on the ellipsoid}

Interesting physical quantities include the 
partition function $\CZ_{\CG_C}$, or more generally expectation values of
supersymmetric loop operators $\CL_\ga$
such as the Wilson- and 't Hooft loops. 
Such quantities are formally defined by the path integral over
all fields on ${\BBE}_{\ep_1,\ep_2}^4$. It was shown in a few
examples for gauge theories from class $\CS$ in \cite{Pe,HH} how to
evaluate this path integral by means of the localization 
technique. A variant of the localization argument was used to 
show that the integral over all fields actually reduces to 
an integral over the locus in field space where the 
scalars $\phi_e$ take constant values $\phi_e=\bar\phi_e\equiv \frac{i}{2}
a_e\si_3={\rm const}$, 
and all other fields vanish. 
This immediately implies that the path integral reduces 
to an ordinary integral over the variables $a_e$. It seems clear
that this argument can be generalized to all theories of class $\CS$.

A more detailed study \cite{Pe,HH} then leads to the conclusion that
the Wilson loop expectation values gave expressions of
the form
\begin{equation}\label{Pestun}
\big\langle \,
W_{e,i}\,\big\rangle_{\BBE_{\ep_1\ep_2}^4}
=\int d\mu(a)\;|\CZ_{\rm inst}(a,m,\tau;\ep_1,\ep_2)|^2\,
2\cosh(2\pi a_e/\ep_i)\,,
\end{equation}
where $i=1,2$. $\CZ_{\rm inst}(a,m,\tau;\ep_1,\ep_2)$ is the so-called
instanton partition function. 
It depends on Coulomb branch moduli $a=(a_1,\dots,a_h)$,
hypermultiplett mass parameters $m=(m_1,\dots,m_n)$, UV gauge coupling 
constants $\tau=(\tau_1,\dots,\tau_h)$, and two 
parameters $\ep_1,\ep_2$. We will briefly summarize some relevant
issues concerning its definition in 
Subsection \ref{sec:inst} below.

A rather nontrivial extension of the method from
\cite{Pe} allows one to treat the case of 't Hooft loops \cite{GOP}
as well, in which case a result of the following form is found
\begin{equation}\label{GOP}
\big\langle \,
T_{e,i}\,\big\rangle_{{\BBE}_{\ep_1,\ep_2}^4}
=\int d\mu(a)\;(\CZ_{\rm inst}(a,m,\tau;\ep_1,\ep_2))^*\,
\CD_{e,i}\cdot\CZ_{\rm inst}(a,m,\tau;\ep_1,\ep_2)\,,
\end{equation}
with $\CD_{e,i}$ being a certain difference operator acting only
on the variable $a_e$, which has coefficients that 
depend only on $a$, $m$ and $\ep_i$, in general.

\subsection{Instanton partition functions - scheme dependence}
\label{sec:inst}

Let us briefly discuss some relevant aspects
concerning the definition 
of $\CZ^{\rm inst}(a,m,\tau;\ep_1,\ep_2)$. This function
is defined in \cite{N} as a partition function 
of a two-parametric deformation $\CG_C^{\ep_1\ep_2}$ 
of $\CG_C$ on $\BR^4$. The theory $\CG_C^{\ep_1\ep_2}$ 
is defined by deforming the Lagrangian of $\CG_C$ by
$(\ep_1,\ep_2)$-dependent terms which break four-dimensional Lorentz
invariance, but preserve one of the supersymmetries of $\CG_C$ on
$\BR^4$. The unbroken
supersymmetry allows one to localize the path integral
defining $\CZ_{\rm inst}(a,m,\tau;\ep_1,\ep_2)$ to a sum over
integrals over the instanton moduli spaces.


Subsequent generalizations to wider classes
of gauge theories from class $\CS$ \cite{AGT,HKS1,HKS2} 
lead to expressions of the following form,
\begin{equation}\label{Zinst:exp}
\CZ^{\rm inst}(a,m,\tau;\ep_1,\ep_2)\,=\,
\CZ_{}^{\rm pert}\,
\sum_{{{\mathbf k}\in(\BZ^{\geq 0})^h}}
\,q_1^{k_1}\cdots q_h^{k_h}\,
\CZ^{\rm inst}_{\mathbf k}(a,m;\ep_1,\ep_2)\,.
\end{equation}
Let us first discuss the 
terms $\CZ^{\rm inst}_{\mathbf k}(a,m;\ep_1,\ep_2)$ 
summed in \rf{Zinst:exp}. These terms can be represented 
as multiple ($h$-fold)
integrals over the moduli spaces $\CM^{\rm inst}_{k,2}$ 
of $SU(2)$-instantons of charge $k$.

\subsubsection{UV issues in the instanton corrections}

It is important to bear in mind that the integrals
defining $\CZ^{\rm inst}_{\mathbf k}(a,m;\ep_1,\ep_2)$ are 
UV divergent due to singularities caused by pointlike, and possibly
colliding instantons, see e.g. \cite{DHKM}. 
Possible IR divergencies are regularized by
the above-mentioned $(\ep_1,\ep_2)$-dependent deformation of the 
Lagrangian \cite{N}.

The explicit formulae for
$\CZ^{\rm inst}_{\mathbf k}(a,m;\ep_1,\ep_2)$ that were used in the
calculations of  expectation values $\big\langle \,
\CL_\ga\,\big\rangle_{{\BBE}_{\ep_1,\ep_2}^4}$ 
performed in \cite{Pe,AGT,GOP,HH} have been obtained using 
particular prescriptions for regularizing the UV-divergencies
which were introduced
in \cite{N,NO} and \cite{NS}, respectively. 
The approach of \cite{N,NO}
uses
a non-commutative deformation of $\CG_C^{\ep_1\ep_2}$ which is known to yield
a smooth resolution of the instanton moduli spaces $\CM^{\rm inst}_{k,2}$
\cite{NS98}.
The second, presented in \cite{NS}, does not work for  
all gauge theories $\CG_C$ from class $\CS$. This 
prescription uses a representation of 
$\CG_C^{\ep_1\ep_2}$ as the limit of a
five-dimensional gauge theory on $\BR^4\times S^1$ 
when the radius of the factor $S^1$ vanishes. It was
shown in \cite{NS} that both prescriptions yield identical results.

These approaches work most straightforwardly for gauge theories 
with gauge group $(U(2))^h$ rather than $(SU(2))^h$. In order to
use the known results for  $(U(2))^h$, the authors of \cite{AGT} 
proposed that the instanton partition functions 
$\CZ^{\rm inst}(a,m,\tau;\ep_1,\ep_2)$ for gauge group $(SU(2))^h$ are
related to their counterparts $\CZ^{\rm inst}_{U(2)}(a,m,\tau;\ep_1,\ep_2)$
defined in theories with gauge group
$(U(2))^h$ by splitting off a ``$U(1)$-factor'',
\begin{equation}\label{U1-factor}
\CZ^{\rm inst}(a,m,\tau;\ep_1,\ep_2)\,=\,
\CZ^{\rm spur}_{U(1)}(m,\tau;\ep_1,\ep_2)\,
\CZ^{\rm inst}_{U(2)}(a,m,\tau;\ep_1,\ep_2)\,.
\end{equation}
Note that the $U(1)$-factor $\CZ^{\rm spur}_{U(1)}(m,\tau;\ep_1,\ep_2)$
does not depend on the Coulomb branch moduli $a$. However, the
precise form of the factor proposed in \cite{AGT} 
was so far mainly motivated by the relations with conformal
field theory discorvered there.

\subsubsection{Non-perturbative scheme dependence ?}\label{NPscheme}

One would expect
that there should be other possibilities for 
regularizing the UV divergencies
in general.
Some examples were explicitly discussed in \cite{HKS1,HKS2}. One may, for
example, use that $Sp(1)\simeq SU(2)$ in order to set up an alternative
scheme for the definition of the instanton partition functions.
It was found
to give an answer $\widetilde{\CZ}^{\rm inst}(a,m,\tau;\ep_1,\ep_2)$ 
that differs from 
$\CZ^{\rm inst}(a,m,\tau;\ep_1,\ep_2)$ by factors that do not depend on 
the Coulomb branch moduli $a$,
\begin{equation}\label{Zinst-altern}
\widetilde{\CZ}^{\rm inst}(a,m,\tilde{\tau};\ep_1,\ep_2)\,=\,
\CZ^{\rm spur}(m,\tau;\ep_1,\ep_2)\,
\CZ^{\rm inst}(a,m,\tau;\ep_1,\ep_2)\,,
\end{equation}
together with a redefinition $\tilde{\tau}=\tilde{\tau}(\tau)$ of the
UV gauge coupling constants.
The possibility to have redefinitions of the UV gauge couplings 
in general is 
suggested by the structure of the 
Uhlenbeck-compactification $\overline{\CM}^{\rm inst}_{k,2}$
of $\CM^{\rm inst}_{k,2}$,
\begin{equation}
\overline{\CM}^{\rm inst}_{k,2}\,=\,
{\CM}^{\rm inst}_{k,2}\cup
\big[{\CM}^{\rm inst}_{k-1,2}\times\BR^4\big]
\cup 
\dots\cup\big[{\rm Sym}^k(\BR^4)\big]\,.
\end{equation}

Interesting for us will in particular be the factors 
$\CZ^{\rm spur}(m,\tau;\ep_1,\ep_2)$ in \rf{Zinst-altern}
which are called spurious 
\cite{HKS1,HKS2}. 
One way to justify this terminology is to note
that such factors will drop out in {\it normalized} expectation
values defined as
\begin{equation}
\big\langle\!\!\big\langle \,
\CL_{\ga}\,\big\rangle\!\!\big\rangle_{{\BBE}_{\ep_1,\ep_2}^4}:=
\big(\big\langle \,
1\,\big\rangle_{{\BBE}_{\ep_1,\ep_2}^4}\big)^{-1}\big\langle \,
\CL_{\ga}\,\big\rangle_{{\BBE}_{\ep_1,\ep_2}^4}\,,
\end{equation}
as follows immediately from the general form of the results for
the expectation values 
quoted in \rf{Pestun} and \rf{GOP}.
The scheme dependence contained in the spurious
factors $\CZ^{\rm spur}(m,\tau;\ep_1,\ep_2)$
should therefore be considered as unphysical.

It would be very interesting to 
understand the issue of the scheme dependence, the
freedom in the choice of UV regularization
used to define ${\CZ}^{\rm inst}(a,m,\tau;\ep_1,\ep_2)$,
more systematically. We will later 
arrive at a precise description of the
freedom left by the approach taken in this paper.

\subsubsection{Perturbative part}\label{pertpart}

The perturbative part $\CZ_{}^{\rm pert}$ in \rf{Zinst:exp} 
factorizes as
$\CZ_{}^{\rm pert}=\CZ_{}^{\rm tree}\,\CZ_{}^{\rm 1-loop}$.

The factor $\CZ^{\rm inst}_{\rm tree}$ represents the tree-level
contribution. It is given by a simple expression proportional
(up to spurious factors) to
\begin{equation}
\CZ_{}^{\rm tree}\,=\,\prod_{e\in\si_1}\,q_e^{a_e^2/\ep_1\ep_2}\,,
\end{equation}
where $\si_1$ is the set of edges of the MS graph $\Ga_{\si}$ 
associated to the pants decomposition $\si$ defining 
the Lagrangian of $\CG_{C}$.

The factor $\CZ^{\rm inst}_{\rm 1-loop}$ is given by certain 
determinants of differential operators. It has the 
following form 
\begin{equation}
\CZ_{}^{\rm 1-loop}\,=\,\prod_{v\in\si_0}\,
\CZ_v^{\rm 1-loop}(a_{e_1(v)},a_{e_2(v)},a_{e_3(v)};\ep_1,\ep_2)\,,
\end{equation}
where $\si_0$ is the set of vertices of the MS graph $\Ga_{\si}$ 
associated to the pants decomposition $\si$, 
and ${e_1(v)},{e_2(v)},{e_3(v)}$ are the edges of $\Ga_{\si}$ that
emanate from $v$. If an edge $e_i(v)$ ends in a boundary component
of $C$, then $a_{e_i(v)}$ will be identified with the mass parameter
associated to that boundary component.

It should be noted that there is a certain freedom 
in the definition of $ \CZ_{v}^{\rm 1-loop}$ due to the regularization 
of divergencies in the infinite products defining $\CZ_v^{\rm 1-loop}$.
This issue has a natural resolution
in the case of partition functions or 
expectation values on ${\BBE}_{\ep_1,\ep_2}^4$ going back to 
\cite{Pe}: what enters into these quantities is 
the absolute value squared
$|\CZ_v^{\rm 1-loop}(a_{e_1(v)},a_{e_2(v)},a_{e_3(v)};\ep_1,\ep_2)|^2$ which 
is unambigously defined \cite{Pe,HH}. There does not seem 
to be a preferred prescription to fix the phase of 
$\CZ_v^{\rm 1-loop}(a_{e_1(v)},a_{e_2(v)},a_{e_3(v)};\ep_1,\ep_2)$,
in general, which can be seen as a part of the perturbative scheme
dependence.

\subsection{Reduction to zero mode quantum mechanics}

We may assign to the expecation values $\langle \CL_{\ga}\rangle$  
an interpretation in terms of
expectation values of operators $\SL_\ga$ which act
on the Hilbert space obtained
by canonical quantization of the gauge theory $\CG_C$ on
the space-time $\BR\times {\BBE}_{\ep_1,\ep_2}^3$, where
${\BBE}_{\ep_1,\ep_2}^3$ is the three-dimensional ellipsoid defined
as
\begin{equation}
{\BBE}_{\ep_1,\ep_2}^3:=\,\{\,(x_1,\dots,x_4)\,|\,
\ep_1^2(x_1^2+x_2^2)+\ep_2^2(x_3^2+x_4^2)=1\,\}\,.
\end{equation}
This is done by interpreting the coordinate $x_0$ for
${\BBE}_{\ep_1,\ep_2}^4$ as Euclidean time. Noting that 
${\BBE}_{\ep_1,\ep_2}^4$ looks near $x_0=0$ as 
$\BR\times {\BBE}_{\ep_1,\ep_2}^3$, we 
expect to be able to represent partition functions 
$\CZ_{\CG_C}({\BBE}_{\ep_1,\ep_2}^4)$ or expectation values
$\big\langle \,
\CL_\ga\,\big\rangle_{\CG_C^{}({\BBE}_{\ep_1,\ep_2}^4)}$
as matrix elements of states in the Hilbert space $\CH_{\CG_C}$ 
defined by
canonical quantization of $\CG_C$ on  
$\BR\times {\BBE}_{\ep_1,\ep_2}^3$.
More precisely
\begin{equation}
\CZ_{\CG_C}({\BBE}_{\ep_1,\ep_2}^4)
\,=\,\langle\,\tau\,|\,\tau\,\rangle\,,\qquad
\big\langle \,
\CL_\ga\,\big\rangle_{{\BBE}_{\ep_1,\ep_2}^4}
\,=\,\langle\,\tau\,|\,\SL_\ga\,|\,\tau\,\rangle\,,
\end{equation}
where $\langle\,\tau\,|$ and $|\,\tau\,\rangle$ are the states 
created by performing the path integral over the 
upper/lower half-ellipsoid
\begin{equation}
{\BBE}_{\ep_1,\ep_2}^{4,\pm}:=\,\{\,(x_0,\dots,x_4)\,|\,x_0^2+
\ep_1^2(x_1^2+x_2^2)+\ep_2^2(x_3^2+x_4^2)=1\;,\;\pm x_0>0\,\}\,,
\end{equation}
respectively, and $\SL_\ga$ is the operator that represents the 
observable $\CL_\ga$ in the Hilbert space 
$\CH_{\CG_C}({\BBE}_{\ep_1,\ep_2}^3)$.

\subsubsection{Localization -- Interpretation 
in the functional Schr\"odinger picture}

The form \rf{Pestun}, \rf{GOP} of the loop operator expectation values
is naturally interpreted in the Hamiltonian framework as follows.
In the functional Schroedinger picture one would represent the 
expecation values 
$\big\langle \,
\CL_\ga\,\big\rangle_{{\BBE}_{\ep_1,\ep_2}^4}$
schematically in the following form
\begin{equation}\label{funSchroe}
\big\langle \,
\CL_\ga\,\big\rangle_{{\BBE}_{\ep_1,\ep_2}^4}
\,=\,\int [\CD\Phi]\;(\Psi[\Phi])^*\,L_{\ga}\Psi[\Phi]\,,
\end{equation}
the integral being extended over all field configuration on the 
three-ellipsoid ${\BBE}_{\ep_1,\ep_2}^3$ at $x_0=0$. The wave-functional
$\Psi[\Phi]$ is defined by means of the path integral 
over the lower half-ellipsoid
${\BBE}_{\ep_1,\ep_2}^{4,-}$ with Dirichlet-type boundary conditions
defined by a field configuration 
$\Phi$ on the boundary ${\BBE}_{\ep_1,\ep_2}^3$ 
of ${\BBE}_{\ep_1,\ep_2}^{4,-}$.  

The fact that the path integral localizes to the locus ${\SL\so\sfc}_C^{}$
defined by
constant values of the scalars,
and zero values for all other fields implies that the path integral
in \rf{funSchroe} can be reduced to an ordinary integral of the 
form
\begin{equation}\label{funSchroe-loc}
\big\langle \,
\CL_\ga\,\big\rangle_{{\BBE}_{\ep_1,\ep_2}^4}
\,=\,\int da\;(\Psi_\tau(a))^*\,L_{\ga}'\Psi_\tau(a)\,,
\end{equation}
with $\CL_{\ga}'$ being the restriction of $\CL_{\ga}$ to $\Loc_C$,
and $\Psi(a)$ defined by means of the path integral 
over the lower half-ellipsoid
${\BBE}_{\ep_1,\ep_2}^{4,-}$ with Dirichlet boundary conditions
$\Phi\in\Loc_C$, $\phi_e=\bar\phi_e=\frac{i}{2}a_e\si_3$. 
The form of the results for expectation values of loop observables
quoted in \rf{Pestun}, \rf{GOP} is thereby naturally explained.

Comparing the results \rf{Pestun} and \rf{GOP} with 
\rf{funSchroe-loc} leads to the conclusion 
that the wave-functions $\Psi_\tau(a)$
appearing in \rf{funSchroe-loc} are represented by the 
instanton partition functions,
\begin{equation}\label{Psi-Zinst}
\Psi_{\tau}(a)\,=\,\CZ^{\rm inst}(a,m,\tau;\ep_1,\ep_2)\,.
\end{equation}
Our goal will be to find an alternative way to characterize
the wave-functions $\Psi_{\tau}(a)$, based on their transformation
properties under electric-magnetic duality.

\subsubsection{Reduction to a subspace of the Hilbert space}

The Dirichlet boundary condition $\Phi\in\Loc_C$, $\phi_e=a_e$ is naturally
interpreted as defining a Hilbert subspace $\CH_0$ within $\CH_{\CG_C}$.
States in $\CH_0$ can, by definition, be represented by wave-functions
$\Psi(a)$, $a=(a_1,\dots,a_h)$.
 
Note that field configurations that satisfy the 
boundary condition $\Phi\in\Loc_C$ 
are annhilated by the supercharge $Q$ 
used in the localization calculations of \cite{Pe,GOP,HH} --
that's just what defined the locus $\Loc_C$ in the first
place. This indicates that the Hilbert subspace $\CH_0$ represents
the cohomology of $Q$ within $\CH_{\CG_C}$.

The algebra of observables acting on $\CH_0$
should contain the supersymmetric Wilson- and 't Hooft loop observables.
The Wilson loops $W_{e,\1}$ 
and $W_{e,\2}$ act diagonally as
operators of multiplication by
$2\cosh(2\pi a_e/\ep_\1)$ and $2\cosh(2\pi a_e/\ep_\2)$, respectively.
The 't Hooft loops will act as certain difference operators.

Let us denote the non-commutative algebra of operators 
generated by polynomial functions of the loop operators
$\SW_{e,i}$ and $\ST_{e,i}$ by $\CA_{\ep_i}$,
where $i=1,2$. We will  denote the
algebra generated by all such supersymmetric loop operators by
$\CA_{\ep_\1\ep_\2}\equiv \CA_{\ep_\1}\times\CA_{\ep_\2}$.

\section{Riemann-Hilbert problem for instanton partition functions}\label{sec:RHgauge}

\setcounter{equation}{0}

The main result of this paper may be summarized in the statement that, 
up to spurious factors, the wave-functions $\Psi_\tau(a)$
in the quantum mechanics of the zero
modes of $\CG_C$ coincide with the Liouville conformal blocks
$\CZ^{\rm Liou}(\be,\al,q;b)$,
\begin{equation}\label{TV}
\Psi_\tau(a)\,\simeq\,
\CZ^{\rm Liou}(\be,\al,q;b)\,.
\end{equation}
The definition of $\CZ^{\rm Liou}(\be,\al,q;b)$ 
will be reviewed and generalized in
Part III below, where we will also spell out the
dictionary between the variables involved.  Combined 
with \rf{Psi-Zinst}, we arrive at the 
relation $\CZ^{\rm inst}(a,m,\tau,\ep_\1,\ep_\2)\simeq
\CZ^{\rm Liou}(\be,\al,q;b)$ proposed in \cite{AGT}.

In this paper we will characterize 
the wave-functions $\Psi_{\tau}(a)$ using the relation between the 
algebra $\CA_{\ep_1\ep_2}$ of supersymmetric loop observables
to the quantized algebras of functions on moduli spaces of 
flat connections. These quantized algebras of functions are
deeply related to Liouville theory, as will be explained in 
Part III of this paper.
Taking into account these
relations will lead to the  relation  \rf{TV}
of  $\Psi_{\tau}(a)$ with the Liouville conformal blocks.

Before we continue to discuss our approach to the 
relation \rf{TV} let us briefly 
review some of the 
known evidence for \rf{TV}, mainly coming from its relation with the
observations of \cite{AGT}.

\subsection{Available evidence}

\label{GTvsQM}

The authors of \cite{AGT} observed in some examples of theories 
from class $\CS$ that one has (up to spurious factors)
an equality of instanton partition functions to the 
conformal blocks $\CZ^{\rm Liou}(\be,\al,\tau;b)$ of
Liouville theory,
\begin{equation}\label{AGT}
\CZ^{\rm inst}(a,m,\tau;\ep_1,\ep_2)\,\simeq\,\CZ^{\rm Liou}(\be,\al,\tau;b)\,,
\end{equation}
assuming a suitable dictionary between the variable involved. 
The results of \cite{AGT} can be generalized to
a subset of the family of theories from class $\CS$ 
called the linear quiver
theories corresponding to surfaces $C$ 
of genus $0$ or $1$ \cite{AFLT}.

For surfaces $C$ of genus 0
we know, on the other hand,  that the Liouville conformal
blocks coincide with certain wave-functions in the quantum theory 
of the Teichm\"uller spaces $\CT(C)$ 
of Riemann surfaces (\cite{T03}, see also Part III of this paper), 
\begin{equation}\label{Psi-ZL}
\CZ^{\rm Liou}(\be,\al,\tau;b)\,=\,\Psi_{\tau}^{\CT}(a)\,
\equiv\langle\,a\,|\,\tau\,\rangle^{}_{\CT(C)}\,.
\end{equation}
The state $\langle\,a\,|$
is an eigenstate of a maximal family of  
commuting geodesic length
operators, while 
$|\,\tau\,\rangle_{\CT(C)}^{}$ is 
defined as an eigenstate of the operators 
obtained in the quantization of certain
complex-analytic coordinates on $\CT(C)$.
The definition of $\Psi_{\tau}^{\CT}(a)$ and the derivation 
of \rf{Psi-ZL}  
will be reviewed and generalized to surfaces $C$ of higher genus
in Part III of our paper.

Combining the observations \rf{Psi-Zinst}, \rf{AGT}
and \rf{Psi-ZL} suggests that 
the quantum mechanics of the zero modes of $\CG_C$ is equivalent 
to the quantum theory 
of the Teichm\"uller spaces, and that we have in particular
\begin{equation}\label{Psi-Psi}
\Psi_{\tau}(a)\,\simeq \,\Psi_{\tau}^{\CT}(a)\,.
\end{equation}

This conclusion was anticipated in \cite{DGOT}, where it was noted
that the existing results on Wilson loop observables
can be rewritten in the 
form
\begin{equation}\label{q-Int0}
\big\langle \,
\CL_\ga\,\big\rangle_{{\BBE}_{\ep_1,\ep_2}^4}
\,=\,\langle\,\tau\,|\,\SL_\ga\,|\,\tau\,\rangle_{\CT(C)}^{}\,,
\end{equation}
using the observations \rf{AGT} and \rf{Psi-ZL} quoted above.
The gauge theoretical calculations leading to \rf{q-Int0} were later
generalized to the case of 't Hooft loops in \cite{GOP}. These results
confirmed the earlier 
proposals made in \cite{AGGTV,DGOT} that the supersymmetric
loop operators in gauge theories $\CG_C$ 
are related to the analogs of 
the Verlinde loop operators in Liouville theory. The 
Verlinde loop operators are further mapped to the geodesic
length operators by the correspondence between Liouville theory
and the quantum Teichm\"uller theory
\cite{T03,DGOT}.

 One should keep in mind that 
the Teichm\"uller spaces $\CT(C)$ are naturally isomorphic to   
the connected components $\CM_{\rm flat}^0(C)$ 
of $\CM_{\rm flat}(C)$.  
Combining all these observations we may 
conclude that for 
surfaces $C$ of genus $0$ 
the expectation values of supersymmetric
loop operators in $\CG_C$ 
can be represented as expectation values of certain
operators in the quantum mechanics obtained by quantizing 
$\CM_{\rm flat}^0(C)$.

Our goal is to understand more directly
why this is so, and to generalize this result to all
theories from class $\CS$.

\subsection{Assumptions}

Our approach for deriving \rf{TV} 
is based on physically motivated assumptions. We will first
formulate the underlying assumptions concisely, and later 
dicuss the underlying motivations. 
\begin{itemize}
\item[(a)] $\Psi_\tau(a)$ can be analytically 
continued with respect to the variables $\tau$ 
to define a multi-valued analytic function on the
coupling constant space $\CM(\CG_C)$.
The boundaries of  $\CM(\CG_C)$, labelled by
pants decompositions $\si$ correspond to weakly-coupled 
Lagrangian descriptions for $\CG_C$.
\item[(b)] The transitions between any two different  weakly-coupled 
Lagrangian descriptions for $\CG_C$ are generated from the
elementary electric-magnetic duality transformations
of the $N_f=4$ and the $\CN=2^*$-theories.
The electric-magnetic 
duality transformations exchange the respective Wilson- and 
't Hooft loop observables.
\item[(c)] The algebra $\CA_{\ep_1\ep_2}$ generated by the supersymmetric
loop observables is isomorphic to the algebra
${\rm Fun}_{\ep_\1}(\CM_{\rm flat}(C))\times 
{\rm Fun}_{\ep_\2}(\CM_{\rm flat}(C))$, where
${\rm Fun}_{\ep}(\CM_{\rm flat}(C))$ is the 
quantized algebra of 
functions on $\CM_{\rm flat}^0(C)\simeq\CT(C)$.
\end{itemize}
Assumptions (a) and (b) can be motivated  
by noting that the theories of class $\CS$ 
are all quiver gauge theories. This combinatorial structure
reduces the S-duality transformations
to those of the building blocks, the $N_f=4$ and the $\CN=2^*$-theories 
\cite{G09}.
The realization of electric-magnetic duality  in these theories has been 
discussed extensively in the literature, going back to the 
works of Seiberg and Witten \cite{SW1,SW2}.

Of particular importance for us is assumption (c). 
Let us first note that this assumption is 
strongly supported by the explicit calculation
of the 't Hooft loop operator expectation values in the $\CN=2^*$-theory
carried out in \cite{GOP}. One finds a precise correspondence between the 
difference operator $\CD_{e,i}$ in \rf{GOP} 
and operator $\SL_t$ representing the trace coordinate $L_t$
in the quantum theory of $\CM_{\rm flat}(C)$ (see 
equation \rf{quantum't Hooft-1,1} below).

It should be possible to verify assumption (c) directly 
by studying the algebra of Wilson-'t Hooft loop operators in 
the theories $\CG_C$ in more generality.
It was proposed in \cite{IOT} that in order
to study the {\it algebra} of supersymmetric loop operators
one may replace the background space-time ${\BBE}_{\ep_1,\ep_2}^4$
by the local model $S^1\times \BR^3$
for the vicinity of the loop operators, taking into 
account the relevant effects of the curvature by a simple twist 
in the boundary conditions. 
This has been used in \cite{IOT} to calculate 
expectation values of supersymmetric loop operators
in several cases. The results give additional support
for the validity of assumption (c). Further development of this
approach may well lead to a derivation of (c) purely within 
four-dimensional gauge theory.

As also pointed out in \cite{IOT}, the  
twisted boundary conditions on $S^1\times \BR^3$
used in this paper are essentially equivalent to the deformation  of
$\CG_C$ studied in \cite{GMN3}.  
Specializing the results of \cite{GMN3} to the $A_1$ theories
of class $\CS$ considered here, one gets a  
non-commutative algebra of observables 
with generators $\BL_{\ga}$ which
can be represented in the form
\begin{equation}\label{GMN-3}
\BL_{\ga}\,=\,\sum_{\eta}\overline{\underline{\Omega}}(\ga,\eta;y)X_\eta\,,
\end{equation}
where $X_\eta$ are generators of the non-commutative algebra
obtained by canonical quantization of the Darboux-coordinates 
studied in \cite{GMN1,GMN}, and the coefficients 
$\overline{\underline{\Omega}}(\ga,\eta;y)$ are indices for certain
BPS states extensively studied in \cite{GMN3}.
It is pointed out in this paper, 
on the one hand, that there is 
a simple physical reason for getting a non-commutative deformation
of the algebra of the Darboux-coordinates generated by the $X_\eta$.
On the other hand it is argued in \cite{GMN3} 
that \rf{GMN-3} coincides with the decomposition 
of geodesic length operators into the (quantized) coordinates
for the Teichm\"uller spaces
introduced by Fock \cite{F97}. It follows that the 
algebra generated by the $\BL_{\ga}$ is isomorphic to the algebra
of geodesic length operators in the quantum Teichm\"uller theory.
This is exactly the algebra ${\rm Fun}_{\ep}(\CM_{\rm flat}(C))$
studied in this paper. We believe that this line of thoughts 
can lead to an insightful derivation of our assumption
(c), but it seems desirable to have a more detailed discussion of 
the applicability of the results of \cite{GMN3} to
our set-up.

Yet another approach towards understanding assumption (c)
starts from a modified set-up in which the gauge theory 
$\CG_C$ is replaced by its Omega-deformed version $\CG_C^{\ep_1\ep_2}$ 
\cite{N,NW}.
In the Omega-deformed theory one may define analogs of the loop
observables $L_\ga$ and wave-functions 
$\Psi_{\tau}^{\rm\sst top}(a)$ in a very similar way as above,
and one has $\Psi_{\tau}^{\rm\sst top}(a)=
\CZ^{\rm inst}(a,m,\tau;\ep_1,\ep_2)$.
Combined with the observation \rf{Psi-Zinst} made above we see that 
\begin{equation}\label{Psi-Psitop}
\Psi_{\tau}(a)\,=\,\CZ^{\rm inst}(a,m,\tau;\ep_1,\ep_2)\,=\,
\Psi_{\tau}^{\rm\sst top}(a)\,.
\end{equation}
This strongly indicates that we may use the
results on the Omega-deformed theory 
$\CG_C^{\ep_1\ep_2}$ from \cite{NW} for the study of 
the gauge theory on ${\BBE}_{\ep_1,\ep_2}^4$. 
In the following Section \ref{NW:sec} we will briefly review the 
argument for (c) in the Omega-deformed theory 
$\CG^{\ep_1\ep_2}_C$ which was given
by Nekrasov and Witten in \cite{NW}.

\subsection{The Riemann-Hilbert problem}\label{RH-def}

The strategy for deriving \rf{TV} may now be outlined as follows.

Assumption (b) implies that the S-duality transformations 
induce a change of representation for the Hilbert space $\CH_{\rm top}$.
Recall that $\Psi_\tau^{\si_1}(a)$ is defined to be 
a joint eigenfunction of the
Wilson loop operators constructed using  
the weakly coupled Lagrangian description associated
to a pants decomposition $\si_1$. 
Considering another pants decomposition $\si_2$ one defines in a similar
manner eigenfunctions  $\Psi_\tau^{\si_2}(a)$ of another family
of operators which are not commuting with the Wilson loop
operators defined from  pants decomposition $\si_1$,
but can be constructed as Wilson loop observables 
using the fields used in the Lagrangian description of 
$\CG_C$ associated to pants decompostion $\si_2$. The 
eigenfunctions $\Psi_\tau^{\si_1}(a)$ and $\Psi_\tau^{\si_2}(a)$
must therefore be 
related by an integral transformations of the 
form
\begin{equation}\label{S-duality2}
\Psi_\tau^{\si_\2}(a_2)\,=\,f_{\si_\2\si_1}(\tau)
\int da_1\;K_{\si_\2\si_\1}(a_2,a_1)\,\Psi_\tau^{\si_1}(a_1)\,.
\end{equation} 
We allow for a spurious prefactor $f_{\si_\2\si_1}(\tau)$ in the sense
explained in Subsection \ref{sec:inst}, as it will 
turn out that we can not eliminate such prefectors by choosing 
an appropriate scheme in general.

Given that we know the data $K_{\si_\2\si_\1}(a_2,a_1)$ and
$f_{\si_\2\si_1}(\tau)$, the 
assumptions (a) - (c)  completely describe of the
analytic properties of $\CZ^{\rm inst}(a,m,\tau,\ep_\1,\ep_\2)$ 
as function on the
coupling constant space $\CM(\CG_C)$. This means  that 
$\CZ^{\rm inst}(a,m,\tau,\ep_\1,\ep_\2)$ can be characterized as the 
solution to a Riemann-Hilbert type problem.
  
A detailed construction of the representation of  $\CA_{\ep_1\ep_2}$ 
on $\CH_0$ will be given in Part II of this paper. 
The main result for our purposes is to show that 
the kernels $K_{\si_\2\si_\1}(a_2,a_1)$ appearing in 
\rf{S-duality2} can be characterized by 
the requirement that this transformation correctly 
exchanges the Wilson-  and 't Hooft loops defined
in the two Lagrangian descriptions associated to $\si_1$ and $\si_2$,
respectively. The technically hardest part 
is to ensure that the 
Moore-Seiberg groupoid of transformations from one 
Lagrangian description 
to another is correctly realized by the 
transformations \rf{S-duality2}.

In Part III we will then show that
this Riemann-Hilbert type problem 
has a solution that is unique up to 
spurious factors as encountered in 
\rf{Zinst-altern}, and given by the Liouville conformal 
blocks appearing on the right
hand side of \rf{TV}. A precise mathematical 
charcterization of the possible spurious factors is 
obtained.

It may be instructive to compare this type of reasoning to the
derivations of exact results for 
prepotentials in supersymmetric gauge theories
pioneered by Seiberg and Witten. The key assumptions made 
in these derivations were the analyticity of the prepotential,
and assumptions on the physical interpretation of its singularities.
Well-motivated assumptions on effective descriptions 
near the singularities of the prepotential $\CF$ lead Seiberg and Witten to 
a characterization of this quantity in terms of 
a Riemann-Hilbert problem. A key assumption was that the
transition between any two singularities of the prepotential
corresponds to electric-magnetic duality.

\section{The approach of Nekrasov and Witten}\label{NW:sec}

\setcounter{equation}{0}

An approach towards understanding the link between the gauge theories
$\CG_C$ and Liouville theory expressed in formula \rf{AGT} 
was proposed in the work \cite{NW} of Nekrasov and Witten. 
This work considers the gauge theory $\CG_C$ 
on four-manifolds $M^{4}$ that have $(U(1))^2$-isometries 
and therefore allow to define the Omega-deformation 
$\CG_C^{\ep_1\ep_2}$ of $\CG_C$. The result may imprecisely be
summarized by saying that the topological sector 
of $\CG_C^{\ep_1\ep_2}$ is represented by the 
quantum mechanics obtained from the quantization of 
$\CM_{\rm flat}(C)$.
The arguments presented in \cite{NW} do not quite suffice to
derive the AGT-correspondence in the strong form \rf{AGT}.

We will argue that one may take the 
arguments of \cite{NW} as a starting point to
reach the more precise
result \rf{AGT}: Certain 
wave-functions in the topologically twisted version 
of $\CG_C^{\ep_1\ep_2}$ considered by \cite{NW} coincide with the 
conformal blocks of
Liouville theory. As the wave-functions in question 
also coincide with the instanton partition functions
(almost by definition), we will thereby get a derivation 
of the AGT-correspondence which is somewhat in the spirit 
of the characterization of the prepotentials that were
pioneered by Seiberg and Witten.

\subsection{The basic ideas}

The approach of Nekrasov and Witten is based on three main ideas:
\begin{itemize}
\item[(i)] The instanton partition functions are defined in  \cite{N} 
as partition functions of $\CG_C^{\ep_1\ep_2}$ on $\BR^{4}$.
The deformation of $\CG_C$ into $\CG_C^{\ep_1\ep_2}$ 
preserves a supersymmetry
which can be used to define a 
topologically twisted version $\CG_C^{\rm top}$ 
of $\CG_C^{\ep_1\ep_2}$. The partition
function of $\CG_C^{\ep_1\ep_2}$ on $\BR^4$ coincides with the partition 
function of $\CG_C^{\ep_1\ep_2}$ on any four-manifold $B^4$ 
with the same topology as $\BR^4$ that has the 
$(S^1)^2$-isometries needed to define the Omega-deformation 
$\CG_C^{\ep_1\ep_2}$ of $\CG_C$ \cite{NW}.

\item[(ii)] The four-manifold $B^4$ may be assumed to 
have a boundary 
$M^3$, and the metric near the boundary may be assumed to 
be the metric on $R\times M^3$. 
Canonical quantization on $\BR\times {M}^{3}$
yields a quantum theory with Hilbert space 
$\CH_{M^3}(\CG_C^{\ep_1\ep_2})$. 
The partition function on 
$B^4$ 
can then be interpreted as a wave-function of the state created
by performing the path integral over $B^4$.

\item[(iii)]
Viewing $S^3$ as a fibration of $(S^1)^2$ over an interval $I$,
one may represent $\CG_C^{\rm top}$ on $R\times S^3$ 
in terms of a topologically twisted 
two-dimensional non-linear sigma model on the world-sheet 
$R\times I$ with target space $\CM_H$, the Hitchin moduli space.
This means that the instanton partition function gets 
re-interpreted as a wave-function of a certain state in the
two-dimensional sigma model on the strip. 
\end{itemize}

Let us consider the
topologically twisted theory $\CG_C^{\rm top}$ on 
$\BR\times {M}^{3}$. 
The topological twist preserves two super-charges
$\SQ$ and $\SQ^{\dagger}$. Choosing $\SQ$ to be the preferred super-charge,
one may identify the Hilbert-space 
$\CH_{\rm top}\equiv \CH_{M^3}^{\rm top}(\CG_C^{\ep_1\ep_2})$ 
of $\CG_C^{\rm top}$ with the $\SQ$-cohomology 
within $\CH_{M^3}(\CG_C^{\ep_1\ep_2})$. 

A few points are clear. 
The Hilbert space $\CH_{\rm top}$ is acted on by the
chiral ring operactors 
\begin{equation}
\su_e:=\,{\rm Tr}(\phi^2_e)\,.
\end{equation}
These operators generate a commutative ring of operators 
acting on $\CH_{\rm top}$. 
It is furthermore argued in \cite[Section 4.9.1]{NW} that 
analogs of the Wilson- and 't Hooft
loop operators can be 
be defined within the gauge theory $\CG_C^{\ep_1\ep_2}$ 
on $\BR\times M^3$  
which commute with $\SQ$,
and therefore define Wilson- and 't Hooft
loop operators $\SW_{e,i}$ and  $\ST_{e,i}$ acting on
$\CH_{\rm top}$. We will  denote the
algebra generated by all such supersymmetric loop operators by
$\CA_{\ep_\1\ep_\2}^{\rm\sst top}\equiv 
\CA_{\ep_\1}^{\rm\sst top}\times\CA_{\ep_\2}^{\rm\sst top}$.


And indeed,
one of the main results of \cite{NW} were the
isomorphisms
\begin{equation}\label{NW-A}
\CA_{\ep_\1}^{\rm\sst top}\times\CA_{\ep_\2}^{\rm\sst top}\simeq
{\rm Fun}_{\ep_\1}(\CM_{\rm flat}(C))\times {\rm Fun}_{\ep_\2}(\CM_{\rm flat}(C))\,,
\end{equation}
where ${\rm Fun}_{q}(\CM_{\rm flat}(C))$ 
is the quantized algebra
of functions on $\CM_{\rm flat}(C)$ that will be defined 
precisely in Part II, together with
\begin{equation}\label{NW}
\CH_{\CG_C}^{\rm top}({M}^{3}_{\ep_1,\ep_2})\,\simeq\,
\CH(\CM_{\rm flat}^{0}(C))\,,
\end{equation}
both sides being understood as
module of ${\rm Fun}_{\ep_\1}(\CM_{\rm flat}(C))\times 
{\rm Fun}_{\ep_\2}(\CM_{\rm flat}(C))$.

\subsection{The effective sigma model description}

It may be instructive to
briefly outline the approach that lead to the
results \rf{NW-A} and \rf{NW}, 
see \cite{NW} for more details.

In order to get a useful effective representation for $\CG_C^{\ep_1\ep_2}$, 
let us note
that we may view three manifolds ${M}^3$ with 
the necessary $(U(1))^2$-isometries  
as a circle fibration $S^1\times S^1\ra I$,
where the base $I$ is an interval.
It was argued in \cite{NW} that the low energy physics of $\CG_C$ can be
represented by a $(4,4)$-supersymmetric sigma model with world-sheet
$\BR\times I$ and
target space being the Hitchin moduli space
$\CM_{\rm H}(C)$. This sigma model can be thought of as being
obtained by compactifying $\CG_C^{\ep_1\ep_2}$ on $S^1\times S^1$.
Due to topological invariance one expects that
supersymmetric observables of $\CG_C^{\ep_1\ep_2}$ get 
represented within the quantum theory of the sigma model. 

An elegant argument for why the sigma model has 
target space $\CM_{\rm H}(C)$ 
can be based on the description of $\CG_C$ as compactification 
of the six-dimensional $(0,2)$-superconformal 
theories of the $A_1$-type on spaces of the form 
$M^4\times C$. If $M^4$ has the structure of a circle fibration,
one expects that the result of compactifying first on 
$C$, then on the circle fibers should be equivalent to 
the result of first compactifying on the circle fibers,
and then on $C$, as far as the resulting topological 
subsector is concerned. If one 
compactifies the six-dimensional $(0,2)$-superconformal
theory on a circle $S^1$, or on $S^1\times S^1$,
the result is a maximally supersymmetric
Yang-Mills theory with gauge group $SU(2)$ on a five-, or four-dimensional 
space-time, respectively.  Minimal energy configurations in the
resulting theories on space-times of the 
form $M\times C$ are represented by solutions of 
Hitchin's equations on $C$ \cite{BJSV}, see also \cite[Subsection 3.1.6]{GMN}.
It follows that the low-energy physics can be effectively 
represented by a sigma model on $M$ which has $\CM_H(C)$
as a target space. This argument has been used in \cite{NW}, see 
also \cite{NRS} for a similar discussion.

The effect of the $\Omega$-deformation is represented within the
sigma-model description by boundary conditions $\CB_{\ep_\1}$ and 
$\CB_{\ep_\2}$ imposed on the sigma model at
the two ends of the interval $I$. It is shown that the 
boundary conditions are represented by the 
so-called canonical co-isotropic branes, see \cite{NW} for 
the definition and further references. The Hilbert space 
$\CH_{M^3}^{\rm top}(\CG_C^{\ep_1\ep_2})$  
thereby gets identified with the space of states 
${\rm Hom}(\CB_{\ep_\1},\CB_{\ep_\2})$ 
of this open two-dimensional
sigma model. 

It was furthermore argued in \cite{NW} that the action 
of the algebra $\CA_{\ep_\1,\ep_\2}^{\rm top}$ of
supersymmetric loop operators on 
$\CH_{M^3}^{\rm top}(\CG^{\ep_\1,\ep_\2}_C)$
gets represented in the sigma model
as the action of the quantized algebra 
of functions on the canonical coisotropic branes via the joining 
of open strings, which defines a natural
left (resp. right) action of $\CA_{\ep_1}(C)\simeq
{\rm Hom}(\CB_{\ep_\1},\CB_{\ep_\1})$
(resp. $\CA_{\ep_2}(C)\simeq {\rm Hom}(\CB_{\ep_\2},\CB_{\ep_\2})$)
on ${\rm Hom}(\CB_{\ep_\1},\CB_{\ep_\2})$. The key result 
obtained in \cite{NW} is then that the algebras 
${\rm Hom}(\CB_{\ep_i},\CB_{\ep_i})$, $i=1,2$, with multiplication
naturally defined by the joining of
strips, are isomorphic to the quantized algebras
of functions ${\rm Fun}_{\ep_i}(\CM_{\rm flat}(C))$ 
on $\CM_{\rm flat}(C)$. The method by which
this conclusion is obtained can be seen as special case 
of a more general framework for producing 
quantizations of algebras of functions on
hyper-K\"ahler manifolds from the canonical coisotropic 
branes of the sigma models on such manifolds \cite{GV}.

\subsection{Instanton partition functions as wave-functions}

Let us extract from  \cite{NW} some implications that will be relevant for us.

Recall that the algebra $\CA_{\ep_\1,\ep_\2}$ is generated by the quantized
counterparts of Wilson- and 't Hooft loop operators. 
It is easy to see from the definitions that
the Wilson loop operators $\SW_{e,i}$ are positive self-adjoint,
and mutually commutative
$[\SW_{e,i},\SW_{e',i}]=0$ for $i=1,2$.
It follows that
there exists a representation 
for $\CH_{\rm top}$
in which the states 
are realized by wave-functions 
$\Psi(a)\equiv \langle a|\Psi\rangle_{\rm top}$, where $a=(a_1,\dots,a_{h})$.

As S-duality exchanges Wilson- and 't Hooft loops, the 't Hooft loops
must also be positive self-adjoint. What is relevant for us is therefore
the subspace of the space of functions on $\CM_{\rm flat}(C)$ characterized
by the positivity of all loop observables. This subspace is 
isomorphic to the space of functions on the Teichm\"uller space
$\CT(C)$, and will be denoted $\CM_{\rm flat}^0(C)$.

Considering the gauge theory $\CG_C^{\ep_1\ep_2}$ on $\BR\times M^3$  
one may naturally consider a state 
$|\,\tau\,\rangle\in\CH_{\CG_C}({M}^{3})$ 
created by performing the path integral over the a Euclidean
four-manifold 
${B}^{4,-}$ with boundary $M^{3}$, 
and its projection 
$|\,\tau\,\rangle_{\rm top}^{}$ to 
$\CH_{\rm top}$. 
We may represent $|\,\tau\,\rangle_{\rm top}^{}$
by its wave-function 
\begin{equation}
\Psi^{\rm \sst top}_\tau(a):=\,\langle\,a\,|\,\tau\,\rangle_{\rm top}^{}\,.
\end{equation}
Note that 
the overlap between an eigenstate $\langle\,a\,|$ 
of all the Wilson loop operators with the state $|\,\tau\,\rangle_{\rm top}^{}$
should be related to the instanton partition function by means of the
metric-independence of the path integrals for $\CG_{C}^{\rm top}$.
This should relate  
$\langle\,a\,|\,\tau\,\rangle_{\rm top}^{}$, given by the path integral
for  $\CG_{C}^{\ep_1\ep_2}$ on ${B}^{4,-}$
to  $\CZ^{\rm inst}(a,m,\tau,\ep_\1,\ep_\2)$ which is defined by
a path integral on $\BR^{4}_{\ep_\1,\ep_\2}$,
\begin{equation}\label{Psi=Z-inst}
\Psi^{\rm \sst top}_\tau(a)\,=\,\CZ^{\rm inst}(a,m,\tau,\ep_\1,\ep_\2)\,.
\end{equation}
The projection 
onto an eigenstate $\langle\,a\,|$ of the Wilson loop operators
is traded for the boundary condition to have fixed scalar
expectation values at the infinity of $\BR^{4}$.

We conclude that the instanton partition functions
$\CZ^{\rm inst}(a,m,\tau,\ep_1,\ep_2)$
represent particular
wave-functions within the quantum theory of $\CM_{\rm flat}^0(C)$.
The isomorphisms \rf{NW-A} and \rf{NW} established in \cite{NW} 
can be taken as the basis for a characterization of the wave-functions
$\Psi^{\rm \sst top}_\tau(a)$ in terms of a Riemann-Hilbert type problem
which will coincide with the one discussed in our previous section 
\rf{sec:RHgauge}. This leads to yet another way to 
find the relation $\Psi_{\tau}(a)=
\Psi_{\tau}^{\rm\sst top}(a)$ that we had pointed out above in 
\rf{Psi-Psitop}. This relation can be understood in a more
phyical way by combining the following 
two observations: On the one hand one may 
note that both in the case of $\CG_C$ on 
${\BBE}_{\ep_1,\ep_2}^{4,-}$, and in the  case of the Omega-deformed
theory $\CG_C^{\ep_1\ep_2}$ on $\BR^4$ the instanton corrections
get localized to the fixed points of the relevant $U(1)\times U(1)$
actions. The two cases are then linked by the key 
observation from \cite{Pe} that the residual 
effect of the curvature of ${\BBE}_{\ep_1,\ep_2}^4$ in the vicinity
of the poles can be modeled by the 
Omega-deformation of \cite{N}.

\newpage
\part{\Large Quantization of $\CM_{\rm flat}^0$ }\label{par:quant}

We are now going to describe the quantum theory of 
$ \CM_{\rm flat}^0(C)\simeq\CT(C)$ in a way that is suitable for
the gauge theoretical applications. This will in particular
lead to 
a precise description of the kernels $K_{\si_\2\si_\1}(a_2,a_1)$
that define the Riemann-Hilbert problem for the 
instanton partition functions.

In Section \ref{q-Mflat} we will explain how the use of pants
decompositions reduces the task to the specification of
a finite set of data. In order to characterize the 
relevant representations of 
the algebra ${\rm Fun}_b(\CM_{\rm flat}(C)$ it suffices to define the 
counterparts fo the Wilson- and 't Hooft loop operators, and
to describe the relations in ${\rm Fun}_b(\CM_{\rm flat}(C)$. 
Transitions between pants decompositions (corresponding to the
S-duality transformations)
can be composed from elementary moves associated to surfaces
of type $C_{0,3}$, $C_{0,4}$ and $C_{1,1}$.
This section summarizes our main results
by listing the explicit formulae for the defining data. 

The rest of Part II of this paper 
(Sections \ref{q-Teich} and \ref{Proofs2})
explains how 
the results summarized in Section \ref{q-Mflat} can be
derived. Our starting point is the quantization of the 
Teichm\"uller spaces constructed in \cite{F97,Ka1,CF,CF2}
which is briefly reviewed in the beginning of Section \ref{q-Teich}. 
The main technical problem is to construct the geodesic length operators,
and to diagonalize 
a maximal commuting set of geodesic length operators
which in our context correspond to the set of Wilson loop operators 
\cite{T05}.
The relevant results from \cite{T05}
are summarized in 
Section \ref{q-Teich}.

Section \ref{Proofs2} describes what remains to be done to 
complete the derivation of the results listed in Section \ref{q-Mflat}. 
An important step, the
explicit calculation of the generators associated to surfaces of genus $0$, 
has recently been taken in \cite{NT}.
A result of particular importance for us is the explicit
calculation of the central extension of the 
representation of the Moore-Seiberg groupoid that is 
canonically associated to the quantum theory of 
$ \CM_{\rm flat}^0(C)\simeq\CT(C)$.

Another approach to the quantization of moduli spaces of flat connections
for noncompact groups
is described in particular in \cite{Gu}, and the case of one-holed
tori was previously discussed in \cite{DG}.

\section{Construction of the quantization
of $\CM_{\rm flat}^0(C)$}\label{q-Mflat}
\renewcommand{\CL}{L}

\setcounter{equation}{0}

An important feature of the description of $\CM_{\rm flat}$ summarised in
Section \ref{sec:2hyp} is the fact that it exhibits a form of 
locality in the sense that the description  
can be reconstructed from the
local pieces isomorphic to $C_{0,4}$ or $C_{1,1}$ 
appearing in pants decompositions. In the relation
with gauge theory one may view this locality as a result of 
the strucure of the MS graph $\Ga_\si$ associated to a pants decompostion:
The Lagrangian includes only couplings between neighboring 
parts of the MS graph.
We are now going to descibe in more
detail how this locality is reflected in the quantum 
theory, and introduce the main data that characterize the quantum 
theory in such a description.

\subsection{Algebra}\label{algebra}

For the case under consideration, the aim is to construct a 
one-parameter family
of non-commutative deformations $\CA_{b}(C)\equiv
{\rm Fun}_b^{\rm alg}(\CM_{\rm flat}(C))$ 
of the Poisson-algebra
of algebraic functions on $\CM_{\rm flat}(C)$. 

For a chosen pants decompostion defined by a 
cut system $\CC$ we will choose as set of generators 
$\{(\CL_s^e,\CL_t^e,\CL_u^e);\ga\in\CC\}\cup
\{\CL_r;r=1,\dots,n\}$. 
The generators $\CL_s^e$, $\CL_t^e$, and 
$\CL_u^e$ are 
associated to
the simple closed curves $\ga_s^e$, $\ga_t^e$, and $\ga_u^e$ 
introduced in Subsection \ref{sec:Genrel}, respectively. 
The generators $\CL_r$ $r=1,\dots,n$ are associated to the 
$n$ boundary components of $C\simeq C_{g,n}$. They will be 
central elements in $\CA_b(C)$.

For each subsurface $C_e\subset C$ associated to a curve $\ga_e$
in the cut system $\CC$ there will be
two types of relations: Quadratic relations of the general form
\begin{equation}\label{q-rels}
\CQ_e(\CL_s^e, \CL_t^e, \CL_u^e) = 0\,,
\end{equation}
and cubic relations
\begin{equation}\label{c-rels}
\CP_e(\CL_s^e, \CL_t^e, \CL_u^e) = 0\,.
\end{equation}
We have not indicated in the notations that the
polynomials $\CQ_e$ and $\CP_e$ may depend
also on the loop variables associated to the boundary components
of $C_e$ in a way that is similar to the classical case described
in Subsection \ref{sec:Genrel}. 
In order to describe the relations it therefore suffices
to specify the polynomials $\CQ_e$ and $\CP_e$
for the two cases $C_e\simeq C_{0,4}$ and $C_e\simeq C_{1,1}$.

\subsubsection{Case $C_e\simeq C_{0,4}\;$:}

Quadratic relation:
\begin{align} \label{CR}
\CQ_{e}(\CL_s, \CL_t, \CL_u):= & \,e^{\pi \textup{i} b^2} \CL_s \CL_t - 
e^{-\pi \textup{i} b^2} \CL_t \CL_s  \\ &
\,- (e^{2\pi \textup{i} b^2} - 
e^{-2\pi \textup{i} b^2}) \CL_u - 
(e^{\pi \textup{i} b^2} - e^{-\pi \textup{i} b^2}) (\CL_1\CL_3+\CL_2\CL_4)\,.
\notag
\end{align}
Cubic relation:
\begin{align}
\CP_{e}(\CL_s, & \CL_t, \CL_u) = \,-e^{\pi\textup{i} b^2} 
\CL_s \CL_t \CL_u \\ 
& + e^{2\pi \textup{i} b^2} \CL_s^2 + 
e^{-2 \pi \textup{i} b^2} \CL_t^2 + e^{2\pi \textup{i} b^2} \CL_u^2\nonumber \\ 
& + e^{\pi\textup{i} b^2} 
\CL_s (\CL_3\CL_4 + \CL_1\CL_2) + e^{- \pi \textup{i} b^2} 
\CL_t (\CL_2\CL_3 + \CL_1\CL_4) + e^{\pi\textup{i} b^2} 
\CL_u (\CL_1\CL_3 + \CL_2\CL_4)  \notag\\
& + 
\CL_1^2+\CL_2^2+\CL_3^2+\CL_4^2+\CL_1\CL_2\CL_3\CL_4 -\big(2\cos\pi b^2)^2 \,.
\nonumber
\end{align}
In the limit $b \rightarrow 0$ it matches \rf{W04}.

\subsubsection{Case $C_e\simeq C_{1,1}\;$:}

Quadratic relation:
\begin{align} \label{CR1,1}
\CQ_{e}(\CL_s, \CL_t, \CL_u):= e^{\frac{\pi \textup{i}}{2} b^2} \CL_s \CL_t - 
e^{-\frac{\pi \textup{i}}{2} b^2} \CL_t \CL_s  
\,- \,(e^{\pi \textup{i} b^2} - 
e^{-\pi \textup{i} b^2}) \CL_u\,.
\end{align}
Cubic relation:
\begin{align}
\CP_{e}(\CL_s,  \CL_t, \CL_u) = 
& \,
e^{\pi \textup{i} b^2}\CL_s^2 + e^{-\pi \textup{i} b^2}\CL_t^2 + 
e^{\pi \textup{i} b^2}\CL_u^2 - e^{\frac{\pi \textup{i}}{2} b^2}\CL_s \CL_t \CL_u 
\notag \\& + L_0-2\cos\pi b^2\,.
\end{align}

The quadratic relations represent the deformation of the 
Poisson bracket \rf{loopPB}, while the cubic relations
will be deformations of the relations \rf{algrel}.

\subsection{Quantization of the Darboux coordinates}

Natural representations $\pi_{\si}$,
of $\CA_b(C)$ by operators on suitable spaces of functions 
can be constructed in terms of the quantum counterparts $\sll_e$, $\sk_e$
of the Darboux variables $l_e$, $k_e$. 
The algebra $\CA_b(C)$ will be represented 
on functions $\psi_\si(l)$ of the tuple $l$ of 
$h=3g-3+n$ variables $l_e$ associated to the edges of $\Ga_\si$.
The  representations $\pi_{\si}$ will be constructed from
operators $\sll_e$, $\sk_e$ which are defined as
\begin{equation}\label{q-Darboux}
\sll_e\,\psi_\si(l):=\,l_e\,
\psi_\si(l)\,,\qquad 
\sk_e\,\psi_\si(l):=\,4\pi b^2
\frac{1}{\textup{i}}\frac{\pa}{\pa l_e}\psi_\si(l)\,.
\end{equation}
We are using the notation $b^2$ for the quantization parameter $\hbar$.

The construction of the representations will reflect the
locality properties emphasized above.
In order to make this visible in the
notations let us introduce the one-dimensional
Hilbert space $\CH_{l_2l_1}^{l_3}$ associated to a hyperbolic
three-holed sphere $C_{0,3}$ with boundary lengths $l_i$, $i=1,2,3$.
We may then identify the Hilbert space $\CH_\si$ of square-integrable
functions $\psi_\si(l)$ on $\BR_+^{h}$ with the direct integral
of Hilbert spaces
\begin{equation}\label{Hilbertsp}
\CH_\si\,\simeq\,\int^{\oplus}_{\BR_+^h}
\prod_{e\in\si_1}
dl_e \;\bigotimes_{v\in\si_0} \CH_{l_2(v),l_1(v)}^{l_3(v)}\,.
\end{equation}
We denoted the set of internal edges of the MS graph $\si$ by $\si_1$,
and the set of vertices by $\si_0$. 

For $C\simeq C_{0,4}$ we may consider, in particular, that pants
decomposition $\si=\si_s$ depicted on the left of Figure \ref{fmove}.
We then have
\begin{equation}
\CH_s^{0,4}:=\CH_{\si_s}\,\simeq\,\int^\oplus dl_e\;
\CH_{l_3l_e}^{l_4}\ot\CH_{l_2l_1}^{l_e}\,.
\end{equation}
Similarly for $C=C_{1,1}$,
\begin{equation}
\CH_s^{1,1}\,\simeq\,\int^\oplus dl_e\;
\CH_{l_0l_e}^{l_e}\,.
\end{equation}

For each edge $e$ of the MS graph $\Ga_{\si}$ associated to  
a pants decomposition $\si$ one has a corresponding subsurface 
$C_e$ that can be embedded into $C$. For any given operator
$\SO$ on $\CH_s^{0,4}$ and any edge $e$ of $\Ga_\si$ such that
$C_e\simeq C_{0,4}$ there is a natural way to 
define an operator $\SO^e$ on $\CH_{\si}$ acting ``locally'' only
on the tensor factors in \rf{Hilbertsp} associated to $C_e$.

More formally one may define $\SO^e$ as follows.
Let $\SO\equiv\SO_{l_4l_3l_2l_1}$ be a family of operators on $\CH_s^{0,4}$.
It can be considered as a function $\SO(\sll_s,\sk_s;l_1,l_2,l_3,l_4)$
of the operators $\sll_s$,
$\sk_s$ that depends parametrically on $l_1,l_2,l_3,l_4$.
Let $\Ga_{\si}$ be an MS graph on $C$. To an edge $e$ of $\Ga_{\si}$
such that
$C_e\simeq C_{0,4}$ let us associate the neighboring edges
$f_i(e)$, $i=1,2,3,4$ numbered according to the convention defined
in Subsection \ref{sec:FNdef}.
We may then use  $\SO_{l_4l_3l_2l_1}$ to
define an operator
$\SO^e$ on $\CH_{\si}$ as
\begin{equation}
\SO^e:=\SO(\sll_e,\sk_e;\sll_{f_1(e)},\sll_{f_2(e)},\sll_{f_3(e)},\sll_{f_4(e)})\,.
\end{equation}
We are using the notation $\sll_f$ for the operators defined above 
if $f$ is an internal edge, and
we identify $\sll_f\equiv l_f$ if $f$ is an edge that ends in 
a boundary component of $C$.
If $C_e\simeq C_{1,1}$ one may associate in a similar
fashion operators $\SO^e$ to families  $\SO\equiv\SO_{l_0}$ 
of operators on $\CH_s^{1,1}$.

It will sometimes be useful to introduce ``basis vectors'' $\langle \,l\,|$ 
for $\CH_\si$, 
more precisely distributions on dense subspaces of $\CH_\si$ such that
the wave-function $\psi(l)$ of a state $|\,\psi\,\rangle$ 
is represented as $\psi(l)\,=\,\langle \,l\,|\,\psi\,\rangle$.
Representing  $\CH_\si$ as in \rf{Hilbertsp} 
one may identify  
\begin{equation}\label{basis}
\langle \,l\,|\,\simeq\,
\bigotimes_{v\in\si_0} v_{l_2(v),l_1(v)}^{l_3(v)}\,,
\end{equation}
where $v_{l_2,l_1}^{l_3}$ is understood as an element of the 
dual $\big(\CH_{l_2,l_1}^{l_3}\big)^{t}$ of the one-dimensional 
Hilbert space $\CH_{l_2,l_1}^{l_3}$.

\subsection{Representations of the trace coordinates}\label{SSec:q-trace}

It suffices to define the operators $\SL_i\equiv \pi_{\si_s}(L_i)$, $i=s,t,u$,
for the two cases $C\simeq C_{0,4}$ and $C\simeq C_{1,1}$. For these
cases we don't need the labelling by edges $e$.
In both cases we will have 
\begin{equation}
\SL_s:=\,2\cosh(\sll_s/2)\,.
\end{equation}
The operators $\SL_i$, $i=t,u$ will be represented
as finite difference operators. 
Considering the operator $\SL_t$ representing 
the 't Hooft loop operator, for example, we will find that it
can be represented in the form
\begin{equation}\label{Diffop}
\SL_t\,\equiv \,\pi_\si(\CL_t)\psi_\si(l)\,=\,\big[ 
D_+(\sll_s)e^{+\sk_s}+D_0(\sll_s)+D_-(\sll_s)e^{-\sk_s}\big]\psi_\si(l)\,,
\end{equation}
with coefficients $D_{\ep}(l)$ that may depend on $l_1,l_2,l_3,l_4$ for 
$C\simeq C_{0,4}$, and on $l_0$ for $C\simeq C_{1,1}$.

\subsubsection{Case $C_e\simeq C_{0,4}$:}

The operators $\SL_t$ and $\SL_u$ are constructed out
of the quantized Darboux coordinates $\sk_s$ and $\sll_s$ as follows
\begin{subequations}\label{rep0,4}\begin{align}\nonumber
\SL_t\,=\,&\frac{1}{ 2(\cosh \sll_s - \cos 2\pi b^2)}
\Big(2\cos\pi b^2(L_2L_3+L_1L_4)+
\SL_s (L_1L_3+L_2L_4)\Big) \\ \label{quantum't Hooft}
& \quad +  \sum_{\ep=\pm 1}
\frac{1}{\sqrt{2\sinh(\sll_{s}/2)}}
{e^{\ep\sk_s/2}}
\frac{\sqrt{c_{12}(\SL_s)c_{34}(\SL_s)}}{2\sinh(\sll_s/2)}
{e^{\ep\sk_s/2}}
\frac{1}{\sqrt{2\sinh(\sll_s/2)}}  
\end{align}
where the notation $c_{ij}(L_s)$ was introduced in \rf{cijdef}.
The operator $\SL_u$ 
is then obtained from $\SL_t$ my means of a simple
unitary operator
\begin{align}
\SL_u\,=\,&  \big[\,\SB^{-1}\cdot\SL_t
\cdot\SB\,\big]_{L_1\leftrightarrow L_2}\,,
\end{align}
\end{subequations}
where we are using the notations 
$L_i:=2\cosh(l_i/2)$, and
\[
\SB:=e^{\pi{\rm i}(\De(\sll_s)-\De(l_\2)-\De(l_\1))}\,,\qquad
\De(l):=\frac{l^2}{(4\pi b)^2}+\frac{1+b^2}{4b}\,.
\] 
The operator $\SB$ will later be recognized as representing the
braiding of holes $1$ and $2$.

\subsubsection{Case $C_e\simeq C_{1,1}$:}

We now find the following expressions for the operators $\SL_t$
and $\SL_u$: 
\begin{subequations}\label{rep1,1}\begin{align}
\SL_t\,=\,& \label{quantum't Hooft-1,1}
\sum_{\ep=\pm 1} \frac{1}{\sqrt{\sinh(\sll_{s}/2)}}
{e^{\ep\sk_s/4}}
\;\sqrt{\cosh\fr{2\sll_s+l_0}{4}\cosh\fr{2\sll_s-l_0}{4}}\;
{e^{\ep\sk_s/4}}
\frac{1}{\sqrt{\sinh(\sll_s/2)}} 
\end{align}
The operator $\SL_u$ can be obtained from $\SL_t$
by means of a unitary operator $\ST$,
\begin{align}
\SL_u\,=\,&  \ST^{-{1}}\cdot\SL_t
\cdot\ST\,,
\end{align}
\end{subequations}
which is explicitly constructed as
\begin{equation}
\ST:=e^{-2\pi i\De(\sll_s)}\,.
\end{equation}
This operator will later be found to represent the Dehn twist.

It is straightforward to check by explicit calculations that the 
relations of  $\CA_b(C)$ are satisfied.
It can furthermore be shown that the representations above are 
unique, see Appendix \ref{uniqueness} for some details.

We furthermore observe that the operators $\SL_i$, 
are positive self-adjoint, but unbounded. 
There is a maximal dense subset 
$\CS_\si$ inside of $\CH_\si$ on which the whole algebra $\CA_b(C)$
of algebraic functions on $\CM_{\rm flat}$ is realized.

\subsection{Transitions between representation}

For each MS graph $\si$ one will get a representation 
$\pi_{\si}$ of the quantized algebra of $\CA_b(C)$ of functions
on $\CM_{\rm flat}(C)$. A natural requirement is that
the resulting quantum theory does not depend on the choice of $\si$ in
an essential way. This can be ensured if there exist unitary
operators $\SU_{\si_2\si_1}$ intertwining between the different 
representations in the sense that
\begin{equation}\label{inter}
\pi_{\si_2}(\CL_\ga)\cdot \SU_{\si_2\si_1}\,=\,
\SU_{\si_2\si_1}\cdot \pi_{\si_1}(\CL_\ga)\,.
\end{equation}
Having such intertwining operators allows one to identify the 
operators $\pi_{\si}(\CL_\ga)$ as different representatives of
one and the same abstract element $\CL_\ga$ 
of the quantized algebra of 
functions  $\CA_b(C)$.
The intertwining property \rf{inter} will turn out to determine the
operators $\SU_{\si_2\si_1}$ essentially uniquely.

It will be found that the operators $\SU_{\si_2\si_1}$
can be represented as integral operators
\begin{equation}
(\SU_{\si_\2\si_\1}\psi_{\si_\1})(l_\2)\,=\,\int dl_\1\;
A_{\si_\2\si_\1}(l_\2,l_\1)\,\psi_{\si_\1}(l_\1)\,.
\end{equation}
This intertwining relation \rf{inter} is then equivalent to 
a system of difference equations for the kernels $A_{\si_\2\si_\1}(l_\2,l_\1)$,
\begin{equation}\label{diffinter}
\pi_{\si_\2}(\CL_\ga)\cdot A_{\si_\2\si_\1}(l_\2,l_\1),
\,=\,A_{\si_\2\si_\1}(l_\2,l_\1)\cdot \overset{\leftarrow}{\pi}_{\si_\1}(\CL_\ga)^t
\,.
\end{equation}
The notation $\overset{\leftarrow}{\pi}_{\si_\2}(\CL_\ga)^t$ indicates that the
transpose of the difference operator ${\pi}_{\si_\1}(\CL_\ga)$ acts on the
variables $l_\1$ from the right. $\pi_{\si_\2}(\CL_\ga)$ acts only
on the variables $l_\2$. The equations \rf{diffinter} represent
a system of difference equations which constrain the kernels
$A_{\si_\2\si_\1}(l_\2,l_\1)$ severely. They will determine the
kernels $A_{\si_\2\si_\1}(l_\2,l_\1)$ essentially uniquely once the 
representations $\pi_\si$ have been fixed.

\subsection{Kernels of the 
unitary operators between different representations}
\label{sec:elmoves}

We now want to list the explicit representations for the generators 
of the 
Moore-Seiberg groupoid in the quantization of $\CM^0_{\rm flat}(C)$.

For many of the following considerations we will find it useful to
replace the variables $l_e$ by 
\begin{equation}
\al_e:=\frac{Q}{2}+\textup{i}\frac{l_e}{4\pi b}\,.
\end{equation}
Using the variables $\al_e$ instead of $l_e$ will in particular help to
compare with Liouville theory.
\subsubsection{B-move}
\begin{equation}
\SB\cdot v_{\al_\2\al_\1}^{\al_\3}\,=\,B_{\al_\2\al_\1}^{\al_\3}
v_{\al_\1\al_\2}^{\al_\3}\,,
\end{equation}
where 
\begin{equation}\label{Bcoeff}
B_{\al_\2\al_\1}^{\al_\3}\,=\,
e^{\pi i(\De_{\al_\3}-\De_{\al_\2}-\De_{\al_\1})}\,.
\end{equation}
\subsubsection{Z-move} 
\begin{equation}
\SZ\cdot v_{\al_\2\al_\1}^{\al_\3}\,=\,
v_{\al_\1\al_\3}^{\al_\2}\,.
\end{equation}
\subsubsection{F-move}
\begin{equation}
\SF\cdot v^{\al_4}_{\al_3\al_s}\ot 
v^{\al_s}_{\al_2\al_1} \;\,=\,
\int^{\oplus}_\BS d\be_t \;\Fus{\al_1}{\al_2}{\al_3}{\al_4}{\be_s}{\be_t}\;
v^{\al_4}_{\be_t\al_1}\ot 
v^{\be_t}_{\al_3\al_2}\,,
\end{equation}
where $\BS=\frac{Q}{2}+i\BR^+$, $Q:=b+b^{-1}$.
The kernel describing the transition between representation $\pi_s$ and 
$\pi_t$ is given in terms of the b-$6j$ symbols as 
\begin{equation} \label{Fvs6j}
\Fus{\al_1}{\al_2}{\al_3}{\al_4}{\be_s}{\be_t}
\,=\,(M_{\be_s}M_{\be_t})^{\frac{1}{2}}\,
\big\{\,{}^{\al_1}_{\al_3}\,{}^{\al_2}_{\al_4}\,{}^{\al_s}_{\al_t}\big\}_b^{}
\,,
\end{equation}
where 
\begin{equation}
M_\be:=|S_b(2\be)|^2=-4\sin\pi (b(2\be-Q))\sin\pi (b^{-1}(2\be-Q))\,,
\end{equation}
and
the b-$6j$ symbols 
$\big\{\,{}^{\al_1}_{\al_3}\;{}^{\al_2}_{\al_4}\;{}^{\al_s}_{\al_t}\big\}_b^{}$
are defined as \cite{PT1,PT2,TV}
\begin{align}\label{6j3}
\big\{\,{}^{\al_1}_{\al_3}\;{}^{\al_2}_{\al_4}\;{}^{\al_s}_{\al_t}\big\}_b^{}
&=\Delta(\al_s,\al_2,\al_1)\Delta(\al_4,\al_3,\al_s)\Delta(\al_t,\al_3,\al_2)
\Delta(\al_4,\al_t,\al_1)\\
&\qquad \times\int\limits_{\CC}du\;
S_b(u-\alpha_{12s}) S_b(u-\al_{s34}) S_b(u -\alpha_{23t}) 
S_b(u-\alpha_{1t4})
\notag \\[-1.5ex] & \hspace{2cm}\times 
S_b( \alpha_{1234}-u) S_b(\alpha_{st13}-u) 
S_b(\alpha_{st24}-u) S_b(2Q-u)\,.
\notag\end{align}
The expression involves the following ingredients:
\begin{itemize}
\item We have used the notations $\al_{ijk}=\al_i+\al_j+\al_k$,
$\al_{ijkl}=\al_i+\al_j+\al_k+\al_l$. 
\item The special function $S_b(x)$ is a variant of the 
non-compact quantum dilogarithm, definition and properties
being collected in Appendix \ref{Qdil},
\item $\Delta(\al_3,\al_2,\al_1)$ is defined as 
\begin{align}
&\Delta(\al_3,\al_2,\al_1)=\bigg(\frac{S_b(\alpha_1+\alpha_2+\alpha_3-Q)}{S_b(\alpha_1+\alpha_2-\alpha_3) 
S_b(\alpha_1+\alpha_3-\alpha_2) S_b(\alpha_2+\alpha_3-\alpha_1)} 
\bigg)^{\frac{1}{2}}\,.
\notag\end{align}
\item The integral is defined in the cases
that $\al_k\in Q/2+\textup{i}\BR$ by a contour $\CC$ which 
approaches $2Q+\textup{i}\BR$ near infinity,
and passes the real axis in the interval $(3Q/2,2Q)$.
\end{itemize}

\subsubsection{S-move}
\begin{equation}
\SS\cdot 
v^{\be_\1}_{\al,\be_\1} \;\,=\,
\int^{\oplus}_\BS d\be_2 \;S_{\be_\1\be_\2}(\al)\;
v^{\be_\2}_{\al,\be_\2}\,,
\end{equation}
where
\begin{align}\label{Skern}
S_{\be_\1\be_\2} (\al_0)\,=\,& \sqrt{2}\,
\frac{\De(\be_\1,\al_0,\be_\1)}{\De(\be_\2,\al_0,\be_\2)}\,
(M_{\be_\1}M_{\be_\2})^{\frac{1}{2}} 
\,\frac{e^{\frac{\pi i}{2}\De_{\al_0}}}{S_b(\al_0)}\times \\
&\times
\int_\BR dt\;e^{2\pi t(2\be_\1-Q)}
\frac{S_b\big(\frac{1}{2}(2\be_\2+\al_0-Q)+it\big)
S_b\big(\frac{1}{2}(2\be_\2+\al_0-Q)-it\big)}
{S_b\big(\frac{1}{2}(2\be_\2-\al_0+Q)+it\big)
S_b\big(\frac{1}{2}(2\be_\2-\al_0+Q)-it\big)}\,.
\notag\end{align}
This ends our list of operators representing the generators 
of the Moore-Seiberg groupoid.

\subsection{Representation of the Moore-Seiberg groupoid}

A projective unitary 
representation of the Moore-Seiberg
groupoid is defined by the family of unitary 
operators $\SU_{\si_\2\si_1}:\CH_{\si_\1}\ra\CH_{\si_\2}$,
$\si_2,\si_\1\in\CM_\0(\Sigma)$ which satisfy the composition
law projectively
\begin{equation}\label{MS-abstr}
\SU_{\si_\3\si_2}\cdot\SU_{\si_\2\si_1}\,=\,\zeta_{\si_\3,\si_\2,\si_\1}\,
\SU_{\si_\3\si_2}\,,
\end{equation}
where $\zeta_{\si_\3,\si_\2,\si_\1}\in\BC$, $|\zeta_{\si_\3,\si_\2,\si_\1}|=1$.
The operators $\SU_{\si_\2\si_1}$ which intertwine
the representations $\pi_{\si}$ according to \rf{inter} will generate
a representation of the Moore-Seiberg groupoid.

\subsubsection{Moore-Seiberg equations}

Let us next list the explicit representations for the relations of the 
Moore-Seiberg groupoid in the quantization of $\CM^0_{\rm flat}(C)$.
In order to state some of them it will be convenient
to introduce the
operator $\ST$ representing the Dehn twist such that
\begin{equation}
\ST\cdot v_{\al_\2\al_\1}^{\al_\3}\,=\,T_{\al_\3}
v_{\al_\2\al_\1}^{\al_\3}\,,
\end{equation}
where 
\begin{equation}\label{g=0,n=3}
T_{\al_2}:=\,B_{\al_\2\al_\1}^{\al_\3}\,B_{\al_\3\al_\2}^{\al_\1}\,=\,
e^{-2\pi i \De_{\al}}
\end{equation}
We claim that the kernels of the operators 
$\SB$, $\SF$, $\SS$ and $\SZ$
defined above satisfy the Moore-Seiberg
equations in the following form:

\begin{subequations}\label{MS-eqns}
{\it Genus zero, four punctures}
\begin{align}\label{hexagon}
\int_{\BS}d\be_t\;&\Fus{\al_1}{\al_2}{\al_3}{\al_4}{\be_s}{\be_t}\,
  B^{\al_4}_{\be_t,\al_1}\,\Fus{\al_2}{\al_3}{\al_1}{\al_4}{\be_t}{\be_u}
\,=\,B^{\be_s}_{\al_2,\al_1}
\,\Fus{\al_2}{\al_1}{\al_3}{\al_4}{\be_s}{\be_u}\,B^{\be_u}_{\al_3,\al_1}\,,\\
\int_{\BS}d\be_2\;& \Fus{\al_1}{\al_2}{\al_3}{\al_4}{\be_1}{\be_2} 
                  \Fus{\al_3}{\al_2}{\al_1}{\al_4}{\be_2}{\be_3}\,=\,
\de_\BS^{}(\be_1-\be_3)\,.
\end{align}
{\it Genus zero, five punctures}
\begin{align}\label{pentagon}
 \int_{\BS}d\be_5\; 
\Fus{\al_1}{\al_2}{\al_3}{\be_{2}}{\be_{1}}{\be_5}
& \Fus{\al_1}{\be_5}{\al_4}{\al_5}{\be_{2}}{\be_{4}}
\Fus{\al_2}{\al_3}{\al_4}{\be_{4}}{\be_5}{\be_{3}} 
=\Fus{\al_1}{\al_2}{\be_3}{\al_{5}}{\be_{1}}{\be_{4}}
\Fus{\,\be_1}{\al_3}{\al_4}{\al_5}{\be_{2}}{\be_{3}}\,.
\end{align}
{\it Genus one, one puncture}
\begin{align}\label{g=1,n=1a}
&\int_{\BS} d\be_\2\;S_{\be_\1\be_\2}(\al)S_{\be_\2\be_\3}(\al)\,=\,
\de_{\BS}^{}(\be_\1-\be_\3)\,(B^{\be_\1}_{\be_\1\al})^{-1}\,,\\
&\int_{\BS} d\be_\2\;S_{\be_\1\be_\2}(\al)\,T_{\be_\2}\, 
S_{\be_\2\be_\3}(\al)\,
=\,e^{6\pi i \chi_b^{}}\,T_{\be_\1}^{-1}\, 
S_{\be_\1\be_\3}(\al)\,
T_{\be_\3}^{-1}\,
\,.
\label{g=1,n=1b}\end{align}
{\it Genus one, two punctures}
\begin{align}\label{g=1,n=2}
S_{\be_\1\be_\2}(\be_\3)\int_{\BS}& d\be_\4\; 
\Fus{\al_\2}{\al_\1}{\be_\2}{\be_\2}{\be_\3}{\be_\4}\,
T_{\be_\4}^{}\,T_{\be_\2}^{-1}\,
\Fus{\be_\2}{\al_\1}{\al_\2}{\be_\2}{\be_\4}{\be_\5}\\
& = \int_{\BS}d\be_\6\;
\Fus{\al_\2}{\al_\1}{\be_\1}{\be_\1}{\be_\3}{\be_\6}
\Fus{\be_\6}{\al_\2}{\al_\1}{\be_\6}{\be_\1}{\be_\5}
S_{\be_\6\be_\2}(\be_\5)\,e^{\pi i(\De_{\al_\1}+\De_{\al_\2}-\De_{\be_\5})}\,.
\notag
\end{align}
\end{subequations}
The delta-distribution $\de_{\BS}(\be_1-\be_2)$ is defined by the ordinary
delta-distribution on the real positive half-line 
$-i(\BS-Q/2)$.


\subsubsection{Mapping class group action}

Having a representation of the Moore-Seiberg groupoid automatically
produces a representation of the mapping class group. 
An element of the mapping class group $\mu$ represents a diffeomorphism
of the surface $C$, and therefore maps any MS graph $\si$ 
to another one denoted $\mu.\si$. Note that the 
Hilbert spaces $\CH_\si$ and $\CH_{\mu.\si}$ are canonically 
isomorphic. Indeed, the Hilbert spaces $\SH_\si$, described
more explicitly in \rf{Hilbertsp}, depend only on the combinatorics
of the graphs $\si$, but not on their embedding into $C$. We may therefore
define an operator $\SM_\si(\mu):\CH_\si\ra\CH_\si$ as
\begin{equation}\label{MCGdef}
\SM_\si(\mu):=\,\SU_{\mu.\si,\si}\,.
\end{equation}
It is automatic that the operators $\SM(\mu)$ define a projective 
unitary representation of the mapping class group ${\rm MCG}(C)$ 
on $\CH_\si$. 

The operators $\SU_{\si_\2,\si_\1}$ intertwine the actions defined thereby,
as follows from 
\begin{equation}\label{MCGinter}
\SM_{\si_\2}(\mu)\cdot\SU_{\si_\2,\si_\1}
=
\eta_{\si_2\si_1}\,\SU_{\mu.\si_\2,\,\mu.\si_\1}\cdot
\SM_{\si_\1}(\mu)
\equiv
\eta_{\si_2\si_1}\,\SU_{\si_\2,\,\si_\1}\cdot
\SM_{\si_\1}(\mu)\,.
\end{equation}
We may therefore naturally 
identify the mapping class group actions
defined on the various $\CH_\si$.

\subsection{Self-duality}

For the application to  gauge theory we 
are looking for a representation 
of {\it two} copies of ${\rm Fun}_{\ep_i}(\CM_{\rm flat}(C))$, $i=1,2$, 
generated from the 
two sets of supersymmetric  Wilson- and
't Hooft loop operators $\ST_{e,i}$, $\SW_{e,i}$
one can define of the four-ellipsoid. The eigenvalues
of the Wilson loop operators $\SW_{e,i}$ 
are $2\cosh(2\pi a_e/\ep_i)$, 
for $i=1,2$, respectively. This can be incorporated 
into the  quantum theory of $\CM^{0}_{\rm flat}(C)$ as
follows.  

Let us identify the quantization parameter $b^2$ with the
ratio of the parameters $\ep_1$, $\ep_2$,
\begin{equation}
b^2\,=\,\ep_1\,/\,\ep_2\,.
\end{equation}
Let us furthermore introduce
the rescaled variables
\begin{equation}
a_e\,:=\,\ep_2\frac{l_e}{4\pi}\,.
\end{equation}
The representations $\pi_{\si}$ on functions $\psi_{\si}(l)$ 
are equivalent to representations on functions $\phi_\si(a)$, 
defined by
\begin{equation}\label{dual-q-Darboux}
\sll_e\,\phi_\si(a):=\,\frac{2\pi a_e}{\ep_2} \,
\psi_\si(l)\,,\qquad 
\sk_e\,\psi_\si(l):=\,
\frac{2}{\textup{i}}\ep_1\frac{\pa}{\pa a_e}\phi_\si(a)\,.
\end{equation}
Let us introduce a second pair of operators
\begin{equation}
\tilde{\sll}_e:=\,\frac{\ep_\2}{\ep_\1}\,\sll_e\,,\qquad
\tilde{\sk}_e:=\,\frac{\ep_\2}{\ep_\1}\,\sk_e\,.
\end{equation}
Replacing in the construction of the operators 
$\SL_{\ga}$ all operators $\sll_e$ by $\tilde{\sll}_e$, 
all $\sk_e$ by $\tilde{\sk}_e$, and all variables $l_i$ by 
$\frac{\ep_\2}{\ep_\1}l_i$ defines operators $\tilde\SL_{\ga}$.
The operators $\tilde\SL_{\ga}$ generate a representation
of the algebra ${\rm Fun}_{b^{-1}}(\CM_{\rm flat}(C))$. It can be checked
that the operatos $\SL_\ga$ commute with the 
operators $\tilde{\SL}_\ga$. Taken together we thereby 
get a representation of ${\rm Fun}_{b}(\CM_{\rm flat}(C))\times
{\rm Fun}_{b^{-1}}(\CM_{\rm flat}(C))$.
The operators $\SL_s^e$ and $\tilde{\SL}_s^e$  correspond to the Wilson loop
operators $\SW_{e,1}$ and $\SW_{e,2}$, respectively.

\subsection{Gauge transformations}\label{gaugetrsf}

Note that the 
requirement that the $\pi_\si(\CL_s^e)$ act as multiplication 
operators leaves a large freedom. 
A gauge transformation 
\begin{equation}\label{gauge}
\psi_\si(l)\,=\,e^{\textup{i}\chi(l)}\psi_\si'(l)\,,
\end{equation}
would lead to a representation $\pi_\si'$ of the form \rf{Diffop}
with $\sk_e$ replaced by
\begin{equation}\label{q-can}
\sk'_e:=\,\sk_e+4\pi b^2\,\pa_{l_e}\chi(l)\,.
\end{equation}
This is nothing but the quantum version of a
canonical transformation $(l,k)\ra (l,k')$ with $k'_e=k_e+f_e(l)$. 
The representation 
$\pi_{\si_s}'(\CL_t)$ may then be obtained from \rf{Diffop}
by replacing 
$D_\ep(l)\ra {D}_\ep'(l)$ with 
\begin{equation}
{D}_\ep'(\sll_s)\,=\,e^{-i\chi(\sll_s)}\,
e^{\ep\sk_s}e^{i\chi(\sll_s)}e^{-\ep\sk_s}
\,D_\ep(\sll_s)\,, \quad\ep=-1,0,1\,. 
\end{equation}

Locality leads to an important restriction
on the form of 
allowed gauge transformations $\chi(l)$. They should preserve the
local nature of the representation $\pi_{\si}^{\chi}$. 
This means that function
$\nu\equiv e^{i\chi}$ must have the form of a product
\begin{equation}
\nu(l)\,=\,\prod_{v\in\si_0} \nu(l_3(v),l_2(v),l_1(v))\,,
\end{equation}
over functions $\nu$
which depend only 
on the variables associated to the vertices $v$ of $\si$. 
This corresponds to replacing  the basis vectors
$v_{l_2,l_1}^{l_3}$ in \rf{basis} by 
${v}_{l_2,l_1}^{l_3}=\nu(l_3,l_2,l_1){v}'{}_{l_2,l_1}^{l_3}$.
We then have, more explicitly,
\begin{equation}\label{D-gaugetr}
{D}_\ep'(l_s)\,=\,d_{43e}^\ep(l_s)d_{e21}^\ep(l_s)\,{D}_\ep(l_s)\,,
\end{equation}
where
\begin{equation}
d_{43e}^\ep(l_s)=
\frac{\nu(l_4,l_3,l_s-4\ep \pi \textup{i}  b^2)}{\nu(l_4,l_3,l_s)}\,,
\qquad d_{e21}^\ep(l_s)=
\frac{\nu(l_s-4\ep \pi\textup{i}  b^2,l_2,l_1)}{\nu(l_s,l_2,l_1)}\,,
\end{equation}
It is manifest that the property of the coefficients $D_\ep(l)$
to depend only on the variables $l_f$ assigned to the nearest
neighbors $f$ of $e$ is preserved by the gauge transformations.

The freedom to change the representations of $\CA_b(C)$ by gauge 
transformations reflects the perturbative scheme dependence mentioned
in Subsection \ref{pertpart}.

\section{Relation to the quantum Teichm\"uller theory}
\label{q-Teich}

\setcounter{equation}{0}

The Teichm\"uller spaces had previously been quantized using 
other sets of coordinates associated to triangulations
of $C$ rather than pants decompositions \cite{F97,CF,Ka1}.
This quantization yields geodesic length operators
quantizing the geodesic length functions on $\CT(C)$ \cite{CF2,T05}.
By diagonalizing the commutative subalgebra generated by
the geodesic length operators associated to a cut system 
one may construct a representation of the Moore-Seiberg 
groupoid \cite{T05}.
We will show that this representation is 
equivalent to the one defined in Section \ref{q-Mflat}.

This section starts by presenting the definitions and results from the 
quantum Teichm\"uller theory that will be needed in this paper.
We will use the formulation introduced by R. Kashaev \cite{Ka1}, 
see also \cite{T05} for a more detailed exposition and a discussion
of its relation to the framework of Fock \cite{F97} and
Chekhov and Fock \cite{CF}. We then review the results from \cite{T05}
on the diagonalization of maximal sets of commuting length 
operators and the corresponding representation of the 
Moore-Seiberg groupoid.

\subsection{Algebra of operators and its representations}

The formulation from \cite{Ka1}
starts from the quantization of a somewhat enlarged 
space $\hat{\CT}(C)$. The usual Teichm\"uller space $\CT(C)$ 
can then be characterized as subspace of $\hat{\CT}(C)$ using 
certain linear constraints. This is motivated by the observation 
that the spaces $\hat{\CT}(C)$ have natural polarizations,
which is far from obvious in the formulation of \cite{F97,CF}.

For a given surface $C$ with 
constant negative curvature metric and at least one puncture one considers
ideal triangulations $\tau$. Such ideal triangulations are defined
by maximal collection of non-intersecting open geodesics which start and
end at the punctures of $C$. We will assume that the triangulations
are decorated, which means that a distinguished corner is chosen in
each triangle.

\begin{figure}[htb]
\epsfxsize5cm
\centerline{\epsfbox{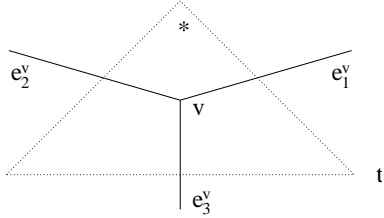}}
\caption{\it Graphical representation of the vertex $v$ dual
to a triangle $t$. The marked corner defines a
corresponding numbering of the edges that emanate at $v$.}
\label{decor}
\end{figure}

We will find it convenient
to parameterize triangulations by their dual graphs which are called
fat graphs $\vf_\tau$. The vertices of $\vf_\tau$ are in one-to-one correspondence
with the triangles of $\tau$, and the edges of $\vf_\tau$ are in 
one-to-one correspondence
with the edges of $\tau$. The relation between a triangle $t$ in 
$\tau$ and the fat graph $\vf_{\tau}$ is depicted in Figure \ref{decor}.
$\vf_\tau$ inherits a natural decoration of its
vertices from $\tau$, as is also indicated in Figure \ref{decor}.

The quantum theory associated to the Teichm\"uller space $\CT(C)$
is defined on the kinematical level by associating to each vertex $v\in\vf_0$,
$\vf_0=\{{\rm vertices\;\,of\;\,}\vf\}$,
of $\vf$ a pair of generators $p_v,q_v$ which are supposed to satisfy
the relations
\begin{equation}
\big[\,p_v\,,\,q_{v'}\,\big]\,=\,\frac{\de_{vv'}}{2\pi i}\,.
\end{equation}
There is a natural representation of this algebra on the 
Schwarz space $\hat{\CS}_\vf(C)$ of rapidly decaying 
smooth functions $\psi(q)$, 
$q:\vf_0\ni v\ra q_v$, generated from 
$\pi_\vf(q_v):=\sq_v$, $\pi_\vf(p_v):=\spp_v$, where
\begin{equation}
\sq_v\,\psi(q):=\,q_v\psi(q)\,,\qquad
\spp_v\,\psi(q):=\,\frac{1}{2\pi i}\frac{\pa}{\pa q_v}\psi(q)\,.
\end{equation}
For each surface $C$ 
we have thereby defined an algebra $\hat{\CA}(C)$ together with 
a family
of representations $\pi_\vf$ of $\hat{\CA}(C)$ on the 
Schwarz spaces $\hat{\CS}_\vf(C)$ which are dense subspaces of
the Hilbert space $\CK(\vf)\simeq L^2(\BR^{4g-4+2n})$. 

The quantized algebra of functions $\CA_{\CT(C)}$ on the Teichm\"uller
spaces is then defined by the quantum version of the Hamiltonian 
reduction with respect to a certain set of constraints. 
To each element $[\ga]$ of the first homology $H_1(C,\BR)$
of $C$ one may associate an operator $\sz_{\vf,\ga}$ that is 
constructed as a linear combination of the operators $p_v$ and $q_v$,
$v\in\vf_0$, see \cite{Ka1,T05} for details. The operators $\sz_{\vf,\ga}$
represent the constraints which can be used to characterize the 
subspace associated to the quantum Teichm\"uller theory within 
$\CK(\vf)$.

\subsection{The projective representation of the 
Ptolemy groupoid on $\CK(\vf)$ \index{Ptolemy groupoid}}
\label{qPtolemy}

The next step is to show that the choice of fat graph $\vf$ is
inessential by constructing unitary operators 
$\pi_{\vf_\2\vf_\1}:\CK(\vf_1)\ra\CK(\vf_2)$
intertwining the 
representations  $\pi_{\vf_1}$ and $\pi_{\vf_\2}$.

The groupoid generated by the transitions $[\vf',\vf]$ from 
a fat graph $\vf$ to $\vf'$ is called the Ptolemy groupoid.
It can be described in terms of generators $\omega_{uv}$, $\rho_u$, $(uv)$ 
and certain relations. The generator $\omega_{uv}$ is the elementary change 
of diagonal in a quadrangle, $\rho_u$ is the clockwise rotation of 
the decoration, and $(uv)$ is the exchange of the numbers associated to 
the vertices $u$ and $v$.
More details 
and further references can be found in \cite[Section 3]{T05}. 

Following \cite{Ka3} closely we shall define a 
projective unitary representation of the Ptolemy groupoid in 
terms of the following set of 
unitary operators 
\begin{equation}\begin{aligned}
\SA_v\;\equiv\;& e^{\frac{\pi i}{3}}
e^{-\pi i (\spp_v+\sq_v)^2}e^{-3\pi i \sq_v^2}\\
\ST_{vw}\;\equiv\;& e_b(\sq_v+\spp_w-\sq_w)e^{-2\pi i\spp_v\sq_w} ,
\end{aligned}\qquad
\text{where}\;\;v,w\in\vf_{\zero}\,.
\label{qPtgens}\end{equation}
The special function $e_b(U)$ can be defined in the strip 
$|\Im z|<|\Im c_b|$, $c_b\equiv i(b+b^{-1})/2$ by means of the 
integral representation
\begin{equation}
\log e_b(z)\;\equiv\;\frac{1}{4}
\int\limits_{i0-\infty}^{i0+\infty}\frac{dw}{w}
\frac{e^{-2{\mathsf i}zw}}{\sinh(bw)
\sinh(b^{-1}w)}.
\end{equation}
These operators are unitary for $(1-|b|)\Im b=0$. They satisfy the 
following relations \cite{Ka3}
\begin{subequations}
\label{q-Ptolemyrels}
\begin{align}
{\rm (i)}& 
\qquad\ST_{vw}\ST_{uw}\ST_{uv}\;=\;\ST_{uv}\ST_{vw},\label{pentrel}\\
{\rm (ii)}& \qquad\SA_{v}\ST_{uv}\SA_{u}\; =\; 
\SA_{u}\ST_{vu}\SA_{v},
\label{symrel}\\
{\rm (iii)}& \qquad\ST_{vu}\SA_{u}\ST_{uv}
\;=\;\zeta\SA_{u}\SA_{v}\SP_{uv},
\label{invrel}\\
{\rm (iv)}& \qquad\SA_u^3\;=\;\id,\label{cuberel}
\end{align}
\end{subequations}
where $\zeta=e^{\pi i c_b^2/3}$, $c_b\df\frac{i}{2}(b+b^{-1})$.
The relations 
\rf{pentrel} to \rf{cuberel} allow us to define a projective 
representation of the Ptolemy groupoid as follows.  
\begin{itemize}
\item Assume that $\om_{uv}\in[\vf',\vf]$.
To $\om_{uv}$ let us associate
the operator
\[ 
\su(\omega_{uv})\,\df\,\ST_{uv}\;:\;\CK(\vf)\ni \fv\;\,\ra\;\,
\ST_{uv}\fv\in\CK(\vf').
\]
\item For each fat graph $\vf$ and vertices $u,v\in \vf_{\zero}$
let us define the following operators
\[ \begin{aligned}
{}& \SA_{u}^{\vf}\;:\;\CK(\vf)\ni \fv\;\,\ra\;\,
\SA_{u}^{}\fv\in\CK(\rho_{u}\circ\vf).\\
{}& \SP_{uv}^{\vf}\;:\;\CK(\vf)\ni \fv\;\,\ra\;\,
\SP_{uv}^{}\fv\in\CK((uv)\circ\vf).
\end{aligned}
\]
\end{itemize}
It follows from \rf{pentrel}-\rf{cuberel}
that the operators $\ST_{uv}$, $\SA_{u}$ 
and $\SP_{uv}$ can be used to generate
a unitary projective representation of the Ptolemy groupoid.

The corresponding automorphisms of the algebra $\CA(C)$ are
\begin{equation}
\sa_{\vf_\2\vf_\1}(\SO):=
\sa\sd[\SW_{\vf_\2\vf_\1}](\SO):=
\SW_{\vf_\2\vf_\1}^{}\cdot\SO\cdot\SW_{\vf_\2\vf_\1}^{}\,.
\end{equation}
The automorphism $\sa_{\vf_\2\vf_\1}$
generate the canonical quantization
of the changes of coordinates for $\hat{\CT}(C)$ 
from one fat 
graph to another \cite{Ka1}.
Let us note that the constraints transform under a
change of fat graph as
$\sa_{\vf_\2\vf_\1}(\sz_{\vf_\1,\ga})=\sz_{\vf_\2,\ga}$.

\subsection{Length operators}

A particularly important class of coordinate functions on the
Teichm\"uller spaces are the geodesic length functions.
The quantization of these observables was studied in 
\cite{CF,CF2,T05}.

Such length operators can be constructed in general 
as follows \cite{T05}. 
We will first define
the length operators for two special cases in which the 
choice of fat graph $\vf$ simplifies the representation of the 
curve $\ga$. 
We then explain how to
generalize the definition to all other cases.
\begin{itemize}
\item[(i)] Let $A_{\ga}$ be an annulus embedded in the surface $C$
containing the curve $\ga$, and let 
$\vf$ be a fat graph which looks inside of $A_{\ga}$ 
as depicted in Figure \ref{annfig}.
\begin{equation}\label{annfig}
\lower.9cm\hbox{\epsfig{figure=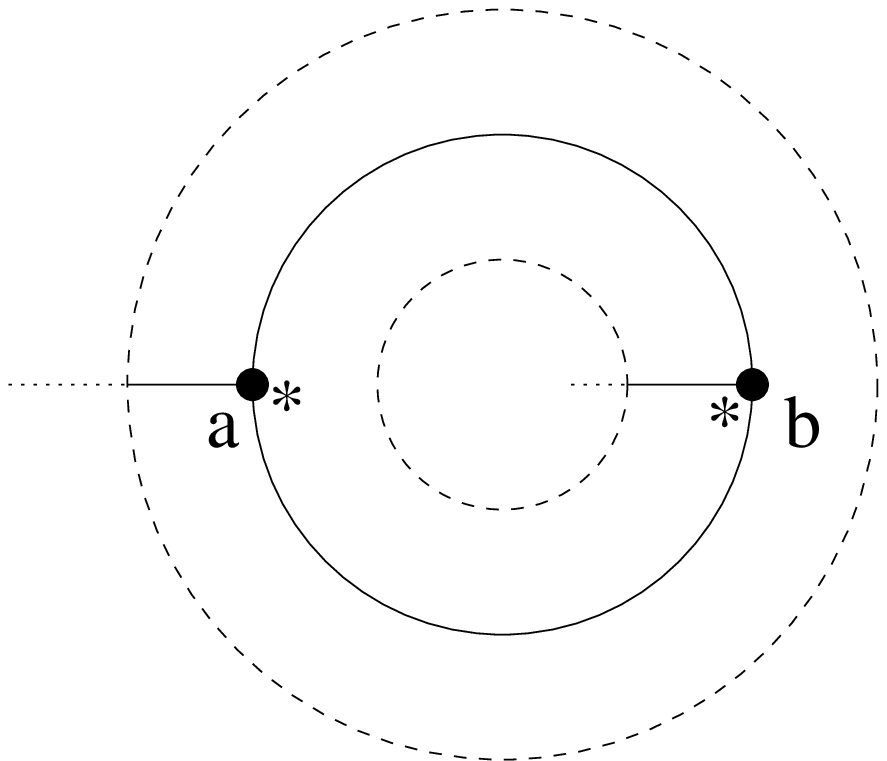,height=3cm}}
\qquad
\begin{aligned}
& \text{Annulus $A_{\ga}$: Region bounded} \\
& \text{by the two dashed circles,}\\
& \text{and part of $\vf$ contained in $A_{\ga}$.}
\end{aligned}
\end{equation}
Let us define the length operators 
\begin{equation}\label{SLsimple}
\begin{aligned}& \SL_{\vf,\ga}:=\,2\cosh 2\pi b\spp_{\ga}+
e^{-2\pi b\sq_{\ga}}\,,\;\;
{\rm where}\\ 
& \spp_{\ga}:=\frac{1}{2}(\spp_a-\sq_a-\spp_b)\,,\quad
\sq_{\ga}:=\frac{1}{2}(\sq_a+\spp_a+\spp_b-2\sq_b)\,.
\end{aligned}
\end{equation}
\item[(ii)] Assume that the curve $\ga\equiv \ga_\3$ 
is the boundary component labelled
by number $3$ 
of a trinion $P_{\ga}$ embedded in $C$ within which the fat graph $\vf$ 
looks as follows:
\begin{equation}\label{trinfig}
\lower.9cm\hbox{\epsfig{figure=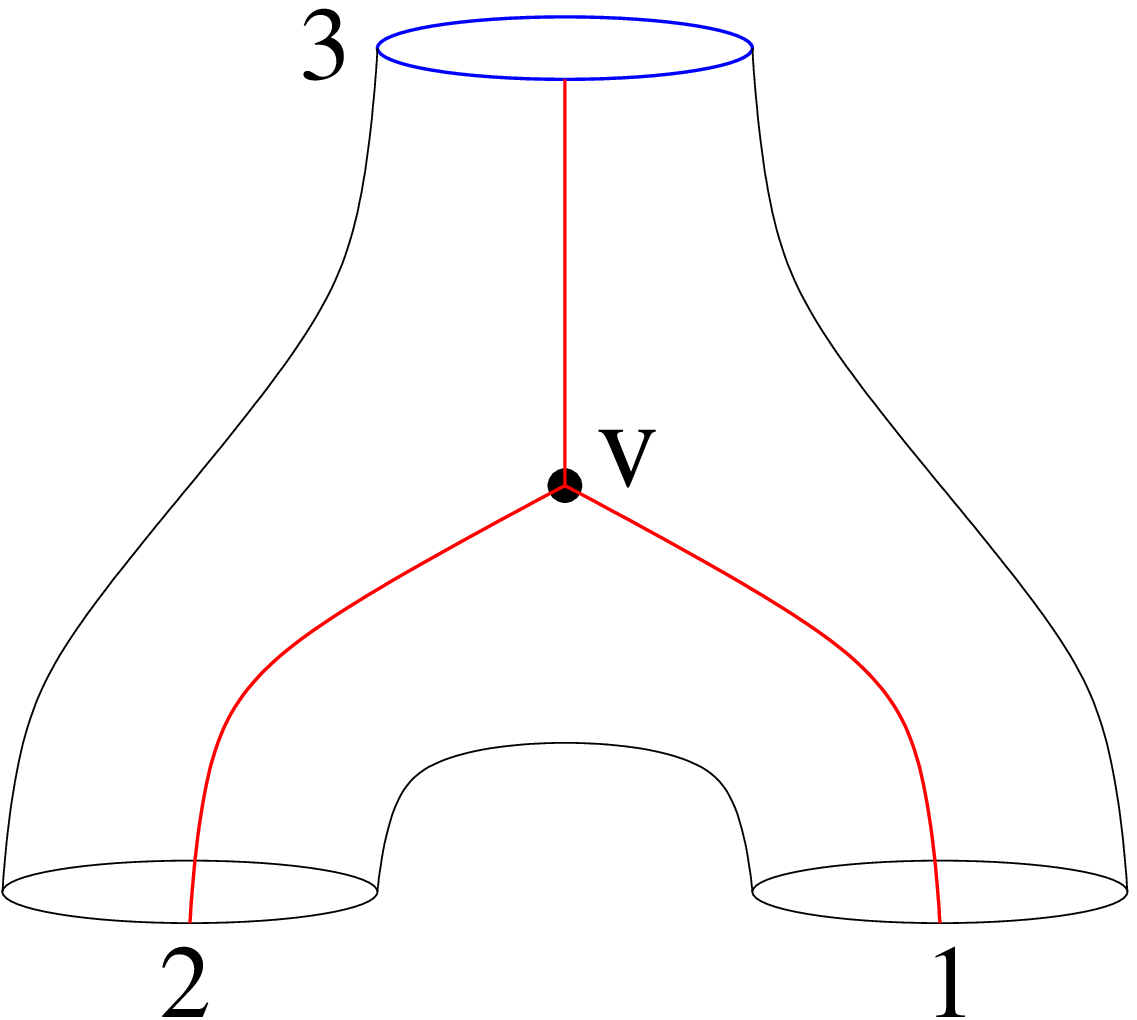,height=3cm}}\,.
\end{equation}
Let $\ga_\ep$, $\ep=1,2$ be the curves
which represent the other boundary components
of $P_{\ga}$ as 
indicated in Figure \ref{trinfig}.
Assume that $\SL_{\ga_\1}$ and $\SL_{\ga_\2}$ are already defined
and define $\SL_{\ga_\3}$ by
\begin{equation}\label{modLlem2}
\SL_{\vf,\ga}\;=\;2\cosh(\sy_v^{\2}+\sy_v^{\1})
+e^{-\sy_v^{{\2}}}\SL_{\ga_\1}^{}
+e^{\sy_v^{\1}}\SL_{\ga_\2}^{}
+e^{\sy_v^{\1}-\sy_v^{\2}}\, ,
\end{equation}
where $\sy_v^{{\ep}}$, $\ep=1,2$
are defined as
$\sy_v^{\2}=2\pi b ({\sq}_{v}+\sz_{\ga_\2})$, 
$\sy_v^{\1}=- 2\pi b ({\spp}_{v}-\sz_{\ga_\1})$.
\end{itemize}
In practise it may be necessary to use part (ii) of the definition 
recursively. 
In all remaining cases we will define the length operator
$\SL_{\vf,\ga}$ as follows: There always exists
a fat graph $\vf_0$ for which one 
of the two definitions above can be used to define 
$\SL_{\vf_0,\ga}$. Let then
\begin{equation}\label{genlength}
\SL_{\vf,\ga}:=\sa_{\vf,\vf_0}^{}(\SL_{\vf_0,\ga})\,.
\end{equation}
It was explicitly 
verified in \cite{NT} that
the definition given above is consistent.
The length operators $\SL_{\vf,\ga}$ are unambigously
defined by (i), (ii) and \rf{genlength} 
above, and we have 
$\SL_{\vf',\ga}=\sa_{\vf',\vf}^{}(\SL_{\vf,\ga})$
if $[\vf',\vf]$ represents an element of the 
Ptolemy groupoid.

The length operators satisfy the following properties:
\begin{enumerate}
\item[(a)] {\bf Spectrum:} $\SL_{\vf,\ga}$ is self-adjoint.
The spectrum of $\SL_{\vf,\ga}$ is simple and equal
to $[2,\infty)$ \cite{Ka4}. This ensures that there exists
an operator $\sll_{\vf,\ga}$ - the {\it geodesic length operator} - 
such that 
$\SL_{\vf,\ga}=2\cosh\frac{1}{2}\sll_\ga$.
\item[(b)] {\bf Commutativity:} 
\[
\big[\,\SL_{\vf,\ga}\,,\,\SL_{\vf,\ga'}\,\big]\,=\,0\quad
{\rm if}\;\; \ga\cap \ga'=\emptyset.
\]
\item[(c)] {\bf Mapping class group invariance:}
\[ 
\sa_\mu(\SL_{\vf,\ga})\,=\,\SL_{\mu.\vf,\ga},
\quad\sa_\mu\equiv\sa_{[\mu.\vf,\vf]},\quad
\text{for all}\;\;\mu\in{\rm MC}(\Sigma).
\]
\end{enumerate}
It can furthermore be shown that this definition reproduces the classical
geodesic length functions on $\CT(C)$ in the classical limit.

\subsection{The Teichm\"uller theory of the annulus}\label{sec:annulus}

As a basic building block let us develop the quantum 
Teichm\"uller theory of an annulus in some detail.
To the simple closed curve $\ga$ that can be embedded into $A$ we associate
\begin{itemize}
\item the constraint 
\begin{equation}
\sz\equiv \sz_{\vf,\ga}:=\frac{1}{2}(\spp_a-\sq_a+\spp_b)\,,
\end{equation}
\item the length operator $\SL\equiv\SL_{\vf,\ga}$,
defined as in \rf{SLsimple}.
\end{itemize}
The operator $\SL$ is positive-self-adjoint, 
The functions
\begin{equation}\label{L1-eigenf}
\phi_{s}(p):=\langle\,p\,
|\,s\,\rangle\,=\,\frac{s_b(s+p+c_b-i0)}{s_b(s-p-c_b+i0)}\,.
\end{equation}
represent the eigenfunctions of the operator $\SL$ with eigenvalue 
$2\cosh 2\pi b s$ in the representation where $\spp\equiv\spp_\ga$ is diagonal
with eigenvalue $p$.
It was shown in \cite{Ka4} that the family of eigenfunctions $\phi_{s}(p)$, 
$s\in\BR^+$,
is delta-function orthonormalized and
complete in $L^2(\BR)$,
\begin{subequations}
\begin{align}
&\int_{\BR_+}dp\;\langle\,s\,|\,p\,\rangle
\langle\,p\,|\,s'\,\rangle\,=\,\de(s^{}-s')\,.\\
&\int_{\BR_+}d\mu(s)\;\langle\,p\,|\,s\,\rangle\langle\,s\,|\,p'\,\rangle\,=\,\de(p^{}-p')\,,
\end{align}
\end{subequations}
where the Plancherel measure $\mu(s)$ is defined as
$d\mu(s)=2\sinh(2\pi bs)2\sinh(2\pi b^{-1}s)ds$.

For later use let us construct the change of representation from 
the representation in which $\spp_a$ and $\spp_b$ are diagonal
to a representation where $\sz$ and $\SL$ are diagonal.
To this aim let us introduce $\sd:=\frac{1}{2}(\sq_a+\spp_a-\spp_b+2\sq_b)$.
We have 
\[
\begin{aligned}
{[}{\sz},\sd]=(2\pi i)^{-1}\,,\\
[\spp,\sq]=(2\pi i)^{-1}\,,
\end{aligned}\qquad
\begin{aligned}
&[\sz,\spp]=0\,,\quad
[\sz,\sq]=0\,,\\
&[\sd,\spp]=0\,,\quad
[\sd,\sq]=0\,.
\end{aligned}
\]
Let $\langle\,p,z\,|$ be an eigenvector of $\spp$ and $\sz$ with
eigenvalues $p$ and $z$, and
$|\,p_a,p_b\,\rangle$ an eigenvector of $\spp_a$ and $\spp_b$
with eigenvalues $p_a$ and $p_b$, respectively.
It follows easily that 
\begin{equation}
\langle\,p,z\,|\,p_a,p_b\,\rangle\,=\,\de(p_b-z+p)e^{\pi i(p+z-p_a)^2}\,.
\end{equation}
The transformation 
\begin{equation}\label{holediag}
\psi(s,z)\,=\,\int_{\BR^2}dpdp_a\;\frac{s_b(s-p+c_b-i0)}{s_b(s+p-c_b+i0)}
e^{\pi i(p+z-p_a)^2}\Psi(p_a,z-p)\,,
\end{equation}
will then map a wave function $\Psi(p_a,p_b)$ in the representation
which diagonalizes $\spp_a$, $\spp_b$ to the 
corresponding wave function $\psi(s,z)$ in the representation
which diagonalizes $\SL$ and $\sz$.

\subsection{Teichm\"uller theory for surfaces with holes}\label{ssec:hole}

The formulation of quantum Teichm\"uller theory introduced above
has only punctures (holes with vanishing geodesic circumference) 
as boundary components. In order to generalize to holes of 
non-vanishing geodesic circumference one may represent each hole
as the result of cutting along a geodesic surrounding a pair of
punctures.
\[
\lower.9cm\hbox{\epsfig{figure=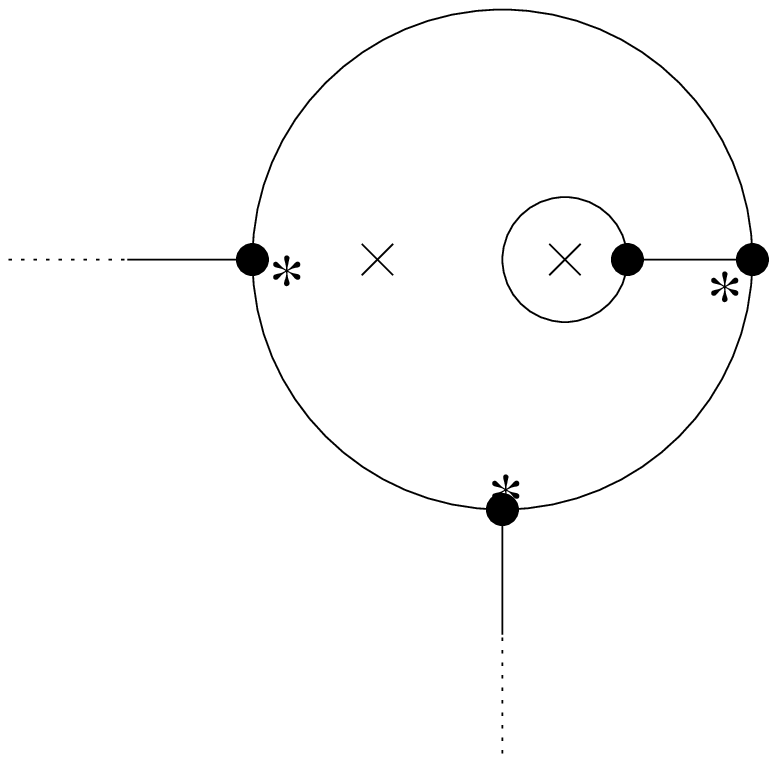,height=3cm}}
\quad
\begin{aligned}
& \text{Example for a} \\
& \text{fat graph in the} \\
& \text{vicinity of two} \\
& \text{punctures (crosses)}\\
& \text{}
\end{aligned}
\qquad
\lower.9cm\hbox{\epsfig{figure=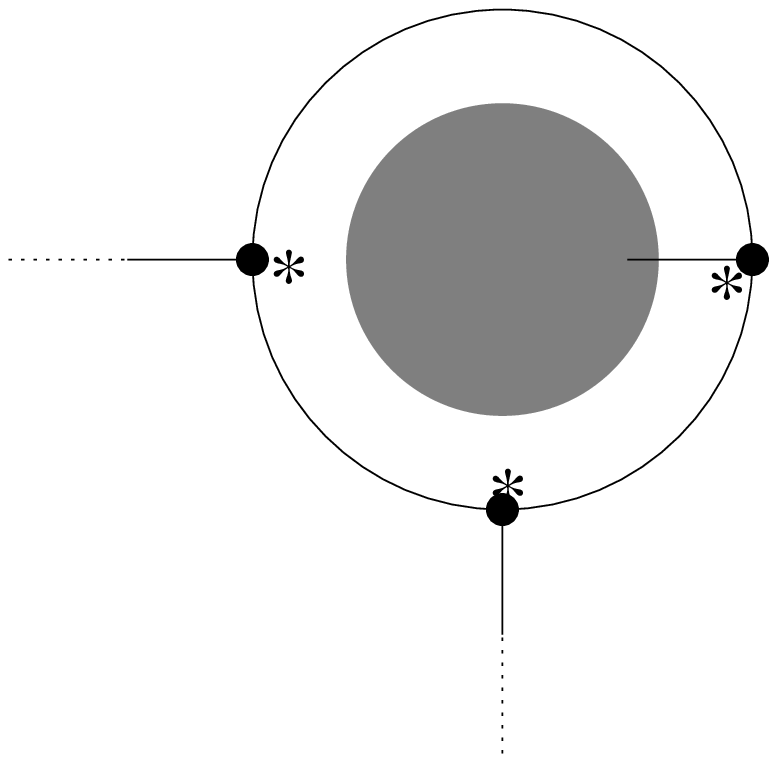,height=3cm}}
\quad
\begin{aligned}
& \text{The same fat graph} \\
& \text{after cutting} \\
& \text{out the hole}\\
& \text{}
\end{aligned}
\]
On a surface $C$ with $n$ holes one may choose $\vf$ to have the following
simple standard form near at most $n-1$ of the holes, which will
be called ``incoming'' in the following:
\begin{equation}\label{holefig}
\lower.9cm\hbox{\epsfig{figure=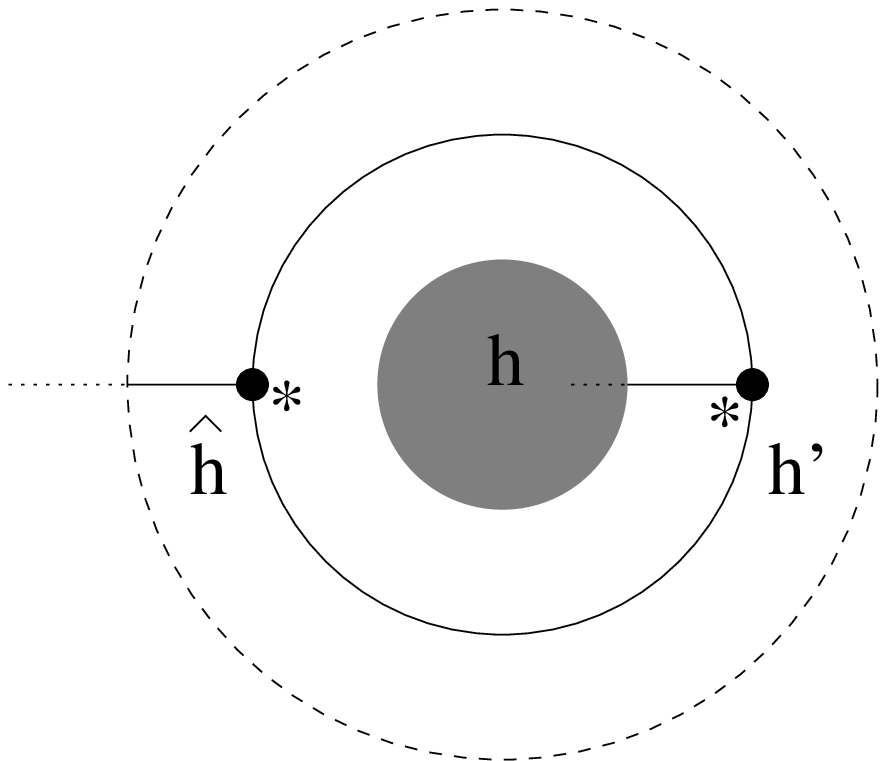,height=3cm}}
\qquad\quad
\begin{aligned}
& \text{Incoming boundary component:} \\
& \text{Hole $h$ (shaded), together with an annular} \\
& \text{neighborhood $A_h$ of $h$ inside $C$, and}\\
& \text{the part of $\vf$ contained in $A_h$.}
\end{aligned}
\end{equation}
The price to pay is a more complicated representation of the 
closed curves which surround the remaining holes.

The simple form of the fat graph near the incoming 
boundary components allows us to use the 
transformation \rf{holediag} 
to pass to a representation where the 
length operators and constraints
associated to these holes are diagonal.
In order to describe the resulting hybrid representation
let us denote by $s_b$ and $z_b$ the assignments of values 
$s_h$ and $z_h$ to each 
incoming hole $h$, while $p$ assigns real numbers $p_v$ to all
vertices $v$ of $\vf$ which do not coincide with any vertex
$\hat{h}$ or $h'$ associated to an incoming hole $h$.
The states will then be described by wave-functions 
$\psi(p;s_b,z_b)$ on which the operators $\SL_{h}$ and $\sz_{h}$
act as operators of multiplication by $2\cosh2\pi bs_h$ and 
$z_h$, respectively. 

For a given hole $h$ one may define a projection $\hat{\CH}(C_{h(s,z)})$ 
of $\hat{\CH}(C)$ to the eigenspace with fixed eigenvalues $2\cosh 2\pi bs$ 
and $z$ of $\SL_h$ and $\sz_h$. States in $\hat{\CH}(C_{h(s,z)})$ 
can be represented by wave-functions $\psi_h(p_h)$, where $p_h$ assigns real
values to all vertices in $\vf_0\setminus\{\hat{h},h'\}$.
The mapping class
action on $\hat{\CH}(C)$ commutes with $\SL_h$ and $\sz_h$. It follows that
the operators $\SM_{\mu}\equiv\SM_{\mu.\vf,\vf}$ representing 
the mapping class group action on $\hat\CH(C)$ project to 
operators  $\SM_{s,z}(\mu)$ generating an action of
${\rm MCG}(C)$ on $\hat\CH(C_{h(s,z)})$.

\subsection{Passage to the length representation}

Following \cite{T05}, we will now describe how to map a maximal commuting 
family of length operators to diagonal form. 
We will start from the hybrid representation described above in which the
length operators and constraints associated to the incoming holes
are diagonal. Recall that states are represented by wave-functions
$\psi(p;s_b,z_b)$ in such a representation, where $p:\tilde{\vf}_0\mapsto\BR$,
and $\tilde{\vf}_0$ is the subset of $\vf_0$ that does not 
contain $\hat{h}$ nor $h'$ for any incoming hole $h$.
A maximal commuting 
family of length operators is associated to a family of simple closed 
curves which define a pants decomposition.

\subsubsection{Adapted fat graphs}

Let
us consider a decorated Moore-Seiberg graph $\si$ on $C$, 
the decoration being the choice of an distinguished boundary
component in each trinion of the pants decomposition defined 
by $\si$. The distinguished boundary component will be called outgoing,
all others incoming.

Such a graph $\si$ allows us to define a cutting of $C$ 
into annuli and trinions. If cutting along a  
curve $\ga$ in the cut system $\CC_\si$ produces two incoming boundary 
components, let $\ga^\pm$ be two curves bounding 
a sufficiently small annular neighborhood $A_{\ga}$ 
of $\ga$ in $C$. Replacing $\ga$ by $\{\ga^+,\ga^-\}$ for
all such curves $\ga$ produces an extended cut system $\hat\CC_\si$
which decomposes $C$ into trinions and annuli. 

Let us call a pants decomposition $\si$ admissible if no
curve $\ga_e\in\CC_\si$ is an outgoing boundary component for the two 
trinions it may separate.  To admissible pants decompostions $\si$
we may associate
a natural fat graph $\vf_\si$  
defined by gluing the following pieces:
\begin{itemize}
\item Annuli: See Figure \rf{annfig}.
\item Trinions: See Figure \rf{trinfig}.
\item Holes: See Figure \rf{holefig}.
\end{itemize}
Gluing these pieces in the obvious way will produce the
connected graph $\vf_\si$ adapted to the 
Moore-Seiberg graph $\si$ we started from. The restriction 
to admissible fat graphs turns out to be inessential \cite{NT}.

\subsubsection{The unitary map to the length representation}

To each vertex 
$v\in\vf_{\si,0}$ assign the length operator $\SL^\2_v$ and $\SL^\1_v$
to the incoming and $\SL_v$ to the outgoing boundary components of the 
pair of pants $P_v$ containing $v$.
The main ingredient will be an operator $\SC_v$ which maps $\SL_v$ to a simple 
standard form,
\begin{equation}
\SC_v\cdot\SL_v\cdot(\SC_v)^{-1}\,=\,2\cosh 2\pi b \spp_v+e^{-2\pi b \sq_v}\,.
\end{equation}
Such an operator can be constructed explicitly as \cite{T05}
\begin{equation}\label{SCdef}
\SC_v:=\,e^{-2\pi i s_\2\sq_v} \,\frac{e_b(\mss^\1_v+\spp_v)}{e_b(\mss^\1_v-\spp_v)} \,
e^{-2\pi i \mss^\1_v\spp_v}\,(e_b(\sq_v-\mss^\2_v))^{-1}\,
e^{-2\pi i(\sz_v^\2\spp_v+\sz_v^\1\sq_v)},
\end{equation}
where $\mss_v^\imath$, $\imath=\1,\2$ are the 
positive self-adjoint operators defined by 
$\SL_v^\imath=2\cosh2\pi b\mss_v^{\imath}$, and $\sz^\2_v$, $\sz^\1_v$
are the constraints associated to the incoming boundary components
of $P_v$. 

The map to the length representation is then constructed 
as follows. Let us first apply the product of the
transformations \rf{holediag} that
diagonalizes the length operators associated to all incoming holes and
embedded annuli. The resulting hybrid
representation has states represented by wave-functions
$\psi(p;s_{\rm\sst A},z_{\rm\sst A})$, where $p$ assigns a real
number $p_v$ to each vertex of $\Ga_\si$, 
whereas $s_{\rm\sst A}$ (resp. $z_{\rm\sst A}$) 
assigns real positive numbers (resp. real numbers) to all\footnote{both embedded annuli and annuli representing incoming boundary components} 
annuli, respectively.

In order to diagonalize all length operators associated to 
all edges of the MS graph $\Ga_\si$ it remains to apply
an ordered product the operators $\SC_v$.
The resulting operator may be represented as the following explicit 
integral transformation: Let $s$ be 
the assignment of real positive numbers $s_e$ to all
edges $e$ of $\Ga_\si$. Define
\begin{align}\label{PF-length}
\Phi(s,z_{\rm\sst A})\,=\,\int\limits_{\BR^{h}} 
\Big({\prod}_{v\in\tilde\vf_0} dp_v\; 
K_{s^{\2}_vs^{\1}_v}^{z^{\2}_vz^{\1}_v}(s_v,p_v)\Big)\,
\psi(p;s_{\rm\sst A},z_{\rm\sst A})\,.
\end{align}
The kernel $K_{s_{\2}s_{\1}}^{z_{\2}z_{\1}}(s,p)$ has the 
following explicit form \cite{NT}
\begin{align}
K_{s_{\2}s_{\1}}^{z_{\2}z_{\1}}(s,p)
& \,=\,\zeta_{\0}\int_\BR dp'\;
{e^{-2\pi i(s_\2-c_b)(s_\2+p'-p+z_\1)}}{e_b(p-z_\1-s_\2-p'+c_b)}
\notag\\
&\hspace{4cm}\times\frac{s_b(s_\1-p'-s_\2)}{s_b(s_\1+p'+s_\2)}
\frac{s_b(s+p'-c_b)}{s_b(s-p'+c_b)}e^{-2\pi i z_\2(2p-z_\1)}
 \,.
\notag\end{align}
The explicit integral transformation \rf{PF-length} defines 
an operator $\hat\SC_\si$. In order to 
get an operator $\SC_\si$ which maps the
representation $\pi_{\vf_\si}^{\CT}$ 
for the quantum Teichm\"uller theory based on the 
Penner-Fock coordinates to the representation
$\pi_{\si}^{}$ defined in this paper
it suffices to compose $\hat\SV_\si$ with the projection 
$\Pi$ defined as
$\phi(s)\equiv(\Pi\Phi)(s):=\Phi(s,0)$.
This corresponds to imposing 
the constraints $\sz_{\vf,\ga}\simeq 0$.

\subsubsection{Changes of MS-graph}

The construction above canonically defines operators $\SU_{\si_\2\si_\1}$ 
intertwining between the representations $\pi_{\si_\1}$ and $\pi_{\si_\2}$ as
\begin{equation}\label{SWtoSU}
\SU_{\si_\2\si_\1}:=\hat\SC_{\si_\2}^{}\cdot\SW_{\vf_{\si_\2}\vf_{\si_\1}}\cdot\hat\SC_{\si_\1}^{-1}\,,
\end{equation}
where $\SW_{\vf_{\si_\2}\vf_{\si_\1}}$ is any operator representing the
move $[\vf_{\si_\2},\vf_{\si_\1}]$ between the fat graph associated to 
$\si_\1$ and $\si_\2$, respectively. In this way one defines
operators  $\SB$, $\SF$, $\SZ$ and $\SS$  associated to the
elementary moves between different MS-graphs. These operators
satisfy operatorial versions of the Moore-Seiberg consistency conditions
\cite{T05,NT}, which follow from the relations of the 
Ptolemy groupoid \rf{q-Ptolemyrels} using \rf{SWtoSU}.

\section{Completing the proofs}\label{Proofs2}

\setcounter{equation}{0}

In order to prove the consistency of the quantization 
of $\CM_{\rm flat}^0(C)$ defined in Section \ref{q-Mflat}
we will take the results of \cite{T05} reviewed in the 
previous Section \ref{q-Teich} as a starting point.
It remains to 
\begin{itemize}
\item[(i)] calculate the kernels of the operators $\SF$, $\SB$, $\SZ$
and $\SS$,
\item[(ii)] calculate the explicit form of the difference operators
$\SL_t^e$ in this representation, and
\item[(iii)] calculate the central extension of the Moore-Seiberg groupoid.
\end{itemize}
The solution of these tasks will be described in this section.

\subsection{The Moore-Seiberg groupoid for surfaces of genus $0$}

To begin, let us note that
the kernels of the operators $\SF$, $\SB$
and $\SZ$ have been calculated in \cite{NT}, giving the results
stated in Section \ref{q-Mflat}.  

The key observation \cite{NT} leading to the explicit
calculation of the kernels of $\SF$, $\SZ$ and $\SB$ 
is the fact that 
the operators $\SC_v$ defined in \rf{SCdef} are closely related 
to the Clebsch-Gordan maps of the modular double of $\CU_q(\fsl(2,\BR))$
\cite{PT2}. This observation
implies directly that the matrix elements of the operator
$\SF$ must coincide with the b-$6j$ symbols of \cite{PT2}.
Fixing a suitable normalization and using the
alternative integral representation found in \cite{TV} 
one gets precisely formula
\rf{Fvs6j}.

One may furthermore use the results of \cite{BT1} to prove that
the operator $\SB$ acts diagonally with eigenvalue given in 
\rf{Bcoeff}. For more details we may refer to \cite{NT}.

\subsection{Preparation I -- Alternative normalizations}\label{Sec:C-norm}

The representation for $\CA_b(C)$ constructed in Section \ref{q-Mflat}
has a severe draw-back: The appearance of square-roots in the 
expressions for the loop operators and for the kernels of $\SU_{\si_\2\si_\1}$
obscures some beautiful and profound analytic properties that
will later be found to have important consequences. We
shall therefore now introduce useful alternative normalizations
obtained by writing 
\begin{equation}\label{v-renorm}
v^{\al_\3}_{\al_\2\al_\1}\,=\,\varrho(\al_3,\al_2,\al_1)
\tilde{v}^{\al_\3}_{\al_\2\al_\1}\,,
\end{equation}
and taking $\tilde{v}^{\al_\3}_{\al_\2\al_\1}$
as the new basis vector for $\CH^{\al_\3}_{\al_\2\al_\1}$. 
It will be useful
to consider vectors  $\tilde{v}^{\al_\3}_{\al_\2\al_\1}$ that may have a norm
different from unity. It will be useful to consider, in particular,
\begin{equation}
\varrho(\al_3,\al_2,\al_1)=\sqrt{C(\bar\al_3,\al_2,\al_1)}\,,
\end{equation}
where 
$\bar\al_3=Q-\al_3$, and $C(\al_3,\al_2,\al_1)$ is the function defined 
as
\begin{align}\label{ZZform}
C(\alpha_1, & \alpha_2,\alpha_3)=\left[\pi\mu\gamma(b^2)b^{2-2b^2}
\right]^{(Q-\sum_{i=1}^3\alpha_i)/b}\times\\
& \times \frac{\Upsilon_0\Upsilon(2\alpha_1)\Upsilon(2\alpha_2)
\Upsilon(2\alpha_3)}{
\Upsilon(\alpha_1+\alpha_2+\alpha_3-Q)
\Upsilon(\alpha_1+\alpha_2-\alpha_3)\Upsilon(\alpha_2+\alpha_3-\alpha_1)
\Upsilon(\alpha_3+\alpha_1-\alpha_2)}.
\notag\end{align}
The
expression on the right hand side of \rf{ZZform} is constructed
out of the
special function $\up(x)$ which is related to the Barnes
double Gamma function $\Ga_b(x)$ 
as $\up(x)=(\Ga_b(x)\Ga_b(Q-x))^{-1}$. The function
$C(\alpha_1,\alpha_2,\alpha_3)$ is known to be the expression for the 
three-point function in Liouville theory, as was conjectured in 
\cite{DO,ZZ}, and derived in \cite{Teschner:2001rv}. 

Note that the gauge transformation defined by \rf{v-renorm}
will modify the kernels representing the elementary moves of the 
MS groupoid. In the representation defined via \rf{v-renorm}
one may represent the F-move, for example, 
by the kernel 
\begin{equation}
\FusC{\al_1}{\al_2}{\al_3}{\al_4}{\be_{1}}{\be_2}
\,=\,\frac{\varrho(\al_4,\al_t,\al_1)\varrho(\al_t,\al_3,\al_2)}{\varrho(\al_4,\al_3,\al_s)\varrho(\al_s,\al_2,\al_1)}
\Fus{\al_1}{\al_2}{\al_3}{\al_4}{\be_{1}}{\be_2}\,.
\end{equation}

We'd like to stress that the appearance of the function 
$C(\alpha_1,\alpha_2,\alpha_3)$ 
can be motivated without any reference to Liouville 
theory by the intention to make important
analytic properties of the kernels representing $\SF$ and $\SS$ 
more easily visible. One may note, in particular, that
$S_b(x)=\Ga_b(x)/\Ga_b(Q-x)$, from which it is easily seen that
the change of normalization removes all square-roots from
the expressions for 
$\Fus{\al_1}{\al_2}{\al_3}{\al_4}{\al_s}{\al_t}$.
The kernel
$\FusC{\al_1}{\al_2}{\al_3}{\al_4}{\al_s}{\al_t}$ is then found to be 
meromorphic in 
all of its arguments. 
A more complete summary of the relevant analytic properties
will be given in the following Subsection \ref{anaprop} below.

\subsection{Preparation II -- Analytic properties} \label{anaprop}

The kernels representing the operators $\SF$ and $\SS$ have remarkable
analytic properties which will later be shown to have 
profound consequences. The origin of the analytic properties 
can be found in the structure of $\CM_{\rm flat}^{\BC}(C)$ as an algebraic variety.
The simple form of the relations describing 
 $\CM_{\rm flat}^{\BC}(C)$ as an algebraic variety implies nice analytic 
properties of the expressions 
for the loop operators in terms of the Darboux coordinates,
and this leads to nice analytic 
properties of the kernels $A_{\si_\2\si_\1}(l_\2,l_\1)$ via 
\rf{diffinter}.

We will here summarize some of
the most important properties.

\subsubsection{Symmetries}

The kernel representing $\SF$ has a large group of 
symmetries. We will state them in the normalization which makes
the realization of the respective symmetries most manifest.
\begin{itemize}
\item {\it Tetrahedral symmetries}:
The coefficients $\big\{\,{}^{\al_1}_{\al_3}\,{}^{\al_2}_{\al_4}\,{}^{\al_s}_{\al_t}\big\}_b^{}$ satisfy the tetrahedral symmetries
\begin{equation}
\big\{\,{}^{\al_1}_{\al_3}\,{}^{\al_2}_{\al_4}\,{}^{\al_s}_{\al_t}\big\}_b^{}
=\big\{\,{}^{\al_2}_{\al_4}\,{}^{\al_1}_{\al_3}\,{}^{\al_s}_{\al_t}\big\}_b^{}
=\big\{\,{}^{\al_2}_{\al_4}\,{}^{\al_s}_{\al_t}\,{}^{\al_1}_{\al_3}\big\}_b^{}
=\big\{\,{}_{\al_1}^{\al_3}\,{}_{\al_2}^{\al_4}\,{}_{\al_t}^{\al_s}\big\}_b^{}\,.
\end{equation}
\item {\it Weyl symmetries}: The kernel 
$\FusC{\al_1}{\al_2}{\al_3}{\al_4}{\al_s}{\al_t}$ is symmetric
under all reflections $\al_i\ra Q-\al_i$, $i\in\{1,2,3,4,s,t\}$.
\end{itemize}
The tetrahedral symmetries are easily read off from the integral representation
\rf{6j3}. The derivation of the Weyl symmetries can be done 
with the help of the alternative integral representation \rf{6j1}.

Similar properties hold for the kernel representing the operator
$\SS$.
\begin{itemize}
\item {\it Permutation symmetry}:
The coefficients $S_{\al_\1,\al_\2}^{}(\al_\0)$ satisfy
\begin{equation}
\De(\al_\1,\al_\0,\al_\1)
S_{\al_\1\al_\2}^{}(\al_\0)\,=\,\De(\al_\2,\al_\0,\al_\2)
S_{\al_\2\al_\1}^{}(\al_\0)\,.
\end{equation}
\item {\it Weyl symmetries}: The kernel 
$S_{\al_\1,\al_\2}^{C}(\al_\0)$ is symmetric
under all reflections $\al_i\ra Q-\al_i$, $i\in\{\0,\1,\2\}$.
\end{itemize}
In other normalizations one will of course 
find a slightly more complicated realization of these symmetries.

\subsubsection{Resonances and degenerate values}\label{Sec:Res}

We will now summarize some of the most important facts concerning
poles, residues and special values of the intertwining kernels. 
Proofs of the statements
below are given in Appendix \ref{6jApp}.

Important simplifications are found for particular values of the arguments. 
Each $\al_i$, $i\in\{1,2,3,4\}$ is member of two out of the four triples
$T_{12s}:=(\al_1,\al_2,\al_s)$, $T_{34s}:=(\al_3,\al_4,\al_s)$,
$T_{23t}:=(\al_2,\al_3,\al_t)$, $T_{14t}:=(\al_1,\al_4,\al_t)$.
We will say that a triple $T_{ijk}$ is resonant if there exist $\ep_i\in\{+1,-1\}$ 
and $k,l\in\BZ^{\geq 0}$
such that
\begin{equation}
\ep_1\big(\al_3-\fr{Q}{2}\big)\,=\,\ep_2\big(\al_2-\fr{Q}{2}\big)+
\ep_3\big(\al_1-\fr{Q}{2}\big)+\fr{Q}{2}+kb+lb^{-1}\,.
\end{equation}
Poles in the variables $\al_i$, $i\in\{1,2,3,4\}$ will occur only if 
one of the triples $T_{12s}$, $T_{34s}$,
$T_{23t}$, $T_{14t}$ is resonant. The location of poles is simplest to describe
in the case of $F^{{\sst C}}$, which has poles in $\al_i$, $i\in\{1,2,3,4\}$
if and only if either $T_{t32}$ or $T_{4t1}$ are resonant. 

Of particular importance will be the 
cases where one of $\al_i$, $i\in\{1,2,3,4\}$
takes one of the so-called degenerate values
\begin{equation}
\al_i\,\in\,\BD\,,\qquad\BD:=\{\,\al_{nm}\,,n,m\in\BZ^{\geq 0}\,\}\,,\qquad
\al_{nm}:=-nb/2-mb/2\,.
\end{equation}
Something remarkable may happen under this
condition if the triple containing both $\al_i$ and $\al_s$
becomes resonant: Let us 
assume that $\al_s\in\BF_{nm}(\al_j)$, where
\begin{equation}
\BF_{nm}(\al_j)\,=\,\big\{\,\al_j-(n-k)\fr{b}{2}-(m-l)\fr{1}{2b}\,,
k=0,2,\dots,2n\,,l=0,2,\dots,2m\,\big\}\,.
\end{equation}
The kernel $F^{C}$ becomes proportional to   
a sum of delta-distributions supported on resonances of the triple
containing both $\al_t$ and $\al_i$ as expressed in the formulae 
\begin{subequations}
\label{degval}
\begin{align}\label{al1deg}
&\lim_{{\al_1\ra\al_{nm}}}
\FusC{\al_1}{\al_2}{\al_3}{\al_4}{\al_s}{\al_t}_{{\al_{21}^s\in
\BF_{nm}(\al_2)}}^{}\,=\,
\sum_{\be_t\in\BF_{nm}(\al_4)}\de(\al_t-\be_t)
\fus{\underline{\al}_1}{\al_2}{\al_3}{\al_4}{\al_s}{\be_t} \\
&\lim_{{\al_2\ra\al_{mn}}}
\FusC{\al_1}{\al_2}{\al_3}{\al_4}{\al_s}{\al_t}_{\al_{21}^s\in\BF_{nm}(\al_1)}^{}\,=\,
\sum_{\be_t\in\BF_{nm}(\al_3)}\de(\al_t-\be_t)
\fus{{\al}_1}{\underline{\al}_2}{\al_3}{\al_4}{\al_s}{\be_t}\,, 
\label{al2deg}\end{align}
\end{subequations}
and similarly for $\al_3$ and $\al_4$. 
The delta-distributions 
$\de(\al_t-\be_t)$
on the right of
\rf{degval} are to be understood as complexified versions 
of the usual delta-distributions.
$\de(\al-\be)$ is defined to be the linear functional
defined on spaces $\CT$ of entire analytic test function $t(\al)$ as
\begin{equation}\label{deltadef}
\big\langle\,\de(\al-\be)\,,\,t\,\big\rangle\,=\,t(\be)\,,
\end{equation}
with $\langle\de(\al-\be),t\rangle:\CT'\times\CT\ra \BC$ being the
pairing between $\CT$ and its dual $\CT'$. The identities \rf{degval}
are likewise understood as identities between distributions on $\CT$.

\subsection{Intertwining property}

In this subsection we are going to describe a quick way
to prove that the unitary operators defined in Subsection 
\ref{sec:elmoves} correctly map the representations 
$\pi_{\si_1}$ to $\pi_{\si_2}$, as expressed in equations 
\rf{inter}. The proof we will give here 
exploits the remarkable analytic properties of the 
fusion coefficients $\Fus{\al_1}{\al_2}{\al_3}{\al_4}{\al_s}{\al_t}$
summarized in Subsection \ref{anaprop}.

One may, in particular, use the relations \rf{al1deg} in order
to derive from the pentagon relation \rf{pentagon} systems
of difference equations relating  
$\Fus{\al_1}{\al_2}{\al_3}{\al_4}{\al_s}{\al_t}$ to the
residues of its poles, like for example
\begin{align}\label{smallpentagon}
 \sum_{\be_5\in\BF_{nm}(\be_2)} 
\fus{\underline{\al}_1}{\al_2}{\al_3}{\be_{2}}{\be_{1}}{\be_5}
& \fus{\underline{\al}_1}{\be_5}{\al_4}{\al_5}{\be_{2}}{\be_{4}}
\Fus{\al_2}{\al_3}{\al_4}{\be_{4}}{\be_5}{\be_{3}} 
=\fus{\underline{\al}_1}{\al_2}{\be_3}{\al_{5}}{\be_{1}}{\be_{4}}
\Fus{\,\be_1}{\al_3}{\al_4}{\al_5}{\be_{2}}{\be_{3}}\,,
\end{align}
valid for $\al_1=\al_{nm}$, $\be_1\in\BF_{nm}(\al_2)$ and 
$\be_4\in\BF_{nm}(\al_5)$. Similar equations can be derived
for $\al_i=\al_{nm}$, $i=2,3,4,5$.

Further
specializing \rf{smallpentagon} to $\al_1=-b$ and 
$\be_1=\al_2$, $\be_4=\al_5$, 
for example, yields a somewhat simpler difference equation
of the form
\begin{align}\label{smallpentagon2}
 \sum_{k=-1}^1 \dus{\,\al_2}{\al_3}{\al_4}{\al_5}{\be_{2}}{\be_{3}}
\,\Fus{\al_2}{\al_3}{\al_4}{\al_5}{\be_2+kb,}{\be_{3}} 
=0\,.
\end{align}
The coefficients $\dus{\,\al_2}{\al_3}{\al_4}{\al_5}{\be_{2}}{\be_{3}}$ 
are given as
\begin{equation}
\dus{\,\al_2}{\al_3}{\al_4}{\al_5}{\be_{2}}{\be_{3}}
=\fus{-b}{\al_2}{\al_3}{\be_{2}}{\al_2}{\be_2+kb}
\fus{-b}{\be_2+kb}{\al_4}{\al_5}{\be_{2}}{\al_5}-
\de_{k,0}\fus{\underline{\al}_1}{\al_2}{\be_3}{\al_{5}}{\al_2}{\al_5}\,.
\end{equation}
By carefully evaluating the relevant residues of 
$\Fus{\al_1}{\al_2}{\al_3}{\al_4}{\al_s}{\al_t}$ (see Appendix \ref{App:res} 
for a list of the relevant results) one may
show that \rf{smallpentagon2} is equivalent to 
the statement \rf{diffinter} that the fusion transformation 
correctly intertwines the representation $\pi_{\si_s}$ 
and $\pi_{\si_t}$ of $\CL_t$ associated
to the pants decompositions $\si_s$ and $\si_t$, respectively.
Alternatively one may use this argument in order to
{\it compute} the explicit form of the operator $\SL_t$ in 
the representation  where $\SL_s$ is diagonal.

\subsection{The S-kernel and the central extension}\label{sec:Skern}

It remains to calculate the kernel of $\SS$  
and the central extension, as parameterized by the 
real number $\chi_b^{}$ depending on the deformation 
parameter $b$ in \rf{g=1,n=2}. One way to calculate the kernel
of $\SS$ directly within quantum Teichm\"uller theory is described 
in Appendix \ref{SfromTeich}. We conclude that 
the operators $\SB$, $\SF$, $\SS$ are all represented by 
kernels that depend {\it meromorphically} on all their variables.

We had noted above that
the operators $\SB$, $\SF$, $\SZ$ and $\SS$ 
satisfy the operatorial form of the Moore-Seiberg consistency 
equations up to 
projective phases \cite{T05}. Being represented by meromorphic
kernels, this implies the validity of \rf{MS-eqns} up to 
projective phases. One may then use special cases  of
\rf{degval}, like
\begin{equation}
\FusC{\al_1}{0}{\al_3}{\al_4}{\al_1}{\be_t}\,=\,\de(\be_t-\al_3)
\end{equation}
in order to 
check that the relations \rf{hexagon}, 
\rf{pentagon} and \rf{g=1,n=2} have to hold {\it identically},
not just up to a phase. All but one of the 
remaining projective phases can be
eliminated by a redefinition of the generators.
We have chosen to parameterize the remaining phase by means of 
the real number $\chi_b^{}$ which appears in the relation 
\rf{g=1,n=1b}. This is of
course conventional, redefining the the kernels
by a phase would allow one to move the phase from relation
\rf{g=1,n=1b} to other relations. Our convention will
turn out to be natural in Part III of this paper. The explicit
formula for the phase $\chi_b^{}$ will be determined below.

In order to derive a formula for $S_{\be_\1\be_\2}(\al_0)$ 
we may then consider 
the relation \rf{g=1,n=2} in the special case $\al_1=\al_2$
and take the limit where 
$\be_1$ and $\be_3$ are sent to zero. The details are somewhat
delicate. We will here give an outline of the argument, with 
more details given in Appendix \ref{App:S}. 
It turns out to be necessary to 
send $\be_1$ and $\be_3$ to zero simultaneously. 
One will find a
simplification of  relation \rf{g=1,n=2} in this limit 
due to the 
relation
\begin{equation}\label{limitF}
\lim_{\ep\downarrow 0}
{{\rm F}_{\ep,\al_\3}^{\rm\sst L} 
\big[ {}_{\ep\;}^{\ep\;} {}_{\al_\1}^{\al_\1} \big]}
\,=\,\de(\al_3-\al_1)\,.
\end{equation}
Using equation \rf{limitF} it becomes straightforward
to take $\be_1=\be_3=\ep$ and send $\ep\ra 0$ 
in the relation \rf{g=1,n=2}, leading to
\begin{equation}\label{S-MSform}\begin{aligned}
F_{0\al}^{\rm\sst L} 
\big[ {}_{\be_\1}^{\be_\1} {}_{\be_\1}^{\be_\1} \big]
& {S_{\be_\1\be_\2}^{\rm\sst L}(\al)} = 
S_{0\be_2}^{\rm\sst L}
\int_\BS d\be_\3 \;e^{-\pi i(2\De_{\be_\2}+2\De_{\be_\1}-2\De_{\be_\3}-\De_\al)}\;
{{\rm F}_{0\be_\3}^{\rm\sst L} \big[ {}_{\be_\2}^{\be_\2} {}_{\be_\1}^{\be_\1} \big]}
{\rm F}_{\be_\3\al}^{\rm\sst L} \big[ {}_{\be_\2}^{\be_\1} {}_{\be_\2}^{\be_\1} \big]
\,,
\end{aligned}\end{equation}
where $S_{0\be}^{\rm\sst L}:=\lim_{\ep\ra 0}S_{\ep\be}^{\rm\sst L}(\ep)$.
This formula determines ${S_{\be_\1\be_\2}^{\rm\sst L}(\al)}/S_{0\be_2}^{\rm\sst L}$ in terms of 
${\rm F}_{\be_s\be_t}^{\rm\sst L} 
\big[ {}_{\be_\4}^{\be_\3} {}_{\be_\1}^{\be_\2} \big]$. 
In Appendix \ref{app:S-kernel} it is shown that
one may evaluate the integral in \rf{S-MSform} explicitly,
leading to the formula  
\begin{align}\label{Skern2}
S_{\be_\1\be_\2}^{\rm\sst L} (\al_0)=\,& S_{0\be_\2}^{\rm\sst L} \,
\frac{N(\be_\1,\al_0,\be_\1)}{N(\be_\2,\al_0,\be_\2)}\,
\frac{e^{\frac{\pi i}{2}\De_{\al_0}}}{S_b(\al_0)}\times \\
&\times
\int_\BR dt\;e^{2\pi t(2\be_\1-Q)}
\frac{S_b\big(\frac{1}{2}(2\be_\2+\al_0-Q)+it\big)
S_b\big(\frac{1}{2}(2\be_\2+\al_0-Q)-it\big)}
{S_b\big(\frac{1}{2}(2\be_\2-\al_0+Q)+it\big)
S_b\big(\frac{1}{2}(2\be_\2-\al_0+Q)-it\big)}\,,
\notag\end{align}
where $N(\al_3,\al_2,\al_1)$ 
is defined as
\begin{align}\label{N-def}
&N(\al_3,\al_2,\al_1)\,=\,\\
&=\,\frac{\Ga_b(2Q-2\al_3)\Ga_b(2\al_2)\Ga_b(2\al_1)\Ga_b(Q)}
{\Ga_b(2Q-\al_1-\al_2-\al_3)\Ga_b(Q-\al_1-\al_2+\al_3)
\Ga_b(\al_1+\al_3-\al_2)\Ga_b(\al_2+\al_3-\al_1)}\,.
\notag\end{align}

It remains to determine $S_{0\be_2}^{\rm\sst L}$. In order to do this, 
let us note (using formula \rf{Ilim1} in Appendix \ref{spp:Sprop}) 
that the expression \rf{Skern2} simplifies for $\al_0\ra 0$
to an expression of the form
\begin{equation}
S_{\be_\1 \be_\2}^{\rm\sst L} :=\lim_{\al_0\ra 0}S_{\be_\1\be_\2}^{\rm\sst L} (\al_0)\,=
\,\frac{S_{0\be_2}^{\rm\sst L}}{|S_b(2\be_\2)|^2}\,
2\cos\left(\pi(2\be_1-Q)(2\be_2-Q)\right)\,.
\end{equation}
It then follows from \rf{g=1,n=1a} that we must have 
\begin{equation}
S_{0\be}^{\rm\sst L}\,=\,\sqrt{2}|S_b(2\be)|^2=
-2^{\frac{5}{2}}\,\sin\pi b (2\be-Q)\,\sin\pi b^{-1} (2\be-Q)\,.
\end{equation}
One may observe an interesting phenomenon: The analytic
continuation of $S_{\be_\1\be_\2}^{\rm\sst L}$ to the value $\be_\1=0$ does not
coincide with the limit 
$S_{0\be}^{\rm\sst L}:=\lim_{\ep\ra 0}S_{\ep\be}^{\rm\sst L}(\ep)$.
This can also be shown directly using the integral representation 
\rf{Skern2}, see Appendix \ref{spp:Sprop}.

Direct calculation using relation \rf{g=1,n=1b} in the special case
$\al=0$ then shows that
$\chi_b^{}$ is equal to
\begin{equation}\label{centext}
\boxed{\quad \chi_b^{}\,=\,\frac{{\mathbf c}}{24}\,,\qquad 
{\mathbf c}\,=\,1+6(b+b^{-1})^2\,.\quad}
\end{equation}
We conclude that the quantization of Teichm\"uller space produces
a projective representation of the Moore-Seiberg groupoid
with central extension given in terms of the Liouville central
charge ${\mathbf c}$, as is necessary for the relation between Liouville theory
and the quantum Teichm\"uller theory to hold in higher genus.

\subsection{General remarks}

It should be possible to verify the consistency of the 
quantum theory of 
$ \CM_{\rm flat}^0(C)\simeq\CT(C)$ defined in Section \ref{q-Mflat}
without 
using the relation with the quantum Teichm\"uller theory 
described in Section \ref{q-Teich} above. However,
the most difficult
statements to prove would then be the consistency 
conditions \rf{MS-eqns}.
We may note, however, that
the relations \rf{hexagon}-\rf{pentagon} can be proven by using the 
relation between the fusion 
coefficients 
$\Fus{\al_1}{\al_2}{\al_3}{\al_4}{\be_s}{\be_t}$ and the 6j-symbols
of the modular double of $\CU_q(\fsl(2,\BR))$ \cite{PT2,NT}, or
with the fusion and braiding coefficients of quantum Liouville theory 
\cite{PT1,Teschner:2001rv,T03a}.

Any proof that the operators defined in Section 
\ref{q-Mflat} satisfy 
the full set of consistency conditions \rf{MS-eqns} could 
be taken as the basis for an alternative approach to 
the quantum Teichm\"uller theory that is
entirely based on the loop coordinates associated to pants 
decompositions rather than triangulations of the Riemann surfaces.

A more direct way to prove the consistency conditions \rf{MS-eqns}
could probably start by demonstrating the fact
that the operators $\SU_{\si_\2\si_\1}$ correctly intertwine
the representations $\pi_\si$ according  to \rf{inter}. 
It follows
that any operator intertwining a representation $\pi_{\si_\1}$ with 
itself like $\SU_{\si_\1\si_\3}\SU_{\si_\3\si_\2}\SU_{\si_\2\si_\1}$ 
acts trivially on all generators $\SL_{\si_\1,\ga}$. This should imply
that such operators must be proportional to the 
identity, from which the validity of the 
consistency conditions \rf{MS-eqns} up to projective phases
would follow.

However, such an
approach would lead into difficulties of functional-analytic nature that
we have not tried to solve. One would need to show, in particular,
that any operator commuting with $\pi_{\si}(\CA_b(C))$ has to be 
proportional to the identity. 

The proof of \rf{MS-eqns} using the  
quantum Teichm\"uller theory described in
Section \ref{q-Teich}
seems to be the most elegant for the time being.

\newpage
\part{\Large Conformal field theory}

We are now going to 
describe an alternative approach to the quantization of $\CM^0_{\rm flat}(C)$,
and explain why it is intimately related to the Liouville theory. 
It will be shown that the conformal blocks, naturally identified with 
certain wave-functions in the quantum theory of  $\CM^0_{\rm flat}(C)$,
represent solutions to the Riemann-Hilbert type problem formulated
in Subsection \ref{RH-def} above.

This will in particular clarify why we need to have the spurious
prefactors $f_{\si_\2\si_1}(\tau)$ in the S-duality transformations 
\rf{S-duality2}, in general. They will be identified with transition
functions of the projective
line bundle which plays an important role in the geometric approach
to conformal field theory going back to \cite{FS}. This observation 
will lead us to the proper geometric characterization of the
non-perturbative scheme dependence observed in Subsection 
\ref{NPscheme}, and will allow us to define natural prescriptions
fixing the resulting ambiguities.

\section{Classical theory}\label{Kaehlerquant}

\setcounter{equation}{0}

\subsection{Complex analytic Darboux coordinates}\label{cplxDarboux}

In order to establish the relation with conformal field theory it
will be useful to consider an alternative quantization scheme
for  $\CM^{0}_{\rm flat}(C)\simeq \CT(C)$ which makes explicit use of the
complex structure on these spaces. In order to do this, it will
first be convenient to identify a natural complexification of the
spaces of interest by representing $\CM^{0}_{\rm flat}(C)$ as a connected
component of the real slice $\CM_{\rm flat}(C)$ within 
$\CM^{\BC}_{\rm flat}(C)$.

Let us begin by recalling 
that natural Darboux coordinates
for an important component of the moduli space of flat ${\rm SL}(2,\BC)$
connections can be defined in terms of
a special class of local systems called
{\it opers}.

\subsubsection{Opers}

In the case $\fg=\fsl_2$ one may define
opers as bundles admitting a connection that locally looks
as
\begin{equation}\label{operconn}
\nabla'\,=\,\frac{\pa}{\pa y}+M(y)\,, \qquad
M(y)\,=\,\bigg(\,\begin{matrix} 0 & -t(y) \\ 1 & 0 \end{matrix}\,\bigg)\,.
\end{equation}
The equation $\nabla' h=0$ for horizontal sections $s=(s_1,s_2)^t$
implies the second order 
differential equation $(\pa_y^2+t(y))s_2=0$.  
Under holomorphic
changes of the local coordinates on $C$,
$t(y)$  transforms as
\begin{equation}\label{opertrsf}
t(y)\;\mapsto\;(y'(w))^2t(y(w))+\frac{c}{12}\{y,w\}\,,
\end{equation}
where $c=c_{\rm\sst cl}:=6$, and the 
Schwarzian derivative $\{y,w\}$ is defined as
\begin{equation}
\{y,w\}\,\equiv\,\left(\frac{y''}{y'}\right)'-
\frac{1}{2}\left(\frac{y''}{y'}\right)^2\,.
\end{equation}
Equation \rf{opertrsf}  
is the transformation law characteristic for {\it projective}
c-connections, which are also called $\fsl_2$-opers, or opers for short.

Let ${\rm Op}(C)$ the space of $\fsl_2$-opers on a
Riemann surface $C$.
Two opers represented by ${{t}}$ and ${{t}'}$, respecticely, 
differ by a holomorphic
quadratic differentials $\vartheta=({{t}}-{{t}'})(dy)^2$.
This implies that the space ${\rm Op}(C_{g,n})$ of $\fsl_2$-opers
on a fixed surface $C_{g,n}$ of genus $g$ with $n$ marked points
is $h=3g-3+n$-dimensional.
Complex analytic coordinates for ${\rm Op}(C_{g,n})$ are
obtained by picking a reference oper ${t}_0$,
a basis ${\vartheta}_1,\dots,{\vartheta}_{h}$ for the vector
space of quadratic differentials,
and writing any other oper as
\begin{equation}\label{accessdef}
{t}(dy)^2\,=\,{t}_0(dy)^2+\sum_{r=1}^{h}h_r\,{\vartheta}_r\,.
\end{equation}

The space of opers forms an affine 
bundle $\CP(C)$ over the Teichm\"uller space
of deformations of the complex structure of $C$.
The monodromy representations $\rho_{{P}}:\pi_1(C_{g,n})\ra {\rm SL}(2,\BC)$
of the connections $\nabla'$ will generate a
$3g-3+n$-dimensional subspace
in  the character variety $\homsl$ of surface group representations.
Varying the complex structure of the underlying surface $C$, too,
we get a subspace of $\homsl$ of complex dimension
$6g-6+2n$. 
It is important that the mapping $\CP(C)\ra \homsl$
defined by associating to the family of opers 
$\pa_y^2+t(y;q)$ its monodromy
representation $\rho_t^{}$ is locally biholomorphic \cite{He,Ea,Hu}.

\subsubsection{Projective structures}\label{sec:projstruct}

A projective structure is a particular atlas of complex-analytic coordinates
on $C$ which is such that the transition functions are all given by Moebius
transformations
\begin{equation}\label{Moebius}
y'(y)\,=\,\frac{ay+b}{cy+d}\,.
\end{equation}
It will be useful to note that there is a natural
one-to-one correspondence between projective structures and opers.
Given an oper, in a patch $\CU\subset C$ 
locally represented by
the differential operator $\pa_y^2+t(y)$, one may construct a
projective structure by taking the ratio 
\begin{equation}
w(y):=f_1(y)/f_2(y)\,,
\end{equation}
of two linearly independent solutions $f_1$, $f_2$ 
of the differential equation $(\pa_y^2+t(y))f(y)=0$
as the new coordinate in $\CU$. The oper 
will be represented by the differential operator $\pa_w^2$ 
in the coordinate $w$, as follows from \rf{opertrsf} observing 
that
\begin{equation}
t(y)\,=\,\frac{c}{2}\{w,y\}\,.
\end{equation}
The bundle $\CP(C)$ may therefore be identified with the space
of projective structures on $C$.

\subsubsection{Complex structure on $\CP(C)$}

The space $\CP(C)$ is isomorphic as a complex manifold to
the holomorphic cotangent bundle $T^*\CT(C)$ over the Teichm\"uller
space $\CT(C)$. In order to indicate how this isomorphism 
comes about, let us recall some basic results from the complex
analytic theory of the Teichm\"uller spaces.\footnote{A standard 
reference is \cite{Na}. A useful
summary and further 
references to the original literature can be found in 
\cite{TT}. The results that are relevant for us
are very concisely summarized in \cite[Section 1]{BMW}.}

Let $\CQ(C)$ be 
the vector space of meromorphic quadratic differentials on $C$ which are
allowed to have poles only at the punctures of $C$. The poles are
required to be of second order, with fixed leading coefficient. A Beltrami differential $\mu$
is a $(-1,1)$-tensor, locally written as $\mu_{\bz}^zd\bz/dz$. 
Let $\CB(C)$ be the space of all measurable Beltrami differentials such that
$\int_C |\mu \vartheta|<\infty$ for all $q\in\CQ(C)$. There is a natural pairing 
between $\CQ(C)$ and $\CB(C)$ defined as
\begin{equation}\label{q-mu-pairing}
\langle\,\vartheta\,,\,\mu\,\rangle:=\,\int_C\mu \vartheta\,.
\end{equation}
Standard Teich\-m\"uller theory establishes the basic isomorphisms of 
vector spaces 
\begin{align}
T\CT(C)\,&\simeq\,\CB(C)/\CQ(C)^{\perp}\,,\label{TTeich}\\
T^*\CT(C)\,&\simeq\,\CQ(C)\,,
\end{align}
where $\CQ(C)^{\perp}$ is the subspace in $\CB(C)$ on which all linear forms 
$f_\vartheta$, $\vartheta\in\CQ(C)$, defined by $f_\vartheta(\mu)\equiv \langle \vartheta,\mu\rangle$
vanish identically.  

The relation between $\CP(C)$ and $T^*\CT(C)$ follows from the
relation between ${\rm Op}(C)$ and the space $\CQ(C)$ of quadratic
differentials explained above. What's not immediately obvious is the fact 
there is a natural complex structure on $\CP(C)$ that makes 
the isomorphism $\CP(C)\simeq T^*\CT(C)$ an isomorphism of complex manifolds.

To see this, the key ingredient is the
existence of a {\it holomorphic section} of the bundle
$\CP(C)\ra\CT(C)$, locally represented by opers $\pa_y^2+t(y;q)$ that 
depend holomorphically on $q$. Such a section 
is provided by the Bers double uniformization.
Given two Riemann surfaces $C_\1$ and $C_\2$ there exists a subgroup 
$\Gamma(C_\1,C_\2)$ of
${\rm PSL}(2,\BC)$ that uniformizes $C_\1$ and $C_\2$ simultaneously in the following
sense: Considering the natural action of $\Gamma(C_\1,C_\2)$ on $\BC$ by 
Moebius transformations,
the group $\Ga$ will have a domain of discontinuity of the form 
$\Omega(C_\1,C_\2)=\Omega_\1\sqcup\Omega_\2$
such that $\Omega_\1/\Ga(C_\1,C_\2)\simeq C_\1$, 
$\Omega_\2/\Ga(C_\1,C_\2)\simeq \bar{C}_\2$, where
$\bar{C}_\2$ is obtained from $C_\2$ by orientation reversal. 
Let $\pi_\1:\Omega_\1\ra C_\1$ be the corresponding
covering map.
The Schwarzian derivatives  $\CS(\pi_\1^{-1})$ and $\CS(\pi_\2^{-1})$ 
then define a families of opers on $C_\1$ and $\bar{C}_\2$, 
respectively. The family of opers  defined by $\CS(\pi_\1^{-1})$
depends holomorphically on the complex structure moduli $q_\2$ of $C_\2$.

\subsubsection{Symplectic structure on $\CP(C)$}

Note furthermore 
that the corresponding mapping $\CP(C)\simeq T^*\CT(C)\ra \homsl$ is
symplectic in the sense that the canonical cotangent
bundle symplectic structure
is mapped to the Atiyah-Bott symplectic structure $\Omega_J$ on the
space of flat complex connections \cite{Ka}.
We may, therefore, choose a set of local coordinates 
${q}=({q}_1,\dots,{q}_{h})$
on $\CT(C_{g,n})$
which are conjugate to the coordinates $h_r$ defined above in the sense that
the Poisson brackets coming from this symplectic structure are
\begin{equation}\label{Poissopers}
\{\,{q}_r\,,{q}_s\,\}\,=\,0\,,\qquad \{\,h_r\,,{q}_s\,\}\,=\,\de_{r,s}\,,\qquad
\{\,h_r\,,h_s\,\}\,=\,0\,.
\end{equation}

Let us note that one may also use non-holomorphic sections $t'(y;q,\bar{q})$ in 
$\CP(C)\ra \CT(C)$ in order to get such Darboux coordinates $(q,h)$.
This amounts to a shift of the variables $h_r$ by a function of the variables
$q$,
\[
h_r'\,=\,h_r+\nu_r(q,\bar{q})\,,
\]
which clearly preserves the canonical form of the Poisson brackets  
\rf{Poissopers}.

\subsection{Twisted cotangent bundle $T^*_{c}\CM(C)$}\label{tw-T*CT}

The affine bundle $\CP(C)$ over $\CT(C)$ descends to a twisted 
cotangent bundle over the moduli space $\CM(C)$ of complex structures
on $C$. To explain what this means let us 
use a covering $\{\CU_\imath;\imath\in\CI\}$
of $\CM(C)$. Within each patch $\CU_\imath$ we may consider local
coordinates $q=(q_1,\dots,q_h)$ for $\CM(C)$, which may be completed to 
a set of local 
Darboux coordinates $(q,h)$ for $\CP(C)$ such that
\[
\Om\,=\,\sum_{r=1}^h dh_r\wedge dq_r\,.
\]
$\CP(C)$ is a {\rm twisted} holomorphic cotangent bundle over $\CM(C)$ if
the Darboux coordinates transform as
\begin{equation}\label{H-trans}
\La=\sum_r h_r^{\imath}\,dq^{\imath}_r\,=\,\sum_r h_r^{\jmath}\,dq^{\jmath}_r-
\chi^{\imath\jmath}\,,
\end{equation}
with $\chi^{\imath\jmath}$ being locally defined holomorphic one-forms on 
$\CU_{\imath\jmath}\equiv\CU_{\imath}\cap\CU_{\jmath}$.
The collection of one-forms defines a 1-cocycle with values in the sheaf of holomorphic one-forms.
We may always write $\chi^{\imath\jmath}=\pa g_{\imath\jmath}$ for locally defined holomorphic
functions $g_{\imath\jmath}$ on $\CU_{\imath\jmath}$. 
The functions $f^{c}_{\imath\jmath}:=e^{2\pi i g_{\imath\jmath}}$ will
then satisfy relations of the form
\begin{equation}\label{triple}
f^{c}_{\imath_\3\imath_\2}\,f^{c}_{\imath_\2\imath_\1}\,=\,\si_{\imath_\3\imath_\2\imath_\1}\,
f^{c}_{\imath_\3\imath_\1}\,,
\end{equation}
where $\si_{\imath_\3\imath_\2\imath_\1}$ is constant 
on the triple overlaps $\CU_{\imath_\1\imath_\2\imath_\3}\equiv 
\CU_{\imath_\1}\cap\CU_{\imath_\2}\cap\CU_{\imath_\3}$.
A collection of functions $f^{c}_{\imath\jmath}$ on $\CU_{\imath\jmath}$ that satisfy \rf{triple}
defines a so-called {\it projective line bundle} \cite{FS}. 
The obstruction to represent it as an ordinary
line bundle is represented by a class in $\check{H}^2(\CM(C),\BC^\ast)$.  


It was pointed out in \cite{FS} that any holomorphic section 
of $\CP(C)\ra \CT(C)$, represented by a family of opers $t\equiv t(y;q)$,  
can be considered as a connection on a certain
{holomorphic projective line bundle} $\CE_c$. 
The connection is locally represented by the one-forms $(\pa_r+A_r)dq_r$ 
on $\CT(C)$ such that
\begin{equation}
A_r(\tau)\,=\,\int_C t \mu_r\,,
\end{equation}
for a collection of Beltrami differentials $\mu_r$ which represent
a basis to the tangent space $T\CT(C)$ dual to the chosen set of  
coordinates $q_r$. 
One may
define a family of local sections 
$\CF_\imath$ of $\CE_c$ which are horizontal 
with respect to the connection $A_t$ 
as solutions to the differential equations
\begin{equation}\label{Ec-sect}
\pa_r \ln \CF_\imath\,=\,-\int_C t \mu_r\,.
\end{equation} 
The transition functions $f^{c}_{\imath\jmath}$ of $\CE_c$ are then defined
by $f^{c}_{\imath\jmath}:=\CF_{\imath}^{-1}\CF_{\jmath}^{}$. In general it 
will not be possible to choose the integration constants in the
solution of \rf{Ec-sect} in such a way that in \rf{triple} we find
$\si_{\imath_\3\imath_\2\imath_\1}=1$ for all nontrivial triple intersections 
$\CU_{\imath_\3\imath_\2\imath_\1}$.
 
The resulting projective line bundle $\CE_c$ 
is uniquely characterized by the real number $c$ if the family
$t$ is regular at the boundary of $\CM(C)$. It was shown in \cite{FS} that 
\begin{equation}
\CE_c\,=\,(\la_{\rm H})^{\frac{c}{2}}\,,
\end{equation}
where $\la_{\rm H}$ is the so-called Hodge line bundle, the determinant bundle
${\rm det}\Omega_{\rm H}\equiv\bigwedge^g\Omega_{\rm H}$ 
of the bundle of rank $g$ over $\CM(C)$ 
whose fiber over a point of $\CM(C)$ is the space of abelian differentials
of first kind on $C$.

\subsection{Projective structures from the gluing construction}

\label{Glueing-2}

In Subsection \ref{sec:glueing}
we have described how to construct local 
patches of coordinates $q=(q_1,\dots,q_h)$ 
for $\CT(C)$  
by means of the gluing construction.  
There is a corresponding natural choice 
of coordinates $H=(h_1,\dots,h_h)$ for $T^*\CT(C)$ defined as follows. 
The choice of the 
coordinates $q$ defines a basis for $T\CT(C)$ 
generated by the tangent vectors $\pa_{q_r}$
which can be represented by Beltrami differentials $\mu_r$ via \rf{TTeich}. 
The dual basis
of quadratic differentials $\vt_r$ is then 
defined by the condition  $\langle\vartheta_r,\,\mu_s\rangle=\de_{r,s}$.
This defines coordinates $h_r$ for  $T^*\CT(C)$.

In order to make the coordinates $(q,h)$ for $T^*\CT(C)$ into
coordinates for $\CP(C)$, one needs to choose a section $\CS:\CT(C)\ra\CP(C)$.
It will be very important to note that the gluing 
construction allows one to define natural
choices for local  sections of $\CP(C)$ as follows.

Let us 
represent the three-punctured
spheres used in the gluing construction
as $C_{0,3}\sim\BP^1\setminus\{0,1,\infty\}\sim
\BC\setminus\{0,1\}$. A natural choice of coordinate
on $C_{0,3}$ is then coming from the coordinate $y$ on the complex 
plane $\BC$. Let us choose the coordinates 
around the punctures $0$, $1$ and $\infty$  
to be $y$, $1-y$ and $1/y$, respectively.
The surfaces $C$ obtained from
the gluing construction will then automatically come with an
atlas of local coordinates which has transition functions always 
represented by
Moebius transformations \rf{Moebius}.
It follows that the gluing construction 
naturally defines families of projective structures over 
the multi-discs $\CU_{\si}$ with coordinates $q$, or equivalently 
according to Subsection \ref{sec:projstruct} a section 
$\CS_\si:\CU_{\si}\ra\CP(C)$. One could replace  the representation
of $C_{0,3}$ as 
$C_{0,3}\sim\BP^1\setminus\{0,1,\infty\}$
by 
$C_{0,3}\sim\BP^1\setminus\{z_1,z_2,z_3\}$,
leading to other sections $\CS:\CU_{\si}\ra\CP(C)$.

We may define such a section $\CS_\si$ for any pants decomposition
$\si$. The sections $\CS_{\si}$ define corresponding 
local trivializations of the projective line bundle $\CE_c$
according to our discussion in Subsection \ref{tw-T*CT}.
The trivializations coming from pants decompositions 
lead to a particularly simple representation for the transition 
functions $f^{c}_{\si_\2\si_\1}$ defining $\CE_c$, which will be calculated
explicitly in the following.

\subsubsection{Transition functions}\label{trans-Ec}

It is enough to calculate the resulting transition functions
for the elementary moves $B$, $F$ and $S$ generating the MS
groupoid. In the case of $B$ and $F$ it suffices to note that the 
gluing of two three-punctured spheres produces a
four-punctured sphere that may be represented
as $C_{0,4}\sim\BP^1\setminus\{0,1,q,\infty\}$, with $q$ being the
gluing parameter. The B-move corresponds to the Moebius transformation
$y'=q-y$ which exchanges $0$ and $q$. Being related by
a Moebius transformation, the projective structures
associated to two pants decompositions $\si_\1$ and $\si_\2$ related
by a B-move must coincide. We may therefore assume that 
$g_{\si_\2\si_\1}=1$ if 
$\si_\1$ and $\si_\2$ differ by a B-move. The F-move corresponds
to $y'=1-y$, so that $g_{\si_\2\si_\1}=1$ if 
$\si_\1$ and $\si_\2$ differ by a F-move. 

The only nontrivial case is the S-move. We 
assume that $C_{1,1}$ is obtained from 
a three-punctured sphere $C_{0,3}\sim\BP^1\setminus\{0,1,\infty\}$
by gluing annular neighborhoods of $0$ and $\infty$. The resulting 
coordinate $y_\si$ on $C_{1,1}$ is coming from the coordinate $y$ on 
$C_{0,3}\sim\BP^1\setminus\{0,1,\infty\}$. 
A nontrivial transition function $g_{\si_\2\si_1}$ will be found if  
$\si_\1$ and $\si_\2$ differ by a S-move since the coordinates 
$y_{\si_\1}$ and $y_{\si_\2}$ are {\it not} related by a Moebius transformation.

In order to see this, it is convenient to introduce 
the coordinate 
$w_\si$ related to the coordinate $y_\si$ on the complex plane 
by $y_\si=e^{w_\si}$. The coordinate $w_\si$ would be the natural
coordinate if we had represented $C_{1,1}$ as 
\[ 
C_{1,1}\sim\big\{\,w\in\BC;\; w\sim w+n\pi+m\pi\tau;\;
n,m\in\BZ\,\big\}\setminus\{0\}\,.
\] 
This corresponds to representing $C_{1,1}$ by gluing the two 
infinite ends of the punctured cylinder 
$\{w\in\BC;\; w\sim w+n\pi;n\in\BZ\}\setminus\{0\}$. The
corresponding alternative 
pants decomposition of $C_{1,1}$ will be denoted
$\tilde{\si}$.

The transition function  $g_{\si_\2\si_\1}$ defined by our conventions
for the gluing construction will then be nontrivial since
the relation $y_\si=e^{w_\si}$ is {\it not} a Moebius transformation.
The relation between the projective
structures associated to pants decompositions $\si$ 
and $\tilde{\si}$ can be calculated from \rf{opertrsf}, 
\begin{equation}
\tilde{t}(w)\,=\,e^{2w}\,t(e^w)-\frac{c}{24}\,.
\end{equation}
We thereby get a nontrivial transition function $g_{\tilde\si \si}$ 
between the trivializations of $\CE_c$ associated to 
$\si$ and $\tilde{\si}$
equal to $\frac{c}{24}\tau$ up to an additive constant. 

Let us assume that $\si_\2$ is obtained from $\si_\1$ by an S-move.
The projective structures associated to the coordinates
$w_{\si_\1}$ and $w_{\si_\2}$ will coincide since the S-move
is represented by the Moebius transformation $w_{\si_\2}=-w_{\si_\1}/\tau$.
The resulting transition function $g_{\tilde\si_\2\tilde\si_\1}=1$ is trivial. 
Taken together we conclude that 
\begin{equation}\label{transfct}
g_{\si_\2\si_\1}\,=\,g_{\si_\2\tilde\si_2}+g_{\tilde\si_\2\tilde\si_\1}+g_{\tilde\si_\1\si_\1}
\,=\,\frac{c}{24}\Big(\tau+\frac{1}{\tau}\Big)+h_{\si_\2\si_\1}\,,
\end{equation}
with $h_{\si_\2\si_\1}$ being constant, if $\si_\1$ and $\si_\2$ are related
by an S-move. These are the only nontrivial transition functions
of $\CE_c$ in the representation associated to pants 
decompositions defined above. The argument above determines $g_{\si_\2\si_\1}$ 
up to an additive ambiguitiy $h_{\si_\2\si_\1}$. Precise normalizations
fixing this ambiguity will be defined next.

\section{The generating functions $\CW$}

\setcounter{equation}{0}

\subsection{Definition}

We have used two radically different representations for the space 
$\CP(C)$: As cotangent bundle $T^*\CT(C)$, on the one hand, and
as character variety $\homsl$ on the other hand. In Section
\ref{sec:2hyp} we had introduced systems of Darboux coordinates 
$(l,k)$ associated to MS-graphs $\si$ for the  character variety $\homsl$.
We had previously introduced Darboux coordinates $(q,h)$ 
with the help of the isomorphism $\CP(C)\simeq T^*\CT(C)$. Important
objects are the generating functions $\CW(l,q)$ that characterize
the transitions between these sets of coordinates.

Let us briefly explain how the functions $\CW(l,q)$ are defined.
The locally defined one-forms $\sum_r k_rdl_r-\sum_r h_rdq_r$ are 
$\pa$-closed since $\sum_r dk_r\wedge dl_r=\sum_r dh_r\wedge dq_r$ \cite{Ka}, 
therefore locally
exact, 
\begin{equation}\label{exW}
\sum_r k_rdl_r-\sum_r h_rdq_r\,=\,
\pa \CW\,.
\end{equation}
It follows that the change of coordinates $(l,k)\ra(q,h)$ can locally be 
described in terms of a generating function $\CW$. Let us start, 
for example, with the coordinates $(q,h)$. 
For fixed  values of $l$, let us define the functions $h_r(l,q)$ 
as the solutions to the system of equations
\begin{equation}\label{monofix}
2\cosh(l_r/2)\,=\,{\rm tr}(\rho_{q,h}(\ga_r))\,,
\end{equation}
where $\rho_{q,h}$ is the monodromy representation of the 
oper $\pa_y^2+t_0(y;q)+\sum_{r}h_r\vartheta_r(y)$. 
Equation \rf{exW} ensures integrability of the equations
\begin{align}
&h_r(l,q)\,=\,-\frac{\pa}{\pa q_r}\CW(l,q)\,,\label{H-Wrel}
\end{align}
which define $\CW(q,l)$ up to a function of $l$. This ambiguity is 
fixed by the equations 
\begin{align}
&k_r(l,q)\,=\,\frac{\pa}{\pa l_r}\CW(l,q)\,,\qquad k_r(l,q)
\equiv k_r(\rho_{q,h(l,q)})\,,
\label{k-Wrel}
\end{align}
following from \rf{exW}, 
where $k_r(\rho)$ is the value of the coordinate $k_r$ on the monondromy $\rho$
as defined in Section
\ref{sec:2hyp}.

Comparing \rf{H-Wrel} with \rf{Ec-sect} we realize $\CF_{\rm cl}(l,q)\equiv
e^{\CW(l,q)}\CF_0(q)$
as the local section of the projective holomorphic line bundle $\CE_c$
that is horizontal with respect to the connection defined 
by the family of opers 
$\pa_y^2+t_0(y;q)+\sum_{r}h_r(q,l)\vartheta_r(y)$.

\subsection{Changes of coordinates}

We have introduced systems of coordinates $(l,k)$ and $(q,h)$ 
that both  depend on the
choice of a pants decomposition $\si$.
In order to
indicate the dependence on the choices of pants decompositions
underlying the defininitions of the coordinates we shall use
the notation 
$\CW_{\si,\si'}(l,q)$ if coordinates  $(l,k)$ were defined using the
pants decomposition $\si$ and if coordinates $(q,h)$ 
were define using the pants decomposition
$\si'$.

\subsubsection{Changes of coordinates $(l,k)$}

Let us compare the functions $ \CW_{\si_\2,\si'}(l,q)$ and
$\CW_{\si_\1,\si'}(l,q)$ associated to two different choices 
of pants decompositions
$\si_\2$ and $\si_\1$, respectively. It is clear that 
there must exist a relation of the form
\begin{equation}\label{kltrans}
\CW_{\si_\2,\si'}(l_\2,q)\,=\,
\CF_{\si_\2\si_\1}(l_\2,l_\1(l_\2,q))+\CW_{\si_\1,\si'}(l_\1(l_\2,q),q)\,,
\end{equation}
where $\CF_{\si_\2\si_\1}(l_\2,l_\1)$ is the generating function
for the change of Darboux coordinates $(k_\2,l_\2)$ associated to
$\si_\2$ to $(k_\1,l_\1)$ associated to $\si_\1$, respectively. 

The generating 
function $\CF_{\si_\2\si_\1}(l_\2,l_\1)$
can be represented up to an additive constant
by choosing a path $\varpi_{\si_\2\si_\1}\in[\si_\2,\si_\1]$ 
connecting $\si_1$ and $\si_2$, 
representing it as sequence of Moore-Seiberg moves
$[m_N\circ m_{N-1}\circ\dots\circ m_1]$, and adding the
generating functions $\CF_{m_i}$ 
representing the changes of Darboux variables
associated to the moves $m_i$. Changes of the 
path $\varpi_{\si_\2\si_\1}\in[\si_\2,\si_\1]$ will change the result
by an additive constant.

The generating 
functions $\CF_{\si_\2\si_\1}(l_\2,l_\1)$ can be identified as the 
semiclassical 
limits of $b^2\log A_{\si_\2\si_\1}(l_\2,l_\1)$, with
$A_{\si_\2\si_\1}(l_\2,l_\1)$ being the kernels of the
operators generating the representation of the Moore-Seiberg groupoid 
constructed in Part II.

\subsubsection{Changes of coordinates $(q,h)$}

It turns out that $\CW_{\si,\si'}(l,q)$, considered as function
of $q$, can be extended to functions on all of $\CT(C)$ by 
analytic continuation\footnote{We don't have a direct proof of this fact
at the moment, but we may infer it indirectly from the corresponding statement 
about the Liouville conformal blocks $\CZ^{\rm L}$ 
together with the fact that the 
$\CW_{\si,\si}(l,q)$ coincide with 
the semiclassical limit $b\ra 0$ of
$b^2\log \CZ^{\rm L}$.}.
We will use the same notation $\CW_{\si,\si'}(l,q)$
for the result of the analytic continuation.

Comparing the transformation 
\rf{H-trans} of the coordinates $h_r$ with \rf{H-Wrel}, we see that 
the functions  $ \CW_{\si,\si'_\2}(l,q)$ and
$\CW_{\si,\si_\1'}(l,q)$ defined by using different
pants decompositions for the definition of coordinates $(q,h)$ 
are related by the transition functions in the projective line 
bundle $\CE_c$, 
\begin{equation}\label{qHtrans}
\CW_{\si,\si'_\2}(l,q)\,=\, g_{\si_\2',\si_\1'}^{}(q)+
\CW_{\si,\si_\1'}(l,q)\,.
\end{equation}
This reflects the changes of coordinates $h_r$ induced by changes of
the sections $\CP(C)\ra\CT(C)$ associated to transitions between
different pants decompositions.

By combining \rf{kltrans} and \rf{qHtrans} one gets, in particular,
\begin{equation}\label{kl-qHtrans}
\CW_{\si_\2,\si_2}(l_\2,q)\,=\,g_{\si_\2,\si_\1}^{}(q)+
\CF_{\si_\2\si_\1}(l_\2,l_\1(l_\2,q))+\CW_{\si_\1,\si_1}(l_\1(l_\2,q),q)\,.
\end{equation}
In order to define $\CF_{\si_\2\si_\1}(l_\2,l_\1(l_\2,q))$
and $g_{\si_\2,\si_\1}^{}(q)$ unambigously one would
need to fix a normalization prescription for
$\SW_{\si,\si'}(l,q)$. 

\subsubsection{Mapping class group action}

Note that in the case $\si'_\2=\mu.\si'$, $\si'_\1\equiv\si'$
we get from \rf{qHtrans}
\begin{equation}\label{qHtrans-M}
\CW_{\si,\mu.\si'}(l,q)\,=\,g_{\mu}^{}(q)+
\CW_{\si,\si'}(l,q)\,.
\end{equation}
We have used the shortened notation
\begin{equation}
g_{\mu}^{}(q):=g_{\mu.\si,\si}(q)\,.
\end{equation}
Taken together we find, in the particular case $\si=\si'$
\begin{equation}\label{cl-MCGact}
\CW_{\mu.\si,\mu.\si}(l,q)\,=\,g_{\mu}^{}(q)+
F_{\mu.\si,\si}(l,\tilde{l}(l,q))+
\CW_{\si,\si}(\tilde{l}(l,q),q)\,.
\end{equation}
Below we will fix a specific
normalization 
for $\CW_{\si,\si}(l,q)$.  
Thanks to the uniqueness of analytic continuation 
the sum of terms $F_{\mu.\si,\si}(l,\tilde{l}(l,q))+
g_{\mu}^{}(q)$ appearing in \rf{cl-MCGact}
will then be uniquely defined.

\subsection{Behavior at the boundaries of $\CT(C)$}

It will be important for us to understand the behavior of the generating 
functions $\CW(l,q)$ 
at the boundaries of the Teichm\"uller spaces $\CT(C)$.
This will in particular allow us to define a natural choice for the
precise normalization of the functions 
$\CW_{\si,\si}(l,q)$.

By means of pants decompositions one may reduce the problem 
to the cases of the four-punctured
sphere $C=C_{0,4}$, and the one-punctured torus $C=C_{1,1}$.

\subsubsection{Genus zero, four punctures, singular term}

Let us first 
consider $C=C_{0,4}=\BP^1\setminus\{z_1,z_2,z_3,z_4\}$.
We may assume that $z_1=0$, $z_3=1$, $z_4=\infty$, and identify 
the complex structure parameter $q$
with $z_2$. The opers on $C$ can be represented in the form 
$\pa_y^2+t(y)$, where
\begin{equation}\label{fouroper}
t(y)=\frac{\de_3}{(y-1)^2}+\frac{\de_1}{y^2}+\frac{\de_2}{(y-q)^2}+\frac{\upsilon}{y(y-1)}+
\frac{q(q-1)}{y(y-1)}\frac{H}{y-q}.
\end{equation}

The relation \rf{H-Wrel} becomes simply
\begin{equation}\label{H-W}
H(l,q)\,=\,-\frac{\pa}{\pa q}\CW(l,q)\,.
\end{equation}
This relation determines $\CW(l,q)$ up to $q$-independent functions of $l$. 
For $q\ra 0$ it may be shown
that $\CW(l,q)$ behaves as
\begin{equation}\label{CWfactor}
\CW(l,q)\,=\,(\de(l_1)+\de(l_2)-\de(l))\log q+
\CW_0(l)+\CO(q)\,,
\end{equation} 
where
$\de(l)=\frac{1}{4}+\left(\frac{l}{4\pi}\right)^2$. 
Indeed, this is equivalent to the statement that $H(l,q)$ behaves as 
\begin{equation}\label{Hasym}
H(l,q)\,\sim\,\frac{\de(l)-\de(l_1)-\de(l_2)}{q}+\CO(q^0)\,,
\end{equation}
for $q\ra 0$. To prove this, let us first calculate the monodromy
of $\pa_y^2+t(y)$ around the pair of points $z_1$ and $z_2$ 
as function of the parameters $q$ and $D:=qH$. 
It is straightforward to show that the differential equation
$(\pa_y^2+t(y))g(y)=0$ will have a solution of the form 
\begin{equation}\label{gseries}
g(y)=y^{\nu}\sum_{l=0}^{\infty}y^l\,g_{l}+\CO(q)\,,
\end{equation}
provided that $\nu$ is one of the two solutions of
\begin{equation}\label{D-eqn}
\nu(\nu-1)+\de(l_1)+\de(l_2)+D\,=\,0+\CO(q)\,.
\end{equation}
The solution \rf{gseries} has diagonal 
monodromy $e^{2\pi i\nu}$ around $(z_1,z_2)\equiv(0,q)$. Note that
$\nu$ and $l$ are related as
$\nu=\frac{1}{2}+i\frac{l}{4\pi}$. The equation \rf{Hasym} follows.

A more detailed analysis of the solutions to the differential equation
$\pa_y^2+t(y)$ 
shows that the expansion of the 
function $\CW(l,q)$ in powers of $q$ 
is fully defined by \rf{H-W}
combined with the boundary condition \rf{CWfactor}
once $\CW_0(l)$ is specified.

\subsubsection{Genus zero, four punctures, constant term}

In order to determine $\CW_0(l)$ let us recall that
the Darboux variable $k$ conjugate to $l$ is
obtained from $\CW(l,q)$ as
\begin{equation}
k\,=\,4\pi i\frac{\pa}{\pa l}\CW(l,q)\,.
\end{equation}
Having fixed a definition for the coordinate $k$ 
by means of \rf{FN-FK}, 
we should therefore be able
to determine $\CW(l,q)$ up to a constant, including the precise
form of $\CW_0(l)$.
The result is the following:
\begin{claim} $\quad$\\
{\it The function $\CW_0(l)$ characterizing the
asymptotics \rf{CWfactor} of $\CW_0(l,q)$ 
is explicitly given as
\begin{equation}\label{CW0}
\CW_0(l)\,=\,\frac{1}{2}(
C^{\rm cl}(l_4,l_3,l)+C^{\rm cl}(-l,l_\2,l_\1))\,,
\end{equation}
where $C^{\rm cl}(l_\3,l_\2,l_\1)$ is explicitly given as
\begin{align}\notag
C^{\rm cl}(l_\3,l_\2,l_\1)\,=\,&
\left(\frac{1}{2}+\frac{i}{4\pi}(l_\3+l_2+l_1)\right)\log(\pi\mu)
-\sum_{i=1}^3\upc\big(1+\fr{i}{2\pi}l_i\big)\\
&+\sum_{s_1,s_2=\pm}\upc\big(\fr{1}{2}+\fr{i}{4\pi}(l+s_\1l_\1+s_\2l_\2+l_\3)\big)
\label{clthree}\end{align}
with function $\upc(x)$ defined as
\begin{equation}
\upc(x)\,=\,\int_{1/2}^{x}du\;\log\frac{\Gamma(u)}{\Gamma(1-u)}\,.
\end{equation}
}\end{claim}

The proof is described in Appendix \ref{sec:W-as}.
A formula for $\CW_0(l)$ that is very similar (but not 
quite identical) to \rf{CW0} was previously proposed in \cite{NRS}.

Let us note that the function $C^{\rm cl}(l_\3,l_\2,l_\1)$
coincides with the classical Liouville action for the three-punctured sphere
\cite{ZZ}.

\subsubsection{Genus one, one puncture}

It remains to discuss the case $C=C_{1,1}$. The discussion is similar,
the results are the following. 
The opers on $C_{1,1}$ can be represented in the form 
$\pa_y^2+t(y)$, where
\begin{equation}\label{oneoper}
t(y)\,=\,\de(l_0)\,\wp(\ln y)+H(l,q)\,, 
\end{equation}
with $\wp(w)$ being the Weierstrass elliptic function
\begin{equation}
\wp(w)=\frac{1}{w^2}+\sum_{(n,m)\neq (0,0)}\left(\frac{1}{(w-\pi n-m\pi\tau)^2}-
\frac{1}{(\pi n+m\pi\tau)^2}\right)\,.
\end{equation}
$\CW(l,q)$ behaves as
\begin{equation}\label{CWfactor,g=1}
\CW(l,q)\,=\,-\de\log q+\CW_0(l)+\CO(q)\,,
\end{equation}
where
\begin{equation}\label{CW0-torus}
\CW_0(l)\,=\,\frac{1}{2}C^{\rm cl}(l,-l,l_0)\,.
\end{equation}
As before we note that \rf{H-Wrel}, \rf{k-Wrel}
determine $\CW(l,q)$ only up to a constant, 
equation \rf{CW0-torus} holds for a particular 
convention fixing this constant.

\subsection{The real slice}

We had pointed out earlier that the monodromy map induces a map
$
\rho:\CP(C)\ra \homsl
$
that is locally biholomorphic. A natural real slice in $\homsl$ 
is $\homslr$, which contains a connected component 
isomorphic to $\CM_{\rm flat}^0(C)$. 
The corresponding slice in $\CP(C)$ can locally
be described
by a family of opers $t(y;q,\bq)$ that is real analytic in $q,\bq$.

Let us consider coordinates $q,\bq$ introduced using a pants 
decomposition $\si$. We will furthermore assume that the local 
coordinates $y$
are coming from the projective structure naturally 
associated to the pants decomposition $\si$.

There exists a real analytic
function $S_{\si}(q,\bar{q})$ on $\CT(C)$ 
such that 
\begin{equation}\label{SL-pot}
t(y)\,=\,\sum_{r=1}^h h_r\,\vt_r\,,
\qquad h_r=-\frac{\pa}{\pa q_r}S_{\si}(q,\bar{q})\,.
\end{equation}
The function $S_{\si}(q,\bar{q})$ is related to the 
generating function $\CW_{\si,\si}(l,q)$ as
\begin{equation}
S_{\si}(q,\bar{q})\,=\,2\,{\rm Re}(\CW_{\si,\si}(l(q,\bar{q}),q))\,,
\end{equation}
where $l_e(q,\bar{q})$ is the length of the geodesic $\ga_e$ in the
hyperbolic metric which corresponds to the complex structure specified
by $q,\bq$.

It is clear that the function $S_{\si}(q,\bar{q})$ represents
a hermitian metric in a (generically) projective line bundle $\CE_c$.
This means more concretely that the mapping class group 
acts on $S_{\si}(q,\bar{q})$ as follows
\begin{equation}\label{SL-trans}
S_{\si}(\mu.q,\mu.\bar{q})\,=\,|f^{c}_\mu(q)|^2\,S_{\si}(q,\bar{q})\,,\qquad
\mu\in{\rm MCG}(C)\,.
\end{equation} 
The functions $f^{c}_\mu(q)$ are
transition functions of the projective line bundle ${\CE}_c$. 

The function $S_{\si}(q,\bar{q})$ is nothing but the classical 
Liouville action. It should be possible to give a direct proof of
this claim along the lines of \cite{TZ87a,TZ87b,TT}. It will follow indirectly
from the relations with quantum Liouville theory to be described later.

\subsection{Scheme dependence}\label{sec:scheme1}

In the above we have given an unambiguous definition of 
the generating functions $\CW_{\si,\si}(l,q)$. One should keep in mind
that the definition was based on the use of the projective
structures that were defined using the gluing constructions of 
Riemann surfaces $C$. This corresponds to choosing 
particular local sections $t_0(y,q)$ 
of $\CP(C)$ in the definition of the 
coordinates $h_r$ via \rf{accessdef}. 

One may, of course, consider other choices for the local sections 
$t_0(y,q)$ than the one chosen for convenience above. This would modify
the coordinates $h_r$ by functions of $q$, leading to 
a modification of $\CW(l,q)$ by some function $\CW_0(q)$ that
depends on $q$ and parameterically on $c$. The dependence of $\CW(l,q)$ on 
the variables $l$ would be unaffected.

\newpage

\section{Quantization}\label{hol-quant}

\setcounter{equation}{0}

Summary:
\begin{itemize}
\item Functions on $\CP(C)$ $\rightsquigarrow$ Ring of holomorphic 
differential operators on $\CT_{g,n}$.
\item Quantization of twisted cotangent bundle $T^*_c\CM(C)$ $\rightsquigarrow$
Eigenstates $v_q$ of operators $\sq_e$: 
Section of holomorphic vector bundle 
$\CW(C)\ot \CE_c$, where
$\CW(C)$: flat projective vector bundle defined from repr. of ${\rm MCG}(C)$ 
defined in  Part \ref{par:quant}.
\item Quantization of generating functions $\CW_\si(l,q)$ 
$\rightsquigarrow$  matrix elements 
$\CF_l^\si(q)\equiv{}^{}_{\si}\!\langle\,v_q\,,\,\de_l^{\si}\,\rangle\!{}^{}_{\si}$.
\item Results of Parts II and III $\rightsquigarrow$ Riemann-Hilbert type problem 
for $\CF_l^\si(q)$.
\end{itemize}

\subsection{Algebra of functions - representations}

\subsubsection{}

We want to describe the quantization of the spaces
$\CM^{0}_{\rm flat}(C)\simeq\CT(C)$ 
in a way that makes explicit use of 
the complex structure on these spaces. In order to do this, 
we find it convenient to represent $\CM^{0}_{\rm flat}(C)$ as a connected
component of the real slice $\CM_{\rm flat}(C)$ within 
$\CM^{\BC}_{\rm flat}(C)$. As a preliminary, we are going to explain 
how such a description works in a simple example.

Let us consider $\BR^2$ with real coordinates $x$ and $p$ and Poisson bracket
$\{x,p\}=1$. Canonical quantization will produce operators $\spp$ and $\sx$
with commutation relations $[p,x]=-i\hbar$, which can be
realized on a space of functions $\psi(x)$ 
of a real variable $x$. This is a simple analog of the 
quantization scheme discussed in Part II.

We now want to use a quantization scheme that makes explicit 
use of the complex structure of $\BR^2\simeq \BC$. In order to do this
let us consider $\BR^2$ as a real slice of the  
space $\BC^2$. One could, of course, use complex coordinates  
$x$ and $p$ for $\BC^2$ with Poisson bracket $\{x,p\}=1$, 
and describe the real slice $\BR^2$ by the requirement 
$x^*=x$, $p^*=p$. Alternatively one may use 
the complex analytic coordinates $a=x+ip$ and $a'=x-ip$ for $\BC^2$
which have Poisson bracket $\{a,a'\}=-2i$. 
The real slice $\BR^2$ is then described 
by the equation $a'=a^*$ which expresses $a'$ as 
a non-holomophic function of the complex analytic coordinate $a$
on the real slice $\BR^2$.

Quantization of the Poisson bracket $\{a,a'\}=-2i$
gives operators $\sa$, $\sa'$ which satisfy $[\sa,\sa']=2\hbar$. 
This algebra can be represented on functions $\Psi(a)$ in terms
of the holomorphic differential operator $\frac{\pa}{\pa a}$. 
If $\sa$ and $\sa'$ were independent variables, we could also realize
the algebra $[\bar\sa,\bar\sa']=-2\hbar$ generated by the hermitian 
conjugate operators on non-holomorphic functions 
$\Psi(a)\equiv\Psi(a,\bar{a})$.

But in the case of interest, $a'$ is a non-holomorphic function of
$a$ by restriction to the real slice. We want to point out that
it is then natural to realize $[\sa,\sa']=2\hbar$ 
on {\it holomorphic} functions $\Psi(a)$, thereby making 
explicit use of the complex structure on the phase space $\BR^2$. 
There is a natural isomorphism with the representation defined on
functions $\psi(x)$ 
of a real variable $x$ which can be described as an integral 
transformation of the form
\begin{equation}
\Psi(a)\,=\,\int dx\;\langle a|x\rangle\Psi(x)\,,
\end{equation}
where the kernel $\langle a|x\rangle$ is the complex conjugate of the
wave-function $\psi_a(x)=\langle x|a\rangle$ of an eigenstate 
of the operator $\sa=\sx+i\spp$ with eigenvalue $a$.

The representation of the Hilbert space using 
{\it holomorphic} functions $\Psi(a)$ is known as the coherent state
representation in quantum mechanics.

\subsubsection{}

In the present case we regard the Darboux coordinates $(l,k)$ as analogs of
the coordinates $(x,p)$, while the coordinates $(q,h)$ take the role of 
$(a,a^*)$.  Both $(k,l)$
and $(q,h)$ form systems of Darboux coordinates for $\CT(C)$. The coordinates 
$q_r$ alone are complex analytic coordinates for $\CQ(C)$, and the coordinates 
$h_r$ are non-holomorphic functions $h_r=h_r(q,\bar{q})$ 
-- this is in exact analogy
to the case of $(a,a^*)$. Important differences will 
follow from the fact that the relation
between $(q,h)$ and $(l,k)$ is much more complicated than the 
relation between $(x,p)$ and $(a,a^*)$. It is no longer true that
$h_r$ is the complex conjugate of $q_r$.

Quantization is canonical on a purely algebraic level: 
We introduce a noncommutative algebra
with generators $\hat{q}=(\hat{q}_1,\dots,\hat{q}_h)$ and $\hat{h}=(\hat{h}_1,\dots,\hat{h}_h)$ and relations
\begin{equation}\label{Hrel}
[\,\hat{h}_r\,,\,\hat{q}_s\,]\,=\,b^2\,\de_{r,s}\,.
\end{equation}
The resulting algebra is the natural quantization of the algebra of 
holomorphic functions on the cotangent bundle $T^*\CT(C)$
which will be denoted as ${\rm Fun}_{b}(T^*\CT(C))$. 

There is an obvious realization of the algebra 
${\rm Fun}_{b}(T^*\CT(C))$
on functions $\Psi(q)$ locally defined on subsets of $\CT(C)$.
The generators $\hat{\sq}_r$ corresponding to the coordinate $q_r$ 
introduced in Section \ref{cplxDarboux}
are represented
as operators of multiplication by $q_r$, and
the generators $\hat h_r$ associated to the conjugate "momenta"
$h_r$  should be represented by the differential operators
$\sh_r\equiv b^2\pa_{q_r}$ in such a representation,
\begin{equation}\label{K-quant}
\sq_r\Psi(q)\,=\,q_r\Psi(q)\,,\qquad\sh_r\Psi(q)\,=\,
b^2\,\frac{\pa}{\pa q_r}\,\Psi(q)\,.
\end{equation}

The resulting representation should be seen as an
analog of the coherent state representation of quantum
mechanics.

\subsubsection{}

As both $(k,l)$
and $(q,h)$ form systems of Darboux coordinates for $\CT(C)$, 
we expect that there exists a unitary equivalence between 
the representations on functions $\psi(l)$ defined in Part II, and the
representation on holomorphic functions $\Psi(q)$ we are constructing 
here. This means in particular that
there should ultimately be a representation of the scalar product in $\CH(C)$
within each of these
representations
\begin{equation}
\langle\,\Psi\,,\,\Psi\,\rangle\,=\,\int d\mu(l)\;|\psi(l)|^2\,=\,
\int_{\CT(C)} d\mu(q,\bar q)\;|\Psi(q)|^2\,. 
\end{equation} 
Normalizability of the wave-functions $\psi(q)$ 
will restrict both the appearance of singularities in the analytic
continuation of $\psi(q)$ over all of $\CT(C)$, and the behavior 
of $\psi(q)$ at the boundaries of $\CT(C)$.
In our case it is not apriori obvious how 
to identify a natural domain for 
the action of the operators $(\sq,\sh)$ which represent
${\rm Fun}_{b}(T^*\CT(C))$ on {\it holomorphic}
wave-functions $\Psi(q)$. However, it is certainly natural to expect that 
$\Psi(q)$ has to be analytic on all of $\CT(C)$. 
It will furthermore be necessary to demand that the behavior
of $\Psi(q)$ at the boundaries of $\CT(C)$ is "regular" in a sense
that needs to be made more precise. A more precise description of 
the space of wave-functions that is relevant here will eventually 
follow from the results to be described below.

It is natural to introduce eigenstates $v_q$ of the 
position operators $\sq_r$
such that
\begin{equation}\label{cohwavefct}
\Psi(q)\,=\,\langle\,v_q\,,\,\Psi\,\rangle\,.
\end{equation}
The definition of the coordinates $q$ will in general require
the consideration of a  local patch $\CU_{\imath}\subset\CT(C)$.
The corresponding wave-functions will be denoted as $\Psi_\imath(q)\equiv
\langle v_q^\imath,\Psi\rangle$. When the coordinates 
$q$ come from the gluing construction we will use the index $\si$
instead of $\imath$.

\subsubsection{}

Important further requirements are motivated by the fact that
the cotangent bundle $T^*\CT(C)$ descends to a twisted cotangent bundle
over $T^*_c\CM(C)$ for which coordinates like $(q,h)$ 
represent local systems of 
coordinates. Recall that the coordinates $H^{\imath}$ and $H^{\jmath}$
associated to different patches 
$\CU_{\imath}$ and $\CU_{\jmath}$ are related via \rf{H-trans},
where 
\begin{equation}
\chi_{\imath\jmath}^{}\,=\,\frac{1}{2\pi i}\pa\log f^{c_{\rm cl}}_{\imath\jmath}\,.
\end{equation}
The relation \rf{H-trans} has a natural quantum counterpart, 
\begin{equation}\label{q-H-trans}
\sum_r dq^{\imath}_r\frac{\pa}{\pa q_r^{\imath}}\Psi_\imath(q)\,=\,
\sum_r dq^{\jmath}_r\frac{\pa}{\pa q_r^{\jmath}}\Psi_\jmath(q)
-\frac{1}{b^2}\,\chi^{\imath\jmath}\,,
\end{equation}
which leads us to require that
\begin{equation}\label{Psi-trans}
\Psi_\imath(q)\,=\,f^{c}_{\imath\jmath}(q)\Psi_\jmath(q)\,,
\end{equation}
where the parameter $c$ will be given by $c_{\rm cl}$ up to 
corrections of order $b^2$ that will be determined later.

The mapping class group ${\rm MCG}(C)$ acts by holomorphic 
transformations on $\CT(C)$. We will use the notation $\mu.\tau$ 
for the image of a point $\tau\in\CT(C)$ under
$\mu\in{\rm MCG}(C)$. 
We require that there is a
representation of ${\rm MCG}(C)$ on $\CH(C)$ which 
is represented on the wave-functions $\Psi_\si(q)$ most naturally as
\begin{equation}\label{MCGact}
(\SM_{\mu}\Psi)_{\mu.\si}(q)\,=\,\Psi_{\si}(\mu.q)\,,\quad{\rm or}\quad
\SM_{\mu}^{-1}v_q^{\mu.\si}\,=\,v_{\mu.q}^{\si}\,.
\end{equation}
This requirement should be understood as one of the properties defining
the representations $\Psi_\si(q)$, or equivalently the eigenstates $v_q^\si$.

\subsection{Relation between length representation and 
K\"ahler quantization}
\label{sec:length-Kahler}

There should exist expansions of  the form
\begin{equation}\label{asym}
\Psi_{\si,\si'}(q)\,=\,\int dl\;
\langle\, v_q^{\si}\,,\, \de_{l}^{\si'}\,\rangle
\langle\,\de_l^{\si'}\,,\,\Psi\,\rangle
\,=\,\int dl\;
\CF_{\si,\si'}(l,q)\,\psi_{\si'}(l)\,.
\end{equation}
The requirement \rf{asym} introduces key objects, 
the eigenfunctions $\Psi^{\si,\si'}_{l}(q)\equiv\CF_{\si,\si'}(l,q)$ 
of the length operators.
We will mostly restrict attention to the diagonal case 
$\si\equiv \si'$ in the following, and denote $ \Psi^{\si}_{l}(q)\equiv
\Psi^{\si,\si}_{l}(q)$.

The wave-functions
$\Psi_{l}^{\si_\1}(q)$ 
and
$\Psi_{l}^{\si_\2}(q)$
associated to different patches 
$\CU_{\si_\1}$ and $\CU_{\si_\2}$ are related
by an integral transformation of the following form:
\begin{equation}\label{VB}
\Psi_{l_\1}^{\si_\1}(q)\,=\,f^{c}_{\si_\1\si_\2}(q)\,\int dl_\2\;U_{\si_\1\si_\2}(l_\1,l_\2)
\Psi_{l_\2}^{\si_\2}(q)\,,
\end{equation}
as follows from
\begin{align}\label{pre-VB}
\langle\,v_q^{\si_\1}\,,\,\de_{l_\1}^{\si_\1}\,\rangle & \,=\,
f^{c}_{\si_\1\si_\2}(q)\,\langle\,v_q^{\si_\2}\,,\,\de_{l_\1}^{\si_\1}\,\rangle\\
& \,=\,
f^{c}_{\si_\1\si_\2}(q)\,\langle\,v_q^{\si_\2}\,,\,
\SU_{\si_\1\si_\2}\de_{l_\1}^{\si_\2}\,\rangle\,.
\notag\end{align}

Let us now consider the wave-function 
$\langle\,v_q^{\mu.\si}\,,\,\de_{l_\1}^{\mu.\si}\,\rangle$,
where $\mu\in{\rm MCG}(C)$. On the one hand, 
\begin{align}\label{pre-RH}
\Psi^{\mu.\si}_l(q) &\,\;=\;
\langle\,v_q^{\mu.\si}\,,\,\de_{l}^{\mu.\si}\,\rangle 
\,=\,\langle\,v_q^{\mu.\si}\,,\,\SM_{\mu}^{}\de_{l}^{\si}\,\rangle\notag\\
& \,{\={MCGact}}\,\langle\,v_{\mu.q}^{\si}\,,\,\de_{l}^{\si}\,\rangle\,=\,
\Psi^{\si}_l(\mu.q)\,.
\notag\end{align}
In the first line we have beeen using that $\SM_{\mu}=\SU_{\mu.\si,\si}$,
in passing to the second the unitarity of $\SM_\mu$ and
our requirement
\rf{MCGact}.
Another way of representing the wave-function 
$\langle\,v_q^{\mu.\si}\,,\,\de_{l_\1}^{\mu.\si}\,\rangle$ is found
by
specializing \rf{VB} to the case that $\si_\1=\mu.\si$, and 
$\si_\2=\si$. Taken together we find
\begin{equation}\label{RH}
\Psi_{l_\1}^{\si}(\mu.q)\,=\,f^{c}_{\mu.\si,\si}(q)\,\int dl_\2\;M_{\mu}(l_\1,l_\2)
\Psi_{l_\2}^{\si}(q)\,.
\end{equation}
Note that one may read \rf{RH} as expression of the fact that
the wave-functions $\Psi^{\si}_{l}(q)$ 
represent sections of the holomorphic vector
bundle $\CV(C):=\CW(C)\ot\CE_{c}$ over $\CM(C)$, 
where $\CW(C)$ is the projective local system 
defined by the projective representation of the 
mapping class group constructed in Part II. For the reader's 
convenience we have reviewed the notion of a projective local system
in Appendix \ref{App:proj}.
It is very important that the holomorphic 
bundle $\CV(C)$ of Hilbert spaces over 
$\CM(C)$ is an ordinary vector bundle as opposed to a projective one,
as the latter can not have any section. 

The kernels $M_{\mu}(l_\1,l_\2)$ in \rf{RH} have been defined
in Part II. The classical limits of
$-b^2 \log M_{\mu}(l_\1,l_\2)$
may be identified with the 
generating functions $F_{\mu.\si,\si}(l_\1,{l}_\2)$ that appear in 
\rf{cl-MCGact}. The transition functions $f^{c}_{\mu.\si,\si}(q)$ in \rf{RH}
may then be identified with $e^{2\pi i g_{\mu.\si,\si}^{c}(q)}$, 
with $g_{\si_\2\si_\1}^{c}(q)$ being the
transition function of $\CE_{c}$ defined via \rf{cl-MCGact}.

Having specified the data $M_{\mu}(l_\1,l_\2)$ and $f^{c}_{\mu.\si,\si}(q)$
defining the vector bundle $\CV(C)$, one may regard 
\rf{VB} as definition of a Riemann-Hilbert type problem
for the wave-functions $\Psi_{l}^{\si}(q)$.
If $\CV(C)$ were a projective 
vector bundle, the Riemann-Hilbert problem \rf{RH} would not have any
solution. 
The fact that it has a solution for
\begin{equation}
c\,=\,c_{\rm cl}+13b^2+6b^4\,,
\end{equation}
will immediately follow from the relation with Liouville theory 
to be exhibited 
in the next section. 
Note that 
\begin{equation}
{\mathbf c}:=\,\frac{c}{b^2}\,=\,13+6\left(b^2+b^{-2}\right)\,,
\end{equation}
coincides with the expression of the 
central extension found in equation \rf{centext} above.

\subsection{Uniqueness and asymptotics}

Uniqueness of the solution to the 
Riemann-Hilbert problem defined above 
can then be shown by a variant
of the argument used in \cite{T}: Any two solutions of the
Riemann-Hilbert problem differ by multiplication with a 
meromorphic function with possible poles at the boundary $\pa\CM(C)$
of $\CM(C)$. In order to fix this ambiguity one needs
to fix the asymptotic behavior at $\pa\CM(C)$. 
Let us consider the component of $\pa\CM(C)$ where the gluing parameter
$q_e$ vanishes. We need to distinguish the cases $C_e\simeq C_{0,4}$
and $C_e\simeq C_{1,1}$, as before. 

Let us consider the case $C_e\simeq C_{0,4}$.
Note that
the functions $\CF_{\si}(l,q)\equiv\CF_{\si,\si}(l,q)$ 
represent the quantum counterparts of
the generating functions $\CW_\si(l,q)$, in the sense that
\begin{equation}
\CW_{\si}(l,q)\,=\,-\lim_{b\ra 0}b^2\log\CF_{\si}(l,q)\,.
\end{equation}
In view of the asymptotic behavior 
\rf{CWfactor} and \rf{CWfactor,g=1} of $\CW_\si(l,q)$
it is therefore natural to require that the functions $\CF_\si(l,q)$
should have asymptotics of the form 
\begin{equation}\label{as-ansatz}
\log\CF_{\si}(l,q)\,=\,
(\De(l_e)-\De(l_\1)-\De(l_\2))\log q_e+\CF_{0,\si}(l)+\CO(q_e)\,.
\end{equation}
The functions $\De(l)$
should coincide with $\frac{1}{b^2} \de(l)$ up to possible quantum 
corrections, 
$b^2\De(l)=\de(l)+\CO(b^2)$. The form \rf{as-ansatz}
of the asymptotic behavior is equivalent to the validity of a
quantized version of the relation \rf{Hasym} 
which takes the following form
\begin{equation}\begin{aligned}
& \Big(b^2\big[(1-\nu)q_e\pa_{q_e}^{}+\nu \pa_{q_e}^{}q_e\big]
+\de(l_\1)+\de(l_\2)-\de(l)\Big)
\CF_{\si}(l,q)\,=\,0\,.
\end{aligned}\end{equation}
On the left hand side we have parameterized the 
ambiguity in the operator ordering using the parameter $\nu\in[0,1]$. 
Consistency with the realization of the B-move, given 
in \rf{Bcoeff}, requires that $\nu=\frac{1}{2}+\frac{b^2}{4}$. 
This determines the possible quantum corrections in the definition
of the function $\De(l)$ to be $\De(l)
=\frac{1}{b^2}\de(l)+\nu$, which gives
\begin{equation}
\De(l)\,=\,
\left(\frac{l}{4\pi b}\right)^2+\frac{Q^2}{4}
\,,\qquad Q=b+b^{-1}\,.
\end{equation}
In a very similar way one may treat the case $C_e\simeq C_{1,1}$. 
Having fixed the asymptotics, the 
solution to the Riemann-Hilbert problem is unique 
up to multiplication 
by a constant.

\subsection{Scheme dependence}

We had noted above in Subsection \ref{sec:scheme1} that the
definition of the observables $h_r$ depends on the  choice of
a projective structure. A similar issue must therefore be
found in the quantum theory concerning the 
definition of the operators $\sh_r$. We have to allow
for redefinitions of the operators $\sh_r$ that correspond to 
redefinitions of the eigenstates $v_q$ by multiplicative
factors which may depend on $q$. 

This freedom is physically irrelevant in the following sense.
What is physically relevant are normalized expectation values of 
observable like
\begin{equation}
\big\langle\!\!\big\langle \,\SO\,\big\rangle\!\!\big\rangle_q:=\,\frac{
\langle\, v_q\,,\SO\,v_q\,\rangle}{\langle\, v_q\,,v_q\,\rangle}\,.
\end{equation}
It is clear that such expectation values are unaffected by
redefinitions of the eigenstates $v_q$ by multiplicative, $q$-dependent
factors. This is how the scheme dependence discussed in Section 
\ref{sec:inst} manifests itself in the quantum theory of 
$\CM^0_{\rm flat}(C)$.

\section{Relation to quantum Liouville theory}

\setcounter{equation}{0}

We will now argue that
the conformal field theory called Liouville theory is mathematically
best interpreted as the harmonic analysis on Teichm\"uller spaces, 
which is another name for the quantum theory defined in the previous section.
This will partly explain why
the Riemann-Hilbert type problems defined in 
Sections \ref{sec:RHgauge} and \ref{hol-quant} are solved
by Liouville theory.

\subsection{Virasoro conformal blocks}\label{Vircfbl}

\renewcommand{\fw}{w}
\renewcommand{\fv}{v}
\renewcommand{\fe}{e}

\subsubsection{Definition of the conformal blocks}

\newcommand{\vir}{{\rm Vir}_{\mathbf c}}
\newcommand{\CFB}{{\SC\SB}}
\newcommand{\HFB}{{\SH\SC\SB}}
\newcommand{\SFB}{{\SS\SC\SB}}

The Virasoro algebra ${\rm Vir}_c$ has generators $L_n$,  $n\in\BZ$,
and relations
\begin{equation}\label{Vir}
[L_n,L_m] = (n-m)L_{n+m}+\frac{\mathbf c}{12}n(n^2-1)\de_{n+m,0}.
\end{equation}
Let $C$ be a Riemann surface $C$ with $n$ marked points 
$P_1,\ldots,P_n$. At 
each of the marked points $P_r$, $r=1,\ldots,n$,  
we choose local coordinates 
$w_r$, which vanish at $P_r$. We will fix a projective
structure on $C$ and assume that the patches around
the points $P_r$ are part of an atlas defining the
projective structure.
We associate highest weight representations
$\CV_r$, of $\vir$ 
to $P_r$, $r=1,\ldots,n$. The representations
$\CV_r$ are generated from highest weight vectors $e_r$ with
weights $\De_r$. 

The conformal blocks
are then defined to be the linear functionals
$\CF:\CV_{[n]}\equiv\otimes_{r=1}^n\CV_r\ra\BC$ that
satisfy the invariance property
\begin{equation}\label{cfblvir}
\CF_C(T[\chi]\cdot v) = 0\quad \forall v\in\CR_{[n]},\quad \forall\chi\in
\FV_{{\rm out}},
\end{equation}
where $\FV_{{\rm out}}$ is the Lie algebra of meromorphic
differential operators on $C$ which may have poles only at
$P_1,\ldots,P_n$. The action of $T[\chi]$ on
$\otimes_{r=1}^n\CR_r\ra\BC$ is defined as
\begin{align}\label{Tdef}
T[\chi] = \sum_{r=1}^n {\rm id}\ot\dots\ot\underset{(\rm r-th)}{L[\chi^{(r)}]}\ot\dots\ot{\rm id},\quad
L[\chi^{(r)}] := \sum_{k\in\BZ} L_k^{} \chi_k^{(r)} \in \vir,
\end{align}
where $\chi_k^{(r)}$ are the coefficients of the 
Laurent expansions of $\chi$ at the points $P_1,\dots P_n$,
\begin{equation}
\chi(z_r) = \sum_{k\in\BZ} \chi_k^{(r)}\,w_r^{k+1} \,\pa_{w_r} \in
\BC(\!(w_r)\!)\pa_{w_r}\,.
\end{equation}
It can be shown that the central extension vanishes on the
image of the Lie algebra $\FV_{{\rm out}}$ in
$\bigoplus_{r=1}^n \vir$, making the definition consistent.
We may refer to \cite{AGMV,W88} for early discussions
of this definition in the physics literature,
and to \cite{BF} for a mathematically rigorous 
treatment.

The vector space
of conformal blocks associated to the Riemann surface $C$
with representations $\CV_r$ associated to the marked
points $P_r$, $r=1,\ldots,n$ will be denoted as
$\CFB(\CV_{[n]},C)$. It is the space of solutions to 
the defining invariance conditions \rf{cfblvir}.
The space $\CFB(\CV_{[n]},C)$ is infinite-dimensional in general. 
Considering 
the case $n=1$, $\De_1=0$ and $g>1$, 
for example, one may see this more explicitly
by noting that for $z_1$ in generic position\footnote{We assume that $z_1$
is not a Weierstrass point.} one may
find a basis for $\FV_{{\rm out}}$ generated by vector fields 
which have a pole of order higher than $3g-3$. This follows from the
Weierstrass gap theorem. The conditions \rf{cfblvir} will then allow us to 
express the values of $\CF$ on arbitrary vectors in $\CV_1$
in terms of the values 
\begin{equation}\label{values}
\CF\big(L_{3-3g}^{k_{3g-3}}\dots L_{-1}^{k_1}e_1\big)\,,\qquad
k_1,\dots,k_{3g-3}\in\BZ^{>0}\,,
\end{equation}
were $e_1$ is the highest weight vector of $\CV_1$. We note
that $\CF$ is completely defined by the values \rf{values}.
$\CFB(\CV_{[n]},C)$ is therefore isomorphic as a vector space to the 
space of {\it formal} power series in $3g-3$ variables.

\subsubsection{Conformal blocks as expectation values of 
chiral vertex operators}

Let us also introduce the notation 
\begin{equation}\label{partdef}
\CZ^{\rm\sst L}(\CF,C)\,=\,\CF(e_1\ot\dots \ot e_n)\,,
\end{equation}
for the value of $\CF$ on the product of highest weight
vectors. $\CZ^{\rm\sst L}(\CF,C)$ can be interpreted as a chiral 
``partition function'' from a physicist's point of view.
It may alternatively be interpreted as an expectation value
of a product of $n$ chiral primary fields inserted into 
a Riemann surface $C$. This interpretation
may be expressed using the notation
\begin{equation}\label{cfblnotation}
\CZ^{\rm\sst L}(\CF,C) = \left\langle\, 
\Phi_n(z_n)\dots\Psi_1(z_1)\,\right\rangle_{\CF}^{}.
\end{equation}
The state-operator correspondence associates chiral vertex operators
$\Phi_r(v_r|z_r)$ to arbitrary vectors $v_r\in\CV_r$. The vertex operators
$\Phi_r(v_r|z_r)$ are called the descendants of $\Phi_r(z_r)$. The
value $\CF(v_1\ot\dots \ot v_n)$ 
is therefore identified with the expectation value
\begin{equation}\label{cfblnotation-2}
\CF(v_1\ot\dots\ot v_n) = \left\langle\, 
\Phi_n(v_n|z_n)\dots\Psi_1(v_1|z_1)\,\right\rangle_{\CF}^{}.
\end{equation}
There are generically many different ways to ``compose'' chiral 
vertex operators. The necessary choices are encoded in the choice of 
$\CF$ in a way that will become more clear in the following.

\subsubsection{Deformations of the complex structure of 
$C$}

A key point that needs to be understood about spaces of
conformal blocks is the dependence on the complex structure
of $C$. There is a canonical way to represent infinitesimal
variations of the complex structure on the spaces of
conformal blocks. By combining the definition of conformal
blocks with the so-called ``Virasoro uniformization'' of
the moduli space ${\CM}_{g,n}$ of complex structures on
$C=C_{g,n}$ one may construct a representation of
infinitesimal motions on ${\CM}_{g,n}$ on the space of
conformal blocks.

The ``Virasoro uniformization'' of the moduli space
${\CM}_{g,n}$  may be formulated as the statement that the
tangent space $T{\CM}_{g,n}$ to ${\CM}_{g,n}$ at $C$ can be
identified with the double quotient
\begin{equation}\label{VirUni}
T{\CM}_{g,n} = \Ga(C\setminus\{x_1,\ldots,x_n\},\Theta_C)
\left\backslash \bigoplus_{k=1}^n \BC(\!(t_k)\!)\pa_k
\right/ \bigoplus_{k=1}^n t_k\BC[[t_k]]\pa_k,
\end{equation}
where  $\BC(\!(t_k)\!)$ and
$\BC[[t_k]]$ are the spaces of formal Laurent and Taylor series
respectively, and $\Ga(C\!\setminus\{x_1,\ldots,x_n\},\Theta_C)$ is the
space of vector fields that are holomorphic on
$C\setminus\{x_1,\ldots,x_n\}$.

Given a tangent vector $\vartheta\in T{\CM}_{g,n}$, 
it follows from the Virasoro uniformization \rf{VirUni} 
that we may find an element
$\eta_{\vartheta}$ of $\bigoplus_{k=1}^n
\BC(\!(t_k)\!)\pa_k$, which represents $\vartheta$ via
\rf{VirUni}.
Let us then consider $\CF(T[\eta_\vartheta] v)$ with $T[\eta]$
being defined in \rf{Tdef} in the case that the vectors
$v_k$ are the highest
weight vectors $e_k$ for all $k=1,\ldots,n$. 
\rf{VirUni} suggests to define the
derivative $\de_{\vartheta} \CF(v)$ of $\CF(v)$ in the direction of 
 $\vartheta\in T{\CM}_{g,n}$ as
\begin{equation}\label{Viract}
\de_{\vartheta} \CF(v) := \CF(T[\eta_{\vartheta}^{}]v),
\end{equation}
 Dropping the condition that $\fv$ is a product of highest
weight vectors one may try to use \rf{Viract} to define
$\de_{\vartheta} \CF$ in general. 
And indeed, it is well-known that \rf{Viract} 
leads to the definition of a canonical connection on 
the space $\CFB(\CV_{[n]},C)$ of conformal blocks which is 
projectively
flat, see e.g. \cite{BF} for more details.

There is no hope to integrate the canonical connection
on  $\CFB(\CV_{[n]},C)$ to produce a bundle over $\CM(C)$ with 
fiber at a Riemann surface $C$ being $\CFB(\CV_{[n]},C)$, in general.

The first problem is that the connection defined by \rf{Viract} is not
flat, but only projectively flat. It can only define a 
connection on the projectivized space $\BP\CFB(\CV_{[n]},C)$,
in general.
For the readers convenience we have
gathered some basic material on connections on bundles 
of projective spaces in Appendix \ref{App:proj}.
As we will see in a little more detail later, 
one may trivialize the curvature 
at least locally, opening the possibility to integrate
\rf{Viract} at least in some local patches $\CU\subset\CM(C)$.

The other problem is simply that  $\CFB(\CV_{[n]},C)$ is way too big,
as no growth conditions whatsoever are imposed on the values 
\rf{values} for general elements $\CF\in\CFB(\CV_{[n]},C)$.
One needs to find interesting subspaces of $\CFB(\CV_{[n]},C)$
which admit useful topologies.

We will later even be able to identify 
natural Hilbert-subspaces $\HFB(\CV_{[n]},C)$ 
of  $\CFB(\CV_{[n]},C)$. The  Hilbert-subspaces $\HFB(\CV_{[n]},C)$ will 
be found to  glue into a bundle of projective vector
spaces $\CW(\CV_{[n]},C)$ over $\CM(C)$ with connection 
defined via \rf{Viract} -- this is the best possible
situation one can hope for in cases where the spaces of 
conformal blocks are infinite-dimensional.

\subsubsection{Propagation of vacua}

The vacuum representation $\CV_0$ which corresponds to
$\De_r=0$ plays a distinguished role. If
$\Phi_0(\fv_0|w_0)$ is the vertex operator associated to
the vacuum representation, we have
\begin{equation}\label{psi0}
\Phi_0(\fe_0|w_0) = {\rm id},\quad
\Phi_0(L_{-2}\fe_0|w_0) = T(w_0),
\end{equation}
where $T(z)$ is the energy-momentum tensor. It can be shown
that the spaces of conformal blocks with and without
insertions of the vacuum representation are canonically
isomorphic, see e.g. \cite{BF} for a proof. The isomorphism between
$\CFB(\CV_0\ot\CV_{[n]},C_{g,n+1})$ and
$\CFB(\CV_{[n]},C_{g,n})$ is simply given by evaluation at
the vacuum vector $\fe_0\in\CV_0$
\begin{equation}\label{vacprop}
\CF'(\fe_0^{}\ot\fv) \equiv
\CF(\fv),\qquad \fv\in\CV_{[n]}\,,
\end{equation}
as is also obvious from \rf{psi0}. This fact is often
referred to as the ``propagation of vacua''.

One may then define the expectation value of the energy momentum
tensor defined by a fixed element $\CF$ 
as follows
\begin{equation}\label{Tvaldef}
T_{\CF}(w_0)\equiv\langle\!\langle\,T(w_0)\,\rangle\!\rangle^{}_{\CF}:=
\CF'(L_{-2}\fe_0\ot\fv) \,/\,
\CF(\fv)\,.
\end{equation} 
We are assuming that the local coordinate $w_0$ is part of an 
atlas defining the chosen projective structure on $C$. It follows
that $T_{\CF}(w_0)$ transforms like a quadratic differential 
when going from one patch of this atlas to another.

The invariance property \rf{cfblvir} allows
us to rewrite $\CF'(L_{-2}\fe_0\ot\fv)$ in the form
\begin{equation}
\CF'(L_{-2}\fe_0\ot\fv)\,=\,\CF'(\fe_0\ot\vartheta_{w_0}^{}\fv)\,,
\end{equation}
with $\vartheta_{w_0}=T[\chi_{w_0}^{}]$, for a 
vector field $\chi_{w_0}^{}$ that has a pole at $w_0$. We may then use
\rf{vacprop} to write $\CF'(\fe_0\ot\vartheta_{w_0}^{}\fv)=
\CF(\vartheta_{w_0}^{}\fv)$. It follows that 
$T_{\CF}(w_0)$ can be expressed in terms of 
$\CF$ as 
\begin{equation}\label{T-def}
T_{\CF}^{}(w_0)\,=\,
\CF(\vartheta_{w_0}^{}\fv)/\CF(\fv)\,.
\end{equation} 
Recalling the definition 
\rf{Viract}, we observe that that the canonical 
connection can be characterized in terms of the
expectation value $T_{\CF}^{}(w_0)$.

\subsubsection{Parallel transport}

Note that the value $\CF(\vartheta_{w_0}^{}\fv)$
in \rf{T-def}, by definition, represents the
action of a differential operator $\CT_{w_0}^{}$ 
corresponding to a tangent vector to $\CM(C)$ 
on $\CF$.
This statement may be expressed in the form of a differential equation
for $\CZ^{\rm\sst L}(\CF,C)$
\begin{equation}\label{transport}
\CT_{w_0}^{}\,\CZ^{\rm\sst L}(\CF,C)\,=\,T_{\CF}(w_0)\,\CZ^{\rm\sst L}(\CF,C)\,.
\end{equation}
The differential equation \rf{transport} may be re-written using local 
coordinates $q=(q_1,\dots,q_h)$ for $\CT(C)$ whose variation
is described by means of Beltrami-differentials $(\mu_1,\dots,\mu_h)$ as
\begin{equation}\label{Cfbltransp}
\big[{\pa}_{q_r}+\CA_r(\CF,q)\big]\,
\CZ^{\rm\sst L}(\CF,C)\,=0\,,\qquad \CA_r(\CF,q):=\int_{C}\mu_r T_{\CF}\,.
\end{equation} 
Our aim is to use \rf{Cfbltransp} to construct 
a family $\CF_q$ of conformal blocks over a neighborhood $\CU$ of $\CM(C)$.
We first need to ensure that the partial derivatives
$\frac{\pa}{\pa q_r}$ whose action is defined via \rf{Cfbltransp}
do indeed commute. This amounts to the trivialization of the  curvature
of the canonical connection within $\CU$. 

One way do this concretely uses the 
atlas of local coordinates produced by the gluing construction 
of Riemann surfaces. One may consider 
Beltrami-differentials $\mu_r$ which are compactly supported in
non-intersecting annular regions $A_r$ on $C$. 
Equation \rf{Cfbltransp} then describes 
the variations of the conformal blocks
with respect to the coordinates $q_r$ for $\CT(C)$
defined by the gluing construction.

Let us assume that $\CF$ is such that \rf{Cfbltransp} can be integrated
to define a function $\CZ^{\rm\sst L}(\CF,q)$ in a neighborhood of a point in 
$\CM$ represented by the surface $C$. Note that the Taylor expansion
of  $\CZ^{\rm\sst L}(\CF,q)$ is completely defined by the conformal
block $\CF\in\CFB(\CV_{[n]},C)$. Derivatives of $\CZ^{\rm\sst L}(\CF,q)$  
are related to the values $\CF(T[\eta_{\vartheta}^{}]v)$ via \rf{Viract}.
These values can be computed in terms of the values \rf{values}
which characterize $\CF$
by using the defining invariance condition \rf{cfblvir}. 
Conversely let us note that the values \rf{values}
characterizing a conformal block can be computed from 
the derivatives of $\CZ^{\rm\sst L}(\CF,q)$ via \rf{Viract}.

It may not be possible to integrate 
\rf{Cfbltransp} for arbitrary $\CF\in\CFB(\CV_{[n]},C)$ as
the numbers \rf{values} which characterize $\CF$ may grow too quickly.
We will denote the subspace of $\CFB(\CV_{[n]},C)$ spanned by
the conformal blocks $\CF$ for which \rf{Cfbltransp} can locally
be integrated to an analytic function $\CZ^{\rm\sst L}(\CF,q)$ by
 $\CFB^{\rm\sst an}_{\rm\sst loc}(\CV_{[n]},C)$.

Let us stress that for any given function
$\CZ^{\rm\sst L}(q)$ which is analytic in a neighborhood of a point $q_0$ in 
$\CM$ represented by the surface $C$ one may define a family of 
conformal blocks
$\CF_q\in\CFB(\CV_{[n]},C_q)$
by using the Taylor expansion of $\CZ^{\rm\sst L}(q)$ around $q$ to define the values 
\rf{values} which characterize the elements $\CF_q\in\CFB(\CV_{[n]},C_q)$.
The conformal blocks $\CF$ 
in $\CFB^{\rm\sst an}_{\rm\sst loc}(\CV_{[n]},C)$  
are therefore 
in one-to-one correspondence with analytic
functions $\CZ^{\rm\sst L}(\CF,q)$ defined locally in open subsets $\CU\subset\CM$.

\subsubsection{Scheme dependence}\label{sec:scheme2}

In the definition 
of the conformal blocks we assumed that a projective 
structure on $C$ had been chosen. 
This allows us in particular to define an
expectation value $T_{\CF}(w_0)$ of the energy-momentum
tensor which transforms as a quadratic differential when 
going from one local coordinate patch on $C$ to another.
In order to define families of conformal blocks 
using the canonical
connection one needs to have {\it families} of projective 
structures over local patches $\CU\subset\CM(C)$ that allow one
to trivialize the curvature of the canonical connection 
locally in $\CU$. Such families certainly exist, we had 
pointed out earlier that the families of projective structures
defined by the gluing construction 
described in Subsection \ref{Glueing-2} do the job.

One may describe changes of the underlying projective
structure by considering the corresponding 
oper $\pa_y^2+t_0(y)$, and modifying $t_0(y)$ 
by addition of a quadratic differential
$\sum_{r=1}^{h}h_r\,{\vartheta}_r$. The parallel transport
defined using the modified projective structure will
remain integrable if there exists a potential $\CZ_0(q)$ 
on $\CU$ such that $h_r=-\pa_{q_r}\CZ_0(q)$.
The result will be a modification of the partition functions
$\CZ^{\rm\sst L}(\CF,q)$ by a universal factor,
a function $\CZ_0(q)$ of $q$ independent of the 
choice of $\CF$. This may be regarded as the conformal 
field theory counterpart of the scheme dependence discussed in 
Subsections \ref{sec:inst} and \ref{sec:scheme1}.

\subsubsection{Mapping class group action}

Let $\CFB^{\rm an}(\CV_{[n]},C)$ be the subspace of 
$\CFB^{\rm\sst an}_{\rm\sst loc}(\CV_{[n]},C)$ which can be analytically
continued over all of $\CT(C)$. 
Note that $T_\CF(w_0)$ defines a projective c-connection on $C$. 
Given a family of conformal blocks $\CF_q$ defined in a 
subset $\CU\subset\CM$ one gets a corresponding family of projective 
connections  $T_{\CF_q}(w_0)$. If $\CF\in\CFB^{an}(\CV_{[n]},C)$
one may analytically continue the family of projective 
connections  $T_{\CF_q}(w_0)$ over all of $\CT(C)$. The
resulting section of $\CP(C)\ra\CT(C)$ may then be used to define  
a family of local sections of the projective line bundle $\CE_c$
as explained in Subsection \ref{tw-T*CT}. It is defined by the equations
\rf{Ec-sect} which coincide with \rf{Cfbltransp} in the present case.

Analytic continuation along closed curves in $\CM(C)$ defines
an action of the mapping class group on $\CFB^{\rm an}(\CV_{[n]},C)$.
We will later define a subspace  $\CFB^{\rm\sst temp}(\CV_{[n]},C)$
of  $\CFB^{\rm an}(\CV_{[n]},C)$
which is closed under this action. It may be characterized by
the condition that the partition functions $\CZ^{\rm\sst L}(\CF,q)$  are
``tempered'' in a sense that will 
be made more precise. The spaces 
$\CFB^{\rm\sst temp}(\CV_{[n]},C_q)$ associated to local families
$C_q$ of Riemann surfaces glue into a projective local 
system $\CW_{\rm L}(C)$ over $\CM(C)$.

A vector bundle that is not projective is \cite{FS}
\begin{equation}
\CV_{\rm L}(C):=\,\CW_{\rm L}(C)\ot \CE_c\,.
\end{equation}
Picking a basis for  $\CFB^{\rm\sst temp}(\CV_{[n]},C_q)$ in some 
$\CU\subset\CM(C)$
one may define a section of $\CV(C)$ by means of 
analytic continuation. Natural bases for  $\CFB^{\rm\sst temp}(\CV_{[n]},C_q)$
can be defined by means of the gluing construction, 
as will be explained next.

\subsection{Gluing construction of conformal blocks}\label{glue}


\subsubsection{Gluing boundary components}

Let us first consider a Riemann surface $C_{21}$
that was obtained by gluing two surfaces $C_2$ and $C_1$
with $n_2+1$ and $n_1+1$ boundary components, respectively.
Given an integer $n$, let sets $I_1$ and $I_2$ be such that
$I_1\cup I_2=\{1,\ldots,n\}$. 
Let us consider conformal
blocks $\CF_{C_i}\in\CFB(\CV^{[n_i]}_i,C_i)$ where
$\CV^{[n_2]}_2=(\otimes_{r\in I_2}\CV_r)\otimes\CV_0$ and
$\CV^{[n_1]}_1=\CV_0\otimes(\otimes_{r\in I_1}\CV_r)$ with
the same representation $\CV_0$ assigned to $z_{0,1}$ and
$z_{0,2}$, respectively. Let
$\langle\,.,.\,\rangle_{\CV_0}$ be the invariant bilinear
form on $\CV_0$. For given $v_2\in\otimes_{r\in I_2}\CV_r$
let $W_{\fv_2}$ be the linear form on $\CV_0$ defined by
\begin{equation}
W_{\fv_2}(\fw) := \CF_{C_2}(\fv_2\ot\fw),\quad\forall
\fw\in\CV_0,
\end{equation}
and let $\SC_\1(q)$ be the family of linear operators
$\CV_{1}^{[n_1]}\ra \CV_0$ defined as
\begin{equation}\label{Vprefactor}
\SC_\1(q)\cdot v_1 := \sum_{e\in B(\CV_0)}
q^{L_0}{e}\;\CF_{C_1}(\check{e}\ot v_1),
\end{equation}
where we have used the notation $B(\CV_0)$ for a basis of
the representation $\CV_0$ and $\check{e}$ for the dual of
an element $e$ of $B(\CV_0)$ defined by $\langle\,
\check{e},e'\,\rangle_{\CV_0} = \de_{e,e'}$. We may then
consider the expression
\begin{equation}\label{Vfactor}
\CF_{C_{21}}(\fv_2\ot\fv_1) := W_{\fv_2}(\SC_\1(q)\cdot\fv_1).
\end{equation}
We have thereby defined a new conformal block associated to
the glued surface $C_{21}$, see \cite{T08} for more
discussion. The insertion of the operator $q^{L_0}$ plays
the role of a regularization. It is not a priori clear that
the linear form $W_{\fv_2}$ is defined on infinite linear
combinations such as $\SC_1(q)\cdot\fv_1$. Assuming
$|q|<1$, the factor $q^{L_0}$ will produce an suppression
of the contributions with large $L_0$-eigenvalue, which
renders the infinite series produced by the definitions
\rf{Vfactor} and \rf{Vprefactor} convergent.

An operation representing 
the gluing of two boundary components of a single Riemann surface
can be defined in a very similar way.

\subsubsection{Gluing from pairs of pants}\label{gluingpants}

One can produce any Riemann surface $C$ by gluing pairs of
pants. The different ways to obtain $C$ in this way are
labeled by pants decompositions $\si$. The elementary building 
blocks are the conformal blocks associated to three-punctured spheres
$C_{0,3}$, which are well-known to be uniquely 
defined  up to normalization
by the invariance property \rf{cfblvir}. We fix the normalization 
such that the value of $\CF_{C_{0,3}}$ on the product of 
highest weight vectors is 
\begin{equation}\CF_{C_{0,3}}(e_3\ot e_2 \ot e_1)\,=\,
\sqrt{C(Q-\al_3,\al_2,\al_1)}\,,
\end{equation}
where $C(\al_3,\al_2,\al_1)$ is the function defined in \rf{ZZform}.

Using the
gluing construction recursively leads to the definition of
a family of conformal blocks
$\CF^{\si}_{\be,q}$
depending on the following set of data:
\begin{itemize}
\item $\si$ is a pants decomposition.
\item $q$ is the coordinate for $\CU_{\si}\subset\CT(C)$ defined 
by the gluing construction.
\item $\be$ is an assignment $\be:e\mapsto \be_{e}\in 
\BS\equiv\frac{Q}{2}+i\BR$, defined
for all edges on $\Ga_{\si}$. 
\end{itemize}
The parameters
$\be_{e}$ determine the Virasoro representations
$\CV_{\Delta_{e}}$ to be used in the gluing
construction of the conformal blocks from pairs of
pants via
\begin{equation}\label{Dedef}
\De_{e} \,=\, \be_e(Q-\be_e)\,,\qquad {\mathbf c}=1+6Q^2\,.
\end{equation}

The partition functions $\CZ^{\rm\sst L}_{\si}(\be,q)$
defined from $\CF^\si_{\be,q}$ via \rf{partdef} are entire
analytic with respect to the variables $\al_r$, meromorphic
in the variables $\be_e$, with poles at
the zeros of the Kac determinant, and it can be argued that
the dependence on the
gluing parameters $q$ is analytic in a open multi-disc $\CU_\si$
of full dimension $3g-3+n$
\cite{T03a,T08}.

\subsubsection{Change of pants decomposition}

It turns out that the conformal blocks
$\CZ^{\rm\sst L}_{\si_1}(\be,q)$ constructed by the gluing
construction in a neighborhood of the asymptotic region of
$\CT(C)$ that is determined by $\si_1$ have an analytic
continuation to
the asymptotic region of $\CT(C)$ determined by a second
pants decomposition $\si_2$. A fact \cite{Teschner:2001rv,T03a,T08}\footnote{A full
proof of the statements made here does not appear in the
literature yet. It can, however, be assembled from building
blocks that are published. By using the groupoid of changes
of the pants decompositions it is sufficient to verify the claim for
the cases $g=0,n=4$ and $g=1,n=1$, respectively. For
$g=0,n=4$ this was done in \cite{Teschner:2001rv}, see also \cite{T03a}.
The case of $g=1,n=1$ was recently reduced to the case
$g=0,n=4$ in \cite{HJS}.}
 of foundational importance for the subject is that
the analytically continued conformal blocks
$\CZ^{\rm\sst L}_{\si_\2}(\be_\2,q)$
can be represented as a linear combination
of the conformal blocks $\CZ^{\rm\sst L}_{\si_\1}(\be_\1,q)$, which takes
the form
\begin{equation}\label{CBTrans}
\CZ^{\rm\sst L}_{\si_\2}(\be_\2,q) \,=\,E_{\si_\2\si_\1}(q)
\int d\mu(\be_\1)\;W_{\si_2\si_1}(\be_\2,\be_\1)\,
\CZ^{\rm\sst L}_{\si_\1}(\be_\1,q)\,.
\end{equation}
The mapping class group acts naturally,
\begin{equation}\label{CB-MCG}
\CZ^{\rm\sst L}_{\mu.\si}(\be,q)\,=\,\CZ^{\rm\sst L}_{\si}(\be,\mu.q)\,.
\end{equation}
Combining \rf{CBTrans} and \rf{CB-MCG} yields a relation of the form
\begin{equation}\label{Mgrptrsf}
\CZ^{\rm\sst L}_{\si}(\be_\2,\mu.q) \,=\,E_{\mu.\si,\si}(q)
\int d\mu(\be_\1)\;W_{\mu.\si,\si}(\be_\2,\be_\1)\,
\CZ^{\rm\sst L}_{\si}(\be_\1,q)\,.
\end{equation}
The transformations \rf{Mgrptrsf} define the 
infinite-dimensional vector bundle 
$\CV_{\rm L}(C)=\CE_c\ot\CW_{\rm L}(C)$. The
constant kernels $W_{\si_2\si_1}(\be_\2,\be_\1)$ represent the transition
functions of $\CW_{\rm L}(C)$, while the prefactors $E_{\si_2\si_1}(q)$
can be identified as transition functions of the 
projective line bundle $\CE_c$.

It suffices to calculate the relations \rf{CBTrans} in the cases
of surfaces $C=C_{0,4}$, and $C=C_{1,1}$. 
This was done in \cite{Teschner:2001rv}
for  $C=C_{0,4}$, where a relation of the form
\begin{equation}\label{fustrsf}
\CZ_{\si_s}^{\rm\sst L}(\be_1,q) \,=\,
\int_\BS d\be_2\;\Fus{\al_1}{\al_2}{\al_3}{\al_4}{\be_{1}}{\be_2}\,
\CZ_{\si_t}^{\rm\sst L}(\be_2,q)\,,
\end{equation}
was found. The pants decompositions $\si_s$ and $\si_t$ are depicted on 
the left and right half of Figure \ref{fmove}, respectively.
Using this result, the case $C=C_{1,1}$ was treated in \cite{HJS},
the result being
\begin{equation}\label{Strsf-L}
\CZ_{\si_s}^{\rm\sst L}(\be_1,q) 
\,=\,e^{\pi i\frac{c}{12}(\tau+1/\tau)}
\int_\BS d\be_2\;S_{\be_1\be_2}(\al_0)\,
\CZ_{\si_t}^{\rm\sst L}(\be_2,q)\,,
\end{equation}
where $q=e^{2\pi i \tau}$, as usual.
The pants decompositions $\si_s$ and $\si_t$ are 
depicted in Figure \ref{smove}.
The prefactor is due to the fact that the conformal blocks
defined according to the gluing construction differ by a factor 
of $q^{\frac{c}{24}}$ from the conformal blocks considered in \cite{HJS}.
It represents the only non-trivial transition functions of $\CE_c$
according to our discussion in Subsection \ref{trans-Ec}.

We should again remember that the definition 
of the partition functions  $\CZ^{\rm\sst L}_{\si}(\be,q)$ was based
on a particular scheme, the choice of the projective 
structure coming from the gluing construction described
above. Using a diffent scheme would modify the 
partition functions by $\beta$-independent functions
of $q$. 

\subsection{Comparison with the K\"ahler quantization of $\CT(C)$}

We had previously identified the space of conformal blocks
$\CFB^{\rm\sst an}_{\rm\sst loc}(\CV_{[n]},C)$ with the space of functions $\CZ(q)$
locally defined on patches $\CU\subset\CT(C)$. This space is naturally
acted  on by the algebra of differential operators ${\SD\SO}(\CT(C))$, 
which is directly related to the action of  ${\SD\SO}(\CT(C))$ on 
spaces of conformal blocks defined by means of the Virasoro algebra
via \rf{VirUni}. These observations already indicate that the space
of wave-functions $\Psi(q)$ that represent the Hilbert space 
$\CH(C)$ 
in the representation coming from the
K\"ahler quantization scheme 
should coincide with a suitable
Hilbert-subspace of $\HFB(\CV_{[n]},C)$ 
of  $\CFB(\CV_{[n]},C)$.

The direct calculations of the kernels $W_{\si_2\si_1}(\be_2,\be_1)$ 
carried out
for the generators $Z,B,F$ in \cite{Teschner:2001rv,T03a},
and for $S$ in \cite{HJS} 
yield results that coincide with the kernels defined
in Subsection \ref{sec:elmoves}.  
It follows that $\CW_{\rm L}$ coincides with the projective local system 
from the quantization of $\CM^0_{\rm flat}(C)$,
\begin{equation}
\CW_{\rm L}(C)\,=\,\CW(C)\,.
\end{equation}
This implies immediately that the conformal blocks 
$\CZ^{\rm\sst L}_{\si}(\be,q)$ represent the solution to the Riemann-Hilbert 
problem that was found to characterize the 
wave-functions $\Psi_l^{\si}(q)$ which describe the relation
between length representation and K\"ahler quantization.

These results imply furthermore 
that there is a natural Hilbert space structure on the spaces of
conformal blocks which is such that the mapping class group action 
becomes unitary. The Hilbert spaces $\HFB(\CV_{[n]},C)$ 
of conformal blocks are 
isomorphic as representations of the Moore-Seiberg 
groupoid to the Hilbert spaces
of states constructed in the quantization of $\CM^0_{\rm flat}(C)$ in Part II.

Within $\HFB(\CV_{[n]},C)$ one may consider the maximal domains of definition 
of the algebras $\CA_b(C)$ of quantized trace functions, which can
be seen as natural analogs $\SFB(\CV_{[n]},C)$ of the Schwarz 
spaces of test functions in distribution
theory. The spaces $\SFB(\CV_{[n]},C)$ are Fr\'echet spaces with 
topology given by the family of semi-norms defined from the
expectation values of the operators representing the 
elements of $\CA_b(C)$ on $\SFB(\CV_{[n]},C)$.
The (topological) dual of $\SFB(\CV_{[n]},C)$ 
is the space of ``tempered'' distributions on $\SFB(\CV_{[n]},C)$,
which will be identified with the subspace 
$\CFB^{\rm\sst temp}(\CV_{[n]},C_q)$
of $\CFB(\CV_{[n]},C)$ spanned by
``tempered'' conformal blocks.

\section{Relation to gauge theory}

\setcounter{equation}{0}

\subsection{The solution to the Riemann-Hilbert problem}

We have seen that the kernels representing S-duality transformations
in the gauge theory coincide with the kernels representing the changes
of pants decomposition in Liouville theory.
Taken together we conclude that 
\begin{equation}\label{AGT-fin}
\CZ^{\rm inst}_\si(a,m,\tau;\ep_1,\ep_2)\,=\,
\CZ^{\rm spur}_\si(\al,\tau;b)\,\CZ^{\rm\sst L}_{\si}(\be,\al,q;b)\,,
\end{equation}
where the following identifications of parameters have been used,
\begin{subequations}\label{paramid}
\begin{align}
&b^2\,=\,\frac{\ep_1}{\ep_2}\,,\qquad
\hbar^2\,=\,\ep_1\ep_2\,,\qquad q=e^{2\pi i\tau}\,,\\
&\be_e\,=\,\frac{Q}{2}+i\frac{a_e}{\hbar}\,,\qquad
\al_r\,=\,\frac{Q}{2}+i\frac{m_r}{\hbar},\,\qquad Q:=b+b^{-1}\,.
\end{align}
\end{subequations}
The factors $\CZ^{\rm spur}_\si(\al,\tau;b)$ 
represents the scheme dependence discussed
previously.
We expect that the possibility to have such factors 
is related to the issues raised by the 
necessity to introduce a UV regularization in 
the study of the gauge theories $\CG_C$ mentioned in 
Subsection \ref{sec:inst}.
It would be very interesting
to investigate the scheme dependence coming from
possible choices of  UV regularizations of the gauge theories
more systematically. 

\subsection{Chiral ring}

Let us recall that there are further supersymmetric
observables which should be realized on 
$\CH_0$ or $\CH_{\rm top}$, respectively: the chiral ring operators
$\su_r:={\rm Tr}(\phi^2_r)$. 
We are going to propose that the operators $\su_r$ are
directly related to the operators $\sh_r$ arising in the
quantum theory of $\CM_{\rm flat}^0(C)$,
\begin{equation}\label{chring}
\su_r\,\simeq\,\ep_2^2\sh_r\,.
\end{equation} 
This is 
nontrivially supported by
the calculations of certain examples in
\cite{LMN,FFMP,FMPT}.

The existence of a relation of the form \rf{chring} is
natural in view of the fact that
the prepotential
\begin{equation}
\CF(a,m,\tau):=\lim_{\ep_1,\ep_2\ra 0}\ep_1\ep_2
\CZ^{\rm inst}_\si(a,m,\tau;\ep_1,\ep_2)\,,
\end{equation}
satisfies Matone type relations of the general form
\begin{equation}\label{Matone}
u_r\,=\,\frac{\pa}{\pa \tau_r}\CF(a,\tau)\,.
\end{equation} 
A proof of the relations \rf{Matone} that is valid 
for all theories of class $\CS$
was given in \cite{GT}. It was based on the observation 
that both the coordinates $(a,a^D)$ describing the 
special geometry underlying Seiberg-Witten theory, and
the coordinates $(\tau,h)$ introduced above can be
seen as systems of Darboux coordinates for the same
space $T^*\CT(C)$. The prepotential $\CF(a,m,\tau)$ is the 
generating function of the change of variables
between $(a,a^D)$ and $(\tau,h)$ \cite{GT}. 

This observation can be obtained
in the limit for $\ep_2\ra 0$ from the 
fact that 
\begin{equation}
\CW(a,m,\tau;\ep_2):=\lim_{\ep_1\ra 0}\ep_1
\CZ^{\rm inst}_\si(a,m,\tau;\ep_1,\ep_2)\,,
\end{equation}
coincides with the generating function $\CW(l,\tau)$ defined above, 
taking into account the identifications \rf{paramid}.
Passing to the limit $\ep_2\ra 0$, we may observe that
\[
\ep_2^2(\pa_y^2+t(y))\,\equiv\, \ep_2^2\pa_y^2-\vt(y)
\] turns into the
quadratic differential $-\vt(y)$ when $\ep_2$ is sent to zero
keeping $\vt(y)$ finite.
Using $\vt(y)$ we define the Seiberg-Witten curve $\Sigma$ 
as usual by
\begin{equation}
\Sigma\,=\,\{\,(v,u)\,|\,v^2=\vt(u)\,\}\,.
\end{equation}
It follows by WKB analysis of the differential equation 
$\ep_2^2\pa_y^2+t(y)\chi=0$ that
the coordinates $l_e$ have asymptotics that can be expressed in terms of
the Seiberg-Witten differential  $\Lambda$ on $\Sigma$ 
defined such that $\Lambda^2=\vt(u) (du)^2$.
We find
\begin{equation}\label{lk/aaD}
\frac{l_e}{2}\,\sim\,\frac{2\pi}{{\ep_\2}}\,a_e\,,\qquad
\frac{\kappa_e}{2}\,\sim\,\frac{2\pi}{{\ep_\2}}\,a_e^{\rm\sst D}\,,
\end{equation}
where $a_e$ and $a_e^{\rm\sst D}$ are periods of the 
Seiberg-Witten differential  $\Lambda$ defined as
\begin{equation}
 a_e:=\int_{\hat\ga_s^e}\La\,, \qquad a_e^{\rm\sst D}:
=\int_{\hat\ga_t^e}\La\,,
\end{equation}
with $\hat\ga_s^e$, $\hat\ga_t^e$
being cycles on $\Sigma$ that project to $\ga^e_s$ and $\ga^e_t$,
respectively.

It may also be interesting to note that the relation \rf{chring}
relates
the scheme dependence in the definition of the conformal
blocks to a possible quantum-field theoretical
scheme-dependence in the definition 
of the chiral ring operators $\su_r$. 

We thereby realize that the quantum theory 
of $\CM^0_{\rm flat}(C)$ studied in this paper
can also be interpreted as the quantization of the
geometrical structure encoding
the low energy physics of the
$A_1$ gauge theories of class $\CS$: Recall that the prepotential 
can be characterized as the generating function for the change of 
Darboux coordinates $(a,a^D)\leftrightarrow(\tau,h)$ for $T^*\CT(C)$ 
\cite{GT}. 
Turning on $\ep_2$ ``deforms'' $(a,a^D)$ into $(k,l)$, see \rf{lk/aaD}. 
The wave-functions $\Psi_\tau(a)$ studied in this paper 
represent the change of coordinates $(k,l)\leftrightarrow(\tau,h)$
on the quantum level. By combining these observations
we realize that the quantum mechanics of scalar zero modes that 
represents the non-perturbative skeleton of $\CG_C$ can be 
obtained from the Seiberg-Witten theory of $\CG_C$ in two steps:
The first is the deformation of the cotangent bundle
$T^*\CT(C)$ representing the Seiberg-Witten theory of $\CG_C$ 
into the twisted cotangent bundle
$T^*_{\ep_2}\CT(C)$ which is isomorphic to $\CM^0_{\rm flat}(C)$.
The second step is the quantization of
$T^*_{\ep_2}\CT(C)\simeq\CM^0_{\rm flat}(C)$. The parameter
$\ep_1$ of the  Omega-deformation plays the role of Planck's constant in the 
second step. The combination of the two steps may be interpreted as
the quantization of the Seiberg-Witten theory of $\CG_C$, with 
quantization parameter $\hbar=\ep_1\ep_2$. One has a certain freedom in 
quantizing $T^*\CT(C)$ which is parameterized by the ``refinement 
parameter'' $b^2=\ep_1/\ep_2$.

 \newpage

\appendix

\part{\Large Appendices}

\section{Uniqueness of the representations}\label{uniqueness}

\setcounter{equation}{0}

Let us look at the question of uniqueness of representation for the 
algebra \rf{CR} with the constraint 
(\ref{c-rels}). Let us write operators $L_t$ and $L_u$ in the following form
\beqa
&& L_t = D_+ e^{+\sk} + D_0 + D_- e^{-\sk}, \nonumber \\ &&
L_u = E_+ e^{+\sk} + E_0 + E_- e^{-\sk},
\eeqa
and substitute these operators into the algebra (\ref{CR}). Considering the coefficient corresponding to different difference operators $e^{+k}, I, e^{-k}$ one finds the following relation between $\underline{E}=\{E_+,E_0,E_-\}$ and $\underline{D}=\{D_+,D_0,D_-\}$ coefficients
\beqa
E_+ &=& e^{-l_s/2} e^{-\pi \textup{i} b^2} D_+ \nonumber \\
E_0 &=& \frac{1}{e^{\pi \textup{i} b^2}+e^{-\pi \textup{i} b^2}} \left( L_s D_0 - L_1 L_3 - L_2 L_4 \right) \nonumber \\
E_- &=& e^{l_s/2} e^{-\pi \textup{i} b^2} D_-,
\eeqa
which is true for the set of $\underline{D}$ and $\underline{E}$ coefficients given in the previous section.

Let us now check which constraints we obtain from (\ref{c-rels}). Again combining the coefficients corresponding to the shift operators $e^{+2\sk}, e^{+\sk}, I, e^{-\sk}, 
e^{-2\sk}$ we see that coefficients behind the shift operators $e^{+2\sk}$ and $e^{-2\sk}$ are trivially zero while the constraint for the coefficients behind $e^{+\sk}$ 
and $e^{-\sk}$ to be zero are equivalent and takes the following form
\beqa \label{relat1}
&& \frac{e^{-\pi \textup{i} b^2} e^l - e^{-3 \pi \textup{i} b^2}}{e^{\pi \textup{i} b^2} + e^{-\pi \textup{i} b^2}} D_0 + \frac{e^{3 \pi \textup{i} b^2} e^{-l} - e^{-3 \pi \textup{i} b^2}}{e^{\pi \textup{i} b^2} + e^{-\pi \textup{i} b^2}} e^{-\sk} D_0 e^{+\sk} \nonumber \\ && = \frac{e^{- \pi \textup{i} b^2} e^{l/2} - e^{\pi \textup{i} b^2} e^{-l/2}}{e^{\pi \textup{i} b^2} + e^{-\pi \textup{i} b^2}} (L_1 L_3 + L_2 L_4) + e^{-\pi \textup{i} b^2} (L_2 L_3 + L_1 L_4),
\eeqa
which satisfies for $D_0$ presented in the previous section. Let us now write the constraint appearing from the trivial shift operator
\beqa \label{relat2}
&& (e^{-2 \pi \textup{i} b^2} - e^{2 \pi \textup{i} b^2} e^l) D_+ e^{+\sk} D_- e^{-\sk} + (e^{-2 \pi \textup{i} b^2} - e^{2 \pi \textup{i} b^2} e^{-l}) D_- e^{-\sk} D_+ e^{+\sk} \nonumber \\ && + \frac{e^{-4 \pi \textup{i} b^2} + 2 e^{-2 \pi \textup{i} b^2} -1 - e^l - e^{-l}}{(e^{\pi \textup{i} b^2}+e^{-\pi \textup{i} b^2})^2} D_0^2 - \frac{(L_1 L_3 + L_2 L_4)^2}{(e^{\pi \textup{i} b^2}+e^{-\pi \textup{i} b^2})^2} \nonumber \\ && +\left( 2 \frac{(e^{l/2}+e^{-l/2})(L_1 L_3 + L_2 L_4)}{(e^{\pi \textup{i} b^2}+e^{-\pi \textup{i} b^2})^2} + e^{-\pi \textup{i} b^2} (L_2 L_3 + L_1 L_4) \right) D_0 \nonumber \\ && + e^{2 \pi \textup{i} b^2} (e^{l/2} + e^{-l/2})^2 - (e^{\pi \textup{i} b^2} + e^{-\pi \textup{i} b^2})^2 + e^{\pi \textup{i} b^2} (e^{l/2} + e^{-l/2}) (L_3 L_4 + L_1 L_2) \nonumber \\ && + L_1^2 + L_2^2 + L_3^2 + L_4^2 + L_1 L_2 L_3 L_4 = 0,
\eeqa
which is satisfied for  (\ref{quantum't Hooft}).

Let us look more closely at the constraint (\ref{relat1}). We already know that there exists one solution $D_0$ but it might happen that there are additional solutions. Imagine that apart from the solution we have there could be additive or multiplicative additional solutions to (\ref{relat1}). 


An additive modification of $D_0$,
$$
D_0 = D_0^{(0)} + D_0^{\rm add},
$$
where $D_0^{(0)}$ is coefficient found previously, 
would have to be a solution to the following equation
$$
\frac{e^{-\pi \textup{i} b^2} e^l - e^{-3 \pi \textup{i} b^2}}{e^{\pi \textup{i} b^2} + e^{-\pi \textup{i} b^2}} D^{\rm add}_0 + \frac{e^{3 \pi \textup{i} b^2} e^{-l} - e^{-3 \pi \textup{i} b^2}}{e^{\pi \textup{i} b^2} + e^{-\pi \textup{i} b^2}} e^{-\sk} D_0^{\rm add} e^{+\sk} = 0,
$$
and is equal  to 
\beq
D_0^{\rm add} = \tilde{D}_0\,e^{-\frac{l^2}{8 \pi \textup{i} b^2}} \frac{S_b(-\frac{l}{2 \pi \textup{i} b} +b) S_b(-\frac{l}{2 \pi \textup{i} b} - b)}{S_b(-\frac{l}{2 \pi \textup{i} b}) S_b(-\frac{l}{2 \pi \textup{i} b} + 2 b)} = e^{-\frac{l^2}{8 \pi \textup{i} b^2}} \frac{\sinh (\frac l2 + \pi \textup{i} b^2) }{\sinh (\frac l2 - \pi \textup{i} b^2)} ,
\eeq
with $\tilde{D}_0$  being an $4\pi i b^2$-periodic functions of $l$.
Any non-vanishing modification of this kind would spoil the reality
of the solution.

For resolving the constraint (\ref{relat2}) we introduce
\beq
E_{-+} \ = \ D_- e^{-k} D_+ e^{+k},
\eeq
and observe that
$$
D_+ e^{+k} D_- e^{-k} = e^{+k} \left( e^{-k} D_+ e^{+k} D_- e^{-k} \right) = e^{+k} E_{-+} e^{-k},
$$
which allows us to rewrite (\ref{relat2}) as
\beqa \label{relat2m}
&& (e^{-2 \pi \textup{i} b^2} - e^{2 \pi \textup{i} b^2} e^l) e^{+\sk} E_{-+} e^{-\sk} + (e^{-2 \pi \textup{i} b^2} - e^{2 \pi \textup{i} b^2} e^{-l}) E_{-+} \nonumber \\ && + \frac{e^{-4 \pi \textup{i} b^2} + 2 e^{-2 \pi \textup{i} b^2} -1 - e^l - e^{-l}}{(e^{\pi \textup{i} b^2}+e^{-\pi \textup{i} b^2})^2} D_0^2 - \frac{(L_1 L_3 + L_2 L_4)^2}{(e^{\pi \textup{i} b^2}+e^{-\pi \textup{i} b^2})^2} \nonumber \\ && +\left( 2 \frac{(e^{l/2}+e^{-l/2})(L_1 L_3 + L_2 L_4)}{(e^{\pi \textup{i} b^2}+e^{-\pi \textup{i} b^2})^2} + e^{-\pi \textup{i} b^2} (L_2 L_3 + L_1 L_4) \right) D_0 \nonumber \\ && + e^{2 \pi \textup{i} b^2} (e^{l/2} + e^{-l/2})^2 - (e^{\pi \textup{i} b^2} + e^{-\pi \textup{i} b^2})^2 + e^{\pi \textup{i} b^2} (e^{l/2} + e^{-l/2}) (L_3 L_4 + L_1 L_2) \nonumber \\ && + L_1^2 + L_2^2 + L_3^2 + L_4^2 + L_1 L_2 L_3 L_4 = 0.
\eeqa

As in the case of constraint for $D_0$ we consider an 
additive deviation to $E^{(0)}_{-+}$, 
$$
E_{-+} = E^{(0)}_{-+} + E^{\rm add}_{-+},
$$
and find the following equation for $E^{\rm add}_{-+}$:
\beq
\left( e^{-2 \pi \textup{i} b^2} - e^{2 \pi \textup{i} b^2} e^l \right) e^{+\sk} 
E^{\rm add}_{-+} e^{-\sk} + \left( e^{-2 \pi \textup{i} b^2} - e^{2 \pi \textup{i} b^2} e^{-l} \right) E^{\rm add}_{-+} = 0,
\eeq
whose solution is
\beq \label{solut2}
E^{\rm add}_{-+} = \tilde{E}_0\,\frac{e^{-\frac{l^2}{8 \pi \textup{i} b^2} - \frac{l}{2}}}{\sinh \frac{l}{2} \sinh(\frac{l}{2} - 2 \pi \textup{i} b^2)},
\eeq
with $\tilde{E}_0$ being $4\pi \ii b^2$-periodic.
Again one sees that solution (\ref{solut2})
would spoil the reality of the solution.

The only freedom we are left with is the gauge transformation since (\ref{relat2m}) fixes only the product (up to the shift) of $D_-$ and $D_+$.
To see more manifestly the conclusion above let us take the classical limit of constraints (\ref{relat1}) and (\ref{relat2}) which become
\beqa \label{relat1_clas}
&& (e^{l/2} - e^{-l/2})^2 D_0 = L_s (L_1 L_3 + L_2 L_4) + 2 (L_2 L_3 + L_1 L_4),
\eeqa
which defines $D_0$ unambiguously. Let us now write the constraint appearing from the trivial shift operator
\beqa \label{relat2_clas}
&& -(e^{l/2} - e^{-l/2})^2 D_+ D_- - \frac 14 (e^{l/2} - e^{-l/2})^2 D_0^2 - \frac 14 (L_1 L_3 + L_2 L_4)^2 \nonumber \\ && +\left( \frac 14 L_s (L_1 L_3 + L_2 L_4) + (L_2 L_3 + L_1 L_4) \right) D_0 + L_s^2 - 4 \nonumber \\ && + L_s (L_3 L_4 + L_1 L_2) + L_1^2 + L_2^2 + L_3^2 + L_4^2 + L_1 L_2 L_3 L_4 = 0,
\eeqa
from which one finds unambiguously $D_+ D_-$. And the only freedom is in multiplying $D_+$ by $e^{\pi \textup{i} \chi(l)}$ and $D_-$ by $e^{-\pi \textup{i} \chi(l)}$, i. e. the gauge freedom.

Let us finally remark that
assuming the cyclic symmetry for algebra of loop 
operators under permutations of two points on a sphere
\beq
L_s \rightarrow_{2\leftrightarrow3} L_t \rightarrow_{1\leftrightarrow2} L_u \rightarrow_{2\leftrightarrow4} L_s
\eeq
one gets the cyclic symmetry for the cubic relation (\ref{c-rels}),
so in a sense the two first lines in (\ref{c-rels}) 
are fixed by cyclic symmetry.

\section{Special functions}\label{Qdil}
\setcounter{equation}{0}

\subsection{The function $\Ga_b(x)$}

The function $\Ga_b(x)$ is a close relative of the double
Gamma function studied in \cite{Ba}. It 
can be defined by means of the integral representation
\begin{equation}
\log\Ga_b(x)\;=\;\int\limits_0^{\infty}\frac{dt}{t}
\biggl(\frac{e^{-xt}-e^{-Qt/2}}{(1-e^{-bt})(1-e^{-t/b})}-
\frac{(Q-2x)^2}{8e^t}-\frac{Q-2x}{t}\biggl)\;\;.
\end{equation}
Important properties of $\Ga_b(x)$ are
\begin{align}
{} \text{functional equation} \quad &
\Ga_b(x+b)=\sqrt{2\pi}b^{bx-\frac{1}{2}}\Ga^{-1}(bx)\Ga_b(x). \label{Ga_feq}\\
{}\text{analyticity}\quad &
\Ga_b(x)\;\text{is meromorphic, it has poles only}\nonumber\\ 
{}& \text{at}\;\,  
x=-nb-mb^{-1}, n,m\in\BZ^{\geq 0}. 
\end{align}
A useful reference for further properties is \cite{Sp}.

\subsection{Double Sine function}\label{Qdil1}

The special functions used in this note are all
build from the so-called double Sine-function. This function
is closely related to the special function here denoted 
$e_b(x)$, which was introduced under the name of {\em quantum dilogarithm}
in \cite{FK2}. These special functions are simply related
to the Barnes double Gamma function \cite{Ba}, and were also introduced 
in studies of quantum groups and integrable models in 
\cite{F2,Ru,Wo,V}.

In the strip $|{\rm Im}(x)| < \fr{Q}{2}$, function $e_b(x)$ has the 
following integral representation
\begin{equation}\label{wint}
 e_b(x)= 
 \exp \Biggl\{ -
 \int\limits_{\BR+\textup{i}0} \frac{dt}{4\, t} \,
 \frac{ e^{-2\textup{i} t x}}{\sinh b t \, \sinh{\frac{t}{b}} } \Biggr\} \,,
\end{equation}
where the integration contour goes around the pole $t=0$ in the 
upper half--plane. 
The function $s_b(x)$ is then 
related to $e_b(x)$ as follows
\begin{equation}
 s_b(x)\,=\, 
 e^{\frac{\textup{i} \pi}{2} x^2 + 
      \frac{\textup{i} \pi}{24}(b^2 + b^{-2})}e_b(x)\,.
\end{equation}
The analytic continuation of $s_b(x)$ to the
entire complex plane is a meromorphic function with the
following properties 
\begin{align}
\text{functional equation}  \label{wfunrel} \quad&
 \frac{s_b(x + \fr{\textup{i}}{2}b^{\pm 1})}{s_b(x - \fr{\textup{i}}{2}b^{\pm 1})} = 
 2 \, \cosh (\pi b^{\pm 1} x) \,, \\[0.5mm]
\text{reflection property} \label{wrefl} \quad&
 s_b(x) \; s_b(-x) = 1  \,, \\[0.5mm]
  \label{wcc} \text{complex conjugation} \quad&
  \overline{s_b(x)} = s_b(-\bar{x}) \,,\\[0.5mm]
 \label{wan} 
 \text{zeros\,/\,poles} \quad& 
 (s_b(x))^{\pm 1} = 0 \ \Leftrightarrow  
  \pm x \in  \big\{ \textup{i}\fr{Q}{2}{+}nb{+}mb^{-1};n,m\in\BZ^{\geq 0}\big\}   
	\,, \\[0.5mm]
\label{wres}
 \text{residue} \quad& 
  \Res_{x=-\textup{i}\frac{Q}{2}} s_b(x)=\frac{\textup{i}}{2\pi}
 \,, \\[0.5mm]
 \text{asymptotics} \quad& s_b(x) \sim 
\left\{
\begin{aligned}
& e^{- \frac{\textup{i} \pi}{2}(x^2 + \frac{1}{12}(b^2+b^{-2}))}\;\;{\rm for}\;\,
|x|\ra\infty,\;\, |{\rm  arg}(x)|<\fr{\pi}{2} \,, \\
 & e^{+ \frac{\textup{i} \pi}{2}(x^2 + \frac{1}{12}(b^2+b^{-2}))}\;\;{\rm for}\;\,
|x|\ra\infty,\;\, |{\rm  arg}(x)|>\fr{\pi}{2} \,.
\end{aligned}\right.
\end{align}
Of particular importance for us is the behavior for $b\ra 0$, which 
is given as 
\begin{equation}
e_b\left(\frac{v}{2\pi b}\right)\,=\,
\exp\bigg(\!\!-\frac{1}{2\pi b^2}\Li_2(-e^v)\bigg)\Big(1+\CO(b^2)\Big)\,.
\end{equation}

In our paper we mainly use 
the special function $S_b(x)$ defined by
\begin{equation}\label{Sbx}
 S_b(x) := s_b(\textup{i}x -\fr{\textup{i}}{2}Q)\,. 
\end{equation}
In terms of $\Ga_b(x)$ the double Sine-function is given as
$$
S_b(x) = \frac{\Ga_b(x)}{\Ga_b(Q-x)}.
$$
We will use the properties
\begin{align}
\label{Sb1}
\text{self--duality} \quad&  
 S_b(x) = S_{b^{-1}}(x) \,, \\
\label{Sb2}
\text{functional equation} \quad&
 S_b(x + b^{\pm 1}) = 2 \, \sin (\pi b^{\pm 1} x) \, S_b(x) \,,\\
\label{Sb3}
\text{reflection property} \quad&
  S_b(x) \, S_b(Q-x) = 1 \,.
\end{align}

\subsection{Integral identities}

We will use the following set of integral identities.
\begin{propn}
\begin{eqnarray}\label{Corollary 1}
\int_{i\BR} 
dz\; \prod_{i=1}^3 S_b(\mu_i-z) 
S_b(\nu_i+z) = \prod_{i,j=1}^3 S_b(\mu_i+\nu_j),
\end{eqnarray}
where the balancing condition is $\sum_{i=1}^3 \mu_i+\nu_i = Q.$
\end{propn}
This identity  was recently understood
as a pentagon identity in \cite{KLV}.

\begin{propn}
\begin{eqnarray}\label{Corollary 2}
\frac 12 \int_{i\BR} 
dz\; \frac{S_b(\mu \pm z) S_b(\nu \pm z)}{S_b(\pm 2 z)} e^{-2 \pi \textup{i} z^2}  = S_b(\mu+\nu) e^{-\frac 12 \pi \textup{i} (\mu-\nu)^2 + \frac 12 \pi \textup{i} Q (\mu+\nu)}.
\end{eqnarray}
The following notation has been used
$
S_b(\alpha \pm u) :=
S_b(\alpha+u)
S_b(\alpha-u).
$
\end{propn}

\begin{propn}
\begin{align}\label{Corollary 3}
& \int_{i\BR} 
 dy\; \prod_{i=1}^3 S_b(\mu_i-y) \prod_{i=1}^2 S_b(\nu_i+y) e^{\pi \textup{i} \lambda y} e^{-\frac 12 \pi \textup{i} y^2}  =  \\[-.5ex] & \hspace{2cm} =\prod_{i=1}^3 S_b(\mu_i+\nu_2) e^{\frac 12 \pi \textup{i} \lambda^2} e^{\frac 18 \pi \textup{i} Q^2} e^{-\frac 12 \pi \textup{i} Q(\lambda+\nu_1)}\notag\\[-.5ex] 
&\hspace{4cm}\times \frac 12 
\int_{i\BR} 
dy\; \frac{\prod_{i=1}^3 S_b(\mu_i+\sigma\pm y) S_b(\nu_1 - \sigma \pm y)}{S_b(\pm 2 y)} e^{-2 \pi \textup{i} y^2} , 
\nonumber \end{align}
where
$$
2 \sigma = Q - \sum_{i=1}^3 \mu_i - \nu_2,
$$
and the following balancing condition is satisfied
$$
\sum_{i=1}^3 \mu_i + \sum_{i=1}^2 \nu_i = \lambda + \frac Q2.
$$
\end{propn}

The proof of the above
Propositions is easily obtained from the reduction of elliptic
hypergeometric integrals to the hyperbolic level \cite{DS} (the details can
be found in \cite{Bu} or in \cite{SV11}).
identity \ref{Corollary 1}, \ref{Corollary 2}
and \ref{Corollary 3} are equivalent to  Theorem 5.6.7,
Theorem 5.6.6 and Theorem 5.6.17 in \cite{Bu}, respectively.

\section{Analytic properties of intertwining kernels}
\label{6jApp}
\setcounter{equation}{0}

\subsection{Preparations}

It will be convenient to factorize the expression for 
$\FusC{\al_1}{\al_2}{\al_3}{\al_4}{\al_s}{\al_t}$ as
\begin{align}\label{F-Ca}
&\FusC{\al_1}{\al_2}{\al_3}{\al_4}{\al_s}{\al_t}
=\frac{\Gamma_b(2Q-2\al_s) \Gamma_b(2\al_s)}{\Gamma_b(2Q-2\al_t) \Gamma_b(2\al_t)}
\times \big\{\,{}^{\al_1}_{\al_3}\,{}^{\al_2}_{\al_4}\,{}^{\al_s}_{\al_t}\big\}_b^{C}\,
\end{align}
with b-$6j$ symbols 
$\big\{\,{}^{\al_1}_{\al_3}\,{}^{\al_2}_{\al_4}\,{}^{\al_s}_{\al_t}\big\}_b^{C}$
in the normalization from Subsection \ref{Sec:C-norm} given by the formula
\begin{align}\label{b-6j-C}
\big\{\,{}^{\al_1}_{\al_3}\,{}^{\al_2}_{\al_4}\,{}^{\al_s}_{\al_t}\big\}_b^{C}:=
\frac{T(\al_t,\al_3,\al_2)
T(\al_4,\al_t,\al_1)}{S(\al_s,\al_2,\al_1)S(\al_4,\al_3,\al_s)}\times\CJ\,,
\end{align}
where
\begin{align}\label{CJdef}
& \CJ:=\,\int\limits_{\CC}du\;
S_b(u-\alpha_{12s}) S_b(u-\al_{s34}) S_b(u -\alpha_{23t}) 
S_b(u-\alpha_{1t4})
\notag \\[-1.5ex] & \hspace{2cm}\times 
S_b( \alpha_{1234}-u) S_b(\alpha_{st13}-u) 
S_b(\alpha_{st24}-u) S_b(2Q-u)\,,
\notag\end{align}
and
\begin{align}
&S(\al_3,\al_2,\al_1)=\Ga_b(2Q-\al_{123})\Ga_b(\al^3_{12})\Ga_b(\al^1_{23})\Ga_b(\al^2_{31})\\
&T(\al_3,\al_2,\al_1)=\Ga_b(\al_{123}-Q)\Ga_b(Q-\al^3_{12})
\Ga_b(Q-\al^1_{23})\Ga_b(Q-\al^2_{31})\,.
\end{align}
We are using the notations $\al_{ijk}=\al_i+\al_j+\al_k$, $\al_{ij}^k=\al_i+\al_j-\al_k$.

\subsection{Resonant values}

Singular behavior of the integral $\CJ$ could be caused by the 
behavior of the integrand at infinity, or by the pinching of the 
contour $\CC$ between poles of the integrand.
It is not hard to check that the integral converges for $u\ra\infty$ for all
values of the variables $\al_i$. It is furthermore straightforward to
check that the pinching of the contour of integration in \rf{CJdef}
only occurs when at least one of the triples $T_{s12}$, $T_{43s}$,
$T_{t32}$, $T_{4t1}$ is resonant, using the terminology from
Subsection \ref{Sec:Res}. Taking into account the poles and zeros of
the prefactors in \rf{F-Ca} 
one easily verifies that the 
 b-$6j$ symbols 
$\big\{\,{}^{\al_1}_{\al_3}\,{}^{\al_2}_{\al_4}\,{}^{\al_s}_{\al_t}\big\}_b^{C}$
are entire in $\al_s$, and have poles iff one of $T_{t32}$, $T_{4t1}$ is resonant.

We are going to consider the 
 b-$6j$ symbols 
$\big\{\,{}^{\al_1}_{\al_3}\,{}^{\al_2}_{\al_4}\,{}^{\al_s}_{\al_t}\big\}_b^{C}$
as distribution on a space $\CT$ of functions $f(\al_t)$ which are (i)
entire, (ii) decay faster than any exponential for $\al_r\ra \infty$ along
the axis $Q/2+i\BR$, and (iii) Weyl-symmetric $f(\al_t)=f(Q-\al_t)$.
For $\al_i\in Q/2+i\BR$, $i=1,2,3,4,s$ one defines
\begin{equation}\label{D-def}
D_{\al_s}\big\{\,{}^{\al_1}_{\al_3}\,{}^{\al_2}_{\al_4}\,\big\}(f)
:=\frac{1}{2}
\int_{Q/2+i\BR} d\al_t\;
\big\{\,{}^{\al_1}_{\al_3}\,{}^{\al_2}_{\al_4}\,{}^{\al_s}_{\al_t}\big\}_b^{C}
f(\al_t)\,.
\end{equation}
Assuming  $\al_i\in Q/2+i\BR$, $i=1,2,3,4$, one easily checks 
that 
$\tilde{f}(\al_s):=D_{\al_s}\big\{\,{}^{\al_1}_{\al_3}\,{}^{\al_2}_{\al_4}\,\big\}(f)$
has the properties (i)-(iii) above. This means that 
the operator $\SF$ maps $\CT$ to itself.

Consider now the analytic continuation of 
$D_{\al_s}\big\{\,{}^{\al_1}_{\al_3}\,{}^{\al_2}_{\al_4}\,\big\}$ with respect to the
parameter $\al_2$. It can always be represented in the form 
\rf{D-def}, but the contour of integration may need to be deformed. 
The result can generically be represented as an integral over
the original contour $Q/2+i\BR$ plus a finite sum over 
residue terms. The residue terms define generalized delta-distributions 
as introduced in \rf{deltadef}.

\subsection{Degenerate values}

We are particularly interested in the case where takes one of the 
degenerate values 
\begin{equation}\label{degval-ex}
\al_2\,=\,-kb/2-lb^{-1}/2\,.
\end{equation} 
Note that this is a necessary condition 
for having a double resonance,
\begin{equation}\label{doubres}
\al_{12}^s\,=\,-k'b-l'b^{-1}\quad\wedge\quad
\al_{s2}^1\,=\,-k''b-l''b^{-1}\,,
\end{equation}
where $k=k'+k''$, $l=l'+l''$. The prefactor in \rf{b-6j-C} proportional to 
$(S(\al_s,\al_2,\al_1))^{-1}$ vanishes in the case of a double resonance.
It follows that only residue terms can appear in the expression 
for $D_{\al_s}\big\{\,{}^{\al_1}_{\al_3}\,{}^{\al_2}_{\al_4}\,\big\}$
at double resonance \rf{doubres}.

So let us look at the residue terms that become relevant in the analytic continuation
from $\Re(\al_2)=Q/2$ to the values \rf{degval-ex}. Relevant are the poles from
the triple $T_{t32}$, in particular the poles at
\begin{align}
&\al_{32}^t\,=\,-k_1b-l_1b^{-1}\,,\\
&\al_{t2}^3\,=\,-k_2b-l_2b^{-1}\,,
\end{align}
where $k=k_1+k_2$, $l=l_1+l_2$. It is for some considerations convenient to 
assume that $\Re(\al_3)=Q/2-\ep+iP_3$ for some small real number $0<\ep<b/2$. 
It follows that the poles 
\begin{align}
&\al_t\,=\,\fr{Q}{2}-\ep+iP_3+(k_1-k_2)\fr{b}{2}+(l_1-l_2)\fr{1}{2b}\,,
\end{align}
with $k_1-k_2\leq 0$ and $l_1-l_2\leq 0$ will have crossed the contour of integration
from the right, 
and the poles
\begin{align}
&\al_t\,=\,\fr{Q}{2}-\ep+iP_3-(k_2-k_1)\fr{b}{2}-(l_2-l_1)\fr{1}{2b}\,,
\end{align}
with $k_2-k_1<0$ and $l_2-l_1<0$ will have crossed the contour of integration
from the left. The form of the distribution given in \rf{al2deg} follows
easily from these observations.

\subsection{Residues}\label{App:res}

We list here some relevant residues.
\begin{align} \label{FM11C}
& \fus{-b}{\al_2}{\al_3}{\al_4}{1}{0} = \frac{1}{\Gamma(-2b^2)} 
 \\ & \qquad\qquad\qquad \times 
\frac{\Gamma(1-2b\al_4) \Gamma(-Qb-b^2+2b\al_4) \Gamma(2-2b\al_2) \Gamma(2Qb-2b\al_2-b^2)}{\Gamma(2Qb-b^2-b\al_{234})) \Gamma(-b^2-b\al_{34}^2) \Gamma(1-b\al_{24}^3) \Gamma(1-b \al_{23}^4 )};
\nonumber\\
 \label{FM21C}
& \fus{-b}{\al_2}{\al_3}{\al_4}{-1}{0} = \frac{1}{\Gamma(-2b^2)} 
\\ & \nonumber \qquad\qquad\qquad \times 
\frac{\Gamma(1-2b\al_4) \Gamma(-Qb-b^2+2b\al_4) \Gamma(2b\al_2-2b^2) \Gamma(2b\al_2-b^2)}{\Gamma(-b^2+b\al_{23}^4) \Gamma(-b^2+b\al_{24}^3)) \Gamma(1-b\al_{34}^2) \Gamma(-Qb-b^2+b\al_{234})};\\
\label{FM31C}
& \fus{-b}{\al_2}{\al_3}{\al_4}{0}{1} = 2 \cos(\pi b^2) \frac{ \Gamma(-2b^2)}{\Gamma(-b^2)^2} \\ \nonumber& \qquad\qquad\qquad \times \frac{\Gamma(-Qb+2b\al_4) \Gamma(-Qb+2b\al_4+b^2) \Gamma(2Qb-2b\al_2) \Gamma(2b\al_2)}{\Gamma(b\al_{34}^2) \Gamma(b\al_{24}^3) \Gamma(Qb-b\al_{23}^4) \Gamma(-Qb+b\al_{234})};\\
\label{FM41C}
& \fus{-b}{\al_2}{\al_3}{\al_4}{0}{-1} = 2 \cos(\pi b^2) \frac{ \Gamma(-2b^2)}{\Gamma(-b^2)^2} \\  \nonumber &\qquad\qquad\qquad \times \frac{\Gamma(Qb-2b\al_4) \Gamma(Qb-2b\al_4+b^2) \Gamma(2Qb-2b\al_2) \Gamma(2b\al_2)}{\Gamma(2Qb-b\al_{234}) \Gamma(Qb-b\al_{24}^3) \Gamma(Qb-b\al_{34}^2) \Gamma(b\al_{23}^4)};
\\ \label{FM51C}
& \fus{-b}{\al_2}{\al_3}{\al_4}{0}{0} = \frac{ \Gamma(-Qb+2b\al_4-b^2) \Gamma(2Qb-2b\al_2)}{\Gamma(-b^2)\Gamma(2b\al_4) \Gamma(1-2b\al_2) } \\ 
\nonumber & \qquad \qquad\qquad\times \left\{ 1 + 2 \cos \pi b^2 \frac{\sin[\pi b(\al_2-\al_3+\al_4)] \sin[\pi b(-Q+\al_2+\al_3+\al_4)]}{\sin[2 \pi b \al_2] \sin[2 \pi b \al_4]} \right\},
\end{align}
where the notation
$\al_{ij}^k=\al_i+\al_j-\al_k$ was used. 
From the above fusion matrices one can 
derive the 't Hooft--Wilson loop intertwining relation.

\section{The kernel for the S-move} \label{App:S}
\setcounter{equation}{0}

We here describe in more detail our derivation of formula \rf{Skern}
for the kernel representing the S-move. As outlined in the main text,
we are using the following strategy:
\begin{itemize}
\item
Definition \rf{SWtoSU} defines operators  $\SB$, $\SF$, $\SZ$ and $\SS$
within the quantum Teichm\"uller theory  
which
satisfy operatorial versions of the Moore-Seiberg consistency conditions
\cite{T05}.
\item The direct calculation of the kernel of the operator 
$\SS$ presented in Subsection \ref{SfromTeich} below
shows that this operator is represented by a kernel
which depends meromorphically on its arguments.
\item It was explained in 
Subsection \ref{sec:Skern}
that this allows us to 
use the Moore-Seiberg equation \begin{align}\label{g=1,n=2-App}
S_{\be_\1\be_\2}(\be_\3)\int_{\BS}& d\be_\4\; 
\Fus{\al_\2}{\al_\1}{\be_\2}{\be_\2}{\be_\3}{\be_\4}\,
T_{\be_\4}^{}\,T_{\be_\2}^{-1}\,
\Fus{\be_\2}{\al_\1}{\al_\2}{\be_\2}{\be_\4}{\be_\5}\\
& = \int_{\BS}d\be_\6\;
\Fus{\al_\2}{\al_\1}{\be_\1}{\be_\1}{\be_\3}{\be_\6}
\Fus{\be_\6}{\al_\2}{\al_\1}{\be_\6}{\be_\1}{\be_\5}
S_{\be_\6\be_\2}(\be_\5)\,e^{\pi i(\De_{\al_\1}+\De_{\al_\2}-\De_{\be_\5})}\,.
\notag
\end{align}
to derive a formula for $S_{\be_\1\be_\2}(\be_\3)$ in terms of the kernel
for $F$. More details are given in Subsection \ref{app:MS-sol} below.
\item The integrals in the resulting formula for $S_{\be_\1\be_\2}(\be_\3)$
will be calculated explicitly in Subsection
\ref{app:S-kernel}, leading to our formula 
\rf{Skern}.
\end{itemize}
A faster way to find the formula \rf{Skern} would be to use 
the intertwining property \rf{diffinter} to derive an difference 
equation for the kernel $S_{\be_\1\be_\2}(\al)$. The problem would then
be to show that the resulting formula solves the Moore-Seiberg 
equations. This is manifest in our approach.

\subsection{Calculation using Teichm\"uller theory}

\label{SfromTeich}

\begin{figure}[htb]
\epsfxsize5cm
\centerline{\epsfbox{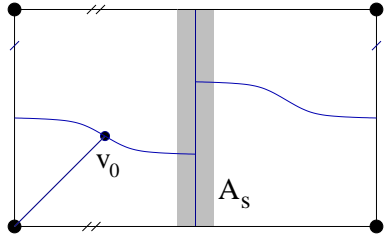}}
\caption{\it Fat graph on a one-holed torus $C_{1,1}$, 
represented as rectangle with
opposite sides identified. The hole sits at the corners
of the rectangle.
The annulus $A_s$ (grey) contains the 
geodesic $\ga_s$ defining a pants decomposition of
$C_{1,1}$.}\label{TorFig}\vspace{.3cm}
\end{figure}

We shall work within the representation for quantum Teichm\"uller theory
associated to the fat graph drawn in Figure \ref{TorFig}. The representation
associated to the annulus $A_s$ in Figure \ref{TorFig} is taken to be 
the one defined in Subsection \ref{sec:annulus}.

For the following 
it will suffice to work in a reduced representation defined by 
setting the constraint $\sz$ to zero. The length operator
$\SL_s$ is then defined by using \rf{SLsimple}. In order to define 
the operator $\SL_0$ representing the length of the hole of
$C_{1,1}$ we may use formula \rf{modLlem2}. The length operator
$\SL_t$ has to be calculated using \rf{genlength}
by finding a fat graph $\vf_0$ which allows
one to use the definition \rf{SLsimple}. 
The resulting formulae for the relevant length operators are 
\begin{align}
&\SL_s\,=\,2\cosh 2\pi b\spp_s+e^{2\pi b \sq_s}\,,\\
&\SL_t\,=\,2\cosh 2\pi b\spp_t
           +e^{2\pi b\sq_t}+e^{-2\pi b\sq_0}e^{\pi b(\sq_t-\spp_t)}\,,\\
& \SL_0\,=\,2\cosh 2\pi b\spp_0+2\cosh (\pi b \spp_0)\SL_se^{-2\pi b\sq_0} 
+e^{-4\pi b\sq_0}\,.
\end{align}
In the expression for $\SL_t$ we have been using the notations
\begin{equation}
\spp_t:=\fr{1}{2}(\sq_s-\spp_s-\spp_0)\,,\qquad
\sq_t:=-\fr{1}{2}(3\spp_s+\sq_s+\spp_0)\,.
\end{equation}
Let us consider eigenstates $|\,a,m\,\rangle_s$
and   $|\,a,m\,\rangle_t$ to the pairs of mutually
commuting operators $(\SL_s,\SL_0)$ and $(\SL_t,\SL_0)$,
respectively
\begin{equation}
\begin{aligned}
&\SL_s\,|\,a,m\,\rangle_s\,=\,2\cosh 2\pi b a\,|\,a,m\,\rangle_s\,,\\
&\SL_t\,|\,a,m\,\rangle_t\,=\,2\cosh 2\pi b a\,|\,a,m\,\rangle_t\,,
\end{aligned}\qquad
\begin{aligned}
&\SL_0\,|\,a,m\,\rangle_s\,=\,2\cosh 2\pi b m\,|\,a,m\,\rangle_s\,,\\
&\SL_0\,|\,a,m\,\rangle_t\,=\,2\cosh 2\pi b m\,|\,a,m\,\rangle_t\,.
\end{aligned}
\end{equation}
We shall work in a representation where the operators $\spp_s$ and $\sq_0$ 
are diagonal. States are represented by 
wave-functions $\phi^s_{a,m}(p_s,q_0):=\langle\,p_s,q_0\,|\,a,m\,\rangle_s$
and $\phi^t_{a,m}(p_s,q_0):=\langle\,p_s,q_0\,|\,a,m\,\rangle_t$.

These wave-functions are related by an integral transformation of the 
form
\begin{equation}\label{Strsf}
\phi^t_{a_t,m}(p_s,q_0)\,=\,\int da_s\;S_{a_ta_s}(m)\,\phi^s_{a_s,m}(p_s,q_0)\,.
\end{equation}
In order to simplify the calculation it helps to consider the
limit $q_0\ra\infty$. Note that $\SL_0$ can be
approximately be represented by $2\cosh 2\pi b \spp_0$ 
in this limit.
Both $\phi^s_{a,m}(p_s,q_0)$ and $\phi^t_{a,m}(p_s,q_0)$
can be normalized to have a leading asymptotic behavior 
for $q_0\ra\infty$ of the form
\begin{align}
& \phi^s_{a,m}(p_s,q_0)\,\sim\,(e^{2\pi im q_0}+R_m^se^{-2\pi im q_0})\psi^s_{a}(p_s)\,,\\
& \phi^t_{a,m}(p_t,q_0)\,\sim\,(e^{2\pi im q_0}+R_m^te^{-2\pi im q_0})\psi^t_{a}(p_s)\,,
\end{align}
where $\psi^s_{a}(p_s)$ and
$\psi^t_{a}(p_t)$ must be eigenfunctions of 
the operators $\SL_s'$ and $\SL_t'$ obtained from 
$\SL_s$ and $\SL_t$ by sending  $q_0\ra\infty$ and 
considering a representations of $(\spp_s,\sq_s)$ on functions
$\psi(p_s)$ of a single variable on which $\spp_s$ acts as multiplication
operator. Equation \rf{Strsf} implies
\begin{equation}\label{Strsf'}
\psi^t_{a_t,m}(p_s)\,=\,\int da_s\;S_{a_ta_s}(m)\,\psi^s_{a_s,m}(p_s)\,.
\end{equation}

The calculation of the kernel $S_{a_ta_s}(m)$ is now straightforward. 
Recall that a complete set of orthonormalized eigenfunction 
of $\SL_s$ is given by the functions defined in \rf{L1-eigenf}.
Note furthermore that
\begin{equation}
\SL_t'\,=\,2\cosh 2\pi b\spp_t
           +e^{2\pi b\sq_t}\,.
\end{equation}
The eigenfunctions of $\SL_t'$ in a representation
in which $\spp_t$ is diagonal are therefore obtained from 
\rf{L1-eigenf} by obvious substitutions. We finally need
that 
$
\langle \,p_s\, | \,p_t\, \rangle\, = \,
e^{\pi \textup{i} (p_s^2+p_t^2)}\, e^{4 \pi \textup{i} p_s p_t} 
e^{-2 \pi \textup{i} m (p_s+p_t)} \,.$
The kernel representing the modular transformation $S$ is then given 
as
\begin{align}\nonumber
&  S_{a_s a_t}(m_0) = 
\langle \,a_s\,|\,a_t\, \rangle\\
&\, =\, 
\int dp_s dp_t\; 
\langle \,a_s\,|\,p_s\, \rangle\, 
\langle\, p_s\,|\,p_t\, \rangle\,
\langle\, p_t\,|\,a_t\, \rangle \\
&\, = \int dp_s\; e^{\pi \textup{i} (p_s-2m)p_s}\frac{s_b(a_{s}-p_s+c_b-i0)}{s_b(a_{s}+p_s-c_b+i0)}
\int dp_t \;
e^{ \pi \textup{i}(p_t- 2m)p_t}\,
\frac{s_b(a_{t}+p_t+c_b-i0)}{s_b(a_{t}-p_t-c_b+i0)}e^{4 \pi \textup{i} p_s p_t} \,.
\nonumber\end{align}
It is easy to see that $S_{a_s a_t}(m_0)$ is meromorphic in
$m_0$, $a_s$ and $a_t$.

\subsection{Solving the
Moore-Seiberg relations for the S-kernel} 
\label{app:MS-sol}

We now want to explain how
to derive the formula 
\begin{equation}\label{S-MSform-App}
F_{0\al}^{\rm\sst L} 
\big[ {}_{\be_\1}^{\be_\1} {}_{\be_\1}^{\be_\1} \big]
{S_{\be_\1\be_\2}^{\rm\sst L}(\al)}=
{S_{0\be_\2}^{\rm\sst L}}
\int d\be_\3 \;e^{-\pi i(2\De_{\be_\2}+2\De_{\be_\1}-2\De_{\be_\3}-\De_\al)}\;
{{\rm F}_{0\be_\3}^{\rm\sst L} \big[ {}_{\be_\2}^{\be_\2} {}_{\be_\1}^{\be_\1} \big]}
{\rm F}_{\be_\3\al}^{\rm\sst L} \big[ {}_{\be_\2}^{\be_\1} {}_{\be_\2}^{\be_\1} \big]
\,.\end{equation}
for ${S_{\be_\1\be_\2}^{\rm\sst L}(\al)}$ from 
equation \rf{g=1,n=2-App}. As explained in the main text,
we mainly need
the identity
\begin{equation}\label{limitF-app}
\lim_{\ep\downarrow 0}
{{\rm F}_{\ep,\al_\3}^{\rm\sst L} 
\big[ \,{}_{\ep\;}^{\ep\;} {}_{\al_\1}^{\al_\1} \big]}
\,=\,\de(\al_3-\al_1)\,.
\end{equation}
Setting $\al_1=\al_2$ and taking $\be_1=\ep,\be_3=\ep$, $\ep\ra 0$ using 
\rf{limitF-app} yields \rf{S-MSform-App}. 

One might be tempted to take $\be_1\ra 0$ first. This turns out not to 
be straightforward, as the convergence of the integrals in 
\rf{g=1,n=2-App} would then be lost. Doing this naively would seem to
lead to an equation similar to \rf{S-MSform-App}, but with 
${S_{0\be_\2}^{\rm\sst L}}$ replaced by 
${\tilde{S}_{0\be_\2}^{\rm\sst L}}:=
\lim_{\be_1\ra 0}{S_{\be_1\be_\2}^{\rm\sst L}}$, which is not 
the same as ${S_{0\be_\2}^{\rm\sst L}}:=\lim_{\ep\ra 0}
S_{\ep,\be_\2}^{\rm\sst L}(\ep)$. The fact that
${S_{0\be_\2}^{\rm\sst L}}\neq {\tilde{S}_{0\be_\2}^{\rm\sst L}}$ 
can be verified explicitly
using equations \rf{Ilim1}, \rf{Ilim2} below.

It remains to prove \rf{limitF-app}.
In order to do this, we will show that 
\begin{equation}\label{limF:ex}
{{\rm F}_{\ep,\al_\3}^{\rm\sst L} 
\big[ {}_{\ep\;}^{\ep\;} {}_{\al_\1}^{\al_\1} \big]}=
{{\rm F}_{0,\al_\3}^{\rm\sst L} 
\big[ {}_{\ep\;}^{\ep\;} {}_{\al_\1}^{\al_\1} \big]}+\CO(\ep)\,,
\end{equation} and 
use the remarkable identity
\begin{equation}\label{F-C}
{{\rm F}_{0\al_\3}^{\rm\sst L} \big[ {}_{\al_\2}^{\al_\2} {}_{\al_\1}^{\al_\1} \big]}
\,=\,\frac{1}{2\pi}
\frac{Z(0)\;Z(\al_\3)}{Z(\al_\2)Z(\al_\1)} C(Q-\al_\3,\al_\2,\al_\1)\,,
\end{equation}
proven below. The function
$C(\al_3,\al_2,\al_1)$ was defined in \rf{ZZform}, and $Z(\al)$ 
is explicitly given as
\begin{equation}
Z(\al)\,=\,
\frac{(\pi\mu\ga(b^2))^{\frac{1}{2b}(Q-2\al)}\,2\pi(Q-2\al)}{\Ga(1+b(Q-2\al))\Ga(1+b^{-1}(Q-2\al))}\,.
\end{equation}
The normalization factor $Z(\al)$ is closely related to the Liouville one-point function on the unit disc \cite{ZZ2}. Note furthermore 
that \cite[Section 4.4]{Teschner:2001rv}
\begin{equation}\label{limitC}
\lim_{\al_2\ra 0}C(Q-\al_3,\al_2,\al_1)\,=\,2\pi\de(\al_3-\al_1)\,.
\end{equation}
The identity \rf{limitF-app}
follows from the combination of \rf{F-C} and \rf{limitC}.

For the calculations necessary to prove \rf{limF:ex} and \rf{F-C} 
we will find it 
convenient to use a further gauge transformation defined by writing 
\begin{equation}\label{v-renorm'}
\tilde{v}^{\al_\3}_{\al_\2\al_\1}\,=\,N(\al_3,\al_2,\al_1)
{w}^{\al_\3}_{\al_\2\al_\1}\,,
\end{equation}
with $N(\al_3,\al_2,\al_1)$ being defined in \rf{N-def}.
The kernels representing the $F$- and $S$-moves in the corresponding 
representation
will be denoted as $\Fusan{\al_1}{\al_2}{\al_3}{\al_4}{\al_s}{\al_t}$
and $S^{\rm\sst PT}_{\be_\1\be_\2}(\al_0)$, respectively. We have
\begin{align} \label{norm}
\FusC{\al_1}{\al_2}{\al_3}{\al_4}{\al_s}{\al_t}
&={\frac{N(\al_s,\alpha_2,\alpha_1)N(\alpha_4,\alpha_3,\al_s)}{N(\al_t,\alpha_3,\alpha_2) N(\alpha_4,\al_t,\alpha_1)} }
\Fusan{\al_1}{\al_2}{\al_3}{\al_4}{\al_s}{\al_t}\,,\\
S^{\rm\sst L}_{\be_\1\be_\2}(\al_0)\,&=\,
\frac{N(\be_\1,\al_0,\be_\1)}{N(\be_\2,\al_0,\be_\2)}\,
S^{\rm\sst PT}_{\be_\1\be_\2}(\al_0)\,.
\label{S-renorm}\end{align}
The kernel  $\Fusan{\al_1}{\al_2}{\al_3}{\al_4}{\al_s}{\al_t}$
can be expressed using the 
formula first derived in 
in \cite{PT2}\footnote{The formula below coincides with equation (228) in \cite{Teschner:2001rv} after shifting $s \rightarrow u-\alpha_s-Q/2$. }, 
\begin{align} \label{6j1}
\Fusan{\al_1}{\al_2}{\al_3}{\bar{\al}_4}{\al_s}{\al_t}= & 
\;\frac{S_b(\alpha_2+\alpha_s-\alpha_1) S_b(\alpha_t+\alpha_1-\alpha_4)}{S_b(\alpha_2+\alpha_t-\alpha_3) S_b(\alpha_s+\alpha_3-\alpha_4)}\,|S_b(2\al_t)|^2 \\
& \times  \int_{\CC}du\; 
S_b(-\alpha_2 \pm (\alpha_1-Q/2)+u) 
S_b(-\alpha_4 \pm (\alpha_3-Q/2) +u) \nonumber \\[-.5ex] &
\hspace{1.125cm}\times  
S_b(\alpha_2 + \alpha_4 \pm (\alpha_t-Q/2) - u) 
S_b(Q \pm (\alpha_s - Q/2) - u) \,.
\notag\end{align}
The following notation has been used
$
S_b(\alpha \pm u) :=
S_b(\alpha+u)
S_b(\alpha-u).
$
The integral in \rf{6j1} will be defined for $\al_k\in Q/2+\textup{i}\BR$ by
using a contour $\CC$ that approaches $Q+\textup{i}\BR$ near infinity,
and passes the real axis in $(Q/2,Q)$, and for other values 
of $\al_k$ by analytic continuation. 
The equivalence between the two different integral representations
of the b-6j symbols was proven in \cite{TV} using methods from \cite{DSV}.

Using the the representation \rf{6j1} and the integral 
identity \rf{Corollary 1} it becomes easy to 
find that
\begin{equation}\label{FPTspecial}
{{\rm F}_{\ep\al_\3}^{\rm\sst PT} \big[ {}_{\al_\2}^{\al_\2} {}_{\al_\1}^{\al_\1} \big]}
=|S_b(2\al_3)|^2
\frac{S_b(\al_1+\al_2+\al_3-Q)}{S_b(\al_2+\al_3-\al_1)S_b(2\al_1)}
(S_b(\ep))^2(1+\CO(\ep))\,,
\end{equation}
from which equation \rf{limF:ex} and 
identity \rf{F-C} follow straightforwardly.

\subsection{Evaluating the integral}\label{app:S-kernel}

We start from equation \rf{S-MSform-App}.
Considering the right hand side, let us represent 
${{\rm F}_{0\be_\3}^{\rm\sst PT} \big[ {}_{\be_\2}^{\be_\2} {}_{\be_\1}^{\be_\1} \big]}$
as
\begin{equation}
{{\rm F}_{0\be_\3}^{\rm\sst PT} \big[ {}_{\be_\2}^{\be_\2} {}_{\be_\1}^{\be_\1} \big]}
\,=\,\lim_{\de\ra 0}{{\rm F}_{\de\be_\3}^{\rm\sst PT} \big[ {}_{\be_\2}^{\be_\2} {}_{\be_\1}^{\be_\1} \big]}\,.
\end{equation}
One may then represent the right hand side of (\ref{S-MSform-App}) 
as the limit $\de\ra 0$ of an expression proportional to 
the following integral:
\beqa
&& I = C \int \frac{e^{-2 \pi \textup{i} (\frac Q2-y)^2} d \ga}{S_b(\pm 2 (\frac Q2-\ga))} \int dx \frac{S_b(-\be_1 \pm (\frac Q2-\be_1) + x) S_b(-Q+\be_2\pm(\frac Q2-\be_2)+x)}{S_b(\frac Q2+\de+x)S_b(- \frac Q2+x) S_b(\be_2-\be_1\pm(\frac Q2-\ga)+x)} \nonumber \\ && \makebox[3em]{} \times \int dy \frac{S_b(-\be_1\pm(\frac Q2-\be_2)+y) S_b(-Q+\be_2\pm(\frac Q2-\be_1)+y)}{S_b(\pm(\frac Q2-\ga)+y) S_b(\be_2-\be_1\pm(\frac Q2-\al_0)+y)},
\eeqa
where
$$
C \simeq 
e^{\frac{\pi \textup{i} Q^2}{2} - \pi \textup{i} (2 \Delta_{\be_2} + 2 \Delta_{\be_1} - \Delta_{\al_0})} \frac{S_b(-Q+\al_0+2\be_2)}{S_b(-Q+2\be_2)S_b(\al_0)} 
S_b(\de).
$$
We use the notation $\simeq$ to indicate equality up to terms that
are less singular 
when $\de\ra 0$. The divergent factor $S_b(\de)$ will be cancelled
by zeros in the prefactors, see \rf{b-6j-C}, so that we only need
to consider the leading singular behavior of the integral $I$ when $\de\ra 0$.

Simplifying the above expression one gets
\beqa \label{Smod2}
&& I \simeq C \int \frac{e^{-2 \pi \textup{i} (\frac Q2-y)^2} d \ga}{S_b(\pm 2 (\frac Q2-\ga))} \int dx \frac{S_b(\frac Q2 - 2\be_1 + x) S_b(-\frac Q2 + x) S_b(-\frac 32 Q + 2 \be_2 +x)}{S_b(\frac Q2+x+\de) S_b(\be_2-\be_1\pm(\frac Q2-\ga)+x)} \nonumber \\ && \makebox[3em]{} \times \int dy \frac{S_b(-\be_1\pm(\frac Q2-\be_2)+y) S_b(-Q+\be_2\pm(\frac Q2-\be_1)+y)}{S_b(\pm(\frac Q2-\ga)+y) S_b(\be_2-\be_1\pm(\frac Q2-\al_0)+y)}.
\eeqa
We may take the integral over the variable $x$ in (\ref{Smod2}) 
using identity \rf{Corollary 1} and get
\beqa \label{Smod3}
&& I \simeq C_1 \int_{-\textup{i} \infty}^{\textup{i} \infty} \frac{e^{-2 \pi \textup{i} (\frac Q2-y)^2} d \ga}{S_b(\pm 2 (\fr{Q}{2}-\ga))}  S_b(\fr{Q}{2} - \be_2 + \be_1 \pm (\fr{Q}{2}-\ga)) \nonumber \\ && \makebox[3em]{} \times \int dy \frac{S_b(-\be_1\pm(\frac Q2-\be_2)+y) S_b(-Q+\be_2\pm(\frac Q2-\be_1)+y)}{S_b(\pm(\frac Q2-\ga)+y) S_b(\be_2-\be_1\pm(\frac Q2-\al_0)+y)},
\eeqa
where
$$
C_1 \simeq C\,S_b(Q-2\be_1) S_b(-Q+2\be_2) S_b(-\de).
$$
Next we take the integral over $\ga$ using identity \rf{Corollary 2} with taking $\nu_2=Q-\be_2+\be_1+(\frac Q2-\al_0)$ (and then apply change of variables $y \rightarrow -y$)
\begin{align} \label{Smod4}
 I \simeq C_2 \int_{-\textup{i} \infty}^{\textup{i} \infty} dy \;
& S_b(\fr{Q}{2}-\be_1-\be_2-y) 
S_b(-\fr{Q}{2} + \be_2-\be_1-y) 
S_b(-\fr{3}{2} Q + \be_1 +\be_2-y) \nonumber \\ 
\times  & S_b(Q-\be_2+\be_1\pm(\fr{Q}{2}-\al_0)+y) 
e^{\pi \textup{i} y (\be_1-\be_2)} e^{-\frac{\pi \textup{i} y^2}{2}},
\end{align}
where
$$
C_2 \simeq e^{-\frac 12 \pi \textup{i} (\frac Q2+\be_2-\be_1)^2} e^{\frac 12 \pi \textup{i} Q (\frac 32 Q-\be_2+\be_1)} C_1.
$$

As a final step we use identity \rf{Corollary 3} with taking $\nu_2=Q-\be_2+\be_1+(\frac Q2-\al_0)$
\beqa \label{Smod5}
&& I \,\simeq\, C_3 \frac 12 \int_{-\textup{i} \infty}^{\textup{i} \infty} dy \frac{S_b(\frac{\al_0}{2} \pm (\frac Q2-\be_1) \pm (\frac Q2-\be_2) \pm y)}{S_b(\pm 2 y)} e^{-2 \pi \textup{i} y^2} dy,
\eeqa
with
$$
C_3 \simeq C_2\,e^{\pi \textup{i} \frac{Q^2}{8}} e^{\pi \textup{i} (\be_1-\be_2)^2} e^{-\frac 12 \pi \textup{i} Q (\frac Q2 + \al_0)} S_b(2Q-2\be_2-\al_0) S_b(Q-\al_0) S_b(2\be_1-\al_0)\,.
$$
We also need 
$\Fusan{\be_1}{\be_1}{\be_1}{\be_1}{\ep}{\al_0}$ for $\ep\ra 0$.
Formula \rf{FPTspecial} 
gives
\beqa
&&  \Fusan{\be_1}{\be_1}{\be_1}{\be_1}{\ep}{\al_0}
\simeq \left( S_b(\epsilon) \right)^2 
\frac{S_b(Q-2\be_1) S_b(-Q+2\be_1+\al_0)}{ S_b(\al_0)}.
\eeqa
By assembling the pieces we come to the following relation
\begin{align}
S_{\be_1 \be_2}^{\rm\sst PT}(\al_0) = &\,\frac 12   S_{0\be_\2}^{\rm\sst PT} \frac{S_b(2\be_1-\al_0) S_b(2Q-2\be_1-\al_0)}{S_b(\al_0)} e^{2 \pi \textup{i} (\be_1-\frac Q2)^2} e^{2 \pi \textup{i} (\be_2-\frac Q2)^2}e^{-\pi \textup{i} (\al_0^2-2 \frac Q4 \al_0)}   \nonumber \\ &  \times \int \frac{S_b(Q-\be_1-\be_2+\frac{\al_0}{2} \pm y) S_b(-Q+\be_1+\be_2+\frac{\al_0}{2} \pm y)}{S_b(\pm 2 y)} e^{-2 \pi \textup{i} y^2} dy \nonumber \\ &\quad\hspace{2.5cm} \times S_b(\be_1-\be_2+\fr{\al_0}{2} \pm y) S_b(-\be_1+\be_2+\fr{\al_0}{2} \pm y)\,.
\end{align}
It remains to apply the following formula  \cite{SV11},
\begin{align}
& \int_{i\BR}dz\; 
\frac{
S_b({Q}/{4} - \mu+{m}/{2} \pm z)}{
S_b({3Q}/{4} - \mu-{m}/{2} \pm z)} e^{4 \pi \textup{i} \xi z} =
\\ & \;\;\;
= \frac 12 e^{2 \pi \textup{i} (\xi^2-(\frac Q4 +\frac m2)^2+\mu^2)} 
S_b(Q/2 - m \pm 2 \xi)
\int_{i\BR} dy\; 
\frac{S_b(\frac{Q}{4}+\frac{m}{2} \pm \mu \pm \xi \pm y)}
{S_b(\pm 2 y)} e^{-2 \pi \textup{i} y^2}\,,
\notag\end{align}
which had been used in this form
in \cite{SV11}, in order to get the desired result, 
\begin{align}\label{Skern3}
& S^{\rm\sst PT}_{\be_\1\be_\2} (\al_0)\,=\,\\ 
& =\,S_{0\be_\2}^{\rm\sst PT} \,
\frac{e^{\frac{\pi i}{2}\De_{\al_0}}}{S_b(\al_0)} 
\int_\BR dt\;e^{2\pi t(2\be_\1-Q)}
\frac{S_b\big(\frac{1}{2}(2\be_\2+\al_0-Q)+it\big)
S_b\big(\frac{1}{2}(2\be_\2+\al_0-Q)-it\big)}
{S_b\big(\frac{1}{2}(2\be_\2-\al_0+Q)+it\big)
S_b\big(\frac{1}{2}(2\be_\2-\al_0+Q)-it\big)}\,.
\notag\end{align}
This is equivalent to formula 
\rf{Skern}, taking into account \rf{S-renorm}.
%

\subsection{Properties of $S_{\be_\1\be_\2}(\al_0)$}\label{spp:Sprop}

In order to derive the key properties of $S_{\be_\1\be_\2}(\al_0)$
let us define
the integral 
$I_{\be_\1\be_\2}^{\al_0}$
\begin{align}
& I_{\be_\1\be_\2}^{\al_0}:=
\frac{1}{S_b(\al_0)}
\int_\BR dt\;e^{2\pi t(2\be_\1-Q)}
\frac{S_b\big(\frac{1}{2}(2\be_\2+\al_0-Q)+it\big)
S_b\big(\frac{1}{2}(2\be_\2+\al_0-Q)-it\big)}
{S_b\big(\frac{1}{2}(2\be_\2-\al_0+Q)+it\big)
S_b\big(\frac{1}{2}(2\be_\2-\al_0+Q)-it\big)}\,.
\notag\end{align}
$I_{\be_\1\be_\2}^{\al_0}$
has the following properties:
\begin{subequations}
\begin{align}\label{Isymm}
 I_{\be_\1\be_\2}^{\al_0}\,&=\,
I_{\be_\2\be_\1}^{Q-\al_0}\,,\\ \label{Isymm2}
 I_{\be_\1,\be_\2}^{\al_0}\,&=\,
I_{\be_\1,Q-\be_\1}^{\al_0}\,=\,I_{Q-\be_\1,\be_\1}^{\al_0}\,,\\
 \lim_{\ep\downarrow 0}I_{\be_\1\be_\2}^{\ep}\,&=\,
\frac{1}{M_{\be_\2}}\,2\cos(\pi(2\be_1-Q)(2\be_2-Q))\,,\label{Ilim1}\\
 \lim_{\ep\downarrow 0}I_{\ep\be}^{\ep}\,&=\,1\,,\qquad 
\lim_{\ep\downarrow 0}I_{\ep\be}^{Q-\ep}\,=\,\frac{M_{\be}}{M_0}\,,
\label{Ilim2}\end{align}
recalling that $M_\be=|S_b(2\be)|^2=-4\sin(\pi b(2\be-Q))
\sin(\pi b^{-1}(2\be-Q))$.
\end{subequations}
Identity \rf{Isymm} follows easily 
from equation (A.31) in \cite{BT2}. 
\rf{Isymm2} is an easy consequence of the symmetry properties 
of the integrand under $t=\ra -t$ and \rf{Isymm}.

In order to derive 
\rf{Ilim1} note that the zero of the prefactor in the definition
of $I_{\be_\1\be_\2}^{\al_0}$ is canceled by a pole of the
integral. This pole results from the fact that
the contour of integration gets pinched between the poles 
at $it=\pm \frac{1}{2}(2\be_\2\pm\al_0-Q)$
in the limit
$\al_0\ra 0$. The residue may be evaluated by deforming the contour into
the sum of two small circles around 
$it=\pm \frac{1}{2}(2\be_\2-Q)+\al_0$ plus some residual contour that
does not get pinched when $\al_0\ra 0$.

In order to prove 
\rf{Ilim2}, one may first use \rf{Isymm}, and then similar arguments 
as used to prove \rf{Ilim1}.

\section{Asymptotics of the generating function $\CW$}

\setcounter{equation}{0}

\label{sec:W-as}

\subsection{Monodromy on nodal surfaces} 

 We need to calculate the monodromy of the
oper $\pa_y^2+t(y)$ on the nodal surface
representing the boundary component of $\CT(C)$ 
corresponding to an pants decomposition $\si$. 
We will need the result to leading order in the gluing
parameters $q_r$. Using the gluing construction one may
represent the nodal surface as union of punctured spheres
and long thin cylinders. Parallel transport along a 
closed curve $\ga$ breaks up into a sequence $M_1,\dots,M_N$
of moves which represent
either the transition $F_{ij}$ from puncture $i$ to puncture $j$ of a
three-punctured sphere, 
the braiding $B_i$
of puncture $i$ on a three-punctured sphere with 
the additional puncture at $y$,
or the propagation $T_e$ along the long thin tube containing the
edge $e$ of $\Ga_\si$. 
To each moves $M_k$ let us associate a 2x2 matrix $m_k$ according to
the following rules:
\begin{subequations}
\begin{itemize}
\item Moves $F_{ij}$: 
\[
\lower1.125cm\hbox{\epsfig{figure=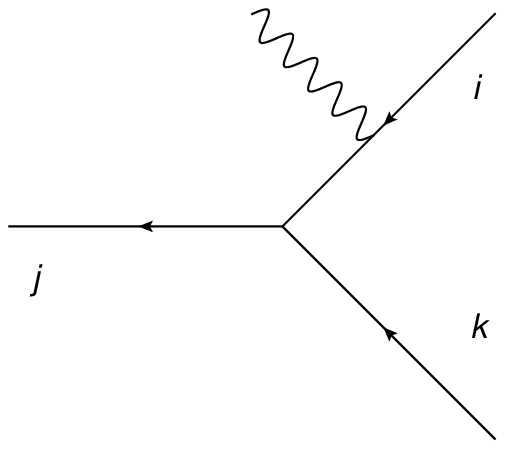,height=2.4cm}}
\quad\overset{F_{ij}}{\longrightarrow}\quad
\lower1.125cm\hbox{\epsfig{figure=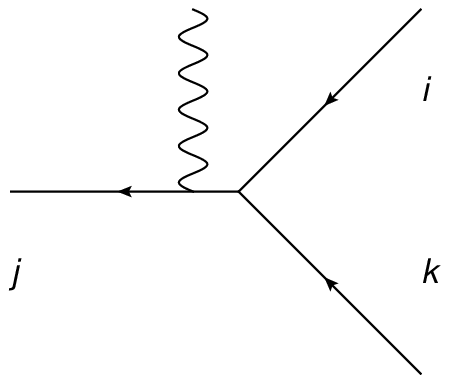,height=2.4cm}}
\]
are represented by the matrix $f^{ij}$ with elements
\begin{equation}
f_{s_1s_2}^{ij}\,\equiv\, f_{s_1s_2}^{}(l_k;l_j,l_i)\,,
\end{equation}
where
\begin{align}
f_{s_1s_2}^{}(l_3;l_2,l_1)
\,=\,\frac{\Ga(1+\ii s_1\frac{l_1}{2\pi})\Ga_b(-\ii s_2\frac{l_2}{2\pi})}
{\prod_{s_3=\pm}
\Ga\big(\frac{1}{2}+\frac{\ii}{4\pi}(s_1l_1-s_2l_2+s_3l_3)\big)}\,. 
\end{align}
\item Moves $B_{i}$: 
\[
\lower1.125cm\hbox{\epsfig{figure=degvert.eps,height=2.4cm}}
\quad\overset{B_{i}}{\longrightarrow}\quad
\lower1.125cm\hbox{\epsfig{figure=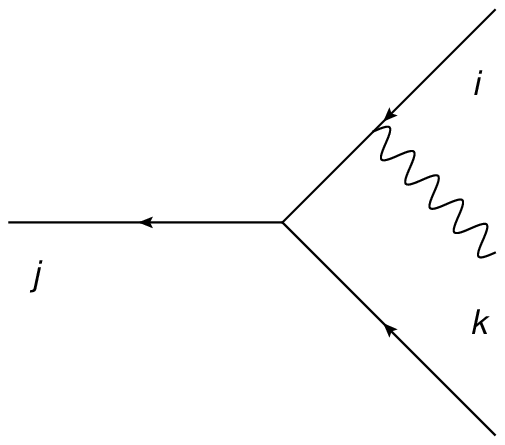,height=2.4cm}}
\]
are represented by the matrix $b^{i}$ with elements
\begin{equation}
b_{ss'}^{i}\,=\,\de_{ss'}^{}\,e^{\frac{\pi {\rm i}}{2}- s\frac{l_i}{2}}\,.
\end{equation}
\item Moves $T_{e}$: 
\[
\lower1cm\hbox{\epsfig{figure=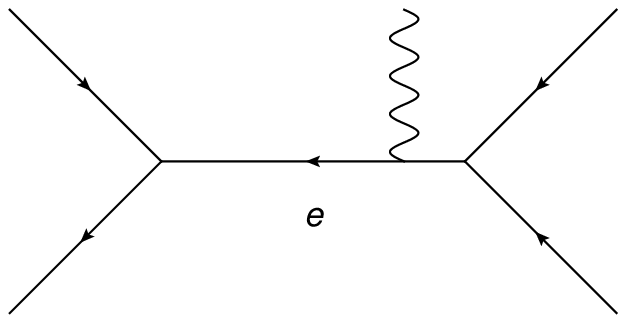,height=2.2cm}}
\quad\overset{T_{e}}{\longrightarrow}\quad
\lower1cm\hbox{\epsfig{figure=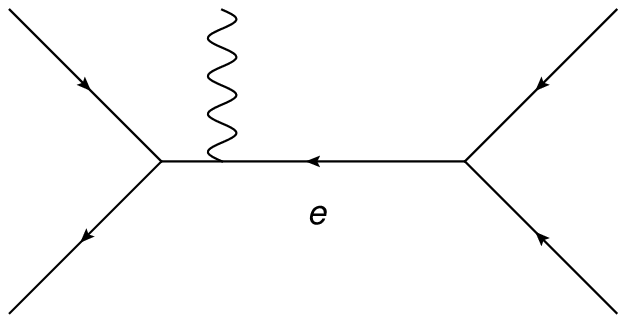,height=2.2cm}}
\]
are represented by the matrix $t^e$ with elements
\begin{equation}
t_{ss'}^{e}\,=\,\de_{ss'}^{}\,q_e^{ik_e/4\pi}\,.
\end{equation}
\end{itemize}
\end{subequations}
If the curve $\ga$ is decribed as the composition
of segments $M_1\circ M_2\circ\dots\circ M_N$,
the trace function $L_{\ga}$ is calculated
as \begin{equation}\label{Tr-rep}
L_{\ga}\,=\,{\rm Tr}(m_1\cdot m_2\cdot\dots\cdot m_N)\,,
\end{equation}
where $m_k$ are the 2x2-matrices representing the moves along the
segments $M_k$.

It is easy to see that the rules above are nothing but the 
limit $b\ra 0$ of the rules defining the Verlinde loop operators
from conformal field theory \cite{AGGTV,DGOT}. This is of course no 
accident. The comparison of the explicit expressions for Verlinde 
loop operators found in \cite{AGGTV,DGOT} with the expressions for
the expressions quoted in Subsection \ref{SSec:q-trace} shows that
the Verlinde loop operators coincide with the quantized trace coordinates,
the respective representations differing only 
by gauge transformations. A more direct explanation of this fact 
will be given elsewhere.

\subsection{Calculation of the constant term}

We may therefore use the results of references \cite{AGGTV,DGOT}.  
This yields, in particular, an expression
for $L_t$ of the form
\begin{subequations}\label{rep0,4-cl}
\begin{align}
& L_{t}\,=\,
D_+^{\rm\sst cl}(l_s)\,e^{+k_s^0}+D_0^{\rm\sst cl}(l_s)+D_-^{\rm\sst cl}(l_s)\,e^{-k_s^0}\,,
\end{align}
where $k_s^0=-\frac{\mathrm i}{2\pi}l_s\log(q)$, and the coefficients 
$D_{\pm}^{\rm\sst cl}(l_s)$ are explicitly given as
\begin{align}
& D_{\pm}^{\rm\sst cl}(l_s) = (2\pi)^4\,
\frac{(\Gamma(1\pm \frac{\mathrm i}{2\pi}l_s)
\Gamma(\pm \frac{\mathrm i}{2\pi}l_s))^2}{\prod\limits_{s,s'=\pm}
\Gamma\big(\frac{1}{2} \pm  \frac{\mathrm i}{4\pi} (l_s+s l_1+s' l_2)\big)
\Gamma\big(\frac{1}{2} \pm  \frac{\mathrm i}{4\pi} b(l_s+s l_3+s' l_4)\big)},
\nonumber\\
&D_0^{\rm\sst cl}(l_s)  =
 \frac{4}{ \cosh l_s - 1}
(\cosh(l_2/2) \cosh(l_3/2) + \cosh(l_1/2) \cosh(l_4/2))\notag\\
&\qquad\quad+\frac{4 \cosh(l_s/2)}{ \cosh l_s - 1}
(\cosh(l_1/2) \cosh(l_3/2) + \cosh(l_2/2) \cosh(l_4/2))\, .\nonumber
 \end{align}
\end{subequations}
This should be compared to \rf{cl-'t Hooft}. 
In the degeneration limit we may use \rf{CWfactor} to represent the leading  
behavior of $k_s$ in the form
\begin{equation}
k_s\,=\,4\pi {\mathrm i}\frac{\pa}{\pa l_s}\CW(l_s,q)\,=
\,k_s^0+4\pi {\mathrm i}\frac{\pa}{\pa l_s}\CW_0(l_s)+\CO(q)\,.
\end{equation}
It follows that we must have 
\begin{equation}
\log D_\pm^{\rm\sst cl}(l_s)\,=\,\log\sqrt{c_{12}(L_s)c_{34}(L_s)}
\pm 4\pi {\mathrm i}\frac{\pa}{\pa l_s}\CW_0(l_s)\,.
\end{equation}
This is a differential equation
for $\CW_0(l)$, solved by \rf{CW0}. \hfill $\square$

\section{Projectively flat connections}\label{App:proj}

\setcounter{equation}{0}

We here want to discuss some generalities on connections on bundles 
of projective spaces and projective line bundles. 
We follow in parts the discussions in \cite{FS,Fe}.

\subsection{Connections on bundles of projective spaces}

Given a holomorphic vector bundle $\CE$ over a complex manifold $X$,
let $\BP(\CE)$ be its projectivization, the bundle of projective spaces 
with fiber at $x\in X$ being the projectivization $\BP(\CE_x)$ of the
fiber $\CE_x$ of $\CE$. A connection on $\BP(\CE)$ is an equivalence
class of locally defined connections $\nabla_\imath$ on $\CE|_{\CU_\imath}$, where 
$\{\CU_{\imath};\imath\in\CI\}$ is a covering of $X$, subject to 
the condition that $a_{\imath\jmath}:=\nabla_\imath-\nabla_{\jmath}$ is a scalar
holomorphic one-form
on the overlaps $\CU_{\imath\jmath}=\CU_{\imath}\cap\CU_{\jmath}$. Two such families of connections
are identified in $\nabla_\imath^{}-\nabla'_{\imath}$ is a scalar 
holomorphic one-form for all $\imath\in\CI$. 

The curvature $F_\imath=\nabla^2_{\imath}$ is a two-form with values in 
${\rm End}(\CE)$ that satisfies $F_{\imath}-F_{\jmath}=da_{\imath\jmath}$ on
overlaps $\CU_{\imath\jmath}$. A connection is flat if $F_\imath$ is 
a scalar, i.e. proportional to the identity in all patches $\CU_{\imath}$.
As $F_{\imath}$ is locally exact, we may always choose a representative 
$\nabla_\imath$ for the equivalence class such that $F_{\imath}=0$ in $\CU_{\imath}$.
Alternatively one may trivialize the scalar one-forms $a_{\imath\jmath}:=\nabla_\imath-\nabla_{\jmath}$ by choosing smooth
scalar one-forms $c_\imath$ such that $a_{\imath\jmath}=c_{\imath}-c_{\jmath}$, 
and considering $\nabla_\imath':=\nabla_{\imath}+c_\imath$ as the preferred
representative for a given equivalence class. The connection $\nabla'_\imath$
is globally defined, but it has non-trivial scalar curvature.

The representation in terms of locally defined flat connections, 
is sometimes referred to as the {\it $\check{\rm C}$ech point of view}.
This point of view will make it clear that the deviation from being a vector
bundle with an ordinary flat connection is controlled by a 
{\it projective holomorphic line bundle}. Such a line bundle $\CL$ is defined
by transitions functions $f_{\imath\jmath}$ defined on 
overlaps $\CU_{\imath\jmath}$ that satisfy 
\[
f_{\imath_\3\imath_\2}\,f_{\imath_\2\imath_\1}\,=\,\si_{\imath_\3\imath_\2\imath_\1}\,
f_{\imath_\3\imath_\1}\,,
\]
on the triple overlaps $\CU_{\imath_\1\imath_\2\imath_\3}\equiv 
\CU_{\imath_\1}\cap\CU_{\imath_\2}\cap\CU_{\imath_\3}$.
The 1-cochain $f_{\imath\jmath}$, $\imath,\jmath\in\CI$, 
defines a class in $\check{H}^2(\Omega^0)$.  
The collection of $f_{\imath\jmath}$
will be called a {\it section} of $\CL$. 
Being one level higher in the $\check{\rm C}$ech-degree than in the case of ordinary line bundles makes 
it seem natural to identify sections with 1-cochains rather than 0-cochains
in the rest of this section.

Having a 
projectively flat vector bundle one gets a
projective {\it line} bundle
by setting
$
f_{\imath\jmath}\equiv e^{2\pi ig_{\imath\jmath}}$,
where the $f_{\imath\jmath}$ are any solutions
of $\pa g_{\imath\jmath}=
\frac{1}{2\pi i}a_{\imath\jmath}$. The collection of one-forms $a_{\imath\jmath}$
defines a $\check{\rm C}$ech-cohomology class in $\check{H}^1(\Omega^1)$, which corresponds 
to a globally defined $(1,1)$-form $\varpi$ by the $\check{\rm C}$ech-Dolbeault isomorphism.
This $(1,1)$-form represents the first Chern class of the projective line bundle
defined by the transition functions $f_{\imath\jmath}$.


\subsection{Projective local systems}

Recall the natural correspondence between
\begin{itemize}
\item[(i)] vector bundles $\CV$ with flat connections $\nabla$,
\item[(ii)] {\it local systems} -- vector bundles with {\it constant} transition functions,
\item[(iii)] representations of the fundamental group
$\rho:\pi_1(X)\ra {\rm End}(V)$ modulo overall conjugation.
\end{itemize}
Indeed, any flat connection $\nabla$ in a vector bundle $\CV$
may be trivialized locally in the patches $\CU_\imath$
by means of gauge transformations. This defines a system of preferred trivializations
for $\nabla$ with constant transition functions. Parallel transport w.r.t. to $\nabla$
defines a representation of the fundamental group. Conversely, given a 
representation of the fundamental group one gets a local system as
$(\tilde{X},V)/\sim$, where $\tilde{X}$ is the universal cover of $X$, and $\sim$ is the 
equivalence relation
\begin{equation}\label{eqrel}
(\,\tilde{x}\,,\,v\,)\,\sim\,(\,\ga\tilde{x}\,,\,\rho(\ga)v\,)\,,\quad
\forall (\,\tilde{x}\,,\,v\,)\,\in\, (\,\tilde{X}\,,\,V\,)\,,\quad
\forall \ga\in\pi_1(X)\,.
\end{equation}
This vector bundle has a canonical flat connection -- the trivial one.

Parallel transport w.r.t. a projectively flat connection defines projective 
representations of the fundamental group $\pi_1(X)$, a map
$\rho:\pi_1(X)$, which assigns to each element $\ga$ of $\pi(X)$ an operator 
$\rho(\ga)\in{\rm End}(E)$ such that
\begin{equation}\label{projrep}
\rho(\ga_1)\cdot\rho(\ga_2)\,=\,e^{2\pi i \chi(\ga_1,\ga_2)}\,
\rho(\ga_1\circ\ga_2)\,.
\end{equation}
The notation anticipates that we will ultimately be interested in 
unitary connections leading to unitary representations of the 
fundamental group.

It is easy to see that there are equally natural correspondences between
\begin{itemize}
\item[(i)] projective vector bundles $\BP(\CV)$ with projectively flat connections $\nabla$,
\item[(ii)] {\it projective local systems} -- projective vector bundles with {\it constant} transition functions,
\item[(iii)] projective representations of the fundamental group
$\rho:\pi_1(X)\ra {\rm End}(V)$.
\end{itemize}
One needs, in particular, to replace the vector space $V$ in the equivalence relation 
\rf{eqrel} by its projectivization. The resulting equivalence relation
makes perfect sense thanks to \rf{projrep}.

\subsection{Riemann-Hilbert type problems} 

It directly follows from the definition of a projectively flat vector bundle that 
an ordinary vector bundle can be obtained by tensoring with a projective 
line bundle. This makes clear how to formulate a suitable generalization of
the Riemann-Hilbert correspondence in case of projectively flat vector bundles.
We need {\it two} pieces of data: 
\begin{itemize}
\item[(a)] a projective representation of the fundamental group
$\rho:\pi_1(X)\ra {\rm End}(V)$, and:
\item[(b)] a holomorphic section of the projective line bundle
canonically associated to $\rho$.
\end{itemize}
We may then ask for $V$-valued holomorphic functions $F(\tilde{x})$ on $\tilde{X}$ that satisfy
\begin{equation}
F(\ga\tilde{x})\,=\,f_\ga(\tilde{x})\,(\rho(\ga) F)(\tilde{x})\,,
\end{equation}
where the functions $f_\ga(\tilde{x})$ represent the 
holomorphic section of the projective line bundle ${\mathcal P}_{\rho}$
canonically associated to $\rho$. 

There is of course an inevitable ambiguity in the solution of this generalized 
Riemann Hilbert problem, represented by the choice of a section of the 
projective line bundle ${\mathcal P}_{\rho}$. This is closely related to 
the issue called scheme dependence in the main text.
A natural point of view is to consider classes
of solutions to the generalized Riemann Hilbert problem which differ
by the choice of a section of ${\mathcal P}_{\rho}$. In our
concrete application we will be able to do slightly better by 
identifying natural choices for the sections of ${\mathcal P}_{\rho}$.

\end{document}